\documentclass[11pt, phd]{osudiss-2} 

\usepackage{notoccite}

\usepackage[math]{blindtext}
\usepackage{slashed}
\usepackage{amsmath,amsfonts,amssymb}
\allowdisplaybreaks[1]
\usepackage{booktabs}
\usepackage{dcolumn}
\usepackage{enumitem}
\usepackage[mathscr,scaled=1.15]{urwchancal}
\DeclareFontFamily{OT1}{pzc}{}
\DeclareFontShape{OT1}{pzc}{m}{it}%
{<-> s * [1.15] pzcmi7t}{}
\DeclareMathAlphabet{\mathpzc}{OT1}{pzc}{m}{it}
\usepackage{graphicx} 
\usepackage{lipsum}
\usepackage{commath}
\usepackage{verbatimbox}
\usepackage[titletoc]{appendix}

\usepackage{bookmark} 
\hypersetup{colorlinks=true,urlcolor=blue,linkcolor=blue, citecolor=blue, breaklinks} 
\usepackage[all]{hypcap}

\usepackage[numbers,square,sort&compress]{natbib}

\usepackage{subcaption}
\usepackage[export]{adjustbox}
\PassOptionsToPackage{obeyspaces}{url}

\newcommand\keywords[1]{\textbf{Keywords}:#1}

\usepackage{thesis-commands}
\graphicspath{{fig/}}

\author{Peng Cheng}
\title{Baryon Spin and Emergent Hadron Mass}

\degree{Doctor of Philosophy}
\graduationyear{November 21, 2023}
\unit{Graduate Program in Physics}

\begin{document}

\frontmatter


\begin{abstract}
One may divide the Standard Model of particle physics into two parts: quantum chromodynamics (QCD) and the unified theory of electroweak interactions. QCD is supposed to be the fundamental theory describing strong interactions whose basic components are gluons and quarks. It is a Poincar\'e invariant quantum gauge field theory built upon the SU(3)-colour gauge group. Fifty years of study suggest that two basic phenomena characterise QCD, namely, colour-field confinement and dynamical chiral symmetry breaking (DCSB). Colour confinement expresses the empirical fact that isolated gluons and quarks have not been captured in detectors.  It means that every observable system is a bound state.  Hence, a solution to QCD must rely heavily on nonperturbative methods. 

This thesis describes the use of Dyson-Schwinger equations (DSEs) to study baryon bound states in QCD.  The DSEs lie within the class of continuum Schwinger function methods (CSMs) and provide a nonperturbative, symmetry-preserving, continuum approach to solving QCD.  In principle, the DSEs are a coupled system of nonlinear integral equations, whose solution delivers results for each of the Schwinger functions (Euclidean space Green functions) that are needed to complete a definition of the theory.  In practice, this tower of equations is truncated, so that only approximate solutions are delivered.  Notwithstanding that, the results deliver far-reaching insights; and with increasing sophistication in the development of nonperturbative truncation schemes, the approximations are becoming increasingly accurate.

In this work, octet baryon axial, induced pseudoscalar, and pseudoscalar form factors are calculated using a symmetry-preserving treatment of a vector $\times$ vector contact interaction (SCI).  The baryons are considered as quark-plus-interacting-diquark bound states, whose structure 
(wave function) is obtained by solving a Poincar\'e-covariant Faddeev equation. 
Since it preserves symmetries, all consequences of partial conservation of the axial current are manifest.  The SCI is characterised by algebraic simplicity, involves no free parameters, and, experience shows, has good predictive power. For instance, one finds that octet baryon axial properties are consistent with only minor violations of SU(3)-flavour symmetry.  This outcome can be interpreted as a dynamical consequence of emergent hadron mass (EHM).  Considering neutral axial currents, the SCI delivers predictions for the flavour separation of octet baryon axial charges and, consequently, produces values for the associated SU(3) singlet, triplet, and octet axial charges. The results indicate that, at the hadron scale $\zeta_{\mathcal{H}}$, valence degrees of freedom carry approximately $50\%$ of an octet baryon's total spin.

This general analysis is followed by a detailed SCI study of proton spin structure, with calculations of the rest-frame quark+diquark angular momentum decomposition of the proton wave function canonical normalisation constant and the proton axial charge.  The SCI analysis of the normalisation yields results that are consistent with more realistic studies, \textit{e.g}., its value is largely determined by quark+diquark $\mathsf S$-wave components, albeit with significant, destructive $\mathsf P$-wave and strong, constructive $\mathsf S \otimes \mathsf P$-wave interference terms.  Moreover, the results for the angular momentum decomposition of the axial charge and its flavour separation are similar to those of the canonical normalisation constant. 
Interestingly, the ratios of $d$ and $u$ quark contributions to the proton axial charge, $g_A^d/g_A^u$, which are computed separately from $\mathsf S$-wave, $\mathsf P$-wave and $\mathsf S \otimes \mathsf P$-wave interference, are roughly the same, \textit{i.e}., all approximately $-0.5$.

Proton structure is one of the principal topics in hadron physics. Its study is expected to reveal key features of both the origin of mass and strong interaction dynamics. This work therefore extended the above analyses to an examination of in-proton parton helicity (spin) distribution functions (DFs).  Basic to its success was exploitation of the existence of an effective charge in QCD which defines an evolution scheme for both unpolarised and polarised parton DFs that is all-orders exact.  Using \emph{Ans\"atze} for hadron-scale proton polarised valence quark DFs, constrained by CSM calculations of the flavour-separated axial charges and insights from perturbative quantum chromodynamics, predictions are delivered for all proton polarised DFs at the measurement scale $\zeta_{\rm C}^2 = 3\,$GeV$^2$.  The pointwise behaviour of the predicted DFs and, consequently, their moments, shows good agreement with results inferred from data.  Notably, flavour-separated singlet polarised DFs are found to be small. On the other hand, the polarised gluon DF, $\Delta G(x;\zeta_{\rm C})$, is large and positive. Based on these results, one finds $\int_{0.05}^1\,dx\,\Delta G(x;\zeta_{\rm C}) = 0.214(4)$ and that experimental measurements of the proton flavour-singlet axial charge should return a value $a_0^{\rm E}(\zeta_{\rm C}) = 0.35(2)$.

\end{abstract}
\clearpage 
\tableofcontents 


\mainmatter
\chapter{Introduction}\label{chapter_background}
Atomic and nuclear physics research spanning over a century has revealed that all matter is made up of particles, including the atoms that make up our bodies, which in turn contain a dense nucleus at their core. This core is made up of nucleons, \textit{i.e}., protons and neutrons, which belong to a larger group of fm-scale particles known as hadrons. In the process of investigating hadrons, it has been discovered that they are complicated bound states of quarks and gluons. The quarks and gluons interact with each other by strong interactions which are described by a Poincar\'e-invariant quantum non-Abelian gauge field theory; namely, quantum chromodynamics (QCD).  While perturbation theory is a crucial tool for studying high-energy processes in the Standard Model, QCD is fundamentally different in that it cannot be studied with perturbation theory when it comes to observable low-energy characteristics of hadrons. The set of experimental and theoretical approaches employed to investigate and map the infrared domain of QCD is known as strong-QCD (sQCD) \cite{Pennington:1996dy} and they must deal with nonperturbative phenomena, such as dynamical chiral symmetry breaking (DCSB) and confinement of quarks and gluons. 

QCD is marked by two emergent phenomena: DCSB and confinement, which have significant implications. These phenomena are particularly apparent in the pion, and the properties of the pion, in turn, indicate a close relationship between confinement and DCSB. As the pion is both a quark-antiquark bound state and a Nambu-Goldstone boson, it occupies a distinct position in Nature. Therefore, understanding the properties of the pion is crucial for revealing some fundamental aspects of the Standard Model. As a result of confinement, colour-charged particles cannot exist in isolation and thus cannot be directly observed. Instead, they combine to form colour-neutral bound states. The phenomenon of confinement is supported by empirical evidence, but lacks a formal mathematical proof. To address this gap, the Clay Mathematics Institute established a ``Millennium Problem'' award of \$1-million for a proof that the $SU_{c}(3)$ gauge theory is mathematically well-defined \cite{Jaffe:2006iao}. One potential outcome of solving this problem would be a resolution of the question of whether the confinement conjecture is correct in pure-gauge QCD. The fundamental challenge of gluon and quark confinement is a pressing issue in modern science because it plays a crucial role in ensuring the absolute stability of the proton. Without confinement, isolated protons would decay, hydrogen atoms would be unstable, nucleosynthesis would be a random occurrence without any long-lasting effects, and the Universe would lack the necessary components for the formation of stars and life. Thus, the existence of the Universe fundamentally depends on the phenomenon of confinement. 

While the realisation and nature of confinement in real-world QCD is still under exploration, the other principal non-perturbative feature of QCD is DCSB, which is better understood.  DCSB leads to the generation of mass from nothing. It is essential to note the term ``dynamical'' as it is different from spontaneous symmetry breaking. The latter typically involves the introduction of a scalar field into the action of the theory under examination, while DCSB arises naturally during the process of quantisation of the classical chromodynamics of massless gluons and quarks. Although there is no simple change of variables in the QCD action that can display DCSB, a large mass-scale is generated by this phenomenon.  DCSB is the most significant mass generating mechanism for visible matter in the Universe, responsible for approximately 98\% of the proton's mass.

The aim of hadron physics is to offer a quantitative understanding of the characteristics of hadrons by solving QCD. Studying the properties and structure of hadron is the central subject of hadron physics. It provides the opportunity to raise and address the fundamental questions in QCD: what is confinement; what is DCSB; and how are they related? The hadron spectrum has long been a research topic in quantum mechanics. The constituent-quark and potential models have extensively studied this problem, see, \textit{e.g}., Refs. \cite{Capstick:2000qj, Aznauryan:2012ba, Crede:2013kia} and related literature. However, they do not provide a unified physical description of light-quark mesons and baryons \cite{Swanson:2012zz}. The former require an accurate representation of DCSB. Although QCD sum rules \cite{Aznauryan:2012ba, Colangelo:2000dp} avoid some of these challenges, they face the issue of having an excessive number of parameters, such as the vacuum condensates. To properly address this issue, it is necessary to compare the inferred values of these parameters with those obtained from realistic nonperturbative analyses of QCD or from experimental data. Recently, there has been a vigorous attempt to combine ideas from the light-front formulation of quantum field theory with models derived from the concept of gauge-gravity duality \cite{Aznauryan:2012ba, Brodsky:2011vgv,Brodsky:2013ar}. Nevertheless, a challenge arises when attempting to connect the many parameters utilised in this approach with the single mass-scale present in QCD.

During the past thirty years, CSMs \cite{Roberts:1994dr, Roberts:2000aa, Maris:2003vk, Bashir:2012fs, Roberts:2012sv}, realised via Dyson-Schwinger equations (DSEs), have grown into a powerful tool for the analysis of hadron physics observables.  Indeed, the DSEs have achieved marked successes, especially during the past decade \cite{Roberts:2015lja, Horn:2016rip, Eichmann:2016yit, Burkert:2017djo, Fischer:2018sdj, Qin:2020rad, Roberts:2020udq, Roberts:2020hiw, Roberts:2021xnz, Roberts:2021nhw, Binosi:2022djx, Papavassiliou:2022wrb, Ding:2022ows, Ferreira:2023fva}, particularly in relation to understanding the causes and various manifestations of emergent hadron mass (EHM). DSEs correlate the features of meson and baryon ground- and excited-states within a single, symmetry-preserving framework. In this context, symmetry-preserving implies adherence to Poincar\'e covariance and satisfies the relevant Ward-Green-Takahashi identities. The challenges surrounding DSE analyses revolve around the fact that the equation for each n-point function is connected to those for higher n-point functions \cite{Roberts:1994dr, Roberts:2000aa}. For instance, the gap equation for the quark 2-point function is linked with that for the gluon 2-point function and the gluon-quark 3-point function. As a result, in order to establish a manageable problem, truncations are required.

The principal focus of this thesis is the study of the properties and structure of octet baryons, especially those of proton, within the DSE framework. A central aim of ongoing theoretical and experimental efforts is to comprehend nucleon structure because the nucleon is the simplest composite object made of three valence light quark \cite{Hofstadter:1955ae}. Nucleon form factors encode its fundamental properties, for example, expressing the charge distribution of its charge carrying constituents. 

Nucleon electromagnetic form factors are quite well known, however, much more needs to be understood about their axial and induced-pseudoscalar form factors, which measure nucleon responses to probes associated with the isovector axial-vector current. Such form factors characterise neutrino-nucleon scattering \cite{Ahrens:1988rr, Kitagaki:1990vs, Bodek:2007vi, Meyer:2016oeg}, exclusive pion electroproduction \cite{Choi:1993vt, Bernard:1994pk, A1:1999kwj,Fuchs:2003vw} (\textit{e.g}., $e^{-} p \rightarrow \pi^{-} p \nu$), radiative muon capture \cite{Hart:1977zz, Jonkmans:1996my, Wright:1998gi}, and ordinary muon capture\cite{Gorringe:2002xx, Bardin:1980mi,
MuCap:2012lei, MuCap:2015boo}. 

At zero momentum transfer, the nucleon axial form factor gives the axial charge $g_A \equiv G_A(0)$, which can be measured with high precision from neutron $\beta$-decay experiments \cite{UCNA:2012fhw, Mund:2012fq, UCNA:2017obv, Darius:2017arh}. 
The nucleon induced pseudoscalar coupling, $g_P^{*}$, can be determined via the ordinary muon capture process, $\mu^{-}+p\to n+\nu_{\mu}$, from the singlet state of the muonic hydrogen atom at the muon capture point $Q^2=0.88 m_\mu^2$ \cite{MuCap:2012lei, MuCap:2015boo, Castro:1977ep, Bernard:2000et, MuCap:2007tkq}, where $m_\mu$ is the muon mass. 
Additionally, kindred decays of hyperons have also attracted much attention \cite{Gaillard:1984ny, Cabibbo:2003cu} because they present an opportunity to shed light on the Cabibbo-Kobayashi-Maskawa (CKM) matrix element $\left|V_{u s}\right|$. Hyperon axial charges are also important in effective field theory of octet baryons\cite{Tiburzi:2008bk, Alexandrou:2009qu} because they enter the expansions of all quantities in chiral perturbation theory. Compared with nucleon, little empirical information is available on hyperon axial charges. 

The flavour-dependent axial charges of the octet baryons, \textit{i.e.}, the singlet axial charge $g_A^{(0)}$, the isovector axial charge $g_A^{(3)}$ and the SU(3) octet axial charge $g_A^{(8)}$ are also fundamental observables in hadron physics. They provide insights into the spin structure and properties of baryons. The charge $g_A^{(0)}$ should indicate the measurable total spin of a given baryon that results from the spins of its valence degrees-of-freedom. Notably in this connection, an angular momentum decomposition of the proton axial charge, $g_{Ap}^{(3)}$, exhibits contributions to its spin that are associated with quark+diquark orbital angular momentum. It is not zero; hence, the proton's spin cannot be attributed solely to the sum of valence quark spins.  Indeed, as is widely known, in quantum field theory, valence quark partons are just one of the factors that contribute to the structure of the proton and the pion. These things highlight the importance of studying the proton's polarised parton distribution functions (DFs): they are an essential part of understanding proton structure. 

This thesis is structured as follows. 
In Chapter\,\ref{chapter2}, I describe the DSE formalism under a symmetry-preserving treatment of a vector $\times$ vector contact interaction (SCI), including the gap equation for the dressed quark propagator; the Bethe-Salpeter equation (BSE) describing two-body problems and the Faddeev equation relevant for the three-body bound states. I also discuss practical calculations in this chapter. 
In Chapter\,\ref{chapter3}, octet baryon axial, induced pseudoscalar, and pseudoscalar form factors are computed using the SCI.  Considering neutral axial currents, I describe the SCI predictions for the flavour separation of octet baryon axial charges and, therefrom, values for the associated SU(3) singlet, triplet, and octet axial charges. 
Based on the work in Chapter\,\ref{chapter3}, I then elucidate the angular momentum decomposition of the proton's axial charges in Chapter\,\ref{chapter4}. The contributions of the various quark+diquark orbital angular momentum components to the canonical normalisation are also obtained with a view to revealing the structure of the wave function. 
In Chapter\,\ref{chapter5}, predictions for all polarised parton DFs are delivered.  The keys to this progress are existence of an effective charge in QCD which defines an evolution scheme for both unpolarised and polarised parton DFs that is all-orders exact and development of well-constrained \textit{Ans\"atze} for the hadron scale in-proton valence quark helicity DFs.
I summarise and provide an outlook in Chapter\,\ref{chapter6}.  
\chapter{Symmetry-preserving treatment of a contact interaction}
\label{chapter2}

\section{Introduction}\label{chap2secintro}

The DSEs are an infinite tower of coupled integral equations relating the Schwinger functions of a quantum field theory to each other.\footnote{\,Schwinger functions may be called the Euclidean Green functions for a theory.  In principle, they are connected to standard Green functions via a straightforward analytic continuation in the time variable.  See Ref.\,\cite{Ding:2022ows} (Sec.\,1) for additional information.}
A quantum field theory is completely defined once all its n-point Schwinger functions are known. 
However, it's impossible for one to solve infinitely many coupled integral equations. 
So it is unavoidable that the tower of coupled equations must be truncated at some point. This implies that the tower of equations has to be truncated at a certain value of n, which corresponds to the maximum number of legs on any Schwinger function that is utilised in the self-consistent solution of the equations. 

A symmetry-preserving truncation scheme applicable to hadrons was introduced in Refs.\,\cite{Munczek:1994zz, Bender:1996bb}. 
That procedure generates a BSE from the kernel of any gap equation for which the diagrammatic content is known. 
In that scheme, all Ward-Green-Takahashi (WGT) identities \cite{Ward:1950xp, Green:1953te, Takahashi:1957xn, Takahashi:1985yz} are preserved, without fine-tuning, and hence, current-conservation and the appearance of Goldstone modes in connection with DCSB are ensured. 
Within DSEs, a symmetry-preserving treatment of vector $\times$ vector contact interaction has proven to be a reliable tool in describing the properties and structure of hadron ground states. So I use it to carry out much of the work in this thesis.

The aim of this thesis is to study the properties and structure of baryons based on DSE, where the one-body gap equation, two-body BSE and three-body Faddeev equation play roles. In this chapter, I will describe them in detail. 
Sec.\,\ref{secsci} introduces the symmetry-preserving treatment of a vector $\times$ vector contact interaction (SCI) used in this thesis. 
Sec.\,\ref{secgap} provides the form of the SCI gap equation, including its regularisation scheme, and explains how to solve it. 
In Sec.\,\ref{secBSE}, I analyse the homogeneous and inhomogeneous vertex BSEs. 
The Faddeev equation is discussed in Sec.\ref{secFaddeev}. 

\section{Contact interaction}\label{secsci}
In the continuum analysis of hadron bound states, the basic element is the quark + anti\-quark scattering kernel. At leading order in the widely-used symmetry-preserving approximation scheme, known as rainbow-ladder (RL) truncation
\cite{Munczek:1994zz,Bender:1996bb}, that kernel can be written as follows:
\begin{subequations}
\label{KDinteraction}
\begin{align}
\mathscr{K}_{\alpha_1\alpha_1',\alpha_2\alpha_2'}  & = {\mathpzc G}_{\mu\nu}(k) [i\gamma_\mu]_{\alpha_1\alpha_1'} [i\gamma_\nu]_{\alpha_2\alpha_2'}\,,\\
{\mathpzc G}_{\mu\nu}(k)  & = \tilde{\mathpzc G}(k^2) T^k_{\mu\nu}\,,
\end{align}
\end{subequations}
where $k = p_1-p_1^\prime = p_2^\prime -p_2$, with $p_{1,2}$, $p_{1,2}^\prime$ being, respectively, the initial and final momenta of the scatterers, and $k^2T_{\mu\nu}^k = k^2\delta_{\mu\nu} - k_\mu k_\nu$.

$\tilde{\mathpzc G}$ serves as the defining element; and it is currently known that, owing to the emergence of a gluon mass-scale \cite{Binosi:2014aea, Gao:2017uox, Cui:2019dwv}, $\tilde{\mathpzc G}$ is nonzero and 
finite at infrared momenta. Hence, when considering long-wavelength hadron properties, it can reasonably be approximated as follows:
\begin{align}
\label{glounP}
\tilde{\mathpzc G}(k^2) & = \frac{4\pi \alpha_{\rm IR}}{m_G^2}\,.
\end{align}
In QCD \cite{Cui:2019dwv}: $m_G \approx 0.5\,$GeV, $\alpha_{\rm IR} \approx \pi$. Following Ref.\cite{Xu:2021iwv}, this value of $m_G$ is retained herein. 
Using Eq.\,\eqref{glounP} and exploiting the fact that a SCI cannot support relative momentum between the constituents of a meson bound-state, the interaction in Eqs.\,\eqref{KDinteraction} can be reduced to:
\begin{align}
\label{KCI}
\mathscr{K}_{\alpha_1\alpha_1',\alpha_2\alpha_2'}^{\rm CI}  & = \frac{4\pi \alpha_{\rm IR}}{m_G^2}
 [i\gamma_\mu]_{\alpha_1\alpha_1'} [i\gamma_\mu]_{\alpha_2\alpha_2'}\,.
\end{align}

The SCI expresses an elementary form of confinement by including an infrared regularising scale, $\Lambda_{\rm ir}$, when defining all integral equations relevant to bound-state problems \cite{Ebert:1996vx}. This expedient removes momenta below $\Lambda_{\rm ir}$, thereby eliminating the thresholds for the quark-antiquark production. The standard choice is $\Lambda_{\rm ir} = 0.24\,$GeV\,$=1/[0.82\,{\rm fm}]$ \cite{Gutierrez-Guerrero:2010waf}, which introduces a confinement length scale roughly comparable to the proton radii \cite{Cui:2022fyr}.

Ultraviolet regularisation is necessary for all integrals in SCI bound-state equations. This process breaks the connection between infrared and ultraviolet scales that is characteristic of QCD. As a result, the associated ultraviolet mass scales, $\Lambda_{\rm uv}$, become physical parameters. These parameters can be interpreted as upper limits for the regions in which distributions within the associated systems are practically momentum-independent.

\section{Gap equation}\label{secgap}

\begin{figure}[htbp]
  \centering
  \includegraphics[width=3.8in]{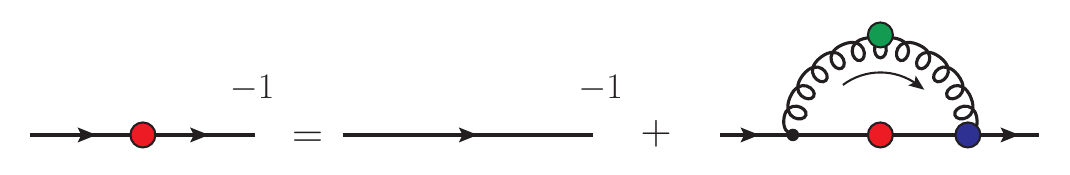}
  \caption{DSE for the quark propagator (gap equation). The lines with red balls are the dressed quark propagators; the curly line with green ball represents the dressed-gluon propagator; the blue ball is the dressed-quark-gluon vertex.}\label{figgap}
\end{figure}

For a quark, the SCI gap equation, depicted in Fig.\ref{figgap}, expressed in Euclidean space (see Appendix~\ref{appEuclid} for the conventions used), is written as:
\begin{subequations}
\label{GapEqn}
\begin{align}
S^{-1}(p) & = S_0^{-1}(p)+\Sigma(p)\,, \\
S_0^{-1}(p) & = i\gamma \cdot p+m\,,    \\
\Sigma(p) & = \frac{16 \pi}{3} \frac{\alpha_{\rm IR}}{m_G^2}
\int \frac{d^4q}{(2\pi)^4} \gamma_\mu S(q) \gamma_\mu\,.
\end{align}
\end{subequations}
where $m$ is the quark current-mass. At zero temperature and chemical potential, the most general Poincar\'e covariant solution of this gap equation involves two scalar function. There are three common, equivalent expressions
\begin{equation}
\label{genS}
S(p)=\frac{1}{i \gamma \cdot p A\left(p^{2}\right)+B\left(p^{2}\right)}=\frac{Z\left(p^{2}\right)}{i \gamma \cdot p+M\left(p^{2}\right)}=-i \gamma \cdot p \sigma_{V}\left(p^{2}\right)+\sigma_{S}\left(p^{2}\right)\,.
\end{equation}
In the second form, $Z(p^2)$ is called the wave-function renormalisation and $M(p^2)$ is the dressed-quark mass function. 

Introducing the first expression of the quark propagator into Eqs.\,\eqref{GapEqn}, one obtains the following
\begin{equation}
\label{GapEqn1}
\begin{aligned}
i \gamma \cdot p A\left(p^{2}\right)+B\left(p^{2}\right) & =i \gamma \cdot p+m+\frac{16 \pi}{3} \frac{\alpha_{\mathrm{IR}}}{m_{\mathrm{G}}^{2}} \int \frac{d^{4} q}{(2 \pi)^{4}} \gamma_{\mu} \frac{1}{i \gamma \cdot q A\left(q^{2}\right)+B\left(q^{2}\right)} \gamma_{\mu} \\
& =i \gamma \cdot p+m+\frac{16 \pi}{3} \frac{\alpha_{\mathrm{IR}}}{m_{\mathrm{G}}^{2}} \int \frac{d^{4} q}{(2 \pi)^{4}} \gamma_{\mu} \frac{-i \gamma \cdot q A\left(q^{2}\right)+B\left(q^{2}\right)}{q^{2} A^{2}\left(q^{2}\right)+B^{2}\left(q^{2}\right)} \gamma_{\mu}\,.
\end{aligned}
\end{equation}
If one then multiplies Eq.\,\eqref{GapEqn1} by ($-i \gamma \cdot p$) and subsequently evaluates a matrix trace over Dirac indices, then one obtains 
\begin{equation}
\label{AEqn}
A\left(p^2\right)=1-\frac{32 \pi}{3} \frac{\alpha_{\mathrm{IR}}}{m_{\mathrm{G}}^{2}} \frac{1}{p^2} \int \frac{d^{4} q}{(2 \pi)^{4}}(p \cdot q) \frac{A\left(q^{2}\right)}{q^{2} A^{2}\left(q^{2}\right)+B^{2}\left(q^{2}\right)}\,.
\end{equation}
It is straightforward to show that $\int d^{4} q(p \cdot q) F\left(\mathrm{p}^{2},\mathrm{q}^{2}\right)=0$ (Appendix \ref{Techn}), and so
\begin{equation}
\label{Ares}
A\left(p^{2}\right)=1\,.
\end{equation}

On the other hand, multiplying Eq.\,\eqref{GapEqn1} by $\mathbf{I}_{D}$ and evaluating a trace over Dirac indices, then
\begin{equation}
\label{BEqn}
B\left(p^{2}\right)=m+\frac{64 \pi}{3} \frac{\alpha_{\mathrm{IR}}}{m_{\mathrm{G}}^{2}} \int \frac{d^{4} q}{(2 \pi)^{4}} \frac{B\left(q^{2}\right)}{q^{2}A^2\left(q^{2}\right)+B^{2}\left(q^{2}\right)}\,.
\end{equation}
Using the result in Eq.\,\eqref{Ares}, one finds
\begin{equation}
\label{BEqn1}
B\left(p^{2}\right)=m+\frac{64 \pi}{3} \frac{\alpha_{\mathrm{IR}}}{m_{\mathrm{G}}^{2}} \int \frac{d^{4} q}{(2 \pi)^{4}} \frac{B\left(q^{2}\right)}{q^{2}+B^{2}\left(q^{2}\right)}\,.
\end{equation}
Since the integral here is $p^{2}-$independent, a $p^{2}-$independent solution must be obtained. Thus, any nonzero solution must be of the form
\begin{equation}
\label{Bres}
B\left(p^{2}\right)=\text { constant }=M\,.
\end{equation}
Inserting Eq.\,\eqref{Bres} into Eq.\,\eqref{BEqn1}, one obtains 
\begin{equation}
\label{MEqn}
\begin{aligned}
M& =m+M \frac{64 \pi}{3} \frac{\alpha_{\mathrm{IR}}}{m_{\mathrm{G}}^{2}} \int \frac{d^{4} q}{(2 \pi)^{4}} \frac{1}{q^{2}+M^{2}}\\
& =m+M \frac{4}{3 \pi} \frac{\alpha_{\mathrm{IR}}}{m_{\mathrm{G}}^{2}} \int_{0}^{\infty} d s s \frac{1}{s+M^{2}}\,,
\end{aligned}
\end{equation}
where $d^{4} q=q^{3} \sin ^{2} \theta_{1} \sin \theta_{2} d q d \theta_{1} d \theta_{2} d \phi$ and $s=q^{2}$ with $d s=2 q d q$.

A heat-kernel-like regularisation procedure is useful in formulating the SCI, \textit{viz}. one writes
\begin{equation}
\label{Regular}
\frac{1}{s+M^{2}}=\int_{0}^{\infty} d \tau e^{-\tau\left(s+M^{2}\right)} \rightarrow \int_{\tau_{u v}^{2}}^{\tau_{i r}^{2}} d \tau e^{-\tau\left(s+M^{2}\right)}=\frac{e^{-\tau_{u v}^{2}\left(s+M^{2}\right)}-e^{-\tau_{i r}^{2}\left(s+M^{2}\right)}}{s+M^{2}}\,,
\end{equation}
where $\tau_{\rm ir,\rm uv}$ are, respectively, infrared and ultraviolet regulators, \textit{i.e.} $\tau_{\rm uv}=1/\Lambda_{\rm uv}$, $\tau_{\rm ir}=1/\Lambda_{\rm ir}$. Using Eq.\,\eqref{Regular}, then Eq.\,\eqref{MEqn} becomes
\begin{equation}
\label{gapactual}
\begin{aligned}
M& =m+M \frac{4 \alpha_{\mathrm{IR}}}{3 \pi m_{\mathrm{G}}^{2}} \mathcal{C}_{0}^{\rm i u}\left(M^{2}\right)\,,
\end{aligned}
\end{equation}
where $\mathcal{C}_{0}^{\rm iu}(\sigma)=\sigma\left[\Gamma(-1, \sigma \tau_{\rm uv}^{2})-\Gamma(-1, \sigma \tau_{\rm ir}^{2})\right]$, with $\Gamma(s, x)$ being the incomplete gamma function
\begin{equation}
\label{ingammafun}
\Gamma(s, x)=\int_{x}^{\infty} d t\, t^{s-1} e^{-t}\,.
\end{equation}
The ``iu'' superscript stresses that the function depends on both the infrared and ultraviolet cutoffs. 

In general, functions of the following form arise in solving SCI bound-state equations: 
\begin{align}
\label{Cn}
  \overline{\cal C}^{\rm iu}_n(\sigma) & = \Gamma(n-1,\sigma \tau_{\textrm{uv}}^{2}) - \Gamma(n-1,\sigma \tau_{\textrm{ir}}^{2}),\; 
  \mathcal{C}_n^{\mathrm{iu}}(\sigma)=\sigma \overline{\mathcal{C}}_n^{\mathrm{iu}}(\sigma),\, n \in \mathbb{Z}^{\geq}\,.
\end{align}
See Appendix~\ref{Techn} for more details. 

Finally, one obtains the SCI dressed-quark propagator:
\begin{equation}
\label{genSinSCI}
S^{-1}(p)=i \gamma \cdot p+M
\end{equation}
with $M$, the dynamically generated dressed-quark mass, obtained by solving Eq.\,\eqref{gapactual}.  

Importantly, there is a critical value of $\alpha_{\mathrm{IR}}/m_{\mathrm{G}}^{2}$ such that, for all values of this ratio which exceed the critical value, a $M\neq 0$ solution exists even when  $m=0$ -- see, \textit{e.g}., Ref.\,\cite{Roberts:2012sv} (Sec.\,2.2).  This is an expression of DCSB in the SCI.

\section{Bethe-Salpeter equation}\label{secBSE}

By projecting the quark-antiquark four-point Schwinger function onto a specified $J^{P}$ quantum number channel, a vertex that satisfies an inhomogeneous BSE can be derived. 
When referring to a vertex equation, one inevitably needs to discuss Ward-Green-Takahashi identities (WGTI), which are basic to preserving symmetries in quantum field theory. 
A meson, characterised by two valence-quarks, with specific $J^{P}$ quantum numbers, appears as a pole in the corresponding projected vertex. 
By equating the residues on both sides of the inhomogeneous BSE, one finds the meson Bethe-Salpeter amplitude by solving the resulting homogeneous BSE. 
It is important to observe that, as a consequence of EHM, any interaction capable of creating mesons as bound states of a quark and antiquark can generate strong colour-antitriplet correlations between any two dressed quarks contained within a hadron. 
These are the so-called diquark correlations.  (See the discussion of Eq.\,\eqref{hbseEqnCIDirdiq}.)

\subsection{Meson and diquark}\label{subsecBSE}

\begin{figure}[!htbp]
  \centering
  \includegraphics[width=3.8in]{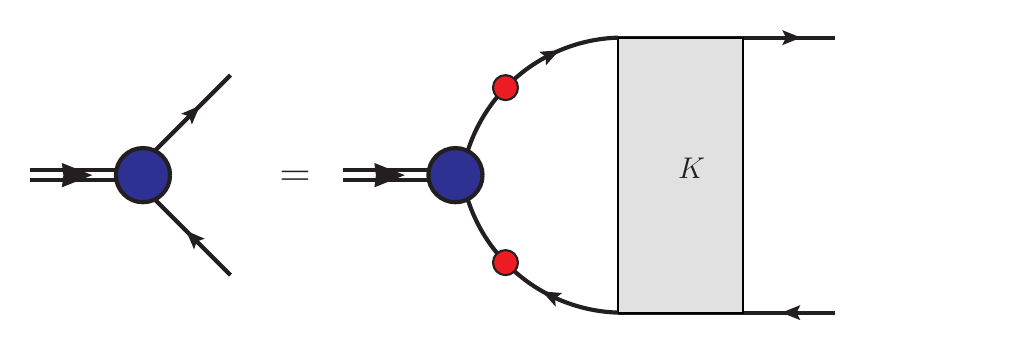}
  \caption{Homogeneous Bethe-Salpeter equation for a quark+antiquark-meson amplitude. Blue ball - the bound state amplitude; and the line with red ball - dressed quark propagator. $\mathrm{K}$ is the two-particle irreducible quark+antiquark scattering kernel.\label{fighbse}}
\end{figure}

In QCD, mesons emerge as bound states seeded by a valence quark + valence antiquark.  
A study of their properties may begin with the solution of a homogeneous BSE, such as that depicted in Fig.\,\ref{fighbse} and represented mathematically as follows:
\begin{equation}
\label{hbseEqn}
[\Gamma(p ; P)]_{t u}=\int \frac{d^{4} t}{(2 \pi)^{4}}[\chi(t ; P)]_{s r} K_{t u}^{r s}(t, p ; P)\,,
\end{equation}
where $\Gamma$ is the bound-state's Bethe-Salpeter amplitude (BSA) and $\chi(t ; P)=S(t+P)\Gamma S(t)$ is its Bethe-Salpeter wave-function, where $S$ is the dressed-quark propagator. 
The $t, u, s$ and $r$ represent colour, flavour and spinor indices, and $K$ is the relevant two-particle irreducible quark+antiquark scattering kernel. This equation possesses solutions for that discrete set of $P^2$-values which correspond to the masses-squared of the bound states. 

In rainbow-ladder truncation and with the interaction in Eq.\,\eqref{KCI}, the homogeneous BSE for a meson with specific $J^P$ quantum numbers, composed of quarks with flavours $f$ and $\bar{g}$, with $\{f,g = u, d, s\}$, \textit{i.e}., lying within SU$(3)$ flavour, can be written as 
\begin{equation}
\label{hbseEqnCI}
^0\Gamma_{f\bar{g}}^{J^P}(p;P) =-\frac{4\pi\alpha_{\mathrm{IR}}}{m_G^2}\int \frac{d^{4} t}{(2 \pi)^{4}} \gamma_{\mu} \frac{\lambda^{a}}{2} S_f(t+P) ^0\Gamma_{f\bar{g}}^{J^P}(t;P) S_g(t) \frac{\lambda^{a}}{2} \gamma_{\mu}\,.
\end{equation}
This equation has a solution for $P^2=-m_{f\bar{g}}^2$. 

The BSA of such a meson, written with full colour, flavour, and spinor structure, takes the form:
\begin{equation}
\label{BSAstruc}
^0\Gamma_{f\bar{g}}^{J^P}(p;P)=\lambda_{c}^{0}
\otimes t_{f\bar{g}}^{e} \otimes
 \Gamma_{f\bar{g}}^{J^P}(p;P)\,,
\end{equation}
where the the colour-singlet character is expressed in $\lambda_{c}^{0}={\rm diag}\{1,1,1\}$ and the flavour matrices are given by

\begin{equation}
\label{flavourmeson}
\begin{array}{ccc}
{\tt t}_{u\bar{d}}^{+} =\left[\begin{array}{ccc}
                           0 & \sqrt{2} & 0 \\
                           0 & 0 & 0 \\
                           0 & 0 & 0
                           \end{array}\right]\,,
&
{\tt t}_{d\bar{u}}^{-} =\left[\begin{array}{ccc}
                           0 & 0 & 0 \\
                           \sqrt{2} & 0 & 0 \\
                           0 & 0 & 0
                           \end{array}\right]\,,
&
{\tt t}_{u\bar{u}-d\bar{d}}^{0} =\left[\begin{array}{ccc}
                                    1 & 0 & 0 \\
                                    0 & -1 & 0 \\
                                    0 & 0 & 0
                                    \end{array}\right]\,,\\[6ex]
{\tt t}_{u\bar{s}}^{+} =\left[\begin{array}{ccc}
                           0 & 0 & \sqrt{2} \\
                           0 & 0 & 0 \\
                           0 & 0 & 0
                           \end{array}\right]\,,
&
{\tt t}_{s\bar{u}}^{-} =\left[\begin{array}{ccc}
                           0 & 0 & 0 \\
                           0 & 0 & 0 \\
                           \sqrt{2} & 0 & 0
                           \end{array}\right]\,,
&
{\tt t}_{u\bar{u}+d\bar{d}-2s\bar{s}}^0 =\frac{1}{\sqrt{3}}
\left[\begin{array}{ccc}
   1 & 0 & 0 \\
   0 & 1 & 0 \\
   0 & 0 & -2
   \end{array}\right]\,,\\[6ex]
{\tt t}_{d\bar{s}}^{0} =\left[\begin{array}{ccc}
                           0 & 0 & 0 \\
                           0 & 0 & \sqrt{2} \\
                           0 & 0 & 0
                           \end{array}\right]\,,
&
{\tt t}_{s\bar{d}}^{0} =\left[\begin{array}{ccc}
                           0 & 0 & 0 \\
                           0 & 0 & 0 \\
                           0 & \sqrt{2} & 0
                           \end{array}\right]\,,
&
{\tt t}_{u\bar{u}+d\bar{d}+s\bar{s}}^0 =
\frac{1}{\sqrt{3}}\left[\begin{array}{ccc}
                     \sqrt{2} & 0 & 0 \\
                     0 & \sqrt{2} & 0 \\
                     0 & 0 & \sqrt{2}
                     \end{array}\right]\,.
\end{array}
\end{equation}

Using $\sum_{a=1}^{8}\left(\frac{\lambda^{a}}{2} \frac{\lambda^{a}}{2}\right)=C_{F} \mathbf{1}=\frac{4}{3} \mathbf{1}$, one can obtain 
\begin{equation}
\label{hbseEqnCIDir}
\begin{aligned}
\Gamma_{f\bar{g}}^{J^P}(p;P) =-\frac{16\pi\alpha_{\mathrm{IR}}}{3 m_G^2}\int \frac{d^{4} t}{(2 \pi)^{4}} \gamma_{\mu} S_f(t+P) \Gamma_{f\bar{g}}^{J^P}(t;P) S_g(t) \gamma_{\mu}.
\end{aligned}
\end{equation}
Since the integrand does not depend on the external relative momentum, $p$, a symmetry preserving regularisation of Eq.\,\eqref{hbseEqnCIDir} results in solutions that are independent of $p$. Here I demonstrate the nature of the BSE through two relevant examples, \textit{viz}.\ the pseudoscalar meson ($J^P=0^{-}$) and the vector meson ($J^P=1^{-}$). 

The solution for a pseudoscalar meson can be written as 
\begin{equation}
\label{psBSA}
\Gamma_{f \bar{g}}^{0^{-}}(P)=\gamma_{5}[ i E_{\left[f \bar{g}\right]}(P)+\frac{\gamma \cdot P}{2M_{f\bar{g}}} F_{\left[f \bar{g}\right]}(P)]\,,
\end{equation}
where $M_{f\bar{g}}=M_f M_{g}/\left[M_f+ M_{g}\right]$. 
If one inserts Eq.\,\eqref{psBSA} into Eq.\,\eqref{hbseEqnCIDir} and employs the symmetry-preserving regularisation of the contact interaction explained below -- see Eqs.\,\eqref{regular} -- one obtains the following explicit form of the homogeneous BSE: 
\begin{equation}
\label{explicithbseps}
\left[\begin{array}{l}
E_{\left[f \bar{g}\right]}\left(P^{2}\right) \\
F_{\left[f \bar{g}\right]}\left(P^{2}\right)
\end{array}\right]=\frac{4 \alpha_{\mathrm{IR}}}{3 \pi m_{G}^{2}} \left[\begin{array}{cc}
\mathcal{K}_{E E}^{\left[f \bar{g}\right]} & \mathcal{K}_{E F}^{\left[f \bar{g}\right]} \\
\mathcal{K}_{F E}^{\left[f \bar{g}\right]} & \mathcal{K}_{F F}^{\left[f \bar{g}\right]}
\end{array}\right]\left[\begin{array}{l}
E_{\left[f \bar{g}\right]}\left(P^{2}\right) \\
F_{\left[f \bar{g}\right]}\left(P^{2}\right)
\end{array}\right]\,,
\end{equation}
with
\begin{subequations}
\label{Kmatps}
\begin{align}
\mathcal{K}_{E E}^{\left[f \bar{g}\right]}& =\frac{3 \pi m_{G}^{2}}{4 \alpha_{\mathrm{IR}}}(1-\frac{m_f+m_g}{M_f+M_g})-
\int_{0}^{1} d \alpha Q^2 \frac{\hat{\alpha}M_{f}+ \alpha M_{g}}{M_f+M_g}\bar{\mathcal{C}}_{1}^{\mathrm{iu}}
\left(\omega_{f \bar{g}}\left(\alpha, P^{2}\right)\right)\,, \\
\mathcal{K}_{E F}^{\left[f \bar{g}\right]}& =\frac{P^{2}}{2 M_{f \bar{g}}} \int_{0}^{1} d \alpha\left[\hat{\alpha} M_{f}+\alpha M_{g}\right] \bar{\mathcal{C}}_{1}^{\mathrm{iu}}\left(\omega_{f \bar{g}}\left(\alpha, P^{2}\right)\right)\,, \\
\mathcal{K}_{F E}^{\left[f \bar{g}\right]}& =\frac{2 M_{f \bar{g}}^{2}}{P^{2}} \mathcal{K}_{E F}^{\left[f \bar{g}\right]}\,, \\
\mathcal{K}_{F F}^{\left[f \bar{g}\right]}& =-\frac{1}{2}\int_{0}^{1} d \alpha\left[M_{f}M_{g}+\hat{\alpha} M_{f}^2+\alpha M_{g}^2\right] \bar{\mathcal{C}}_{1}^{\mathrm{iu}}\left(\omega_{f \bar{g}}\left(\alpha, P^{2}\right)\right)\,,
\end{align}
\end{subequations}
where $w_{f \bar{g}}(\alpha,P)= \hat{\alpha} M_{f}^2+\alpha M_{g}^2+\alpha \hat{\alpha} P^2 (\hat{\alpha}=1-\alpha)$. 
Eq.\eqref{explicithbseps} is a eigenvalue problem: it has a solution at a single value of $P^2 = - m_{f\bar g}^2< 0$, at which point the eigenvector describes the meson's Bethe-Salpeter amplitude. 

In the computation of observables, one must use the canonically-normalised Bethe-Salpeter amplitude, \textit{i.e.}, the amplitude $\Gamma_{f \bar{g}}^{0^-}$ is 
rescaled so that
\begin{equation}
\label{norps}
P_{\mu}=N_{c} \operatorname{tr}_{D} \int \frac{d^{4} t}{(2 \pi)^{4}} \bar{\Gamma}_{f \bar{g}}^{0^-}(-P) \frac{\partial}{\partial P_{\mu}} S_{f}(t+P) \Gamma_{f \bar{g}}^{0^-}(P) S_{g}(t)\,,
\end{equation}
where $N_{c}=3$ and $\bar{\Gamma}_{f \bar{g}}(P)=C^{\dagger} \Gamma_{f \bar{g}}(P)^{T} C$. Using $\frac{\partial}{\partial P_{\mu}}=2 P_{\mu} \frac{d}{d P^{2}}$, one has 
\begin{subequations}
\label{norpsEqn}
\begin{align}
1& =\left[\frac{d}{d P^{2}} \Pi(K, P)\right]_{K=P}\,, \\
\Pi(K, P)& =6 \operatorname{tr}_{D} \int \frac{d^{4} t}{(2 \pi)^{4}} \bar{\Gamma}_{f \bar{g}}(-K) S_{f}(t+P) \Gamma_{f \bar{g}}^{0^{-}}(K) S_{g}(t)\,.
\end{align}
\end{subequations}

With of the canonically normalised Bethe-Salpeter in hand, the leptonic decay of a pseudoscalar meson is described by the following matrix element:
\begin{equation}
\label{leptonicdecayps}
\begin{aligned}
\left\langle 0\left|\bar{q}_{g} \gamma_{\mu} \gamma_{5} q_{f}\right| P^{f \bar{g}}(P)\right\rangle:& = f_{f\bar{g}}^{0^{-}} P_{\mu} \\
& =N_{c} \int \frac{d^{4} t}{(2 \pi)^{4}} \operatorname{tr}_{D}\left[\gamma_{5} \gamma_{\mu} S_{f}(t+P) \Gamma_{f\bar{g}}^{0^{-}}(P) S_{g}(t)\right]\,.
\end{aligned}
\end{equation}
Straightforward algebra now yields the following result:
\begin{equation}
\label{resfps}
f_{f\bar{g}}^{0^{-}}=\frac{N_{c}}{4 \pi^{2}} \frac{1}{M_{f \bar{g}}}\left[\mathcal{K}_{F E}^{\left[f\bar{g}\right]}E_{\left[f\bar{g}\right]}+\mathcal{K}_{F F}^{\left[f\bar{g}\right]}F_{\left[f\bar{g}\right]}\right]\,.
\end{equation}

In the same manner, for a vector meson, one has
\begin{equation}
\label{vcBSA}
\Gamma_{\mu \left(f \bar{g}\right)}^{1^{-}}(P)=\gamma_{\mu}^{\perp} E_{\{f \bar{g}\}}(P)\,,
\end{equation}
where
\begin{subequations}
\label{PTofgamma}
\begin{align}
\gamma_{\mu}^{\perp}& =\gamma_{\mu}-\frac{\gamma \cdot P}{P^{2}} P_{\mu} \\
\gamma_{\mu}^{\|}& =\frac{\gamma \cdot P}{P^{2}} P_{\mu}\,,
\end{align}
\end{subequations}
and $\gamma_{\mu}^{\perp}+\gamma_{\mu}^{\|}=\gamma_{\mu}$, so $P_{\mu} \gamma_{\mu}^{\perp}=0$ and $P_{\mu} \gamma_{\mu}^{\|}=\gamma \cdot P$. Using steps similar to those above, one arrives at the vector meson mass equation
\begin{equation}
\label{explicithbsevc}
1-K^{\{f\bar{g}\}}\left(-m_{\left(f \bar{g}, 1^{-}\right)}^{2}\right)=0\,,
\end{equation}
where
\begin{equation}
\label{Kvc}
K^{\{f\bar{g}\}}\left(P^{2}\right)= \frac{2 \alpha_{\mathrm{IR}}}{3 \pi m_{\mathrm{G}}^{2}} \int_{0}^{1} d \alpha \left[M_{f}M_{g}-\hat{\alpha} M_{f}^2-\alpha M_{g}^2- 2 \alpha \hat{\alpha}P^2\right] \bar{\mathcal{C}}_{1}^{i u}\left(\omega_{f\bar{g}}\right)\,.
\end{equation}

In this case the canonical normalisation condition can be written
\begin{equation}
\label{norvc}
T_{\alpha \beta}(P) P_{\mu}=N_{c} \operatorname{tr}_{D} \int \frac{d^{4} t}{(2 \pi)^{4}} \bar{\Gamma}_{\alpha(f \bar{g})}^{1^-}(-P) \frac{\partial}{\partial P_{\mu}} S_{f}(t+P) \Gamma_{\beta(f \bar{g})}^{1^-}(P) S_{g}(t)\,.
\end{equation}
Summing over the Lorentz indices, Eq.\,\eqref{norvc} can be rewritten in the form 
\begin{equation}
\label{norvc1}
P_{\mu} =\frac{1}{3} \frac{\partial}{\partial P_{\mu}}\left[N_c \operatorname{tr}_{D} \int \frac{d^{4} t}{(2 \pi)^{4}} \bar{\Gamma}_{\alpha (f \bar{g})}^{1^-}(-K) S_{f}(t+P) \Gamma_{\alpha (f \bar{g})}^{1^-}(K) S_{g}(t)\right]_{K=P}\,. 
\end{equation}
So, the canonical normalisation condition is readily expressed as follows:
\begin{equation}
\label{norvc2}
\frac{1}{E_{\{f \bar{g}\}}^{2}}=-\left.9 \frac{m_{\mathrm{G}}^{2}}{4 \pi \alpha_{\mathrm{IR}}} \frac{d}{d P^{2}} K^{\{f \bar{g}\}}\left(P^{2}\right)\right|_{P^{2}=-m_{\left(f \bar{g}, 1^{-}\right)}^{2}}\,,
\end{equation}
where $m_{\left(f \bar{g}1^{-}\right)^2}$ is the solution of Eq.\,\eqref{explicithbsevc}.

The leptonic decay of a vector meson with total momentum $P$ and polarisation $\lambda$ is described by
\begin{equation}
\label{leptonicdecayvc}
\begin{aligned}
\left\langle 0\left|\bar{q}_{g} \gamma_{\mu} q_{f}\right| V^{f \bar{g}}(P,\lambda)\right\rangle:&= f_{\left(f \bar{g}, 1^{-}\right)} m_{\left(f \bar{g}, 1^{-}\right)} \epsilon_{\mu}^{\lambda} \\
&=N_{c} \int \frac{d^{4} t}{(2 \pi)^{4}} \operatorname{tr}_{D}\left[\gamma_{\mu} S_{f}(t+P) \epsilon^{\lambda} \cdot \Gamma_{f\bar{g}}^{1^{-}}(P) S_{g}(t)\right]\,,
\end{aligned}
\end{equation}
where $\{\epsilon_{\mu}^{\lambda} | \lambda = -, 0, +\}$ are the three possible polarisation vectors.
Contracting with $\epsilon_{\mu}^{\lambda}$ and using standard properties of such vectors, Eq.\,\eqref{leptonicdecayvc} can be written in the form
\begin{equation}
\label{leptonicdecayvc1}
f_{\left(f \bar{g}, 1^{-}\right)} m_{\left(f \bar{g}, 1^{-}\right)} =\frac{N_{c}}{3} \int \frac{d^{4} t}{(2 \pi)^{4}} \operatorname{tr}_{D}\left[\gamma_{\mu} S_{f}(t+P) \Gamma_{\mu(f\bar{g})}^{1^{-}}(P) S_{g}(t)\right]\,.
\end{equation}
Then, one readily obtains the following result 
\begin{equation}
\label{resfvc}
f_{\left(f \bar{g}, 1^{-}\right)}m_{\left(f \bar{g}, 1^{-}\right)}=\frac{3 N_{c} m_{G}^2}{8 \pi \alpha_{IR}}  K^{\{f \bar{g}\}}\left(P^2\right)E_{\{f \bar{g}\}}\,.
\end{equation}

Based on the SCI analysis of pseudoscalar mesons in Ref.\,\cite{Xu:2021iwv}, the SCI analysis in Ref.\,\cite{Roberts:2011wy} is improved by keeping all light-quark parameter values but fixing the s-quark current mass $m_s$, and $K$-meson ultraviolet cutoff, $\Lambda_{\rm uv}^K$, through a least-squares fit to measured values of $m_K$, $f_K$, whilst imposing the following relation: 
\begin{equation}
\label{alphaLambda}
\alpha_{\rm IR}(\Lambda_{\rm uv}^{K}) [\Lambda_{\rm uv}^{K}]^2 \ln\frac{\Lambda_{\rm uv}^{K}}{\Lambda_{\rm ir}}
=
\alpha_{\rm IR}(\Lambda_{\rm uv}^{\pi}) [\Lambda_{\rm uv}^{\pi}]^2 \ln\frac{\Lambda_{\rm uv}^{\pi}}{\Lambda_{\rm ir}}\,.
\end{equation}
This procedure eliminates one parameter by imposing the physical constraint that any increase in the momentum-space extent of a hadron wave function should correspond to a reduction in the effective coupling between the constituents. Only $u/d$, $s$ quarks are considered herein.  

In Ref.\,\cite{Xu:2021iwv}, the procedure was also applied to the $c$-quark/$D$-meson and $\bar b$-quark/$B$-meson.  The complete set of results is reproduced in Table~\ref{Tab:DressedQuarks}.
The evolution of $\Lambda_{\rm uv}$ with $m_P$ is described by the following interpolation $(s=m_{P}^2)$:
\begin{equation}
\label{LambdaIRMass}
\Lambda_{\rm uv}(s) = 0.306 \ln [ 19.2 + (s/m_\pi^2-1)/2.70]\,.
\end{equation}

\begin{table}[t]
\caption{\label{Tab:DressedQuarks}
Couplings, $\alpha_{\rm IR}/\pi$, ultraviolet cutoffs, $\Lambda_{\rm uv}$, and current-quark masses, $m_f$, $f=u/d,s,c,b$, that deliver a good description of flavoured pseudoscalar meson properties, along with the dressed-quark masses, $M$, and pseudoscalar meson masses, $m_{P}$, and leptonic decay constants, $f_{P}$, they produce; all obtained with $m_G=0.5\,$GeV, $\Lambda_{\rm ir} = 0.24\,$GeV.
Empirically, at a sensible level of precision \cite{ParticleDataGroup:2020ssz}:
$m_\pi =0.14$, $f_\pi=0.092$;
$m_K=0.50$, $f_K=0.11$;
$m_{D} =1.87$, $f_{D}=0.15$;
$m_{B}=5.30$, $f_{B}=0.14$.
(Isospin symmetry is assumed and dimensioned quantities are listed in GeV.)}
\begin{center}
\begin{tabular*}
{\hsize}
{
l@{\extracolsep{0ptplus1fil}}|
c@{\extracolsep{0ptplus1fil}}|
c@{\extracolsep{0ptplus1fil}}
c@{\extracolsep{0ptplus1fil}}
c@{\extracolsep{0ptplus1fil}}
|c@{\extracolsep{0ptplus1fil}}
c@{\extracolsep{0ptplus1fil}}
c@{\extracolsep{0ptplus1fil}}}\hline
& quark & $\alpha_{\rm IR}/\pi\ $ & $\Lambda_{\rm uv}$ & $m$ &   $M$ &  $m_{P}$ & $f_{P}$ \\\hline
$\pi\ $  & $l=u/d\ $  & $0.36\phantom{2}$ & $0.91\ $ & $0.0068_{u/d}\ $ & 0.37$\ $ & 0.14 & 0.10  \\\hline
$K\ $ & $\bar s$  & $0.33\phantom{2}$ & $0.94\ $ & $0.16_s\phantom{7777}\ $ & 0.53$\ $ & 0.50 & 0.11 \\\hline
$D\ $ & $c$  & $0.12\phantom{2}$ & $1.36\ $ & $1.39_c\phantom{7777}\ $ & 1.57$\ $ & 1.87 & 0.15 \\\hline
$B\ $ & $\bar b$  & $0.052$ & $1.92\ $ & $4.81_b\phantom{7777}\ $ & 4.81$\ $ & 5.30 & 0.14
\\\hline
\end{tabular*}
\end{center}
\end{table}

Using the SCI, colour-antitriplet quark+quark correlations (diquarks) with spin-parity $J^P$, constituted from quarks with flavour $f$ and $g$, are described by a homogeneous BSE that is readily inferred from the BSE for the meson with spin-parity $J^{-P}$, \textit{i.e.}, Eq.\,\eqref{hbseEqnCI}:
\begin{equation}
\label{hbseEqnCIdiq}
^d\Gamma_{f g}^{J^{-P}}(Q) =-\frac{4\pi\alpha_{\mathrm{IR}}}{m_G^2} \int \frac{d^{4} t}{(2 \pi)^{4}} \gamma_{\mu} \frac{\lambda^{a}}{2} S_f(q+Q) ^d\Gamma_{f g}^{J^{-P}}(Q) [S_g(-q)]^{T}\left[\frac{\lambda^{a}}{2}\right]^{T} \left[\gamma_{\mu}\right]^{T}\,,
\end{equation}
where the superscript ``T'' indicates matrix transpose. The flipping of the sign in parity occurs because fermions and antifermions have opposite parity. 

Similar to a meson, the BSA of a diquark can be written as:
\begin{equation}
\label{BSAstrucdiq}
^d\Gamma_{fg}^{J^{-P}}(K)=H_{\bar 3_c}^d \otimes \underline\Gamma_{fg}^{J^{-P}}(K)= H_{\bar 3_c}^d \otimes t^{J}_{fg}\otimes
\Gamma_{fg}^{J^{-P},C}(K)C\,,
\end{equation}
where: the colour-antitriplet character is expressed in $\{H_{\bar 3_c}^d,d=1,2,3\}=\{i\lambda^7, -i\lambda^5, i\lambda^2\}$, with $\{\lambda^k,k=1,...,8\}$ being Gell-Mann matrices, so $(H^{c_3})_{c_1 c_2}=\epsilon_{c_1 c_2 c_3}$ with the Levi-Civita tensor $\epsilon_{c_1 c_2 c_3}$; the flavour structure is expressed via
\begin{equation}
\label{flavourdiq}
\begin{array}{ccc}
{\tt t}_{ud}^0 = \left[\begin{array}{ccc}
                    0 & 1 & 0 \\
                    -1 & 0 & 0 \\
                    0 & 0 & 0
                    \end{array}\right]\,,
&
{\tt t}_{us}^0 = \left[\begin{array}{ccc}
                    0 & 0 & 1 \\
                    0 & 0 & 0 \\
                    -1 & 0 & 0
                    \end{array}\right]\,,
&
{\tt t}_{ds}^0 = \left[\begin{array}{ccc}
                    0 & 0 & 0 \\
                    0 & 0 & 1 \\
                    0 & -1 & 0
                    \end{array}\right]\,,\\[6ex]
{\tt t}_{uu}^1 = \left[\begin{array}{ccc}
                    \sqrt{2} & 0 & 0 \\
                    0 & 0 & 0 \\
                    0 & 0 & 0
                    \end{array}\right]\,,
&
{\tt t}_{ud}^1 = \left[\begin{array}{ccc}
                    0 & 1 & 0 \\
                    1 & 0 & 0 \\
                    0 & 0 & 0
                    \end{array}\right]\,,
&
{\tt t}_{us}^1 = \left[\begin{array}{ccc}
                    0 & 0 & 1 \\
                    0 & 0 & 0 \\
                    1 & 0 & 0
                    \end{array}\right]\,,\\[6ex]
{\tt t}_{dd}^1 = \left[\begin{array}{ccc}
                    0 & 0 & 0 \\
                    0 & \sqrt{2} & 0 \\
                    0 & 0 & 0
                    \end{array}\right]\,,
&
{\tt t}_{ds}^1 = \left[\begin{array}{ccc}
                    0 & 0 & 0 \\
                    0 & 0 & 1 \\
                    0 & 1 & 0
                    \end{array}\right]\,,
&
{\tt t}_{ss}^1 = \left[\begin{array}{ccc}
                    0 & 0 & 0 \\
                    0 & 0 & 0 \\
                    0 & 0 & \sqrt{2}
                    \end{array}\right]\,.
\end{array}
\end{equation}
Here, $C=\gamma_2 \gamma_4$ is the charge-conjugation matrix. 

Using the following properties of Dirac matrices, $C \gamma_{\mu}^{T} C^{\dagger}=-\gamma_{\mu}$, and of Gell-Mann matrices, $\sum_{i=1}^{3}H_{i} \left[\lambda^{a}\right]^{T} \left[H_i\right]^{T}=-\lambda^{a}$, then it is straightforward to show that 
\begin{equation}
\label{hbseEqnCIDirdiq}
\Gamma_{f g}^{J^{-P},C}(p;P) =-\frac{1}{2} \frac{16\pi\alpha_{\mathrm{IR}}}{3 m_G^2}\int \frac{d^{4} t}{(2 \pi)^{4}} \gamma_{\mu} S_f(t+P) \Gamma_{f g}^{J^{-P},C}(t;P) S_g(t) \gamma_{\mu}\,.
\end{equation}
This equation explains the observation that an interaction capable of binding mesons also generates strong diquark correlations in the colour-$\bar{3}$ channel. 
It follows, moreover, that one may obtain the mass and Bethe Salpeter amplitude for a diquark with spin-parity $J^{-P}$ from the equation for a $J^P$-meson in which the only change is a halving of the interaction strength. 

It is now easy to obtain the masses and amplitudes of the $0^{+}$ and the $1^{+}$ diquarks: 
\begin{subequations}
\label{diqBSA}
\begin{align}
\Gamma_{f g}^{0^{+}, C}(P)& =\gamma_{5}\left[ i E_{\left[f g\right]}(P)+\frac{\gamma \cdot P}{M_{fg}} F_{\left[f g\right]}(P)\right]\,, \\
\Gamma_{\mu\left(f g\right)}^{1^{+}, C}(P)& =\gamma_{\mu}^{\perp} E_{\{fg\}}(P)\,,
\end{align}
\end{subequations}
where $M_{f g}=M_f M_{g}/\left[M_f+ M_{g}\right]$.  The masses and associated amplitude functions satisfy simply modified versions of Eqs.\,\eqref{explicithbseps}, and Eqs.\,\eqref{Kvc}, \eqref{norvc2}.

\begin{table}[htbp]
\caption{\label{qqBSAsolutions}
Masses and canonically normalised correlation amplitudes obtained by solving the diquark BSEs.  Recall that isospin-symmetry is assumed.
(Masses listed in GeV.  Amplitudes are dimensionless.)}
\begin{center}
\begin{tabular*}
{\hsize}
{
c@{\extracolsep{0ptplus1fil}}
c@{\extracolsep{0ptplus1fil}}
c@{\extracolsep{0ptplus1fil}}|
c@{\extracolsep{0ptplus1fil}}
c@{\extracolsep{0ptplus1fil}}
c@{\extracolsep{0ptplus1fil}}}\hline
$m_{[ud]}$ & $E_{[ud]}$ & $F_{[ud]}\ $ & $m_{[us]}$ & $E_{[us]}$ & $F_{[us]}\ $ \\
$0.78$ & $2.71$ &  $0.31\ $ &  $0.94$  & $2.78$ & $0.37\ $
\end{tabular*}
\begin{tabular*}
{\hsize}
{
c@{\extracolsep{0ptplus1fil}}
c@{\extracolsep{0ptplus1fil}}
|c@{\extracolsep{0ptplus1fil}}
c@{\extracolsep{0ptplus1fil}}
|c@{\extracolsep{0ptplus1fil}}
c@{\extracolsep{0ptplus1fil}}}\hline
$m_{\{uu\}}$ & $E_{\{uu\}}\ $ & $m_{\{us\}}$ & $E_{\{us\}}\ $ & $m_{\{ss\}}$ & $E_{\{ss\}}\ $ \\
$1.06$ & $1.39\ $ &  $1.22$ &  $1.16\ $  & $1.33$ & $1.10\ $\\\hline
\end{tabular*}
\end{center}
\end{table}

Following the same procedures used for mesons, one can solve the diquark BSEs and calculate the corresponding normalisation constants. As initially observed in Ref.\,\cite{Cahill:1987qr}, owing to similarities between their respective Bethe-Salpeter equations, one may consider a colour-antitriplet $J^P$ diquark as the partner to a colour-singlet $J^{-P}$ meson. Thus, the BSEs for $J^P$ diquark are solved using the dressed-quark propagators and the values of $\Lambda_{\rm uv}$ associated with the $J^{-P}$ mesons \cite{Yin:2019bxe, Yin:2021uom}. The calculated diquark mass-scales and canonically normalised amplitudes are listed in Table~\ref{qqBSAsolutions}. (As explained in Appendix~C of Ref.\,\cite{Chen:2012txa}, when using the SCI, a  slight modification of the canonical normalisation procedure for a given diquark correlation amplitude is necessary, resulting in a $\lesssim 4$\% recalibration, which is already included in Table~\ref{qqBSAsolutions}.)

\subsection{Ward-Green-Takahashi identities}\label{subsecIBSE}
In any study of low-energy hadron observables, it is critical that the relevant WGT identities \cite{Ward:1950xp, Green:1953te, Takahashi:1957xn} be satisfied. 
Failing this, it is impossible, \textit{e.g}., to preserve the pattern of chiral symmetry breaking in QCD and hence a veracious understanding of hadron mass splittings is not achievable. 
WGT identities are also very important for preserving the conserved vector current (CVC) and partially conserved axial-vector current (PCAC). 
For instance, if one dose not ensure satisfaction of the vector WGT identity when computing the pion elastic electromagnetic form factor, it cannot be guaranteed that the pion will have unit charge \cite{Roberts:1994hh}. 

The axial-vector WGT identity (AVWGTI), which expresses chiral symmetry and its breaking pattern is
\begin{equation}
Q_{\mu} \Gamma_{5 \mu}^{f g}(Q)+i\left[m_{f}+m_{g}\right] \Gamma_{5}^{f g}(Q)=S_{f}^{-1}\left(k_{+}\right) i \gamma_{5} + i \gamma_{5} S_{g}^{-1}\left(k\right)\,,
\label{AVWGTI}
\end{equation}
where $Q$ is the incoming momentum of the vertex, and $k_{+}$ and $k$ are the quark's outgoing and incoming momentum. In order to focus on basics, I ignore the flavour matrix of the vertex here. 

The two vertices in Eq.\,\eqref{AVWGTI} are determined by inhomogeneous BSEs, \textit{viz}.\ in RL truncation: 
\begin{subequations}
\label{ihBSEsavps}
\begin{align}
\Gamma_{5 \mu}^{f g}(Q)& =\gamma_{5} \gamma_{\mu}-\frac{16 \pi}{3} \frac{\alpha_{\mathrm{IR}}}{m_{G}^{2}} \int \frac{d^{4} t}{(2 \pi)^{4}} \gamma_{\alpha} S_{f}(t_{+}) \Gamma_{5 \mu}^{f g}(Q) S_{g}(t) \gamma_{\alpha}\,, \label{ihBSEsavps1}\\
\Gamma_{5}^{f g}(Q)& =\gamma_{5}-\frac{16 \pi}{3} \frac{\alpha_{\mathrm{IR}}}{m_{G}^{2}} \int \frac{d^{4} t}{(2 \pi)^{4}} \gamma_{\alpha} S_{f}(t_{+}) \Gamma_{5}^{f g}(Q) S_{g}(t) \gamma_{\alpha}\,.    \label{ihBSEsavps2}
\end{align}
\end{subequations}
One must therefore implement regularisations of these inhomogeneous BSEs that maintain Eqs.\,\eqref{ihBSEsavps}. 

To see what this entails, contract Eq.\,\eqref{ihBSEsavps1} with $Q_\mu$ and combine it with the AVWGTI in Eq.\,\eqref{AVWGTI} and the gap equation in Eq.\,\eqref{gapactual}.  This yields the following two identities: 
\begin{subequations}
\label{WTIdeduc}
\begin{align}
(M_{f}-m_{f})+(M_{g}-m_{g})& =\frac{64 \pi}{3} \frac{\alpha_{\mathrm{IR}}}{m_{G}^{2}} \int_{q} \left[\frac{M_{g}}{q^{2}+M_{g}^{2}}+\frac{M_{f}}{(q+Q)^{2}+M_{f}^{2}}\right]\,, \\
0& =\int_{q} \left[\frac{Q \cdot q}{q^{2}+M_{g}^{2}}-\frac{Q \cdot(q+Q)}{(q+Q)^{2}+M_{f}^{2}}\right]\,.
\end{align}
\end{subequations}

Re-expressing the integrals using Feynman parameterisation, one can arrive at 
\begin{equation}
\label{WTIdeducFeyn1}
\begin{aligned}
(M_{f}-m_{f})+(M_{g}-m_{g})&= \frac{64 \pi}{3} \frac{\alpha_{\mathrm{IR}}}{m_{G}^{2}} \int_{q} \int_{0}^{1} d \alpha \left[\frac{M_{f}+M_{g}}{\left[q^{2}+w_{fg}\left(\alpha,Q\right)\right]}- \right.\\
&\left.\frac{(\hat{\alpha} M_{f}-\alpha M_{g})(M_{f}^2-M_{g}^2+(\hat{\alpha}-\alpha)Q^2)}{\left[q^{2}+w_{fg}\left(\alpha,Q\right)\right]^{2}}\right]\,,
\end{aligned}
\end{equation}
and 
\begin{equation}
\label{WTIdeducFeyn2}
0 =\int_{q}^{\Lambda} \int_{0}^{1} d \alpha \frac{\frac{1}{2} q^{2}+w_{fg}(\alpha,Q)}{\left[q^{2}+w_{fg}(\alpha,Q)\right]^{2}}\,,
\end{equation}
where $w_{fg}(\alpha,Q)=\hat{\alpha} M_{f}^2 +\alpha M_{g}^2+\alpha\hat{\alpha} Q^2$. One must enforce Eqs.\,\eqref{WTIdeducFeyn1} and \eqref{WTIdeducFeyn2} whenever possible in the following. In doing so, one is defining a regularisation scheme for the model that ensures preservation of the AVWGTI. Taking into account the expressions in Appendix\,\ref{Techn}, one finds that this entails
\begin{subequations}
\label{regular}
\begin{align}
(M_{f}-m_{f})+(M_{g}-m_{g})=& \frac{4\alpha_{\mathrm{IR}}}{3\pi m_{G}^{2}} \int_{0}^{1} d \alpha \bigg[\mathcal{C}^{i u}(w_{fg}(\alpha,Q))-(\hat{\alpha} M_{f}-\alpha M_{g})\nonumber\\
&(M_{f}^2-M_{g}^2+(\hat{\alpha}-\alpha)Q^2)\bar{\mathcal{C}}_{1}^{i u}(w_{fg}(\alpha,Q))\bigg]\,, \label{regular1}\\
0=& \int_{0}^{1} d \alpha \mathcal{C}^{i u}(w_{fg}(\alpha,Q))+\mathcal{C}_{1}^{i u}(w_{fg}(\alpha,Q))\,. \label{regular2}
\end{align}
\end{subequations}

With such a symmetry-preserving treatment of the contact interaction, the solutions of Eqs.\eqref{ihBSEsavps} have the general form 
\begin{subequations}
\label{formofavps}
\begin{align}
\Gamma_{5 \mu}^{f g}(Q)& =\gamma_{5}\gamma_{\mu}^{T} P_{T}^{f g}(Q^{2})+\gamma_{5}\gamma_{\mu}^{L} P_{1 L}^{f g}(Q^{2})+i Q_{\mu} \gamma_{5} P_{2 L}^{f g}(Q^{2})\,, \label{formofavps1}\\
i \Gamma_{5}^{f g}(Q)& =i\gamma_{5}E^{f g}(Q^{2})+ \gamma_{5} \frac{\gamma \cdot Q}{2 M_{f g}}  F^{f g}(Q^{2})\,, \label{formofavps2}
\end{align}
\end{subequations}
where $Q_{\mu} \gamma_{\mu}^{T}=0, \gamma_{\mu}^{T}+\gamma_{\mu}^{L}=\gamma_{\mu}$. 

If one inserts Eq.\,\eqref{formofavps2} into Eq.\,\eqref{ihBSEsavps2} and employs the symmetry-preserving regularisation explained in Eqs.\eqref{regular}, then one arrives at the following algebraic equation for the two terms in the pseudoscalar vertex: 
\begin{subequations}
\label{iBSEpsEqn}
\label{algeps}
\begin{align}
&\left[\begin{array}{l}
E^{f g}\left(Q^2\right) \\
F^{f g}\left(Q^2\right)
\end{array}\right]=[I-\mathcal{K}]^{-1}\left[\begin{array}{l}
1 \\
0
\end{array}\right]\,,\\
&I-\mathcal{K}=\left[\begin{array}{ll}
1 & 0 \\
0 & 1
\end{array}\right]-\frac{4 \alpha_{\mathrm{IR}}}{3 \pi m_{G}^{2}}\left[\begin{array}{cc}
\mathcal{K}_{E E}^{f g} & \mathcal{K}_{E F}^{f g} \\
\mathcal{K}_{F E}^{f g} & \mathcal{K}_{F F}^{f g}
\end{array}\right]\,,
\end{align}
\end{subequations}
where the kernel elements are given in Eq.\,\eqref{Kmatps}.  (Owing to a slight notation change in treating the inhomogeneous BSE in this subsection, one must here identify $g$ with $\bar g$ in Eq.\,\eqref{Kmatps}.) As promised, a straightforward calculation reveals the presence of a pole at the lightest $f\bar{g}$-pseudoscalar state ($0^-$). From the term $K_{EE}$, it is easy to obtain the following relation: 
\begin{equation}
\left(m_{f}+m_{g}\right) E^{f g}\left(Q^{2}=0\right)=M_{f}+M_{g}.
\label{EfgRelation}
\end{equation}

Now, returning to the axial-vector vertex. Substituting Eq.\,\eqref{formofavps1} into Eq.\,\eqref{ihBSEsavps1} and drawing on Eq.\,\eqref{AVWGTI}, one finds 
\begin{equation}
\label{P1Lav}
P_{1 L}^{f g}\left(Q^{2}\right) \equiv 1-\frac{m_{f}+m_{g}}{2 M_{f g}} F^{f g}\left(Q^{2}\right)\,.
\end{equation}

One can determine $P_{T}^{f g}(Q^{2})$ by contracting Eq.\,\eqref{formofavps1} with the transverse projection operator
$\gamma_{5}T_{\mu \nu}(Q)=\gamma_{5}(\delta_{\mu \nu}-Q_{\mu} Q_{\nu} / Q^{2})$:
\begin{subequations}
\label{PTav}
\begin{align}
P_{T}^{f g}\left(Q^{2}\right)& =\frac{1}{1+K_{AV}^{f g}\left(Q^{2}\right)}\,, \\
K_{AV}^{f g}\left(Q^{2}\right)& =\frac{2 \alpha_{\mathrm{IR}}}{3 \pi m_{G}^{2}} \int_{0}^{1} d \alpha\left[M_{f}M_{g}+\hat{\alpha} M_{f}^2\right. \\ \nonumber
&\left.+\alpha M_{g}^2 +2 \alpha \hat{\alpha} Q^{2}\right] \bar{\mathcal{C}}_{1}^{\mathrm{iu}}\left(\omega_{f g}\left(\alpha, Q^{2}\right)\right)\,.
\end{align}
\end{subequations}
The transverse part of the axial-vector vertex exhibits a pole at the mass of the lightest $f\bar{g}$ axial-vector state ($1^{+}$). 

Only the computation of $P_{2 L}^{f g}(Q^{2})$ remains. 
This may be accomplished by, first, contracting Eq.\,\eqref{formofavps1} with $i\gamma_{5} Q_{\mu}$ and then, using Eq.\,\eqref{AVWGTI} and the gap equations, one arrives at 
\begin{equation}
\label{P2Lav}
Q^{2} P_{2 L}^{f g}(Q^{2})=(M_{f}+M_{g})-(m_{f}+m_{g}) E^{f g}(Q^{2})\,.
\end{equation}
According to Eq.\,\eqref{EfgRelation}, it is clear that $\lim _{Q^{2} \rightarrow 0} Q^{2} P_{2 L}^{f g}\left(Q^{2}\right)=0$. 
Consequently, $P_{2 L}^{f g}\left(Q^{2}\right)$ is regular at $Q^{2}=0$. 

Using Eq.\,\eqref{EfgRelation} in solving Eqs.\,\eqref{iBSEpsEqn}, one can obtain 
\begin{equation}
\label{Fps0}
F^{fg}(Q^2=0)=2\frac{M_{f g}}{m_{f}+m_{g}} \frac{\mathcal{K}_{F F}^{f g}(Q^2=0)}{\mathcal{K}_{F F}^{f g}(Q^2=0)-1}.
\end{equation}
Inserting $F^{fg}(Q^2=0)$ into Eq.\,\eqref{P1Lav}, one finds
\begin{equation}
\label{P1Lav0}
P_{1L}^{f g}\left(Q^{2}=0\right)=1+\frac{\mathcal{K}_{F F}^{f g}(Q^2=0)}{1-\mathcal{K}_{F F}^{f g}(Q^2=0)}=\frac{1}{1-\mathcal{K}_{F F}^{f g}(Q^2=0)}.
\end{equation}
Comparing $K_{AV}^{f g}\left(Q^{2}\right)$ in Eq.\,\eqref{PTav} with $\mathcal{K}_{F F}^{f \bar{g}}\left(Q^{2}\right)$ in Eq.\,\eqref{Kmatps}, one reads
$K_{AV}^{f g}\left(Q^{2}=0\right)=-\mathcal{K}_{F F}^{f g}\left(Q^{2}=0\right)$. 
Thus, 
\begin{equation}
\label{P1LPTav0}
P_{T}^{f g}\left(Q^{2}=0\right)=\frac{1}{1+K_{AV}^{f g}\left(Q^{2}=0\right)}=P_{1L}^{f g}\left(Q^{2}=0\right).
\end{equation}
This establishes that the axial-vector vertex is regular at $Q=0$ when the current-quark masses are nonzero.

Evidently, a symmetry-preserving calculation of the axial-vector and pseudoscalar vertices has succeeded. 

Following similar procedures, one can develop a symmetry-preserving calculation of the vector and scalar vertices. Detailed discussions are contained in Refs.\,\cite{Xu:2021iwv, Chen:2012txa, Xing:2022sor}.  

\section{Faddeev equation}\label{secFaddeev}
\begin{figure}[htbp]
\centering
  \includegraphics[width=3.8in]{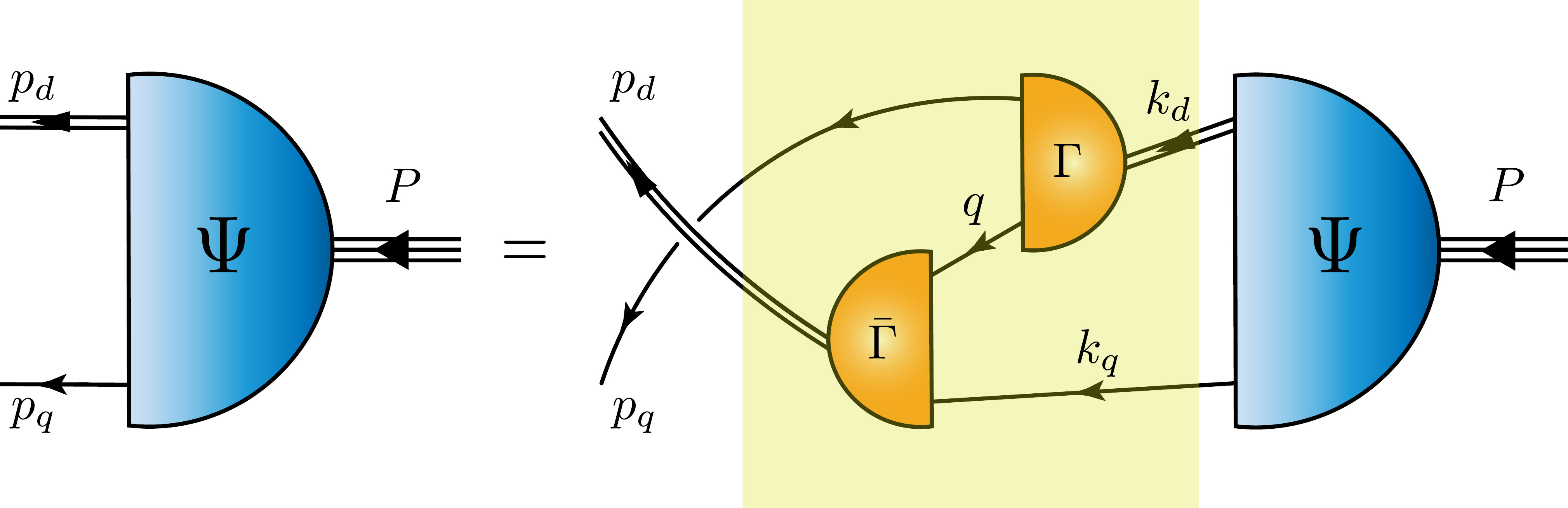}
\caption{\label{figFaddeev}
Integral equation for the Poincar\'e-covariant matrix-valued function $\Psi$, the Faddeev amplitude for a baryon with total momentum $P=p_q+p_d=k_q+k_d$ constituted from three valence quarks, two of which are always contained in a nonpointlike, interacting diquark correlation. $\Psi$ describes the relative momentum correlation between the dressed-quarks and -diquarks.
Legend. \emph{Shaded rectangle} -- Faddeev kernel;
\emph{single line} -- dressed-quark propagator [Section\,\ref{secgap}];
$\Gamma$ -- diquark correlation amplitude [Section\,\ref{subsecBSE}] and \emph{double line} -- diquark propagator. Ground-state $J=1/2^+$ baryons contain both flavour-antitriplet--scalar and flavour-sextet--axial-vector diquarks.}
\end{figure}

Herein, each baryon's dressed-quark core is described via the solution of a Poincar\'e-covariant Faddeev equation, like that depicted in Fig.\,\ref{figFaddeev}. 
The approach is built upon the quark--plus--interacting-diquark picture of baryon structure introduced in Refs.\,\cite{Cahill:1988dx, Reinhardt:1989rw, Efimov:1990uz}.
An updated perspective on this picture is provided in Refs.\,\cite{Barabanov:2020jvn, Lu:2022cjx, Liu:2022ndb, Eichmann:2022zxn}. 

The Faddeev equation is derived following the observation, mentioned above, that an interaction which binds mesons also generates strong diquark correlations in the colour-$\bar{3}$ channel. 
The validity of the diquark approximation in the quark-quark scattering kernel is supported, \textit{e.g}., by the explicitly three-body analyses in Ref.\cite{Eichmann:2011vu}. 
In the quark+diquark  approach, there are two contributions that bind three valence quarks within a baryon \cite{Segovia:2015ufa}. 
One part is expressed in the formation of tight (but not pointlike) quark+quark correlations. 
The other is the attraction generated by the quark exchange depicted in the shaded area of Fig.\,\ref{figFaddeev}, which ensures the diquark correlations within the baryon are fully dynamical. 
Namely, no quark is special because each one participates fully in all diquarks allowed by its quantum numbers. 
The continual rearrangement of the quarks ensures, \emph{inter alia}, that the baryon's dressed-quark wave function adheres to Pauli statistics. 

\subsection{General structure of the Faddeev amplitudes}\label{genralStruofFad} 
The Faddeev amplitude of a spin-$1/2$ baryon may be represented as follows:
\begin{equation}
\label{PsiBaryon}
\Psi = \Psi_1 + \Psi_2 + \Psi_3  \,,
\end{equation}
where the subscript identifies the bystander quark and, \textit{e.g}., $\Psi_{1,2}$ are obtained from $\Psi_3$ through a cyclic permutation of all quark labels. 
I employ the simplest realistic representation of $\Psi$, where an octet baryon is comprised of a sum of scalar and axial-vector diquark correlations: 
\begin{equation}
\label{Psioctet} 
\Psi_3(p_j,\alpha_j,\varphi_j) = {\cal N}_{3}^{0^+}(p_j,\alpha_j,\varphi_j) + {\cal N}_{3}^{1^+}(p_j,\alpha_j,\varphi_j)\,,
\end{equation}
with $(p_j,\alpha_j,\varphi_j)$ being the momentum, spin and flavour labels of the quarks constituting the bound state, and $P=p_1+p_2+p_3$ is the system's total momentum. 

It is plausible that pseudoscalar and vector diquarks may contribute to the Faddeev amplitude of a ground-state octet baryon. However, they have opposite parity compared to the ground-state baryon and hence can only appear in concert with nonzero quark angular momentum. Given that the ground-state baryon is expected to have minimal quark orbital angular momentum, and considering that these diquark correlations are considerably more massive than the scalar and axial-vector diquarks, they can safely be ignored when computing the properties of the ground state.  This is confirmed in Refs.\,\cite{Eichmann:2016hgl, Lu:2017cln, Chen:2017pse}.

The scalar diquark piece in Eq.\,\eqref{Psioctet} takes the form
\begin{equation}
\label{Psi0plus}
\begin{aligned}
\mathcal{N}_3^{0^{+}}\left(p_i, \alpha_i, \tau_i\right)& =\left[^d\Gamma_i^{0^{+}}
\left(\frac{1}{2} p_{[12]} ; K\right)\right]_{\alpha_1 \alpha_2}^{\tau_1 \tau_2} \Delta^{(i,0^{+})}(K)[\Psi^{\mathcal{S},i}(\ell ; P)]_{\alpha_3}^{\tau_3}\,, \\
\Psi^{\mathcal{S},i}(\ell ; P)& =\mathcal{S}^i(\ell ; P) u(P)\,,
\end{aligned}
\end{equation}
where: $K= p_1+p_2=: p_{\{12\}}$, $p_{[12]}= p_1 - p_2$, $\ell := (-p_{\{12\}} + 2 p_3)/3$;
\begin{equation}
\label{scalarqqprop}
\Delta^{(i,0^+)}(K) = \frac{1}{K^2+m_{(i,0^+)}^2}\,,
\end{equation}
is a propagator for the scalar diquark formed from quarks 1 and 2, with $m_{0^+}$ the mass-scale associated with this $0^+$ diquark; $^d\Gamma_i^{0^+}$ is the canonically-normalised Bethe-Salpeter amplitude describing the relative momentum correlation between the quarks, given in Eq.\,\eqref{BSAstrucdiq}; ${\cal S}^i$ is a $4\times 4$ Dirac matrix describing the relative quark-diquark momentum correlation within the baryon, also including the colour-singlet matrix $\lambda_c^0/\sqrt{3}$; and the spinor satisfies
\begin{equation}
(i\gamma\cdot P + M_B)\, u(P) =0= \bar u\, (i\gamma\cdot P + M_B)\,,
\end{equation}
with $M_B$ the mass obtained by solving the Faddeev equation. The spinor, $u$, is normalised such that $\bar{u}(P) u(P)=2 M_B$, and 
\begin{equation}
\label{noru}
2 M_B \Lambda_+(P) = \sum_{\sigma=\pm}u(P;\sigma)\bar u(P;\sigma) = M_B-i\gamma\cdot P\,,
\end{equation}
where in this line I have explicitly indicated the spin label. (See Appendix~A in Ref.\,\cite{Chen:2012qr} for more details.)  I note that $u$ also possesses another column-vector degree of freedom, \textit{viz}.
\begin{subequations}
\label{halfnucleon}
\begin{align}
u_p & =
\left[
\begin{array}{ll}
{\rm r}_1 & u[ud]   \\
{\rm r}_2 &  d\{uu\}    \\
{\rm r}_3 &  u\{ud\}    \\
\end{array} \right]\,,  \\
u_n & =
\left[
\begin{array}{ll}
{\rm r}_1 & d[ud]   \\
{\rm r}_2 &  u\{dd\}    \\
{\rm r}_3 &  d\{ud\}    \\
\end{array} \right]\,,  \\
u_\Lambda & =
\frac{1}{\sqrt{2}}\left[
\begin{array}{ll}
{\rm r}_1 & -\sqrt{2}s[ud]           \\
{\rm r}_2 & u[ds]-d[us]     \\
{\rm r}_3 & u\{ds\} -d\{us\}
\end{array}
\right]\,,  \label{SpinFlavourLambda}\\
u_{\Sigma^+} & =
\left[
\begin{array}{ll}
{\rm r}_1 & u[us]   \\
{\rm r}_2 & s\{uu\} \\
{\rm r}_3 & u\{us\}\\
\end{array}
\right]\,, \label{SpinFlavourSigma}\\
u_{\Xi^0} & =
\left[
\begin{array}{ll}
{\rm r}_1 & s[us]   \\
{\rm r}_2 & s\{us\} \\
{\rm r}_3 & u\{ss\}\\
\end{array}
\right]\,. \label{BSXi0}
\end{align}
\end{subequations}
Since this work assumes isospin symmetry, the unlisted octet charge states may be obtained from those above by applying an isospin-lowering operator. These states are
mass degenerate with those written explicitly. The colour antisymmetry of $\Psi_3$ is implicit in $^d\Gamma^{0^+}$. 

The axial-vector part of Eq.\,\eqref{Psioctet} is
\begin{equation}
\label{Psi1plus}
\begin{aligned}
\mathcal{N}^{1^{+}}\left(p_i, \alpha_i, \tau_i\right) & =\left[^d\Gamma_{i,\mu}^{1^{+}}
\left(\frac{1}{2} p_{[12]} ; K\right)\right]_{\alpha_1 \alpha_2}^{\tau_1 \tau_2} 
\Delta_{\mu \nu}^{(i,1^{+})}(K)\left[\Psi_\nu^{\mathcal{A},i}(\ell ; P)\right]_{\alpha_3}^{\tau_3}\,, \\
\Psi_\nu^{\mathcal{A},i}(\ell ; P) & =\mathcal{A}_\nu^i(\ell ; P) u(P)\,,
\end{aligned}
\end{equation}
where
\begin{equation}
\label{avqqprop}
\Delta_{\mu \nu}^{\left(i, 1^{+}\right)}(K)=\frac{1}{K^2+m_{\left(i, 1^{+}\right)}^2}\left(\delta_{\mu \nu}+\frac{K_\mu K_\nu}{m_{\left(i, 1^{+}\right)}^2}\right)
\end{equation}
is a propagator for the axial-vector diquark formed from quarks 1 and 2 and the remaining elements in Eq.\,\eqref{Psi1plus} are straightforward generalisations of those in Eq.\,\eqref{Psi0plus}. 

For completeness I note that since it is not possible to combine an isospin-0 diquark with an isospin-$1/2$ diquark to obtain isospin-$3/2$, the spin- and isospin-$3/2$ decuplet baryons contain only axial-vector diquark correlations. (Isospin-$3/2$ vector diquarks play practically no role \cite{Liu:2022ndb}.)  This establishes the pattern for the remaining decuplet baryons, allowing them to be expressed via 
\begin{equation}
\label{DecupletFA}
\Psi_3^{10}(p_i,\alpha_i,\varphi_i) = {\cal D}_{3}^{1+}(p_j,\alpha_j,\varphi_j)\,,
\end{equation}
with
\begin{equation}
\label{Psi10}
\begin{aligned}
\mathcal{D}_3^{1+}(p_j,\alpha_j,\varphi_j)& =\left[^d\Gamma_{i,\mu}^{1^{+}}\left(\frac{1}{2} p_{[12]} ; K\right)\right]_{\alpha_1 \alpha_2}^{\tau_1 \tau_2} \Delta_{\mu \nu}^{\left(i, 1^{+}\right)}(K)\left[\Psi_{\nu}^{\mathcal{D},i}(\ell ; P)\right]_{\alpha_3}^{\tau_3},\\
\Psi_{\nu}^{\mathcal{D},i}(\ell ; P)& =\mathcal{D}_{\nu \rho}^i(\ell ; P) u_{\rho}(P),
\end{aligned}
\end{equation}
where $u_\rho(P)$ is a Rarita-Schwinger spinor and, similar to octet baryons, in constructing the Faddeev equations one may focus on that member of each isospin multiplet which has maximum electric charge, \textit{viz.}
\begin{equation}
\label{halfDelta}
\begin{aligned}
u_\Delta= \left[
\begin{array}{c}
\{uu\} u \\
\end{array} \right],\; &
u_{\Sigma^\ast}=
\left[ \begin{array}{c}
\{uu\} s \\
\{us\} u
\end{array} \right],\;
u_{\Xi^\ast}=
\left[
\begin{array}{c}
\{us\} s \\
\{ss\} u
\end{array}\right],\;&
u_\Omega=
\left[
\begin{array}{c}
\{ss\}s
\end{array}
\right]\,.
\end{aligned}
\end{equation}

Although simple to understand visually, the flavour structure expressed in Eq.\,\eqref{halfnucleon} is not that best suited for calculations. 
Instead, it is more convenient to proceed by associating a flavour-space column vector with the baryon spinor: so that, \emph{e.g}., Eq.\,\eqref{BSXi0} is re-expressed as follows:
\begin{equation}
\label{FspaceC}
\underline{\Psi}_{B=\Xi^0} =
[\Psi_{\Xi^0}^{{\mathcal{S}}_{[us]}} {\mathpzc f}_s
+\Psi_{\Xi^0}^{{\mathcal{A}}_{\{us\}}} {\mathpzc f}_s
+\Psi_{\Xi^0}^{{\mathcal{A}}_{\{ss\}}} {\mathpzc f}_u] \otimes \frac{\lambda_c^0}{\sqrt{3}}\,,
\end{equation}
where ${\mathpzc f}_u={\rm column}[1,0,0]$, ${\mathpzc f}_d={\rm column}[0,1,0]$ and ${\mathpzc f}_s={\rm column}[0,0,1]$.The column vector that should be used is determined by the baryon, $B$, and the specified diquark.

The general forms of the matrices ${\cal S}^{i}(\ell;P)$, ${\cal A}^{i}_\nu(\ell;P)$ and ${\cal D}^{i}_{\nu\rho}(\ell;P)$, which characterise the momentum-space correlation between the quark and diquark in the octet and decuplet baryons, respectively, are described in Refs.\,\cite{Oettel:1998bk, Cloet:2007pi, Segovia:2014aza}. 
The requirement that ${\cal S}^{i}(\ell;P)$ represent a positive energy baryon, \textit{i.e.}, an eigenfunction of $\Lambda_+(P)$, entails 
\begin{subequations}
\label{SexpI}
\begin{align}
&{\cal S}^{i}(\ell;P) = s^{i}_1(\ell;P)\,S^1(\ell;P) + s^{i}_2(\ell;P)\,S^2(\ell;P)\,, \\
&S^1(\ell;P)=\mathbf{I}_{\rm D}, S^2(\ell;P)=\left(i\gamma\cdot \hat\ell - \hat\ell \cdot \hat P\, \mathbf{I}_{\rm D}\right)\,, \label{SfunctionsI}
\end{align}
\end{subequations}
where $(\mathbf{I}_{\rm D})_{rs}= \delta_{rs}$, $\hat \ell^2=1$, $\hat P^2= - 1$.  In the baryon rest frame, $s^{i}_{1,2}$ describe, respectively, the upper, lower component of the bound-state baryon's spinor. 

Placing the same constraint on the axial-vector component, one obtains 
\begin{equation}
\label{AexpI}
 {\cal A}^{i}_\nu(\ell;P) = \sum_{n=1}^6 \, p_n^i(\ell;P)\,\gamma_5\,A^n_{\nu}(\ell;P)\,,
\end{equation}
where ($ \hat \ell^\perp_\nu = \hat \ell_\nu + \hat \ell\cdot\hat P\, \hat P_\nu$, $ \gamma^\perp_\nu = \gamma_\nu + \gamma\cdot\hat P\, \hat P_\nu$)
\begin{equation}
\label{AfunctionsI}
\begin{array}{lll}
A^1_\nu= \gamma\cdot \hat \ell^\perp\, \hat P_\nu \,,\; &
A^2_\nu= -i \hat P_\nu \,,\; &
A^3_\nu= \gamma\cdot\hat \ell^\perp\,\hat \ell^\perp\,,\\
A^4_\nu= i \,\hat \ell_\mu^\perp\,,\; &
A^5_\nu= \gamma^\perp_\nu - A^3_\nu \,,\; &
A^6_\nu= i \gamma^\perp_\nu \gamma\cdot\hat \ell^\perp - A^4_\nu\,.
\end{array}
\end{equation}
Finally, because ${\cal D}^{i}_{\nu\rho}(\ell;P)$ is also an eigenfunction of $\Lambda_+(P)$, one obtains
\begin{equation}
\label{DeltaFA}
{\cal D}^{i}_{\nu\rho}(\ell;P) = {\cal S}^{i}(\ell;P) \, \delta_{\nu\rho} + \gamma_5{\cal A}^{i}_\nu(\ell;P) \,\ell^\perp_\rho \,,
\end{equation}
with ${\cal S}^{i}$ and ${\cal A}^{i}_\nu$ given by obvious analogues of Eqs.\,\eqref{SexpI} and \eqref{AexpI}, respectively. 

With detailed forms available for the dressed-quark propagators, diquark Bethe-Salpeter amplitudes, and diquark propagators -- such as those determined by the SCI -- one can now formulate Faddeev equations for the baryons. As illustrated in Fig.\,\ref{figFaddeev}, the kernels of these equations involve the breakup and reformation of diquarks through the exchange of a dressed-quark. Thus, the Faddeev equation for an octet baryon, satisfied by $\Psi_3$, is
\begin{equation}
\label{FaddeevEqn}
\left[\begin{array}{c}
\mathcal{S}^m(k ; P) u(P) \\
\mathcal{A}_\mu^i(k ; P) u(P)
\end{array}\right]=-4 \int \frac{d^4 l}{(2 \pi)^4} \mathcal{M}(k, l ; P)\left[\begin{array}{c}
\mathcal{S}^n(l ; P) u(P) \\
\mathcal{A}_\nu^j(l ; P) u(P)
\end{array}\right]
\end{equation}
where one factor of ``2'' appears because $\Psi_3$ is coupled symmetrically to $\Psi_1$ and $\Psi_2$, and I have evaluated the colour factor ``-2''  (see Appendix\,\ref{CFcoeffi}). 

The kernel in Eq.\,\eqref{FaddeevEqn} is
\begin{equation}
\label{Faddeevkernel}
\mathcal{M}(k, l ; P)=\left[\begin{array}{cc}
\mathcal{M}_{00}^{mn} & \left(\mathcal{M}_{01}\right)_\nu^{mj} \\
\left(\mathcal{M}_{10}\right)_\mu^{in} & \left(\mathcal{M}_{11}\right)_{\mu \nu}^{i j}
\end{array}\right]
\end{equation}
with
\begin{equation}
\label{kernelelements}
\begin{aligned}
\mathcal{M}_{00}^{mn} & =\Gamma_n^{0^{+}}\left(k_{-l}; l_{q q}\right) S^T\left(l_{q q}-k_q\right) \bar{\Gamma}_m^{0^{+}}\left(l_{-k} ;-k_{q q}\right) S\left(l_q\right) \Delta^{\left(n, 0^{+}\right)}\left(l_{q q}\right)\,, \\
\left(\mathcal{M}_{01}\right)_\nu^{mj} & =\Gamma_{j,\mu}^{ 1^{+}}\left(k_{-l}; l_{q q}\right) S^T\left(l_{q q}-k_q\right) \bar{\Gamma}_m^{ 0^{+}}\left(l_{-k} ;-k_{q q}\right) S\left(l_q\right) \Delta_{\mu \nu}^{\left(j, 1^{+}\right)}\left(l_{q q}\right)\,, \\
\left(\mathcal{M}_{10}\right)_\mu^{in} & =\Gamma_n^{ 0^{+}}\left(k_{-l} ; l_{q q}\right) S^T\left(l_{q q}-k_q\right) \bar{\Gamma}_{i,\mu}^{1^{+}}\left(l_{-k};-k_{q q}\right) S\left(l_q\right) \Delta^{\left(n, 0^{+}\right)}\left(l_{q q}\right)\,, \\
\left(\mathcal{M}_{11}\right)_{\mu \nu}^{i j} & =\Gamma_{j,\rho}^{1^{+}}\left(k_{-l} ; l_{q q}\right) S^T\left(l_{q q}-k_q\right)\bar{\Gamma}_{i,\mu}^{1^{+}}\left(l_{-k} ;-k_{q q}\right) S\left(l_q\right) \Delta_{\rho \nu}^{\left(j, 1^{+}\right)}\left(l_{q q}\right)\,,
\end{aligned}
\end{equation}
where 
$k_{-l}=k_q-\frac{1}{2} l_{q q}$, $l_{-k}=l_q-\frac{1}{2} k_{q q}$,
$\ell_q=\ell, k_q=k, \ell_{q q}=-\ell+P, k_{q q}=-k+P$. The decuplet baryons' Faddeev equations are similar to those of the octet baryons, but simpler.

In proceeding, I follow Ref.\,\cite{Roberts:2011cf} and implement a simplification; namely, in the Faddeev equation for a baryon of type $B$, I represent the quark exchanged between the diquarks as 
\begin{equation}
\label{staticapproximation}
S^{\mathrm{T}}(k) \rightarrow \frac{g_B^2}{M}\,,
\end{equation}
where $g_B=g_8=1.18$ for octet baryons and $g_B\to g_{10}=1.56$ for decuplet baryons \cite{Chen:2012qr}.  This is a variant of the so-called ``static approximation'', originally introduced in Ref.\,\cite{Buck:1992wz}, and since then it has been employed in studies of various nucleon properties -- see, \textit{e.g}., Refs.\,\cite{Yin:2019bxe, Yin:2021uom, Wilson:2011aa, Chang:2012cc, Segovia:2013rca, Raya:2021pyr, Gutierrez-Guerrero:2021rsx, Cheng:2022jxe}. 
In conjunction with the diquark correlations produced by Eq.\,\eqref{KCI}, whose Bethe–Salpeter amplitudes are momentum-independent, Eq.\,\eqref{staticapproximation} produces Faddeev equation kernels that are also momentum-independent. Consequently, Eqs.\,\eqref{SexpI} and \eqref{AexpI} simplify dramatically, containing only those terms that are independent of the relative momentum: 
\begin{equation}
\label{basisoctet}
\begin{aligned}
\mathcal{S}^{i}(\ell ; P) & \rightarrow \mathcal{S}^{i}(P)=s^{i}(P) \mathbf{I}_{\mathrm{D}}\,, \\
\mathcal{A}_v^{i}(\mu ; P) & \rightarrow \mathcal{A}_\mu^{i}(P)=a_1^{i}(P) i \gamma_5 \gamma_\mu+a_2^{i}(P) \gamma_5 \hat{P}_\mu\,.
\end{aligned}
\end{equation}

I would like to emphasise that the utilisation of Eq.\,\eqref{staticapproximation} is a matter of convenience rather than necessity. It is employed because it allows us to present algebraic formulas that reveal  qualitative characteristics of the Faddeev equation and enable reliable insights into the mechanisms of bound-state formation and level ordering.  Eliminating this simplification introduces significant additional complexity without material gains in either insight or quantitative agreement with observation \cite{Xu:2015kta}.

\subsection{Explicit example: $\Lambda$ baryon}\label{ExplicitExp} 
Here I illustrate the construction of the Faddeev equation by considering the $\Lambda$ baryon. The $\Lambda$ baryon is an isospin-0, $J^P=(1/2)^+$ state composed of a single quark from each flavour, resulting in a somewhat complicated spin-flavour amplitude. 

Considering Eq.\,\eqref{SpinFlavourLambda}, the ground-state $\Lambda$ may be formed by five possible diquark combinations: 
\begin{equation}
\label{strcutureLam}
[ud]_{0^+}s,\; [us]_{0^+}d, \;[ds]_{0^+}u,\; \{us\}_{1^+}d,\; \{ds\}_{1^+}u\,.
\end{equation}
One can immediately see that $[ud]_{0+}s$ has $I=0$ whilst the others do not possess good isospin. This results in a mixing effect that make it difficult to distinguish between the $\Lambda$ and $\Sigma^0$ isospin-eigenstates. Consequently, constructing the flavour structure of the Faddeev kernel becomes a complex task. 

States with good isospin can be constructed in the following manner: with 
\begin{equation}
\label{VOmatrices}
\begin{array}{cc}
 V=\left(
  \begin{array}{c}
   [ud]s\\[0.7ex]
   [us]d\\[0.7ex]
   [ds]u\\[0.7ex]
   \{us\}d\\[0.7ex]
   \{ds\}u
  \end{array}
  \right)\,, &
 O=\left(
  \begin{array}{ccccc}
   1 & 0 & 0 & 0 & 0\\[0.7ex]
   0 & \frac{1}{\sqrt{2}} & -\frac{1}{\sqrt{2}} & 0 & 0\\[0.7ex]
   0 & \frac{1}{\sqrt{2}} & \frac{1}{\sqrt{2}} & 0 & 0\\[0.7ex]
   0 & 0 & 0 & \frac{1}{\sqrt{2}} & -\frac{1}{\sqrt{2}}\\[0.7ex]
   0 & 0 & 0 & \frac{1}{\sqrt{2}} & \frac{1}{\sqrt{2}}
  \end{array}
  \right) \,,
 \end{array}
\end{equation}
then each of the entries in the new column vector 
\begin{equation}
\label{goodV}
 \tilde{V}=OV=\frac{1}{\sqrt{2}}
 \left(
  \begin{array}{cc}
   \sqrt{2} [ud]s\,, & I=0\\[0.7ex]
   [us]d-[ds]u\,, & I=0\\[0.7ex]
   [us]d+[ds]u\,, & I=1\\[0.7ex]
   \{us\}d-\{ds\}u\,, & I=0 \\[0.7ex]
   \{us\}d+\{ds\}u\,, & I=1
  \end{array}
  \right)\,,
\end{equation}
has good isospin, with the isospin indicated. 

According to Fig.\,\ref{figFaddeev}, one can establish that the column vector $V$ satisfies a Faddeev equation of the form $V=K_{uds} V$, which may be written explicitly as follows: 
\begin{equation}
\label{Vfaddeev}
\left(
  \begin{array}{c}
   [ud]s\\[0.7ex]
   [us]d\\[0.7ex]
   [ds]u\\[0.7ex]
   \{us\}d\\[0.7ex]
   \{ds\}u
  \end{array}
  \right)
=
\left(
\begin{array}{ccccc}
0 & K_{[ud],[us]} & K_{[ud],[ds]} & K_{[ud],\{us\}} & K_{[ud],\{ds\}} \\
K_{[us],[ud]} & 0 & K_{[us],[ds]} & 0 & K_{[us],\{ds\}} \\
K_{[ds],[ud]} & K_{[ds],[us]} & 0 & K_{[ds],\{us\}} & 0 \\
K_{\{us\},[ud]} & 0 & K_{\{us\},[ds]}  & 0 & K_{\{us\},\{ds\}} \\
K_{\{ds\},[ud]} & K_{\{ds\},[us]} & 0 & K_{\{ds\},\{us\}} & 0
\end{array}
\right)
\left(
  \begin{array}{c}
   [ud]s\\[0.7ex]
   [us]d\\[0.7ex]
   [ds]u\\[0.7ex]
   \{us\}d\\[0.7ex]
   \{ds\}u
  \end{array}
  \right)\,,
\end{equation}
where, e.g. $K_{[ud],[us]}$ depicts the disintegration of a $[us]$ scalar diquark through emission of a dressed $u$-quark, which subsequently combines with the $d$-quark to form a $[ud]$ scalar diquark, leaving the $s$-quark as a bystander. In the kernel, the repeated flavour label always indicates the exchanged quark. 

For the convenience of later statements, I introduce a new notation to represent the flavour structure in Eq.\,\eqref{flavourdiq}:

{\allowdisplaybreaks
\begin{equation}
\label{flavourarrays}
\begin{array}{ccc}
{\tt t}^{1=[ud]} = \left[\begin{array}{ccc}
                    0 & 1 & 0 \\
                    -1 & 0 & 0 \\
                    0 & 0 & 0
                    \end{array}\right]\,,
&
{\tt t}^{2=[us]} = \left[\begin{array}{ccc}
                    0 & 0 & 1 \\
                    0 & 0 & 0 \\
                    -1 & 0 & 0
                    \end{array}\right]\,,
&
{\tt t}^{3=[ds]} = \left[\begin{array}{ccc}
                    0 & 0 & 0 \\
                    0 & 0 & 1 \\
                    0 & -1 & 0
                    \end{array}\right]\,,\\[6ex]
{\tt t}^{4=\{uu\}} = \left[\begin{array}{ccc}
                    \sqrt{2} & 0 & 0 \\
                    0 & 0 & 0 \\
                    0 & 0 & 0
                    \end{array}\right]\,,
&
{\tt t}^{5=\{ud\}} = \left[\begin{array}{ccc}
                    0 & 1 & 0 \\
                    1 & 0 & 0 \\
                    0 & 0 & 0
                    \end{array}\right]\,,
&
{\tt t}^{6=\{us\}} = \left[\begin{array}{ccc}
                    0 & 0 & 1 \\
                    0 & 0 & 0 \\
                    1 & 0 & 0
                    \end{array}\right]\,,\\[6ex]
{\tt t}^{7=\{dd\}} = \left[\begin{array}{ccc}
                    0 & 0 & 0 \\
                    0 & \sqrt{2} & 0 \\
                    0 & 0 & 0
                    \end{array}\right]\,,
&
{\tt t}^{8=\{ds\}} = \left[\begin{array}{ccc}
                    0 & 0 & 0 \\
                    0 & 0 & 1 \\
                    0 & 1 & 0
                    \end{array}\right]\,,
&
{\tt t}^{9=\{ss\}} = \left[\begin{array}{ccc}
                    0 & 0 & 0 \\
                    0 & 0 & 0 \\
                    0 & 0 & \sqrt{2}
                    \end{array}\right]\,.
\end{array}
\end{equation}
}

Using the notation above, one may write
\begin{equation}
\label{KmatrixFla}
K_{\text {uds }}=\left(\begin{array}{ccccc}
0 & K_{12} & K_{13} & K_{16} & K_{18} \\
K_{21} & 0 & K_{23} & 0 & K_{28} \\
K_{31} & K_{32} & 0 & K_{36} & 0 \\
K_{61} & 0 & K_{63} & 0 & K_{68} \\
K_{81} & K_{82} & 0 & K_{86} & 0
\end{array}\right)\,,
\end{equation}
where, \textit{e.g}., $K_{12}=K_{[ud],[us]}$.

Calculating the flavour factor (see Appendix\,\ref{CFcoeffi}) and using isospin symmetry, one arrives at 
\begin{equation}
\label{Kmatrixmxing}
K_{\text {uds }}=\left(\begin{array}{ccccc}
0 & \mathcal{K}_{12} & -\mathcal{K}_{13} & -\mathcal{K}_{16} & \mathcal{K}_{16} \\
\mathcal{K}_{21} & 0 & \mathcal{K}_{23} & 0 & \mathcal{K}_{28} \\
-\mathcal{K}_{21} & \mathcal{K}_{23} & 0 & \mathcal{K}_{28} & 0 \\
-\mathcal{K}_{61} & 0 & \mathcal{K}_{62} & 0 & \mathcal{K}_{68} \\
\mathcal{K}_{61} & \mathcal{K}_{62} & 0 & \mathcal{K}_{68} & 0
\end{array}\right)\,.
\end{equation}
Explicit forms of the entries here are presented in Appendix~\ref{FaddeevLambda}.

Owing to the presence of mixing in the $K_{uds}$ kernel, which obscures the $\Lambda$ and $\Sigma^0$ isospin-eigenstate baryons, I will proceed by utilising the matrix $O$ in Eq.\,\eqref{VOmatrices} to construct a non-mixing kernel:
\begin{equation}
\label{KmatrixLam}
\mathcal{K}_{u d s}=O K_{u d s} O^{\mathrm{T}}=\left(\begin{array}{ccccc}
0 & \sqrt{2} \mathcal{K}_{12} & 0 & -\sqrt{2} \mathcal{K}_{16} & 0 \\
\sqrt{2} \mathcal{K}_{21} & -\mathcal{K}_{23} & 0 & -\mathcal{K}_{28} & 0 \\
0 & 0 & \mathcal{K}_{23} & 0 & \mathcal{K}_{28} \\
-\sqrt{2} \mathcal{K}_{61} & -\mathcal{K}_{63} & 0 & -\mathcal{K}_{68} & 0 \\
0 & 0 & \mathcal{K}_{63} & 0 & \mathcal{K}_{68}
\end{array}\right)\,.
\end{equation}
According to Eq.\,\eqref{goodV}, rows $1$, $2$, $4$ map $I=0$ into itself, whereas rows $3$, $5$ do the same for $(I,I_z)=(1,0)$.  

Focusing on the $I = 0$ sector, one arrives at the following Faddeev equation for the $\Lambda$ baryon: 
$\tilde{V}_\Lambda = {\cal K}^\Lambda_{\,uds}
\tilde{V}_\Lambda$, \textit{i.e.}, explicitly,
\begin{equation}
\label{FaddeevLam}
\frac{1}{\sqrt{2}}\left(\begin{array}{c}
\sqrt{2} [u d] s \\
{[u s] d-[d s] u} \\
\{u s\} d-\{\{d s\} u
\end{array}\right)=\left(\begin{array}{ccc}
0 & \sqrt{2} \mathcal{K}_{12} & -\sqrt{2} \mathcal{K}_{16} \\
\sqrt{2} \mathcal{K}_{21} & -\mathcal{K}_{23} & -\mathcal{K}_{28} \\
-\sqrt{2} \mathcal{K}_{61} & -\mathcal{K}_{63} & -\mathcal{K}_{68}
\end{array}\right) \frac{1}{\sqrt{2}}\left(\begin{array}{c}
\sqrt{2} [u d] s \\
{[u s] d-[d s] u} \\
\{u s\} d-\{d s\} u
\end{array}\right)\,.
\end{equation}
Substituting Eqs.\,\eqref{FaddeevEqn} and \eqref{basisoctet} into Eq.\,\eqref{FaddeevLam}, the specific Faddeev equation of the $\Lambda$-baryon can be obtained:
\begin{equation}
\label{eigenequLam}
\left[\begin{array}{l}
s^1(P) \\
s^{[2,3]}(P) \\
a_1^{[6,8]}(P) \\
a_2^{[6,8]}(P)
\end{array}\right]=\left(\begin{array}{cccc}
0 & \sqrt{2} \mathcal{K}_{12}^{00} & -\sqrt{2} \mathcal{K}_{16_1}^{01} & -\sqrt{2} \mathcal{K}_{16_2}^{01} \\
\sqrt{2} \mathcal{K}_{21}^{00} & -\mathcal{K}_{23}^{00} & -\mathcal{K}_{28_1}^{01} & -\mathcal{K}_{28_2}^{01} \\
-\sqrt{2} \mathcal{K}_{6_1 1}^{10} & -\mathcal{K}_{6_1 3}^{10} & -\mathcal{K}_{6_1 8_1}^{11} & -\mathcal{K}_{6_1 8_2}^{11} \\
-\sqrt{2} \mathcal{K}_{6_2 1}^{10} & -\mathcal{K}_{6_2 3}^{10} & -\mathcal{K}_{6_2 8_1}^{11} & -\mathcal{K}_{6_2 8_2}^{11}
\end{array}\right)\left[\begin{array}{l}
s^1(P) \\
s^{[2,3]}(P) \\
a_1^{[6,8]}(P) \\
a_2^{[6,8]}(P)
\end{array}\right]\,.
\end{equation}
Again, compact algebraic expressions for each of the entries in the kernel matrix are listed in Appendix~\ref{FaddeevLambda}. 

Equation \eqref{eigenequLam} is an algebraic eigenvalue problem, the solution to which yields the mass of the dressed-quark-core of the $\Lambda$-resonance and the associated Faddeev amplitude.  The specific Faddeev equations of other baryons can be obtained in a similar way.  They are all algebraic and this enables, \textit{e.g}., a clear understanding of the dynamical character of quark exchange in baryon bound states.

All the elements necessary to construct the baryon Faddeev kernels are now in hand. The value of $\Lambda_{uv}$ in each Faddeev equation is selected as the scale linked to the lightest diquark in the bound state, because this is always the smallest value and, hence, the dominant regularising influence. 

Solving the Faddeev equations, one obtains the masses and amplitudes listed in Table~\ref{SolveFaddeev}. 
The row labels in the table correspond to those identified in Eqs.\,\eqref{halfnucleon}. 
Regarding the masses, I note that the values are deliberately $0.20(2)\,$GeV above experiment \cite{ParticleDataGroup:2020ssz} because Fig.\,\ref{figFaddeev} describes the \emph{dressed-quark core} of each baryon. 
To obtain a complete baryon, resonant contributions should be incorporated into the Faddeev kernel. 
Such ``meson cloud''  effects are known to lower the mass of octet baryons by $\approx 0.2$\,GeV \cite{Burkert:2017djo, Liu:2022ndb, Chen:2017pse, Liu:2022nku, Hecht:2002ej, 
Sanchis-Alepuz:2014wea}. 
(Similar effects are reported in quark models \cite{Garcia-Tecocoatzi:2016rcj, Chen:2017mug}.)  Their impact on baryon structure can be estimated using dynamical coupled-channels models \cite{Aznauryan:2012ba, Julia-Diaz:2007qtz, Suzuki:2009nj, Ronchen:2012eg, Kamano:2013iva}, but that is beyond the scope of the present Faddeev equation analyses. 

\begin{table}[t]
\caption{\label{SolveFaddeev} Masses and unit normalised Faddeev amplitudes obtained by solving the octet baryon Faddeev equations defined by Fig.\,\ref{figFaddeev}.The row label superscript refers to Eqs.\,\eqref{halfnucleon}: for the $\Lambda$-baryon, $r_2$ is a scalar diquark combination; otherwise, it is axial-vector. Canonically normalised amplitudes, explained in connection with Eq.\,\eqref{CanonicalFA}, are obtained by dividing the amplitude entries in each row by the following numbers: ${\mathpzc n}_c^{p,n}=0.157$, ${\mathpzc n}_c^{\Lambda}=0.177$, ${\mathpzc n}_c^{\Sigma}=0.190$, ${\mathpzc n}_c^{\Xi}=0.201$.
(Masses listed in GeV. Amplitudes are dimensionless. Recall that isospin-symmetry is assumed.)}
\begin{center}
\begin{tabular*}
{\hsize}
{
c@{\extracolsep{0ptplus1fil}}
|c@{\extracolsep{0ptplus1fil}}
c@{\extracolsep{0ptplus1fil}}
c@{\extracolsep{0ptplus1fil}}
c@{\extracolsep{0ptplus1fil}}
c@{\extracolsep{0ptplus1fil}}
c@{\extracolsep{0ptplus1fil}}
c@{\extracolsep{0ptplus1fil}}}\hline
  & mass
  & $s^{r_{1}}$ & $s^{r_{2}}$ & $a_{1}^{r_{2}}$ & $a_{2}^{r_{2}}$ & $a_{1}^{r_{3}}$ & $a_{2}^{r_{3}}$\\\hline
$p\ $ & $1.15$ & $\phantom{-}0.88$ & & $-0.38$ & $-0.063$ & $\phantom{-}0.27$ & $\phantom{-}0.044\ $  \\
$n\ $ & $1.15$ & $\phantom{-}0.88$ & & $\phantom{-}0.38$ & $\phantom{-}0.063$ & $-0.27$ & $-0.044\ $  \\
$\Lambda\ $ & $1.33$ & $\phantom{-}0.66$ & $0.62$ &  & & $-0.41$ & $-0.084\ $  \\
$\Sigma\ $ & $1.38$ & $\phantom{-}0.85$ & &$-0.46$  & $\phantom{-}0.15\phantom{3}$  & $\phantom{-}0.22$ & $\phantom{-}0.041\ $\\
$\Xi\ $ & $1.50$ & $\phantom{-}0.91$ & &$-0.29$  & $\phantom{-}0.021$  & $\phantom{-}0.29$ & $\phantom{-}0.052\ $
\\\hline
\end{tabular*}
\end{center}
\end{table}

Notwithstanding these considerations, the quark+diquark picture of baryon structure yields a $\Sigma-\Lambda$ mass splitting that aligns well with experiment. This is because the $\Lambda$ is primarily a scalar diquark system, whereas the $\Sigma$ possesses more axial-vector strength: scalar diquarks are lighter than axial-vector diquarks.

The Faddeev amplitudes in Table~\ref{SolveFaddeev} are unit normalised. When calculating observables, one must use the canonically normalised amplitude, which is defined via the baryon's Dirac form factor in elastic electromagnetic scattering, $F_1(Q^2=0)$.  To wit, for a baryon $B$, with $n_u$ $u$ valence-quarks, $n_d$ $d$ valence-quarks and $n_s$ $s$ valence-quarks, one decomposes the Dirac form factor as follows:
\begin{equation}
\label{CanonicalFA} 
F_1^B(Q^2=0) = n_u e_u F_1^{Bu}(0) + n_d e_d F_1^{Bd}(0)+n_s e_s F_1^{Bs}(0)\,,  
\end{equation}
where $e_{u,d,s}$ represent the quark electric charges, expressed in units of the positron charge. It is subsequently straightforward to calculate the single constant factor that, when used to rescale the unit-normalised Faddeev amplitude for $B$, ensures $F_1^{Bu}(0)=1=F_1^{Bd}(0)=F_1^{Bs}(0)$. So long as a symmetry-preserving treatment is employed for the elastic scattering problem, this single factor guarantees that all three flavour-separated electromagnetic form factors are unity at $Q^2=0$. Detailed examples are provided elsewhere \cite{Wilson:2011aa}.  
\chapter{Octet baryon axialvector and pseudoscalar form factors}
\label{chapter3}
\section{Introduction}\label{chap3secintro}
The proton, $p$, is the only stable hadron and the best known bound state in the baryon octet. Except for protons, all the other octet baryons decay. In many respects, these baryons' semileptonic decays are theoretically the easiest to understand because only one strongly interacting particle is involved in the initial and final states. The neutron, $n$, $\beta$-decay: $n \to p  e^- \bar\nu_e$ is the archetypal process, the study of which has a long history \cite{Pauli:1930pc, Fermi:1934hr}.  Despite that, related decays of hyperons have also attracted much attention \cite{Gaillard:1984ny, Cabibbo:2003cu}, in part because they present an opportunity to shed light on the Cabibbo-Kobayashi-Maskawa (CKM) matrix element $|V_{us}|$ and thereby complement that provided by $K_{\ell 3}$ decays (see Sec. 12.2.2 in Ref.\,\cite{ParticleDataGroup:2020ssz}). It is also of great importance to understand the octet baryons' axial-vector form factors, since they provide insights into the pattern of flavour SU(3) symmetry breaking.

Within the Standard Model, the semileptonic decay $B \to B^\prime \ell^- \nu_\ell$, where $B$ and $B^\prime$ are the initial and final octet baryons and $\ell$ denotes a lepton, involves a valence-quark $g$ in $B$ transforming into a valence-quark $f$ in $B^\prime$. The associated axial-vector transition matrix element are determined by two Poincar\'e-invariant form factors:
\begin{subequations}
\label{jaxdq0}
\begin{align}
\label{jaxdq}
& J^{B^\prime B}_{5\mu}(K,Q)
:= \langle B^\prime(P^\prime)|{\mathcal{A}}^{fg}_{5\mu}(0)|B(P)\rangle \\
\label{jaxdqb}
& =\bar{u}_{B^\prime}(P^\prime) \gamma_5  \bigg[ \gamma_\mu G_A^{B^\prime B}(Q^2) +\frac{iQ_\mu}{2M_{B^\prime B}}G_P^{B^\prime B}(Q^2) \bigg]\,u_B(P)\,. 
\end{align}
\end{subequations}
Here
$G_A^{B^\prime B}(Q^2)$ is the axial form factor and $G_P^{B^\prime B}(Q^2)$ is the induced pseudoscalar form factor; $P$ and $P^\prime$ are, respectively, the momenta of the initial- and final-state baryons, defined such that the on-shell conditions are fulfilled, $P^{(\prime)}\cdot P^{(\prime)}=-m_{B,B^\prime}^2$, with $m_{B,B^\prime}$ being the baryon masses -- again, I work with the Euclidean metric conventions explained in Appendix~\ref{appEuclid});
$M_{B^\prime B} = (m_{B^\prime}+m_B)/2$;
and $u_{B,B^\prime}(P)$ are the associated Euclidean spinors.  (I have suppressed the spin label. See Appendix\,B in Ref.\,\cite{Segovia:2014aza} for details.) Furthermore, $K=(P+P^\prime)/2$ is the average momentum of the system and $Q=P^\prime-P$ is the transferred momentum between initial and final states:
\begin{subequations}
\begin{align}
 -K^2  = \tfrac{1}{2}(m_{B^\prime}^2+m_B^2) + \tfrac{1}{4}Q^2 & =: \tfrac{1}{2}\Sigma_{B^\prime B}+\tfrac{1}{4}Q^2 \,, \\
 -K\cdot Q = \tfrac{1}{2}(m_{B^\prime}^2-m_B^2) & =: \tfrac{1}{2}\Delta_{B^\prime B}\,.
\end{align}
\end{subequations}

Once more, I work in the isospin symmetry limit $m_u=m_d=:m_l$, \emph{i.e}., assume degenerate light-quarks, and treat the $s$ valence quark as approximately twenty-times more massive \cite{ParticleDataGroup:2020ssz}, \emph{viz}.\ $m_s \approx 20\, m_l$.
The overall flavour structure is described by the Gell-Mann matrices $\{\lambda^j|j=1,\ldots,8\}$ so that the flavour-nonsinglet axial current operator can be written
\begin{equation}
\label{jaxx}
{\mathcal{A} }_{5\mu}^{fg}(x) = \bar{\mathpzc q}(x) {\cal T}^{fg} \gamma_5 \gamma_\mu {\mathpzc q}(x)\,,
\end{equation}
where ${\mathpzc q} = {\rm column}[u,d,s]$
and ${\cal T}^{fg}$ is the valence-quark flavour transition matrix.  Hence, for example, the $s\to u$ transition is described by ${\cal T}^{us} = (\lambda^4+i\lambda^5)/2$.

A related form factor, $G_5^{B^\prime B}(Q^2)$, is associated with a kindred pseudoscalar current
\begin{subequations}
\label{jpsdq0}
\begin{align}
\label{jpsdq}
 J^{B^\prime B}_{5}(K,Q) &:=
\langle B^\prime(P^\prime)|{\mathcal{P}}^{fg}_5(0)|B(P)\rangle \\
&=\bar{u}_{B^\prime}(P^\prime) 
\gamma_5\,G_5(Q^2)\,u_B(P)\,,
\label{G5FF}
\end{align}
\end{subequations}
where ${\mathcal{P}}^{fg}_5(x) = \bar{\mathpzc q}(x){\cal T}^{fg}\gamma_5 {\mathpzc q}(x)$ is the flavour-nonsinglet pseudoscalar current operator. This form factor is important because, amongst other things, owing to DCSB, a corollary of EHM \cite{Roberts:2020udq, Roberts:2020hiw, Roberts:2021xnz, Roberts:2021nhw, Binosi:2022djx, Papavassiliou:2022wrb}, one has a partial conservation of the axial current (PCAC) relation for each baryon transition $(2 \mathpzc{m}_{fg}=m_f+m_g)$:
\begin{subequations}
\label{PCAC}
\begin{align}
& 0 = Q_\mu J^{B^\prime B}_{5\mu}(K,Q) + 2i {\mathpzc m}_{f g}  J^{B^\prime B}_{5}(K,Q) \\
\Rightarrow &\;
G_A^{B^\prime B}(Q^2) - \frac{Q^2}{4 M_{B^\prime B}^2} G_P^{B^\prime B}(Q^2) = \frac{{\mathpzc m}_{f g}}{M_{B^\prime B}} G_5^{B^\prime B}(Q^2) \,.
\end{align}
\end{subequations}
Note that the product ${\mathpzc m}_{fg} G_5^{B^\prime B}(Q^2)$ is renormalisation point invariant; neither of these two factors alone possesses this property.

It should be pointed out that PCAC is an operator relation and thus the identities in Eqs.\,\eqref{PCAC} are satisfied for all $Q^2$. They state that the longitudinal component  of the axial-vector current is fully determined by the related pseudoscalar form factor and its intensity is modulated by the ratio of the sum of current-quark masses participating in the transition to the sum of the masses of the involved baryons. The former are determined by Higgs boson couplings into QCD, whereas the latter are largely determined by the scale of EHM.  Therefore, this $Q$-divergence serves as a measure of the interplay between nature's two known mass-generating mechanisms.

Focusing on the case of neutron $\beta$ decay, Eqs.\,\eqref{PCAC} lead to the well-known Gold\-berger-Treiman relation and establish the validity of the pion pole dominance approximation for $G_P^{pn}$. Considering instead a prominent hyperon decay, \emph{e.g}., $\Lambda \to p e^- \bar\nu_e$, one recognises that $G_5^{p \Lambda}$ exhibits a pole at the mass of the charged kaon, \emph{i.e}., when $Q^2+m_K^2=0$. $G_A^{p \Lambda}$ is regular in the neighbourhood of $m_K^2$ because it is tied to the transverse part of the axial current. Then $G_P^{p \Lambda}$ also has a pole at $m_K$. Further, one can define a $Kp\Lambda$ form factor as follows:
\begin{equation}
G_5^{p \Lambda}(Q^2) =: \frac{m_K^2}{Q^2+m_K^2} \frac{2 f_K}{m_u+m_s}G_{Kp\Lambda}(Q^2)\,,
\label{CouplingKpLambda}
\end{equation}
where $f_K$ is the kaon leptonic decay constant, then Eqs.\,\eqref{PCAC} entail
\begin{equation}
G_A^{p\Lambda}(0) = \frac{2 f_K}{m_p+m_\Lambda}G_{Kp\Lambda}(0)\,,
\end{equation}
providing an estimate of the $Kp\Lambda$ coupling in terms of the $\Lambda\to p$ transition's axial-vector form factor at the maximum recoil point. As will become apparent, this relation holds true with an accuracy better than 1\%.

The evidently diverse physics relevance of octet baryon axial-vector transitions emphasises the importance of calculating the associated form factors. However, despite their being some of the simplest baryonic processes, this does not imply that their calculation is simple.  Studies of meson semileptonic transitions \cite{Xu:2021iwv, Chen:2012txa, Xing:2022sor, Yao:2021pyf, Yao:2021pdy} have shown that delivering predictions for the required processes demands dependable calculations of the Poincar\'e-covariant hadron wave functions and the related axial-vector interaction currents and careful symmetry-preserving treatments of the involved matrix elements.

Considering their significance in understanding modern neutrino experiments \cite{Mosel:2016cwa, NuSTEC:2017hzk, Hill:2017wgb, Novario:2020dmr, Lovato:2020kba}, weak interactions and parity violation experiments, the nucleon axial and pseudoscalar form factors have recently attracted a lot of attention, in studies using continuum and lattice methods, \emph{e.g}., Refs.\,\cite{Anikin:2016teg, Chen:2020wuq, Chen:2021guo, Chen:2022odn, Alexandrou:2017hac, Jang:2019vkm}.
Regarding hyperon semileptonic decays, analyses using an array of tools may be found, \emph{e.g}.,
in Refs.\,\cite{Faessler:2008ix, Ledwig:2014rfa, Yang:2015era, Ramalho:2015jem, Yang:2018idi, Qi:2022sus, Erkol:2009ev, Green:2017keo, Bali:2022qja}.
Here, I use CSMs \cite{Eichmann:2016yit, Burkert:2017djo, Qin:2020rad, Roberts:2020udq, Roberts:2020hiw, Roberts:2021xnz, Roberts:2021nhw, Binosi:2022djx, Papavassiliou:2022wrb, Ding:2022ows, Ferreira:2023fva} to extend this body of work on octet baryon axial-vector transitions.
Namely, I construct approximations to the transition matrix elements based on solutions of a symmetry-preserving collection of integral equations for the relevant $n$-point Schwinger functions, $n=2-6$. This approach has become feasible owing to the recent development of a realistic axial current for baryons \cite{Chen:2020wuq, Chen:2021guo}.

In Refs.\,\cite{Chen:2020wuq, Chen:2021guo, Chen:2022odn}, the so-called QCD-kindred framework was used  to compute all form factors associated with the nucleon axial and pseudoscalar currents. 
A straightforward thought is that one could extend it to hyperons. 
However, that would require significant effort. 
An expeditious alternative is to simplify the analysis by using the SCI, introduced in Refs.\,\cite{Gutierrez-Guerrero:2010waf, Roberts:2011wy, Chen:2012qr} and detailed above. 
As already remarked, this approach ensures algebraic simplicity; moreover, very importantly, it provides for the parameter-free unification of octet baryon axial-vector transitions with an array of other baryon properties \cite{Yin:2021uom, Wilson:2011aa, Raya:2021pyr, Xu:2015kta} and studies of the semileptonic decays of pseudoscalar mesons \cite{Xu:2021iwv, Chen:2012txa, Xing:2022sor}. By adopting this approach, one benefits from numerous studies \cite{Yin:2019bxe, Yin:2021uom, Raya:2021pyr, Gutierrez-Guerrero:2021rsx, Xu:2015kta, Wang:2013wk, Bedolla:2015mpa, Bedolla:2016yxq, Serna:2017nlr, Raya:2017ggu, Zhang:2020ecj, Lu:2021sgg} which have shown that SCI predictions provide a valuable quantitative guide when interpreted judiciously. In fact, SCI results often provide both a useful first estimate of an observable and a means of checking
the validity of algorithms used in calculations that rely (heavily) upon high-performance computing.

In Sec.\,\ref{SecTwo}, I sketch the current which ensures preservation of all PCAC identities when the baryons involved are described by the Faddeev equation illustrated in Fig.\,\ref{figFaddeev}.
The sketch is complemented by Appendix~\ref{AppendixCurrents}, which is an extensive description of results for elements that appear in the currents but were not mentioned in the previous chapter.
Using that information, Sec.\,\ref{SecThree} presents and analyses SCI predictions for the axial, induced pseudoscalar, and pseudoscalar transition form factors of the octet baryon.
This is followed by a discussion of the flavour separation of octet baryon axial charges and their connection to the fraction of baryon spin carried by valence degrees of freedom in Sec.\,\ref{SecFour}. Section~\ref{epilogue} provides a summary.

\section{Baryons' axial current}
\label{SecTwo}
\begin{figure}[!t]
  \centering
  \includegraphics[width=3.8in]{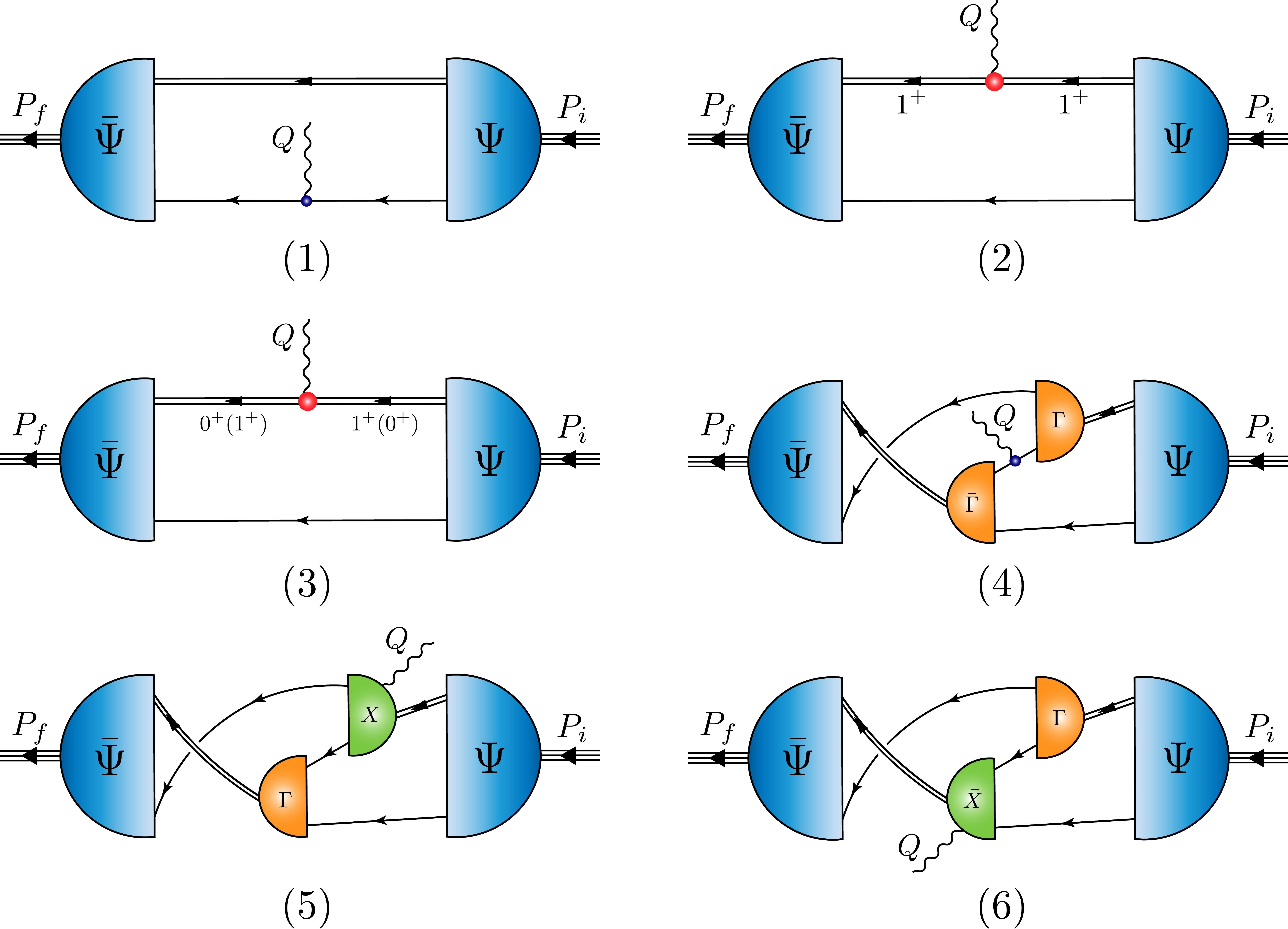}
  \caption{\label{figcurrent} Currents that ensure PCAC for on-shell baryons which are described by the Faddeev amplitudes produced by the equation depicted in Fig.\,\ref{figFaddeev}: \emph{single line}, dressed-quark propagator; \emph{undulating line}, axial or pseudoscalar current; $\Gamma$,  diquark correlation amplitude; \emph{double line}, diquark propagator; and $\chi$, seagull terms.
  A legend is provided in Table~\ref{DiagramLegend} with details in Appendix\,\ref{AppendixCurrentsBaryon}.}
\end{figure}

My analyses of octet baryon axial-vector transition form factors are based on solutions of the Poincar\'e-covariant Faddeev equation depicted in Fig.\,\ref{figFaddeev}. When inserted into the diagrams drawn in Fig.\,\ref{figcurrent}, these solutions deliver a result for the current in Eq.\,\eqref{jaxdq0}, which ensures Eqs.\,\eqref{PCAC}, and all their corollaries for each transition. Details can be found in Refs.\,\cite{Chen:2020wuq, Chen:2021guo}. For subsequent reference, Table~\ref{DiagramLegend} provides a useful separation of the current in Fig.\,\ref{figcurrent}.

\begin{table}[b!]
\caption{\label{DiagramLegend}
Enumeration of terms in the current drawn in Fig.\,\ref{figcurrent}.}
\begin{enumerate}
\item Diagram~1, two distinct terms: $\langle J \rangle^{S}_{\rm q}$ -- probe strikes dressed-quark with scalar diquark spectator; and $\langle J \rangle^{A}_{\rm q}$ -- probe strikes dressed-quark with axial-vector diquark spectator.
\item Diagram~2: $\langle J \rangle^{AA}_{\rm qq}$ -- probe strikes axial-vector diquark with dressed-quark spectator.
\item Diagram~3: $\langle J \rangle^{\{SA\}}_{\rm qq}$ -- probe mediates transition between scalar and axial-vector diquarks, with dressed-quark spectator.
\item Diagram~4, three terms:
    $\langle J \rangle_{\rm ex}^{SS}$ -- probe strikes dressed-quark ``in-flight'' between one scalar diquark correlation and another;
    $\langle J \rangle_{\rm ex}^{\{SA\}}$ -- dressed-quark ``in-flight'' between a scalar diquark correlation and an axial-vector correlation;
    and $\langle J \rangle_{\rm ex}^{AA}$ -- dressed-quark ``in-flight'' between one axial-vector correlation and another.
\item Diagrams~5 and 6 -- seagull diagrams describing the probe coupling into the diquark correlation amplitudes: $\langle J\rangle_{\rm sg}$.  There is one contribution from each diagram to match every term in Diagram 4.
\end{enumerate}
\end{table}

The first step in this analysis of octet baryon transitions is the SCI calculation of every line, amplitude and vertex in Figs.\,\ref{figFaddeev} and \ref{figcurrent}. 
The calculations of dressed quark propagators, diquark correlation amplitudes and Faddeev amplitudes were described in Chapter\,\ref{chapter2}. 
The remaining calculations are detailed in Appendix~\ref{AppendixCurrentsDiquark}. 
By combining the results and employing appropriately selected projection operators, predictions for the baryon axial and pseudoscalar form factors in Eqs.\,\eqref{jaxdqb} and \eqref{G5FF} are readily obtained. 
It is worth noting that Eq.\,\eqref{jaxdq} entails that $G_A^{B^\prime B}$ is entirely determined by the $Q$-transverse part of the baryon axial current \cite{Chen:2021guo}.

\section{Calculated form factors}
\label{SecThree}

\subsection{Axial form factors}
\label{GAsection}
In the isospin-symmetry limit, there are six distinct charged current semileptonic transitions between octet baryons. 
My predictions for the corresponding $G_A(Q^2=0)$ values are recorded in Table~\ref{GAzero}.  In the Cabibbo model of such transitions, which assumes SU$(3)$-flavour symmetry, the couplings in Table~\ref{GAzero} are described by only two distinct parameters (see Table 1 in Ref. \cite{Cabibbo:2003cu}): $D$ and $F$. In these terms, one finds
\begin{equation}
\label{DFvalues}
D = 0.78\,, \; F = 0.43\,,\; F/D = 0.56\,,
\end{equation}
via a least-squares fit to the SCI results, with a mean absolute relative error between SCI results and Cabibbo fit of just 3(2)\%. 
Clearly, the SCI predicts that the violation of SU$(3)$ symmetry in these transitions is small, which confirms the conclusion of many studies. 
This is also shown in the comparison between $n\to p$ and $\Xi^0\to \Sigma^+$. 
The former corresponds to a $d\to u$ transition, and the latter to a $s\to u$ transition; yet, in the Cabibbo model, $G_A^{\Sigma^+\Xi^0}(0)=G_A^{pn}(0)$, and this identity holds true with an accuracy of 4\% in the SCI calculation. 
Similarly, it is observed in experimental results.

It is valuable to provide supplementary context for the results in Eq.\,\eqref{DFvalues}. So I  note that a covariant baryon chiral perturbation theory analysis of semileptonic hyperon decays yields $D=0.80(1)$, $F=0.47(1)$ and $F/D=0.59(1)$ \cite{Ledwig:2014rfa};
and a three-degenerate-flavour lattice QCD (lQCD) computation yields $F/D = 0.61(1)$ \cite{Bali:2022qja}.

\begin{table}[t]
\caption{\label{GAzero}
SCI predictions for $g_A^{B^\prime B}=G_A^{B^\prime B}(Q^2=0)$ compared with experiment \cite{ParticleDataGroup:2020ssz} and other calculations:
Lorentz covariant quark model \cite{Faessler:2008ix};
covariant baryon chiral perturbation theory \cite{Ledwig:2014rfa};
and a lQCD study \cite{Erkol:2009ev}, which used large pion masses ($m_\pi = 0.55\,$--$\,1.15\,$GeV). Quoted error estimates are primarily statistical.
%
}
\begin{center}
\begin{tabular*}
{\hsize}
{
l@{\extracolsep{0ptplus1fil}}
|c@{\extracolsep{0ptplus1fil}}
|c@{\extracolsep{0ptplus1fil}}
|c@{\extracolsep{0ptplus1fil}}
|c@{\extracolsep{0ptplus1fil}}
|c@{\extracolsep{0ptplus1fil}}
|c@{\extracolsep{0ptplus1fil}}}\hline
%
& $n\to p\ $ & $\Sigma^- \to \Lambda\ $ & $\Lambda \to p\ $ & $\Sigma^- \to n\ $ & $\Xi^0 \to \Sigma^+\ $ & $\Xi^- \to \Lambda\ $ \\\hline
SCI\; & $1.24\phantom{(3)}\ $ & $0.66\phantom{(3)}\ $ & $-0.82\phantom{(2)}\ $ & $0.34\phantom{(2)}\ $ & $1.19\phantom{(5)}\ $ & $0.23\phantom{(6)}\ $\\   \hline
\cite{ParticleDataGroup:2020ssz}\; & $1.28\phantom{(3)}\ $ & $0.57(3)\ $ & $-0.88(2)\ $ & $0.34(2)\ $ & $1.22(5)\ $ & $0.31(6)\ $\\
\cite{Faessler:2008ix} & $1.27\phantom{(3)}\ $ & $0.63\phantom{(2)}\ $ & $-0.89\phantom{(2)}\ $ & $0.26\phantom{(2)}\ $ & $1.25\phantom{(4)}\ $ & $0.33\phantom{(4)}\ $\\
\cite{Ledwig:2014rfa} & $1.27\phantom{(3)}\ $ & $0.60(2)\ $ & $-0.88(2)\ $ & $0.33(2)\ $ & $1.22(4)\ $ & $0.21(4)\ $\\
\cite{Erkol:2009ev} & $1.31(2)\ $ & $0.66(1)\ $ & $-0.95(2)\ $ & $0.34(1)\ $ & $1.28(3)\ $ & $0.27(1)\ $\\
\hline
\end{tabular*}
\end{center}
\end{table}

When considering the empirical fact of approximate SU$(3)$-flavour symmetry in the values of octet baryon axial transition charges, one should note that it does not result directly from any basic symmetry. Hence, the apparent near symmetry is actually a dynamical outcome.
The underlying source of any SU$(3)$-flavour symmetry breaking is the Higgs-boson-generated splitting between the current masses of the $s$ and $l=u,d$ valence quarks. However, as mentioned earlier, $m_s/m_l \approx 20$. Therefore, there must be something strongly suppressing the expression of this difference in observable measurements.

The responsible agent is EHM \cite{Roberts:2020udq, Roberts:2020hiw, Roberts:2021xnz, Roberts:2021nhw, Binosi:2022djx, Papavassiliou:2022wrb}. For instance, leptonic weak decays of pseudoscalar mesons proceed via the axial current and $f_K/f_\pi \approx 1.2$. These decay constants serve as order parameters for chiral symmetry breaking, with this effect being primarily dynamical for Nature's three lighter quarks (see Fig.\,2.5 in Ref.\cite{Roberts:2021nhw}). Similarly, one finds SU$(3)$-flavour symmetry breaking on the order of 10\% \cite{Xing:2022sor} in the axial form factors for semileptonic decays of heavy+light pseudoscalar mesons to light vector meson final states. Finally, comparing the hadron-scale valence-quark distribution functions of the kaon and pion, one learns that the $u$ quark carries 6\% less of the kaon's light-front momentum than does the $u$-quark in the pion \cite{Cui:2020dlm, Cui:2020tdf}.

Focusing on the case in this thesis, \emph{i.e}., octet baryon semileptonic transitions, $m_s/m_l \approx 20$ leads to a dressed-quark mass ratio $M_s/M_l \approx 1.4$ -- Table~\ref{Tab:DressedQuarks}; namely, a huge suppression caused by EHM. In turn, this leads to a $\sim 14$\% difference in diquark masses, smaller differences in diquark correlation amplitudes, and, consequently, differences of even smaller magnitude ($\sim 3$\%) between the leading scalar-diquark components of the Faddeev amplitudes of the baryons involved.
In addition, Tables~\ref{interpolatorcoefficientsdu} and \ref{interpolatorcoefficientsus} reveal that the $s\to u$ and $d\to u$ quark-level weak transitions are similar in strength. This is not surprising, since these axial vertices are obtained by solving BSEs which are similar to those that yield the diquark correlation amplitudes. 
Finally, therefore, regarding the $n\to p$\,:\,$\Xi^0\to \Sigma^+$ comparison, \emph{e.g}., Table~\ref{GA0diagrams} reveals that the transition 
is dominated by the scalar diquark components; hence, these transitions should have similar strengths.

\begin{table}[t]
\caption{\label{GA0diagrams}
Diagram separation of octet baryon axial transition charges, presented as a fraction of the total listed in Table~\ref{GAzero}\,--\,Row~1 and made with reference to Fig.\,\ref{figcurrent}.
}
\begin{center}
\begin{tabular*}
{\hsize}
{
l@{\extracolsep{0ptplus1fil}}
|l@{\extracolsep{0ptplus1fil}}
l@{\extracolsep{0ptplus1fil}}
l@{\extracolsep{0ptplus1fil}}
l@{\extracolsep{0ptplus1fil}}
l@{\extracolsep{0ptplus1fil}}
l@{\extracolsep{0ptplus1fil}}
l@{\extracolsep{0ptplus1fil}}}\hline
 & $\langle J \rangle^{S}_{\rm q}$  & $\langle J \rangle^{A}_{\rm q}$ &$\langle J \rangle^{AA}_{\rm qq}$ & $\langle J \rangle^{\{SA\}}_{\rm qq}$
 & $\langle J \rangle_{\rm ex}^{SS}$
 & $\langle J \rangle_{\rm ex}^{\{SA\}}$
 & $\langle J \rangle_{\rm ex}^{AA}$  \\\hline
$g_A^{pn}$ & $0.29\ $ & $\phantom{-}0.013\ $ & $\phantom{-}0.072\ $ & $0.35\ $ & $\phantom{-}0.19\ $ & $\phantom{-}0.051\ $& $0.028\ $\\
$g_A^{\Lambda \Sigma^-}$ & $0.27\ $ & $\phantom{-}0.016\ $ & $\phantom{-}0.023\ $ & $0.42\ $ & $\phantom{-}0.28\ $ & $-0.008\ $ & \\
$g_A^{p \Lambda}$ & $0.45\ $ & & $\phantom{-}0.083\ $ & $0.33\ $ & $\phantom{-}0.082\ $ & $\phantom{-}0.044\ $& $0.013\ $\\
$g_A^{n\Sigma^-}$ & & $\phantom{-}0.13\ $ & $-0.051\ $ & $0.57\ $ & $\phantom{-}0.42\ $ & $-0.076\ $& $0.008\ $\\
$g_A^{\Sigma^+ \Xi^0}$ & $0.41\ $ & $\phantom{-}0.011\ $ & $\phantom{-}0.064\ $ & $0.36\ $ & $\phantom{-}0.12\ $ & $\phantom{-}0.020\ $& $0.013\ $\\
$g_A^{\Lambda \Xi^-}$ & $1.02\ $ & $-0.072\ $ & $\phantom{-}0.12\ $ & $0.12\ $ & $-0.28\ $ & $\phantom{-}0.023\ $ & $0.076\ $\\
\hline
\end{tabular*}
\end{center}
\end{table}

Table~\ref{GA0diagrams} highlights a curious aspect of the quark+diquark picture; namely, the $s\to u$ quark transition $\Sigma^-\to n$ does not receive a contribution from Diagram~1 in Fig.\,\ref{figcurrent} because the only scalar diquark component in $\Sigma^-$ is $d[ds]$ and the neutron does not contain a $[ds]$ diquark. Nevertheless, scalar diquarks continue to be the dominant contributors to $g_A^{n\Sigma^-}$ through Diagrams~3 and 4. Additionally, it is worth recalling that axial form factors derive solely from $Q$-transverse pieces of the baryon current \cite{Chen:2021guo}, resulting in no seagull contributions to $G_A^{B^\prime B}$.

Despite the dominance of scalar diquark components, Table~\ref{GA0diagrams} indicates that axial-vector correlations also play a material role in the transitions.  For example, $\langle J \rangle_{\rm qq}^{\{SA\}}$ is large in all cases but would vanish if axial-vector diquarks were ignored in forming the picture of baryon structure. Their influence is further emphasised below.

\begin{table}[thb]
\caption{\label{InterpolationsGA}
Interpolation parameters for octet baryon axial transition form factors, Eq.\,\eqref{SCIinterpolationsGA}.
(Every form factor is dimensionless; so each coefficient in Eq.\,\eqref{SCIinterpolationsGA} has the mass dimension necessary to cancel that of the associated $s ({\rm GeV}^2)$ factor.)}
\begin{center}
\begin{tabular*}
{\hsize}
{
l@{\extracolsep{0ptplus1fil}}
|c@{\extracolsep{0ptplus1fil}}
c@{\extracolsep{0ptplus1fil}}
c@{\extracolsep{0ptplus1fil}}
c@{\extracolsep{0ptplus1fil}}
c@{\extracolsep{0ptplus1fil}}}\hline
 & $g_0\ $ & $g_1\ $ & $g_2\ $ & $l_1\ $ & $l_2\ $ \\\hline
$\phantom{-}G_A^{pn}\ $ & $1.24\ $ & $1.97\ $ & $\phantom{-}0.29\phantom{0}\ $ & $2.44\ $ & $1.12\ $ \\
$\phantom{-}G_A^{\Lambda \Sigma^- }\ $ & $0.66\ $ & $1.19\ $ & $\phantom{-}0.16\phantom{0}\ $ & $2.73\ $ & $1.48\ $ \\
$-G_A^{p \Lambda}\ $ & $0.82\ $ & $1.00\ $ & $\phantom{-}0.074\ $ & $1.80\ $ & $0.68\ $ \\
$\phantom{-}G_A^{n\Sigma^- }\ $ & $0.34\ $ & $0.43\ $ & $\phantom{-}0.093\ $ & $1.86\ $ & $0.75\ $ \\
$\phantom{-}G_A^{\Sigma^+\Xi^0 }\ $ & $1.19\ $ & $3.28\ $ & $\phantom{-}0.33\phantom{0}\ $ & $3.35\ $ & $1.82\ $ \\
$\phantom{-}G_A^{\Lambda\Xi^- }\ $ & $0.23\ $ & $0.90\ $ & $-0.011\ $ & $4.42\ $ & $2.14\ $ \\
\hline
\end{tabular*}
\end{center}
\end{table}

\begin{figure}[htbp]
\vspace*{2ex}

\leftline{\hspace*{4.5em}{\large{\textsf{A}}}}
\vspace*{-4ex}
\centerline{\includegraphics[width=3.4in]{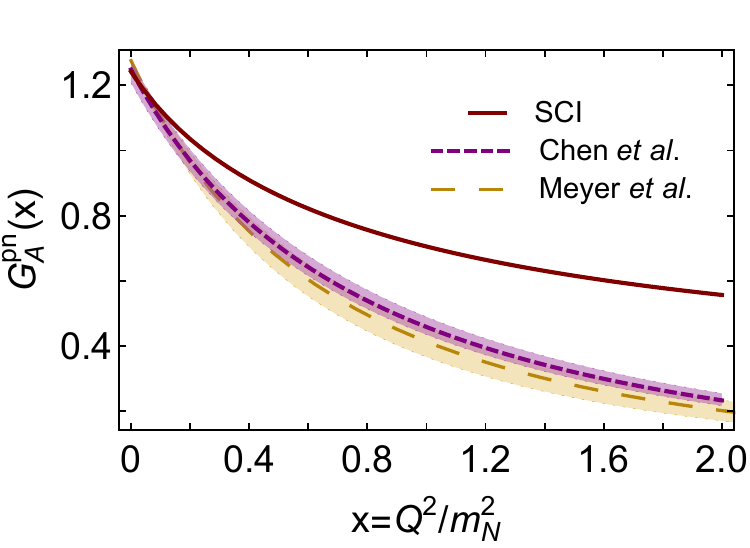}}
\vspace*{1ex}

\leftline{\hspace*{4.5em}{\large{\textsf{B}}}}
\vspace*{-4ex}
\centerline{\includegraphics[width=3.4in]{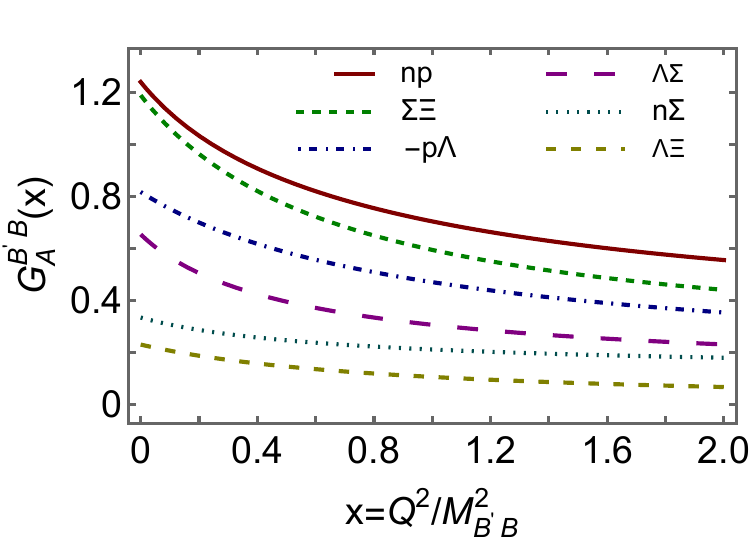}}
\vspace*{1ex}

\leftline{\hspace*{4.5em}{\large{\textsf{C}}}}
\vspace*{-4ex}
\centerline{\includegraphics[width=3.4in]{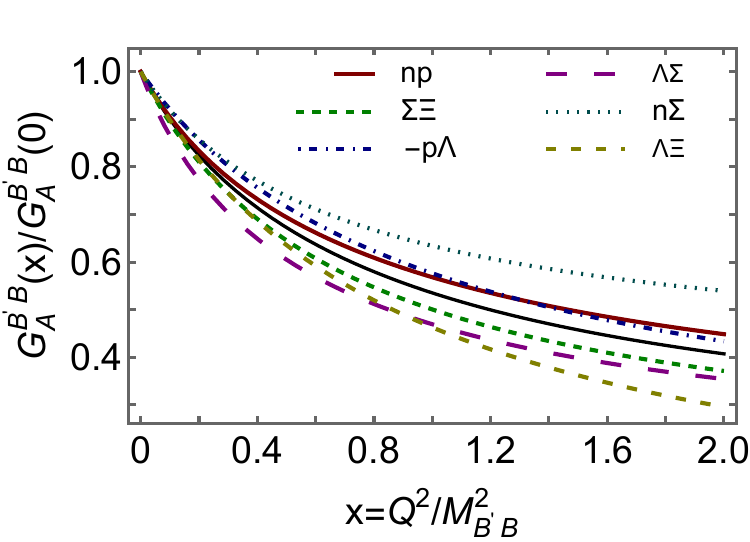}}
\caption{\label{FigGAx}
{\sf Panel A}.
$G_A^{pn}(x=Q^2/m_N^2)$: SCI result computed herein -- solid red curve; prediction from Ref.\,\cite{Chen:2021guo} -- short-dashed purple curve within like-coloured band; and dipole fit to data \cite{Meyer:2016oeg} -- long-dashed gold curve within like-coloured band.
{\sf Panel B}.  Complete array of SCI predictions for octet baryon axial transition form factors: $G_A^{B^\prime B}(x=Q^2/M_{B^\prime B}^2)$.
{\sf Panel C}.  As in Panel B, but with each form factor normalised to unity at $x=0$.  The thinner solid black curve is a pointwise average of the six transition form factors.
}
\end{figure}

On $t=Q^2 \in (-m_{P_{fg}}^2, 2M_{B^\prime B}^2)$, the calculated SCI result for $G_A^{pn}(Q^2= x m_N^2)$ is reliably interpolated using the following function
\begin{equation}
G_A^{B^\prime B}(s) = \frac{g_0+g_1 s+g_2 s^2}{1+l_1 s + l_2 s^2}\,,
\label{SCIinterpolationsGA}
\end{equation}
with the coefficients listed in Table~\ref{InterpolationsGA}. It is drawn in Fig.\,\ref{FigGAx}A and compared with both the CSM prediction from Ref.\,\cite{Chen:2021guo},  which is produced using QCD-kindred momentum dependence for all elements in Figs.\,\ref{figFaddeev} and \ref{figcurrent}, and a dipole fit to low-$Q^2$ data \cite{Meyer:2016oeg}.
As typically found with the SCI: the $Q^2 \lesssim M_l^2$ results are quantitatively sound ($M_l$ is the dressed-mass of the lighter quarks -- Table~\ref{Tab:DressedQuarks}); but the evolution of form factor with increasing $Q^2$ is too slow \cite{Gutierrez-Guerrero:2010waf, Roberts:2011wy, Chen:2012qr}, \emph{i.e}., SCI form factors are too hard at spacelike momenta.

The full set of ground-state octet baryon axial transition form factors is depicted in Fig.\,\ref{FigGAx}B. Interpolations of these functions are achieved using Eq.\,\eqref{SCIinterpolationsGA} along with the relevant coefficients from Table~\ref{InterpolationsGA}.

Fig.\,\ref{FigGAx}C displays the curves from Fig.\,\ref{FigGAx}B renormalised to unity at $x=0$ along with the pointwise average of the renormalised functions. Introducing a dimensionless radius squared associated with the curves drawn, \emph{viz}.
\begin{equation}
(\hat r_A^{B^\prime B})^2 = -6 M_{B^\prime B} \frac{d}{dQ^2}[G_A^{B^\prime B}(Q^2)/G_A^{B^\prime B}(0)],
\end{equation}
in terms of which the usual radius is $r_A^{B^\prime B} = \hat r_A^{B^\prime B}/M_{B^\prime B}$, one can obtain the following comparisons:
\begin{equation}
\begin{array}{ccccc}
\hat r_A^{\Lambda \Sigma^-}/\hat r_A^{pn} & \hat r_A^{p \Lambda}/\hat r_A^{pn} & \hat r_A^{n \Sigma^-}/\hat r_A^{pn} &
\hat r_A^{\Sigma^+ \Xi^0}/\hat r_A^{pn} & \hat r_A^{\Lambda \Xi^-}/\hat r_A^{pn} \\
1.22 & 0.89 & 0.90 & 1.05 & 1.00
\end{array}\,, \label{radiiratio}
\end{equation}
which provide a quantitative representation of the pattern that can be read ``by eye'' from Fig.\,\ref{FigGAx}C.  Evidently, removing the $M_{B^\prime B}$ kinematic factor has exposed a fairly uniform collection of axial transition form factors: the mean value of the ratio in Eq.\,\eqref{radiiratio} is $1.01(13)$.
Considering that SCI form factors are typically hard, the individual SCI radii are likely underestimated; nevertheless, their size relative to $\hat r_A^{pn}$ can serve as a reliable guide. So, for a physical interpretation of these ratios, comparing the SCI result for $\hat r_A^{pn}$ with that in Ref.\,\cite{Chen:2021guo}, one has $\hat r_{A\,{\rm SCI}}^{pn}/\hat r_{A\,\mbox{\footnotesize\cite{Chen:2021guo}}}^{pn}=0.76$ and $\hat r_{A\,\mbox{\footnotesize\cite{Chen:2021guo}}}^{pn}=3.40(4)$.  The dipole fit to data in Fig.\,\ref{FigGAx}A yields $r_{A\,\mbox{\footnotesize\cite{Meyer:2016oeg}}}^{pn}=3.63(24)$.

Considering the $x$-dependence of the axial transition form factors presented in Fig.\,\ref{FigGAx}C, it is noteworthy that at $x=2$ the mean absolute value of the relative deviation from the average curve is 16(8)\%.  Evidently, the magnitude of SU$(3)$-flavour symmetry breaking increases with $Q^2$, \emph{i.e}., as details of baryon structure are probed with higher precision. This may also be highlighted by comparing the $x=2$ values of the $n\to p$ and $\Xi^0\to \Sigma^+$ curves in Fig.\,\ref{FigGAx}C: at $x=2$, the ratio is $\approx 1.2$. In the case of SU(3)-flavour symmetry, it would be unity.

\subsection{Induced pseudoscalar form factors}
\label{SecGP}
\begin{figure}[htbp]
\vspace*{2ex}

\leftline{\hspace*{4.5em}{\large{\textsf{A}}}}
\vspace*{-4ex}
\centerline{\includegraphics[width=3.4in]{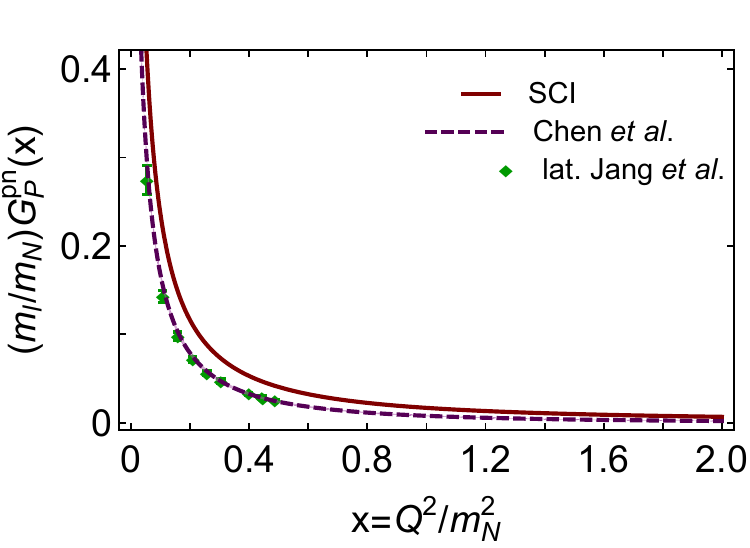}}
\vspace*{1ex}

\leftline{\hspace*{4.5em}{\large{\textsf{B}}}}
\vspace*{-4ex}
\centerline{\includegraphics[width=3.4in]{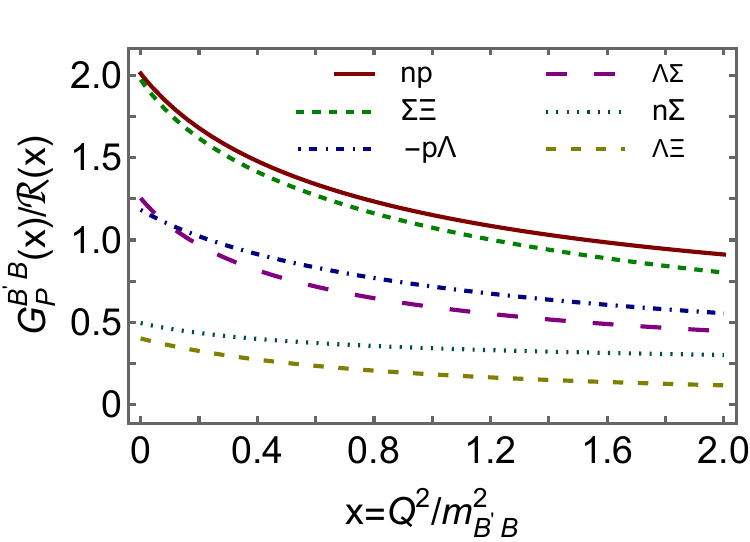}}
\vspace*{1ex}

\leftline{\hspace*{4.5em}{\large{\textsf{C}}}}
\vspace*{-4ex}
\centerline{\includegraphics[width=3.4in]{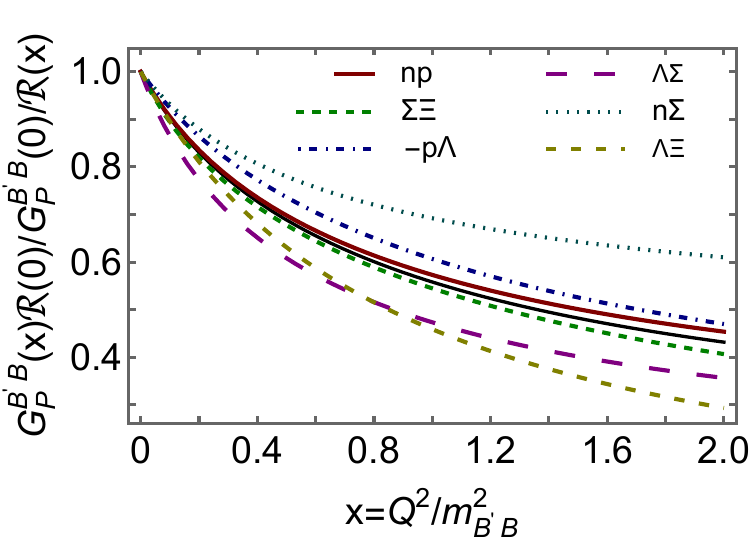}}
\caption{\label{FigGPx}
{\sf Panel A}.
$(m_l/M_N)G_P^{pn}(x=Q^2/m_N^2)$: SCI result computed herein -- solid red curve; prediction from Ref.\,\cite{Chen:2021guo} -- short-dashed purple curve within like-coloured band; and lQCD results \cite{Jang:2019vkm} -- green points.
{\sf Panel B}.  Complete array of SCI predictions for octet baryon axial transition form factors: $G_P^{B^\prime B}(x=Q^2/M_{B^\prime B}^2)/{\cal R}(x)$, Eqs.\,\eqref{SCIinterpolations}.
{\sf Panel C}.  As in Panel B, but with each form factor normalised to unity at $x=0$.  The thinner solid black curve is a pointwise average of the other six curves.
}
\end{figure}

The SCI result for the $n\to p$ induced pseudoscalar transition form factor, $G_P(x)$, is reliably interpolated using the following function:
\begin{subequations}
\label{SCIinterpolations}
\begin{align}
G_{P}^{B^\prime B}(s) & = \frac{g_0+g_1 s+g_2 s^2}{1+l_1 s + l_2 s^2} {\cal R}(s)
\label{SCIinterpolationsGP}\\
{\cal R}(s) & = \frac{m_{P_{fg}}^2}{s+m_{P_{fg}}^2}
\frac{M_{B^\prime B}}{m_{fg}}\,, \label{SCIinterpolationsGP5R}
\end{align}
\end{subequations}
with the coefficients listed in Table~\ref{InterpolationsGP}.  It is drawn in Fig.\,\ref{FigGPx}A and compared with both the CSM prediction from Ref.\,\cite{Chen:2021guo}, which is produced with QCD-kindred momentum dependence for all elements in Figs.\,\ref{figFaddeev} and \ref{figcurrent}, and results from a numerical simulation of lQCD \cite{Jang:2019vkm}. Evidently, there is fair agreement between the SCI result and calculations  that have a closer connection to QCD.

\begin{table}[thb]
\caption{\label{InterpolationsGP}
Interpolation parameters for octet baryon induced pseudoscalar transition form factors, Eq.\,\eqref{SCIinterpolationsGP}.
(Every form factor is dimensionless; so each coefficient in Eq.\,\eqref{SCIinterpolationsGP} has the mass dimension necessary to cancel that of the associated $s\, ({\rm GeV}^2)$ factor.)}
\begin{center}
\begin{tabular*}
{\hsize}
{
l@{\extracolsep{0ptplus1fil}}
|c@{\extracolsep{0ptplus1fil}}
c@{\extracolsep{0ptplus1fil}}
c@{\extracolsep{0ptplus1fil}}
c@{\extracolsep{0ptplus1fil}}
c@{\extracolsep{0ptplus1fil}}}\hline
 {\sf B}  & $g_0\ $ & $g_1\ $ & $g_2\ $ & $l_1\ $ & $l_2\ $ \\\hline
$\phantom{-}G_P^{pn}\ $ & $2.01\ $ & $4.22\ $ & $\phantom{-}0.70\phantom{0}\ $ & $2.96\ $ & $1.57\ $ \\
$\phantom{-}G_P^{\Lambda \Sigma^- }\ $ & $1.25\ $ & $2.09\ $ & $\phantom{-}0.24\phantom{0}\ $ & $2.59\ $ & $1.25\ $ \\
$-G_P^{p \Lambda}\ $ & $1.18\ $ & $1.91\ $ & $\phantom{-}0.15\phantom{0}\ $ & $2.18\ $ & $0.80\ $ \\
$\phantom{-}G_P^{n\Sigma^- }\ $ & $0.50\ $ & $0.44\ $ & $\phantom{-}0.061\ $ & $1.39\ $ & $0.29\ $ \\
$\phantom{-}G_P^{\Sigma^+\Xi^0 }\ $ & $1.97\ $ & $2.38\ $ & $\phantom{-}0.060\ $ & $1.84\ $ & $0.43\ $ \\
$\phantom{-}G_P^{\Lambda\Xi^- }\ $ & $0.40\ $ & $1.34\ $ & $-0.014\ $ & $3.91\ $ & $1.88\ $ \\
\hline
\end{tabular*}
\end{center}
\end{table}

The induced pseudoscalar charge can be determined by muon capture experiments,  $\mu\,+\,p\,\to\,\nu_\mu\,+\,n$:
\begin{equation}
g_p^\ast = \frac{m_\mu}{2m_N} G_p(Q^2 = 0.88\,m_\mu^2)\,,
\end{equation}
where $m_\mu$ is the muon mass. The SCI yields $g_p^\ast=10.3$. For comparison, I record that Ref.\,\cite{Chen:2021guo} predicts $g_p^\ast=8.80(23)$,
the MuCap Collaboration reports $g_p^\ast=8.06(55)$\,\cite{MuCap:2012lei, MuCap:2015boo},
and the world average value is $g_p^\ast=8.79(1.92)$\,\cite{Bernard:2001rs}.  Consequently, one might infer that the SCI result is possibly overestimated by around $15$\%. When evaluating this outcome, it is worth recalling that our SCI analysis is largely algebraic and parameter-free.

Referring to Fig.\,\ref{figcurrent}, a diagram breakdown of $G_{P}^{B^\prime B}(0)$ is presented in Table~\ref{GP0diagrams}. Once again, it is evident that scalar-diquark correlations and $0^+\leftrightarrow 1^+$ transitions play a dominant role in forming the induced pseudoscalar transition charges. In these cases, however, each form factor also receives seagull contributions. They are largest in the case of $\Xi^- \to \Lambda$, where the final state has all three possible types of scalar-diquark correlation. Here, the seagull terms must offset the large contribution from Diagram~1. Significant seagull contributions are also observed in $\Lambda\to p$ and $\Sigma^- \to n$: in the former transition, they exhibit constructive interference with Diagram 4 to compensate for a large contribution from Diagram~1; in the latter, they interfere destructively with Diagram~4. These effects are required by PCAC and guaranteed by our SCI analysis.

\begin{table}[t]
\caption{\label{GP0diagrams}
Diagram separated contributions to $Q^2=0$ values of octet baryon induced pseudoscalar transition form factors, $G_P^{B^\prime B}$, presented as a fraction of the total listed in Table~\ref{InterpolationsGP}\,--\,Column~1 and made with reference to Fig.\,\ref{figcurrent}.
}
\begin{center}
\begin{tabular*}
{\hsize}
{
l@{\extracolsep{0ptplus1fil}}
|c@{\extracolsep{0ptplus1fil}}
c@{\extracolsep{0ptplus1fil}}
c@{\extracolsep{0ptplus1fil}}
c@{\extracolsep{0ptplus1fil}}
c@{\extracolsep{0ptplus1fil}}
c@{\extracolsep{0ptplus1fil}}}\hline
 & $\langle J \rangle^{S}_{\rm q}$  & $\langle J \rangle^{A}_{\rm q}$ &$\langle J \rangle^{AA}_{\rm qq}$ & $\langle J \rangle^{\{SA\}}_{\rm qq}$
 & $\langle J \rangle_{\rm ex}$
 & $\langle J \rangle_{\rm sg}$ \\\hline
$g_P^{pn}$ & $0.54\ $ & $\phantom{-}0.051\ $ & $\phantom{-}0.072\ $ & $0.35\phantom{0}\ $ & $\phantom{-}0.018\ $ & $-0.039\ $ \\
$g_P^{\Lambda \Sigma^-}$ & $0.43\ $ & $\phantom{-}0.054\ $ & $\phantom{-}0.023\ $ & $0.42\phantom{0}\ $ & $\phantom{-}0.023\ $ & $\phantom{-}0.051\ $ \\
$g_P^{p \Lambda}$ & $0.81\ $ & & $\phantom{-}0.073\ $ & $0.32\phantom{0}\ $ & $-0.064\ $ & $-0.14\phantom{0}\ $\\
$g_P^{n\Sigma^-}$ & & $\phantom{-}0.46\phantom{0}\ $ & $-0.047\ $ & $0.56\phantom{0}\ $ & $-0.19\phantom{0}\ $ & $\phantom{-}0.22\phantom{0}\ $ \\
$g_P^{\Sigma^+ \Xi^0}$ & $0.66\ $ & $\phantom{-}0.036\ $ & $\phantom{-}0.061\ $ & $0.33\phantom{0}\ $ & $-0.048\ $ & $-0.033\ $\\
$g_P^{\Lambda \Xi^-}$ & $1.57\ $ & $-0.23\phantom{0}\ $ & $\phantom{-}0.10\phantom{0}\ $ & $0.091\ $ & $\phantom{-}0.13\phantom{0}\ $ & $-0.66\phantom{0}\ $ \\
\hline
\end{tabular*}
\end{center}
\end{table}
The full set of induced pseudoscalar transition form factors for ground-state octet baryons is plotted in Fig.\,\ref{FigGPx}B. 
Division by the factor ${\cal R}(x)$, defined in Eq.\,\eqref{SCIinterpolationsGP5R}, removes kinematic differences associated with quark and baryon masses and pseudoscalar meson poles.
Interpolations of these functions are provided by Eq.\,\eqref{SCIinterpolationsGP} with the appropriate coefficients from Table~\ref{InterpolationsGP}. 
Fig.\,\ref{FigGPx}C redraws these curves renormalised to unity at $x=0$ along with the pointwise average of the rescaled functions. Within the displayed range, the average is similar to the $n\to p$ curve; and at $x=2$, the mean absolute value of the relative deviation from the average curve is 20(14)\%. Once again, these panels reveal that the size of SU$(3)$-flavour symmetry breaking increases with $Q^2$. In this case, comparing the $x=2$ values of the $n\to p$ and $\Xi^0\to \Sigma^+$ curves in Fig.\,\ref{FigGPx}C, the ratio is $\approx 1.2$, which is alike in size with that for the axial transition form factors.

\subsection{Pseudoscalar form factors}

Similar to Eq.\,\eqref{CouplingKpLambda}, the $\pi N N$ form factor is defined via the pseudoscalar current in Eq.\,\eqref{jpsdq0}:
\begin{equation}
G_{\pi NN}(Q^2) \frac{f_\pi}{m_N}  \frac{m_\pi^2}{Q^2+m_\pi^2}= \frac{m_l}{m_N} G_5^{pn}(Q^2)\,.
\end{equation}
In this context, the Goldberger-Treiman relation reads:
\begin{equation}
G_{A}^{pn}(0) = \frac{m_l}{m_N}G_5^{pn}(0)\,.
\end{equation}
Reviewing Eqs.\,\eqref{PCAC} and Table~\ref{InterpolationsGA}, it becomes evident that the relation is satisfied in the SCI. Furthermore, one can obtain the value of the $\pi NN$ coupling constant from the residue of $G_5^{pn}(Q^2)$ at $Q^2+m_\pi^2 = 0$:
\begin{subequations}
\begin{align}
g_{\pi NN}\frac{f_\pi}{m_N} & = \lim_{Q^2+m_\pi^2\to 0}(1+Q^2/m_\pi^2) \frac{m_l}{m_N}G_5^{pn}(Q^2) \label{CouplingpiNN}\\
& \stackrel{\rm SCI}{=} 1.24\,.
\end{align}
\end{subequations}
The SCI prediction is in fair agreement with the results obtained using QCD-kindred momentum dependence for all elements in Figs.\,\ref{figFaddeev} and \ref{figcurrent}, \emph{viz}.\ $1.29(3)$ \cite{Chen:2021guo}; extracted from pion-nucleon scattering data \cite{Baru:2011bw}, $1.29(1)$; inferred from the Granada 2013 $np$ and $pp$ scattering database \cite{NavarroPerez:2016eli}, $1.30$; and determined in a recent analysis of nucleon-nucleon scattering using effective field theory and related tools \cite{Reinert:2020mcu}, $1.30$.

\begin{table}[t]
\caption{\label{PCouplings}
Row~1. Pseudoscalar transition couplings defined by analogy with Eq.\,\eqref{CouplingpiNN}.
Row~2. Value of this quantity at $t=0$ instead of at $t=-m_{P_{fg}}^2$.
Row~3. Relative difference between Rows~1 and 2.
}
\begin{center}
\begin{tabular*}
{\hsize}
{
l@{\extracolsep{0ptplus1fil}}
|c@{\extracolsep{0ptplus1fil}}
c@{\extracolsep{0ptplus1fil}}
c@{\extracolsep{0ptplus1fil}}
c@{\extracolsep{0ptplus1fil}}
c@{\extracolsep{0ptplus1fil}}
c@{\extracolsep{0ptplus1fil}}}\hline
& $\pi p n\ $ & $\pi \Lambda \Sigma\ $ & $K p \Lambda\ $ & $K n \Sigma\ $ & $K \Sigma \Xi\ $ & $K \Lambda \Xi\ $ \\\hline
$g_{P_{fg}B^\prime B} \frac{f_{P_{fg}}}{M_{B^\prime B}}$ &
$1.24\ $ & $0.66\ $ & $-0.83\ $ & $0.34\ $ & $1.21\ $ & $0.25\ $ \\
%
$t=0\ $ & $1.24\ $ & $0.66\ $ & $-0.82\ $ & $0.34\ $ & $1.19\ $ & $0.23\ $ \\\hline
\% difference &
$0.16\ $ & $0.15\ $ & $\phantom{-}1.5\phantom{0}\ $ & $1.8\phantom{0} $ & $1.7\phantom{0}\ $ & $9.1\phantom{0}\ $ \\
\hline
\end{tabular*}
\end{center}
\end{table}

Couplings for all pseudoscalar transitions, defined analogously to Eq.\,\eqref{CouplingpiNN}, are listed in Table~\ref{PCouplings}. Clearly, $G_{P_{fg}B^\prime B}(0) (f_{P_{fg}}/M_{B^\prime B})$ serves as a good approximation to the on-shell value of the coupling in all cases except the $\Lambda \Xi^-$ transition, which is somewhat special owing to the spin-flavour structure of the $\Lambda$, Eq.\,\eqref{SpinFlavourLambda}.  This was emphasised in relation to Table~\ref{GP0diagrams}. Nevertheless, even in this case, the $t=0$ value provides a reasonable guide.

The values presented  in Table~\ref{PCouplings} can be compared with the results from the quark-soliton model (see Table 3 in \cite{Yang:2018idi}).  Converted using empirical baryon masses and meson decay constants, the mean value of $\delta_r^g:=\{|g_{P_{fg} B^\prime B}^{\rm SCI}/g_{P_{fg} B^\prime B}^{\mbox{\rm\footnotesize \cite{Yang:2018idi}}}-1|\}$ is
$0.18(17)$.

\begin{table}[t]
\caption{\label{G50diagrams}
Diagram separated contributions to $Q^2=0$ values of octet baryon pseudoscalar transition form factors, presented as a fraction of the total listed in Table~\ref{InterpolationsG5}\,--\,Column~1 and made with reference to Fig.\,\ref{figcurrent}.
}
\begin{center}
\begin{tabular*}
{\hsize}
{
l@{\extracolsep{0ptplus1fil}}
|c@{\extracolsep{0ptplus1fil}}
c@{\extracolsep{0ptplus1fil}}
c@{\extracolsep{0ptplus1fil}}
c@{\extracolsep{0ptplus1fil}}
c@{\extracolsep{0ptplus1fil}}
c@{\extracolsep{0ptplus1fil}}}\hline
 & $\langle J \rangle^{S}_{\rm q}$  & $\langle J \rangle^{A}_{\rm q}$ &$\langle J \rangle^{AA}_{\rm qq}$ & $\langle J \rangle^{\{SA\}}_{\rm qq}$
 & $\langle J \rangle_{\rm ex}$
 & $\langle J \rangle_{\rm sg}$ \\\hline
$g_5^{pn}$ & $0.51\ $ & $\phantom{-}0.048\ $ & $\phantom{-}0.083\ $ & $0.38\ $ & $\phantom{-}0.017\ $ & $-0.039\ $ \\
$g_5^{\Sigma^- \Lambda}$ & $0.40\ $ & $\phantom{-}0.050\ $ & $\phantom{-}0.025\ $ & $0.44\ $ & $\phantom{-}0.039\ $ & $\phantom{-}0.048\ $ \\
$g_5^{p \Lambda}$ & $0.71\ $ & & $\phantom{-}0.094\ $ & $0.35\ $ & $-0.032\ $ & $-0.12\phantom{0}\ $\\
$g_5^{n\Sigma^-}$ & & $\phantom{-}0.36\phantom{0}\ $ & $-0.057\ $ & $0.59\ $ & $-0.068\ $ & $\phantom{-}0.18\phantom{0}\ $ \\
$g_5^{\Sigma^+ \Xi^0}$ & $0.57\ $ & $\phantom{-}0.028\ $ & $\phantom{-}0.073\ $ & $0.38\ $ & $-0.015\ $ & $-0.028\ $\\
$g_5^{\Lambda \Xi^-}$ & $1.49\ $ & $-0.20\phantom{0}\ $ & $\phantom{-}0.14\phantom{0}\ $ & $0.13\ $ & $\phantom{-}0.040\ $ & $-0.60\phantom{0}\ $ \\
\hline
\end{tabular*}
\end{center}
\end{table}

Similar comparisons can be made with the couplings used in phenomenological hyperon+nucleon potentials \cite{Haidenbauer:2005zh, Rijken:2010zzb}, obtaining $\delta_r^g=0.21(17)$ and $0.15(14)$, respectively.
The dynamical coupled channels study of nucleon resonances in Ref.\,\cite{Kamano:2013iva} uses SU$(3)$-flavour symmetry to express hyperon+nucleon couplings in terms of $g_{\pi NN}$. Compared to those couplings, one finds $\delta_r^g=0.17(15)$. Rescaling the value of $g_{\pi NN}$ used therein to match the SCI prediction, then $\delta_r^g=0.16(14)$. In this last case, the non-zero difference is an indication of the size of SU$(3)$-flavour symmetry violation in $\{g_{P_{fg}B^\prime B}\}$.
These comparisons with phenomenological potentials indicate that the SCI predictions for the couplings in Table~\ref{PCouplings} could serve as useful constraints in refining such models.

\begin{figure}[htbp]
\vspace*{2ex}

\leftline{\hspace*{4.5em}{\large{\textsf{A}}}}
\vspace*{-4ex}
\centerline{\includegraphics[width=3.4in]{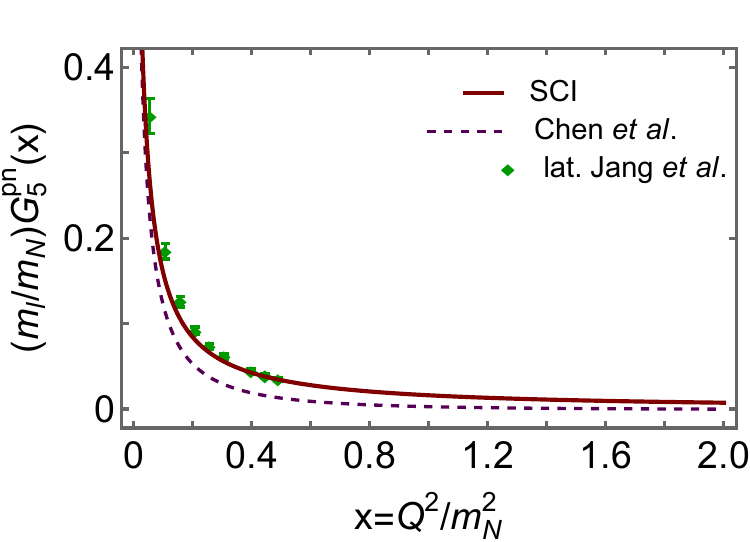}}
\vspace*{1ex}

\leftline{\hspace*{4.5em}{\large{\textsf{B}}}}
\vspace*{-4ex}
\centerline{\includegraphics[width=3.4in]{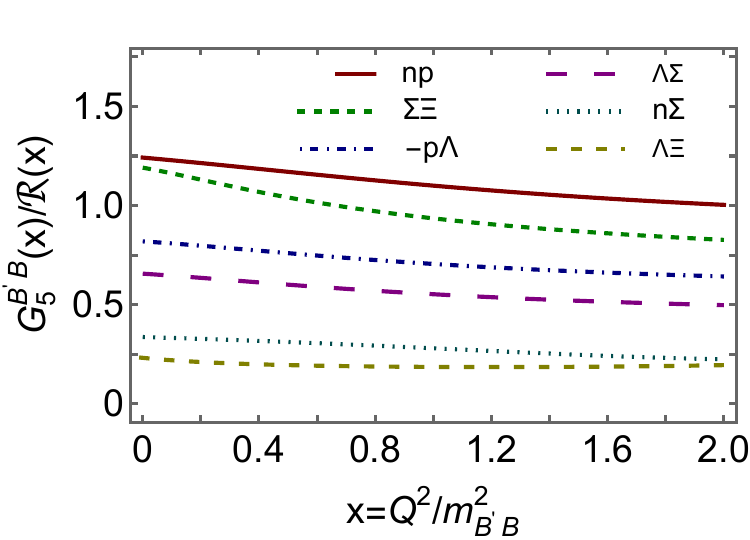}}
\vspace*{1ex}

\leftline{\hspace*{4.5em}{\large{\textsf{C}}}}
\vspace*{-4ex}
\centerline{\includegraphics[width=3.4in]{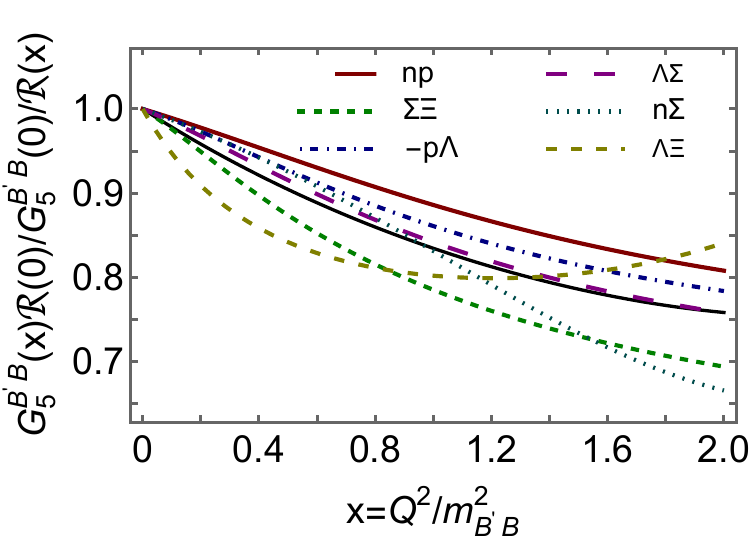}}
\caption{\label{FigG5x}
{\sf Panel A}.
$(m_l/M_N)G_5^{pn}(x=Q^2/m_N^2)$: SCI result -- solid red curve; prediction from Ref.\,\cite{Chen:2021guo} -- short-dashed purple curve within like-coloured band; and lQCD results \cite{Jang:2019vkm} -- green points.
{\sf Panel B}.  Complete array of SCI predictions for octet baryon axial transition form factors: $G_P^{B^\prime B}(x=Q^2/M_{B^\prime B}^2)/{\cal R}(x)$, Eqs.\,\eqref{SCIinterpolations}.
{\sf Panel C}.  As in Panel B, but with each form factor normalised to unity at $x=0$.  The thinner solid black curve is a pointwise average of the other six curves.
}
\end{figure}

Referring to Fig.\,\ref{figcurrent}, a diagram breakdown of $G_{5}^{B^\prime B}(0)$ is presented in Table~\ref{G50diagrams}. Once again, it will be observed that scalar diquark correlations are predominant and $0^+\leftrightarrow 1^+$ transitions play a significant role in building the pseudoscalar transition charges. Moreover, the pattern of diagram contributions is similar to what is observed in $G_{P}^{B^\prime B}(0)$, again largely as a consequence of Eq.\,\eqref{PCAC}: recall, seagulls play no role in $G_{A}^{B^\prime B}(0)$.

\begin{table}[!htbp]
\caption{\label{InterpolationsG5}
Interpolation parameters for octet baryon pseudoscalar transition form factors, Eq.\,\eqref{SCIinterpolationsG5}.
(Every form factor is dimensionless; so each coefficient in Eq.\,\eqref{SCIinterpolationsG5} has the mass dimension necessary to cancel that of the associated $s\, ({\rm GeV}^2)$ factor.)}
\begin{center}
\begin{tabular*}
{\hsize}
{
l@{\extracolsep{0ptplus1fil}}
|c@{\extracolsep{0ptplus1fil}}
c@{\extracolsep{0ptplus1fil}}
c@{\extracolsep{0ptplus1fil}}
c@{\extracolsep{0ptplus1fil}}
c@{\extracolsep{0ptplus1fil}}}\hline
{\sf C}  & $g_0\ $ & $g_1\ $ & $g_2\ $ & $l_1\ $ & $l_2\ $ \\\hline
$\phantom{-}G_5^{pn}\ $ & $1.24\ $ & $\phantom{-}0.13\phantom{0}\ $ & $\phantom{-}0.12\phantom{0}\ $ & $\phantom{-}0.19\ $ & $\phantom{-}0.13\phantom{0}\ $ \\
$\phantom{-}G_5^{\Lambda \Sigma^- }\ $ & $0.66\ $ & $\phantom{-}0.19\phantom{0}\ $ & $\phantom{-}0.075\ $ & $\phantom{-}0.36\ $ & $\phantom{-}0.18\phantom{0}\ $ \\
$-G_A^{p \Lambda}\ $ & $0.82\ $ & $\phantom{-}0.26\phantom{0}\ $ & $\phantom{-}0.14\phantom{0}\ $ & $\phantom{-}0.39\ $ & $\phantom{-}0.25\phantom{0}\ $ \\
$\phantom{-}G_5^{n\Sigma^- }\ $ & $0.34\ $ & $-0.13\phantom{0}\ $ & $\phantom{-}0.019\ $ & $-0.30\ $ & $\phantom{-}0.050\ $ \\
$\phantom{-}G_5^{\Sigma^+\Xi^0 }\ $ & $1.19\ $ & $\phantom{-}1.10\phantom{0}\ $ & $\phantom{-}0.26\phantom{0}\ $ & $\phantom{-}1.03\ $ & $\phantom{-}0.42\phantom{0}\ $ \\
$\phantom{-}G_5^{\Lambda\Xi^- }\ $ & $0.23\ $ & $\phantom{-}0.097\ $ & $-0.014\ $ & $\phantom{-}0.73\ $ & $-0.12\phantom{0}\ $ \\
\hline
\end{tabular*}
\end{center}
\end{table}

The SCI result for the $n \to p$ pseudoscalar transition form factor, $G_5(x)$, is reliably interpolated using the following function:
\begin{equation}
G_{5}^{B^\prime B}(s) = \frac{g_0+g_1 s+g_2 s^2}{1+l_1 s + l_2 s^2} {\cal R}(s),
\label{SCIinterpolationsG5}
\end{equation}
with ${\cal R}(s)$ given in Eq.\,\eqref{SCIinterpolationsGP5R} and the coefficients listed in Table~\ref{InterpolationsG5}. It is plotted in Fig.\,\ref{FigG5x}A and compared with both the CSM prediction from Ref.\,\cite{Chen:2021guo}, which is obtained by using QCD-kindred momentum dependence for all elements in Figs.\,\ref{figFaddeev} and \ref{figcurrent}, and results from a numerical simulation of lQCD \cite{Jang:2019vkm}. The SCI prediction is in fair agreement with that from a numerical simulation of lQCD. The SCI result is harder than the CSM prediction in Ref.\,\cite{Chen:2021guo}, which should be closer to reality. Therefore, it is worth considering the possibility that the lQCD result may also be too hard.

Figure~\ref{FigG5x}B illustrates the complete set of pseudoscalar transition form factors for ground-state octet baryons, each divided by the factor ${\cal R}(x)$ to remove kinematic differences associated with pseudoscalar meson poles and masses. Interpolations of these functions are indicated by Eq.\,\eqref{SCIinterpolationsG5} with the appropriate coefficients listed in Table~\ref{InterpolationsG5}.
In Fig.\,\ref{FigG5x}C, I present  each of the curves from Fig.\,\ref{FigG5x}B after renormalising them to unity at $x=0$ alongside the pointwise average of the renormalised functions. At $x=2$, the mean absolute value of the relative deviation from the average curve is 7(5)\%. Focusing on the $\Xi^0\to \Sigma^+$ curves in Fig.\,\ref{FigG5x}C: at $x=2$, the ratio is $\approx 1.2$, similar to that observed with $G^{\Sigma^+ \Xi^0}_{A,P}$. It is worth mentioning that although $G^{\Lambda\Xi^-}_5(x)/{\cal R}(x)$ does not exhibit a strictly decreasing trend with increasing $x$ within the displayed domain, $G^{\Lambda\Xi^-}_5(x)$ itself does.

\section{Valence spin fraction}
\label{SecFour}
The axial-vector current considered above involves three distinct isospin multiplets and a singlet, which in the isospin symmetry limit may be characterized by the following four baryons: $p$, $\Sigma^+$, $\Xi^-$ and $\Lambda$. Following Ref.\,\cite{Chen:2022odn}, I consider neutral-current processes and carry out a flavour separation of $G_A^{B}$ in each case. The obtained results at $Q^2=0$ define a flavour separation of octet baryon axial charges:
\begin{subequations}
\label{EqFlavour}
\begin{align}
g_{A}^{p} & = \phantom{-} g_{Au}^{p} - g_{Ad}^{p}\,, \\
g_{A}^{\Sigma^+} & = \phantom{-} g_{Au}^{\Sigma^+} - g_{As}^{\Sigma^+}\,, \\
g_{A}^{\Xi^-} & = -g_{Ad}^{\Xi^-} - g_{As}^{\Xi^-}\,, \\
g_{A}^{\Lambda} & =\phantom{-} g_{Au}^\Lambda + g_{Ad}^{\Lambda} - g_{As}^{\Lambda}\,.
\end{align}
\end{subequations}

The flavour-separated charges are of particular interest because $g_{Ah}^{B}$ measures the contribution of the valence-$h$-quark to the light-front helicity of baryon $B$, \emph{i.e}., the difference between the light-front number density of $h$ quarks with helicity parallel to that of the baryon and the corresponding density with helicity antiparallel. Subsequently, one can define the singlet, triplet, and octet axial charges for each baryon, respectively:
\begin{subequations}
\label{axialcharges}
\begin{align}
a_0^B & = g_{Au}^{B} + g_{Ad}^{B} + g_{As}^{B} \,, \label{chargea0}\\
a_3^B & = g_{Au}^{B} - g_{Ad}^{B}  \,,\label{chargea3}\\
a_8^B & = g_{Au}^{B} + g_{Ad}^{B} - 2 g_{As}^{B} \label{chargea8} \,.
\end{align}
\end{subequations}
$a_0^B$ indicates the spin of baryon B arising from the spin of the valence quarks \cite{Deur:2018roz}. Computed within the SCI framework, this quantity is associated with the hadron scale, $\zeta_{\cal H} = 0.33\,$GeV \cite{Cui:2020dlm, Cui:2020tdf, Cui:2021mom, Cui:2022bxn}, whereat all properties of the hadron are carried by valence degrees of freedom.  Consequently, any difference between the SCI value of $a_0^B$ and unity should measure the fraction of the baryon's spin attributed to quark+diquark orbital angular momentum.

The information presented in Appendix~\ref{AppendixCurrentsDiquark} is sufficient to complete the calculation of the charges in Eq.\,\eqref{EqFlavour}.  Table~\ref{FlavourDiagramSeparated} reports the contributions to each charge from the diagrams in Fig.\,\ref{figcurrent}.  Qualitatively, the results are easily understood using the legend in Table~\ref{DiagramLegend}, the spin-flavour structure of each baryon specified in Eqs.\,\eqref{halfnucleon}, and the Faddeev amplitudes in Table~\ref{SolveFaddeev}. For example, regarding the $s$-quark in the $\Lambda$: the $s[ud]$ quark+diquark combination is strong in the Faddeev amplitude, so $g_{As}^{\Lambda}$ receives a dominant  contribution from Diagram~1 scalar diquark bystander; the valence $s$ quark is never isolated alongside an axial-vector diquark, hence $\langle J \rangle^{A}_{\rm s}\equiv 0$; and Diagram~3 provides the other leading contribution, which is fed by the strong $u[ds]-d[us]$ combination transforming into $u\{ds\}-d\{us\}$. Concerning $u$ and $d$ quarks in the $\Lambda$: 
the $u\leftrightarrow d$ antisymmetry of the amplitude's spin-flavour structure entails that whatever contribution $g_{Au}^\Lambda$ receives, $-g_{Ad}^\Lambda$ will be of the same size with opposite sign (weak charges of the $u$ and $d$ quarks are equal and opposite); 
and diagrams involving scalar diquarks must dominate because such diquarks are the most prominent components in the Faddeev amplitude.

Notwithstanding the dominance of scalar diquark contributions in all cases, axial-vector diquarks also play a significant role. I highlighted this with the importance of $u[ds]-d[us]\leftrightarrow u\{ds\}-d\{us\}$ in the $\Lambda$; and it is also worth emphasising the size of the ${\langle J\rangle}_{\rm q}^A$ contribution, which for singly represented valence quarks in $p$, $\Sigma^+$ is much larger in magnitude and has the opposite sign to that connected with the doubly-represented quark.

\begin{table}[t]
\caption{\label{FlavourDiagramSeparated}
With reference to Fig.\,\ref{figcurrent}, diagram contributions to flavour separated octet baryon axial charges, Eq.\,\eqref{EqFlavour}.  ``$0$'' entries are omitted.
Naturally, in the isospin-symmetry limit, the results for $\Sigma^-$ are obtained by making the replacement $g_{Au}^{\Sigma^+} \to g_{Ad}^{\Sigma^-}$; and for the $\Xi^0$, via $g_{Ad}^{\Xi^-} \to g_{Au}^{\Xi^0}$.
}
\begin{center}
\begin{tabular*}
{\hsize}
{
l@{\extracolsep{0ptplus1fil}}
|c@{\extracolsep{0ptplus1fil}}
c@{\extracolsep{0ptplus1fil}}
c@{\extracolsep{0ptplus1fil}}
c@{\extracolsep{0ptplus1fil}}
c@{\extracolsep{0ptplus1fil}}
c@{\extracolsep{0ptplus1fil}}
c@{\extracolsep{0ptplus1fil}}}\hline
 & $\langle J \rangle^{S}_{\rm q}$  & $\langle J \rangle^{A}_{\rm q}$ &$\langle J \rangle^{AA}_{\rm qq}$ & $\langle J \rangle^{\{SA\}}_{\rm qq}$
 & $\langle J \rangle_{\rm ex}^{SS}$
 & $\langle J \rangle_{\rm ex}^{\{SA\}}$
 & $\langle J \rangle_{\rm ex}^{AA}$  \\\hline
$\phantom{-}g_{Au}^{p}$ & $\phantom{-}0.36\ $ & $-0.016\ $ & $\phantom{-}0.11\phantom{0}\ $ & $\phantom{-}0.22\ $ & & $\phantom{-}0.13\phantom{0}\ $& $\phantom{-}0.028\ $ \\
$-g_{Ad}^{p}$ & & $\phantom{-}0.031\ $ & $-0.022\ $ & $\phantom{-}0.22\ $ & $0.24\phantom{0}\ $ & $-0.064\ $& $\phantom{-}0.007\ $\\\hline
$\phantom{-}g_{Au}^{\Sigma^+}$ & $\phantom{-}0.40\ $ & $-0.008\ $ & $\phantom{-}0.15\phantom{0}\ $ & $\phantom{-}0.14\ $ & & $\phantom{-}0.15\phantom{0}\ $& $\phantom{-}0.023\ $ \\
$-g_{As}^{\Sigma^+}$ & & $\phantom{-}0.064\ $ & $-0.014\ $ & $\phantom{-}0.17\ $ & $0.085\ $ & $-0.014\ $& $\phantom{-}0.001\ $\\\hline
$-g_{Ad}^{\Xi^-}$ & & $\phantom{-}0.013\ $ & $-0.020\ $ & $\phantom{-}0.20\ $ & $0.24\ $ & $-0.044\ $& $\phantom{-}0.005\ $ \\
$-g_{As}^{\Xi^-}$ & $-0.61$ & $\phantom{-}0.019\ $ & $-0.066\ $ & $-0.24\ $ & & $-0.026\ $& $-0.005\ $\\\hline
$\phantom{-}g_{Au}^{\Lambda}$ & $\phantom{-}0.086\ $ & $-0.014\ $ & $\phantom{-}0.019\ $ & $-0.087\ $ & $-0.17\phantom{0}\ $ & $\phantom{-}0.035\ $& \\
$-g_{Ad}^{\Lambda}$ & $-0.086\ $ & $\phantom{-}0.014\ $ & $-0.019\ $ & $\phantom{-}0.087\ $ & $\phantom{-}0.17\phantom{0}\ $ & $-0.035\ $& \\
$-g_{As}^{\Lambda}$ & $-0.36\phantom{0}\ $ & & $-0.044\ $ & $-0.21\phantom{0}\ $ & $-0.038\ $ & $-0.016\ $& $-0.003\ $\\
\hline
\end{tabular*}
\end{center}
\end{table}

The summed results for each $g_{Af}^B$ and the corresponding  singlet, triplet, and octet axial charges are listed in Table~\ref{AllCharges}: the pattern of the SCI predictions is similar to that in a range of other studies (see Table~III in Ref.\,\cite{Qi:2022sus}). Based on this information, I first present the following axial charge ratios for each baryon:
\begin{equation}
\begin{array}{cccc}
g_{Ad}^p/g_{Au}^p & g_{As}^{\Sigma^+}/g_{Au}^{\Sigma^+} & g_{Ad}^{\Xi^-}/g_{As}^{\Xi^-} & g_{A(u+d)}^{\Lambda}/g_{As}^{\Lambda}\\
-0.50 & -0.34 &  -0.43 & -0.40
\end{array}\,.\label{Ratioaxial}
\end{equation}
Evidently, if one considers $u+d$ as effectively the singly represented quark in the $\Lambda$, the ratio of axial charges for singly and doubly represented valence quarks is roughly the same in each baryon, \emph{viz}.\ $-0.42(7)$. Further, the magnitude of the ratio is smallest when the singly represented quark is heavier than the doubly represented quark.

It is also worth recalling that the SCI yields results that are in agreement with only small violations of SU$(3)$-flavour symmetry, Eq.\,\eqref{DFvalues}. Thus, one may compare the proton results in Table~\ref{AllCharges} with the following flavour-symmetry predictions:
\begin{equation}
\frac{g_{Ad}^p}{g_{Au}^p}=\frac{F-D}{2F} = -0.39\,, \;
a_8^p=\frac{3 F-D}{F+D}=0.43\,.
\end{equation}
There is a reasonable degree of consistency.

Such agreement is significant as textbook-level analyses yield $g_{Ad}^p/g_{Au}^p = -1/4$ in nonrelativistic quark models with uncorrelated wave functions. The enhanced magnitude of the SCI result can be attributed to the presence of axial-vector diquarks in the proton. Namely, the fact that the Fig.\,\ref{figcurrent}\,--\,Diagram~1 contribution arising from the $\{uu\}$ correlation, in which the probe strikes the valence $d$ quark, is twice as strong as that from the $\{ud\}$, in which the probe strikes the valence $u$ quark. The relative negative sign means this increases $|g_A^d|$ at a cost to $g_A^u$. Consequently, the highly correlated proton wave function, obtained as a solution of the Faddeev equation in Fig.\,\ref{figFaddeev}, lodges a significantly larger fraction of the proton's light-front helicity with the valence $d$ quark.

\begin{table}[t]
\caption{\label{AllCharges}
Net flavour separated and SU$(3)$ baryon axial charges obtained by combining the entries in Table~\ref{FlavourDiagramSeparated} according to Eqs.\,\eqref{EqFlavour}, \eqref{axialcharges}.
``$0$'' entries are omitted.
Recall that these results are for the elastic/neutral processes; hence, the $a_3^B$ entries need not exactly match those in Row~1 of Table~\ref{GAzero}.
}
\begin{center}
\begin{tabular*}
{\hsize}
{
l@{\extracolsep{0ptplus1fil}}
|c@{\extracolsep{0ptplus1fil}}
c@{\extracolsep{0ptplus1fil}}
c@{\extracolsep{0ptplus1fil}}
c@{\extracolsep{0ptplus1fil}}
c@{\extracolsep{0ptplus1fil}}
c@{\extracolsep{0ptplus1fil}}
c@{\extracolsep{0ptplus1fil}}}\hline
$B \ $ & $g_{Au}^B\ $ & $g_{Ad}^B\ $ & $g_{As}^B\ $ & $a_{0}^B\ $ & $a_{3}^B\ $ & $a_{8}^B\ $ \\\hline
$p\ $ & $\phantom{-}0.83\ $ & $-0.41\ $ & & $0.42\ $ & $1.24\ $ & $\phantom{-}0.42\ $ \\
$\Sigma^+\ $ & $\phantom{-}0.85\ $ & & $-0.29\ $ & $0.56\ $ & $0.85\ $ & $\phantom{-}1.42\ $ \\
$\Xi^-\ $ & & $-0.40\ $ & $\phantom{-}0.93\ $ & $0.53\ $ & $0.40\ $ & $-2.26\ $ \\
$\Lambda\ $ & $-0.13\ $ & $-0.13\ $ & $\phantom{-}0.67\ $ & $0.41\ $ & & $-1.61\ $ \\
\hline
\end{tabular*}
\end{center}
\end{table}

The enhancement remains when all elements in Figs.\,\ref{figFaddeev} and \ref{figcurrent} exhibit QCD-kindred momentum dependence, but with a diminished magnitude \cite{Chen:2021guo}: $g_{Ad}^p/g_{Au}^u = - 0.32(2)$. Compared to that analysis, the larger size of the SCI result is likely attributable to the momentum independence of the Bethe-Salpeter and Faddeev amplitudes it generates. This limits the suppression of would-be soft contributions, \emph{e.g}., the two-loop $\langle J \rangle^{SS}_{\rm ex}$ contribution in row~2 of Table~\ref{FlavourDiagramSeparated} is roughly five-times larger than the analogous term in Ref.\,\cite{Chen:2021guo}, significantly enhancing the magnitude of $g_{Ad}^p$.

Referring to Table~\ref{AllCharges}, $a_3^B$ and $a_8^B$ are conserved charges, \emph{i.e}., they are the same at all resolving scales, $\zeta$. This does not hold for the individual terms in their definitions, Eqs.\,\eqref{chargea3} and \eqref{chargea8}: the flavour-separated valence quark charges $g_{Au}^B$, $g_{Ad}^B$ and $g_{As}^B$ evolve with $\zeta$ \cite{Deur:2018roz}. Consequently, the value of $a_0^B$, which is identified with the quark spin contribution to the baryon's total $J=1/2$, changes with scale -- it diminishes slowly with increasing $\zeta$; and as noted above, the SCI predictions in Table~\ref{AllCharges} are made with respect to the hadron scale $\zeta=\zeta_{\cal H} = 0.33\,$GeV \cite{Cui:2020dlm, Cui:2020tdf, Cui:2021mom, Cui:2022bxn}.

Textbook-level analyses yield $a_0^B=1$ in nonrelativistic quark models with uncorrelated wave functions. Thus, in such pictures, all the baryon's spin entirely originates from that of its constituent valence quarks. Herein, on the other hand, considering the hadron scale, then the valence degrees-of-freedom in octet baryons carry roughly one-half the total spin. The mean is
\begin{equation}
\bar a_0^B = 0.50(7)\,.
\end{equation}
Given that there are no other degrees-of-freedom at this scale and considering that the Poin\-car\'e-covariant baryon wave function derived from the Faddeev amplitude discussed in Section~\ref{secFaddeev} properly describes a $J=1/2$ system, the remainder of the total-$J$ must reside in the quark+diquark orbital angular momentum. In keeping with such a picture, this remainder is largest in systems with the lightest valence degrees-of-freedom: $a_0^p \approx a_0^\Lambda < a_0^{\Sigma} \approx a_0^{\Xi}$.  

\section{Summary}
\label{epilogue}
The SCI was used to make predictions for the axial, induced-pseudoscalar, and pseudoscalar transition form factors of 
ground-state octet baryons. This advance contributes to the ongoing efforts of unifying an array of baryon properties \cite{Yin:2019bxe, Yin:2021uom, Wilson:2011aa, Chang:2012cc, Segovia:2013rca, Raya:2021pyr, Gutierrez-Guerrero:2021rsx, Cheng:2022jxe} with analogous treatments of semileptonic decays of heavy+heavy and heavy+light pseudoscalar mesons to both pseudoscalar and vector meson final states \cite{Xu:2021iwv, Xing:2022sor}.
The study required an extensive set of calculations, involving the solution of a collection of integral equations for an array of relevant $n=2$-$6$--point Schwinger functions, \emph{e.g}., gap, Bethe-Salpeter, and, of special importance, Faddeev equations, which describe octet baryons as quark-plus-interacting-diquark bound-states.
Naturally, owing to its symmetry-preserving nature, all mathematical and physical expressions of PCAC are manifest.

Our implementation of the SCI has four parameters, \emph{viz}.\ the values of a mass-dependent quark+antiquark coupling strength chosen at the current-masses of the $u/d$, $s$, $c$, $b$ quarks. Since their values were fixed elsewhere \cite{Xu:2021iwv}, the predictions for octet baryons presented herein are parameter-free. The SCI possesses several merits, including its algebraic simplicity, paucity of parameters, simultaneous applicability to a wide variety of systems and processes, and potential to provide insights that connect and explain numerous phenomena.

Regarding the axial transition form factors of octet baryons, $G_A$, SCI results exhibit agreement with a small violation of SU$(3)$-flavour symmetry [Sec.\,\ref{GAsection}]; and our analysis revealed that this outcome arises as a dynamical consequence of EHM. Namely, the generation of a nuclear size mass scale in the strong interaction sector of the Standard Model serves to mask the impact of Higgs-boson generated differences between the current masses of lighter quarks.
Moreover, the spin-flavour structure of the Poincar\'e-covariant baryon wave functions, described in the presence of both flavour-antitriplet scalar diquarks and flavour-sextet axial-vector diquarks, plays a key role in determining the axial charges and form factors. Significantly, although scalar diquark contributions are dominant, axial-vector diquarks still play a material role, particularly evident in the values of the flavour-separated charges. Therefore, as observed with numerous other quantities \cite{Lu:2022cjx, Cui:2021gzg, Chang:2022jri}, a sound description of observables requires the inclusion of axial-vector correlations in the wave functions of ground-state octet baryons.

The SCI also provides a satisfactory description of the induced-pseudoscalar transition form factors, $G_P$, for octet baryon [Sec.\,\ref{SecGP}]. Qualitatively, similar formative elements are involved in both $G_P$ and $G_A$. The material difference lies in the contribution of seagull terms to the current [Fig.\,\ref{figcurrent}]. $G_A$ is associated with the transverse component of the baryon axial current and thus does not receive any seagull contributions. Conversely, seagull terms play a role in all calculated induced pseudoscalar form factors, being particularly significant for $\Xi^- \to \Lambda$, $\Lambda\to p$ and $\Sigma^- \to n$. Each $G_P(Q^2)$ displays a pole at $Q^2=-m_{\cal P}^2$, where $m_{\cal P}= m_\pi, m_K$, the pion or kaon mass, depending on whether the underlying weak quark transition is $d\to u$ or $s\to u$.

Owing to the partial conservation of the axial current, which entails that the longitudinal part of the axial-vector current is entirely determined by the corresponding pseudoscalar form factor, there exists an intimate connection between the induced pseudoscalar and pseudoscalar transition form factors, $G_{P,5}$, in every case. Therefore, viewed from the correct perspective, all that is mentioned about $G_P$ applies equally to $G_5$. A new feature is the correlation between $G_5$ and various meson+baryon couplings, which can be inferred from the residue of $G_5$ at $Q^2+m_{\cal P}^2=0$ [Table~\ref{PCouplings}]. As computed, the SCI prediction for the $\pi pn $ coupling aligns reasonably well with other calculations and phenomenology.

Working with neutral axial currents, SCI predictions were obtained for the flavour separation of octet baryon axial charges and, consequently, values for the associated SU$(3)$-flavour singlet, triplet and octet axial charges [Sec.\,\ref{SecFour}]. The singlet charge corresponds to the fraction of a baryon's total angular momentum carried by its valence quarks. The SCI predicts that, at the hadron scale, $\zeta_{\cal H}=0.33\,$GeV, this fraction of proton is approximately 42\%. As there are no other degrees of freedom at $\zeta_{\cal H}$, the remainder can be attributed to quark+diquark orbital angular momentum.

Numerous analyses have consistently demonstrated that, when viewed prudently, SCI results serve as a valuable quantitative guideline. Notwithstanding this, it is worth verifying the predictions described herein by employing the QCD-kindred framework, which has been employed widely in studying properties of the nucleon, $\Delta$-baryon, and their low-lying excitations \cite{Liu:2022ndb, Chen:2020wuq, Chen:2021guo, Chen:2022odn, Chen:2018nsg, Lu:2019bjs, Cui:2020rmu}. This holds particularly true for the results concerning octet baryon spin structure.
Furthermore, with continuing progress in the development of the \emph{ab initio} Poincar\'e-covariant three-body Faddeev equation approach to baryon structure \cite{Eichmann:2009qa, Eichmann:2011pv, Wang:2018kto, Qin:2019hgk}, it is expected that octet baryon axial and pseudoscalar current form factors can soon be delivered independently of the quark+diquark scheme. Comparisons between the results obtained from the various frameworks can contribute to the improvement of both approaches. Additionally, it would be valuable to extend the analyses herein to baryons containing one or more heavy quarks; especially, \emph{e.g}., given the role that $\Lambda_b \to \Lambda_c e^- \bar\nu_e$ may play in testing lepton flavour universality \cite{Li:2021qod}.
\chapter{Angular momentum decomposition of proton axial charge}
\label{chapter4}
\section{Introduction}\label{chap4secintro}
Questions concerning the composition of baryons have been a subject of inquiry for approximately a century. The (constituent) quark model \cite{Gell-Mann:1972iag}, along with its subsequent three-body potential models \cite{Capstick:2000qj, Crede:2013kia, Eichmann:2022zxn, Giannini:2015zia,
Plessas:2015mpa}, has provided answers that exhibit an appealing simplicity within the framework of quantum mechanics. In these models, baryons, formed from combinations of $u$, $d$, $s$ valence quark flavours, can be categorised into multiplets of $\mathrm{SU}(6) \otimes \mathrm{O}(3)$, labelled by their flavour content, spin, and orbital angular momentum. From this perspective, the proton, constituted from two $u$ valence quarks and a $d$ valence quark, is viewed as an \textsf{S}-wave ground-state. 

Quark models have widespread practical applications; however, when it comes to spectra, they typically yield masses for radial excitations of the ground state that are notably higher compared to the lowest-mass orbital angular momentum excitation, see Sec.\,15 in Ref.\,\cite{ParticleDataGroup:2020ssz}. 
The best-known example is the Roper resonance, $N(1440){\frac{1}{2}}^+$, discussed elsewhere \cite{Burkert:2017djo}, which is predicted to lie above the nucleon's parity partner, $N(1535){\frac{1}{2}}^-$, contrary to experimental observations. Potential models face challenges from quantum chromodynamics, which demands a Poincar\'e covariant depiction of baryon structure, leading to a Poincar\'e invariant explanation of their properties \cite{Brodsky:2022fqy}. Additionally, while the total angular momentum of a bound state remains Poincar\'e-invariant, any separation into spin and orbital angular momentum components carried by the system's identified constituents is not \cite{Coester:1992cg}. Consequently, potential model wave functions may only offer a basic reference for baryon structure, particularly evident when it comes to assignments to  $\mathrm{SU}(6) \otimes \mathrm{O}(3)$ multiplets. 

CSMs provide a good alternative for studies of baryon composition \cite{Roberts:2015lja, Horn:2016rip, Eichmann:2016yit, Burkert:2017djo, Fischer:2018sdj, Qin:2020rad, Roberts:2020udq, Roberts:2020hiw, Roberts:2021xnz, Roberts:2021nhw, Binosi:2022djx, Papavassiliou:2022wrb, Ding:2022ows, Ferreira:2023fva}.
As seen above, in this framework a baryon is described by using a Poincar\'e-covariant three-body Faddeev equation, whose solution provides the mass and bound state amplitudes. 
It is worth noting here that baryon properties have been calculated \cite{Eichmann:2011vu,
Eichmann:2009qa,Wang:2018kto,Qin:2019hgk} at leading-order using a systematic, symmetry-preserving truncation scheme \cite{Qin:2013mta, Qin:2014vya, Binosi:2016rxz}; and ongoing efforts aim to implement more sophisticated truncations \cite{Qin:2020jig}. 

Meanwhile, the simplified quark-plus-interacting-diquark picture of baryons continues to be successfully employed, 
As described in Chapter\,\ref{chapter2}, the approximation is efficacious because any interaction capable of forming mesons as dressed-quark + antiquark bound states must also lead to strong colour-antitriplet correlations between any two dressed quarks within a hadron. It is worth reiterating that the diquark correlations discussed herein are completely dynamical, appearing in a Faddeev kernel that necessitates their continuous breakup and reformation. Therefore, they are fundamentally distinct from the point-like, static diquarks introduced over fifty years ago \cite{Anselmino:1992vg} to address the so-called ``missing resonance'' problem \cite{Aznauryan:2011ub}. The highly active nature of valence quarks within diquarks entails that the spectrum produced by Fig.\,\ref{figFaddeev} exhibits a richness beyond the explanation of two-body models, something also observed in numerical simulations of lattice-regularised QCD \cite{Edwards:2011jj}. 

A quark+diquark Faddeev equation analysis of the four lowest-lying ($I=\frac{1}{2}$, $J^P={\frac{1}{2}}^{\pm}$) baryons, where, as usual, $I$ is isospin and $J^P$ is spin-parity, is presented in the Ref.\,\cite{Chen:2017pse}. 
These states included the nucleon and its lightest excitations. 
It was found that, projected into the rest frame, the nucleon wave functions have significant $\mathsf S$-wave components; yet they also contain material $\mathsf P$-wave structures and the canonical normalisation receives measurable $\mathsf S \otimes \mathsf P$-wave interference contributions. This result is consistent with that in the previous chapter: the quark+diquark orbital angular momentum must contribute to its spin. So, in this chapter, using the SCI, I study the angular moment decomposition of proton's axial charge as a continuation and supplement to the work in Chapter\,\ref{chapter3}.
The analysis sheds light on the amount of the proton's spin that may be attributed to quark+diquark orbital angular momentum. 

The procedure for a partial wave decomposition of ($\frac{1}{2}$,${\frac{1}{2}}^+$)-baryon bound-state wave functions is outlined in Sec.\ref{secpartialwave}. Solutions for contributions from the various quark+diquark orbital angular momentum components to the canonical normalisation of the Poincar\'e-covariant proton wave function and the proton axial charges are described and analysed in Sec.\ref{secAMD}. Section \ref{cha4review} provides a summary. 

\section{Partial wave decomposition}\label{secpartialwave}

In Sec.\ref{secFaddeev}, I provided a detailed introduction to the baryon Faddeev equation. 
The Faddeev amplitudes, $\Psi^{J^P}$, can be obtained by solving the Faddeev equation, \textit{i.e.}, Eq.\eqref{FaddeevEqn}. 
Then, crucially for what follows in connection with angular momentum decompositions of baryon properties, the (unamputated) Faddeev wave function, $\Phi^{J^P}$, can be computed from the amplitude $\Psi^{J^P}$ simply by attaching the appropriate dressed-quark and -diquark propagators. 

Adapting Eqs.\,\eqref{Psi0plus}, \eqref{Psi1plus}, the wave function of a  $J^P={\frac{1}{2}}^+$ state can be expressed in terms of the following matrix-valued functions:
\begin{subequations}
\label{DecompositionFV}
\begin{align}
\underline{\Phi}^{0^{+}}(\ell ; P)& = S(\ell+P/3) \Delta^{0^{+}}(2 P/3-\ell) \Psi^{{\mathpzc S}=0^{+}}(\ell ; P)\nonumber \\
& =\sum_{k=1}^2 \tilde{s}_k\left(\ell^2, \ell \cdot P\right) S^k(\ell ; P)u(P)\,, \\
\underline{\Phi}_\mu^{1^{+}}(\ell ; P)& =S(\ell+P/3) \Delta_{\mu \nu}^{1^{+}}(2 P/3-\ell) \Psi_\nu^{{\mathpzc A}=1^{+}}(\ell ; P) \nonumber\\
& =\sum_{k=1}^6 \tilde{a}_k\left(\ell^2, \ell \cdot P\right) \gamma_5 A_\mu^k(\ell ; P)u(P)\,,
\end{align}
\end{subequations}
where the forms of $S^k$ and $A_\nu^k$ have been exhibited in Eqs.\eqref{SfunctionsI} and \eqref{AfunctionsI}. It is only when working with the \textit{wave function} that meaningful angular momentum decompositions become available. No information on bound state angular momentum is directly available from a bound state amplitude.

According to Ref.\,\cite{Oettel:1998bk}, the rest frame orbital angular momentum operator is given by 
\begin{equation}
\mathcal{L}^i=\frac{1}{2} \epsilon_{i j k} L^{j k}\,,
\label{OAMoperator}
\end{equation}
whose square characterises the orbital angular momentum,
\begin{equation}
\mathbf{L}^2 \Phi=L(L+1) \Phi\,.
\label{OAMoperatorsquare}
\end{equation}
Here, $\Phi$ is the spinor wave function with positive parity and positive energy and the tensor $L^{j k}$ reads
\begin{equation}
L^{j k}=\sum_{a=1}^3(-i)\left(p_a^j \frac{\partial}{\partial p_a^k}-p_a^k \frac{\partial}{\partial p_a^j}\right)\,,
\label{OAMtensor}
\end{equation}
where $a = 1, 2, 3$ runs over the three valence quark degrees-of-freedom.

In the wave function of Eqs.\,\eqref{DecompositionFV}, the only relative momentum is that between the quark and diquark, $\ell$. Hence, the operator $\mathbf{L}^2$ now takes the form
\begin{equation}
\mathbf{L}^2=\left(2 \ell^i \frac{\partial}{\partial \ell^i}-\boldsymbol{\ell}^2 \nabla^2_\ell+\ell^i \ell^j \frac{\partial}{\partial \ell^i} \frac{\partial}{\partial \ell^j}\right)\,.
\label{OAMoperatorsquarerest}
\end{equation}
By applying the operator $\mathbf{L}^2$ to the basic Dirac components of the functions in Eqs.\,\eqref{DecompositionFV}, one obtains the orbital angular momentum of each component.

Here, in order to enable comparisons with typical formulations of constituent quark models, I list the set of baryon rest-frame quark+diquark angular momentum identifications:
\begin{subequations}
\label{listofOAM}
\begin{align}
&\mathsf{S}: \quad S^1, A_\nu^2,B_\nu^1=\left(A_\nu^3+A_\nu^5\right)\,, \\
&\mathsf{P}: \quad S^2, A_\nu^1,B_\nu^2=\left(A_\nu^4+A_\nu^6\right),C_\nu^2=\left(2 A_\nu^4-A_\nu^6\right) / 3\,, \\
&\mathsf{D}: \quad C_\nu^1=\left(2 A_\nu^3-A_\nu^5\right) / 3\,;
\end{align}
\end{subequations}
\textit{viz}.\ the scalar functions associated with these combinations of Dirac matrices in a Faddeev wave function possess the identified angular momentum correlation between the quark and
diquark. Those functions are:
\begin{subequations}
\begin{align}
&\mathsf{S}: \quad \tilde{s}_1, \tilde{a}_2,\left(\tilde{a}_3+2 \tilde{a}_5\right) / 3\,; \\
&\mathsf{P}: \quad \tilde{s}_2, \tilde{a}_1,\left(\tilde{a}_4+2 \tilde{a}_6\right) / 3, \left(\tilde{a}_4-\tilde{a}_6\right)\,; \\
&\mathsf{D}: \quad\left(\tilde{a}_3-\tilde{a}_5\right)\,.
\end{align}
\label{coefficientofOAM}
\end{subequations}

\section{Solutions and their features}\label{secAMD}

\subsection{Canonical normalisation}\label{subsecAMDnormalisation}

\begin{figure}[t]
\centerline{\includegraphics[width=3.8in]{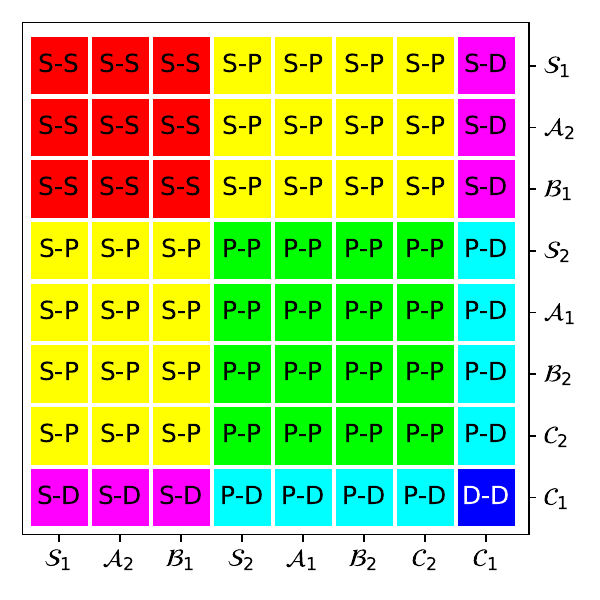}}
\caption{\label{Legend} Legend for interpretation of Figs.\,\ref{fignormalisationAMD} --\ref{figgaudAMD}, identifying interference between the various identified orbital angular momentum basis components in the baryon rest frame.}
\end{figure}

\begin{figure}[!htbp]
\centerline{\includegraphics[width=3.8in]{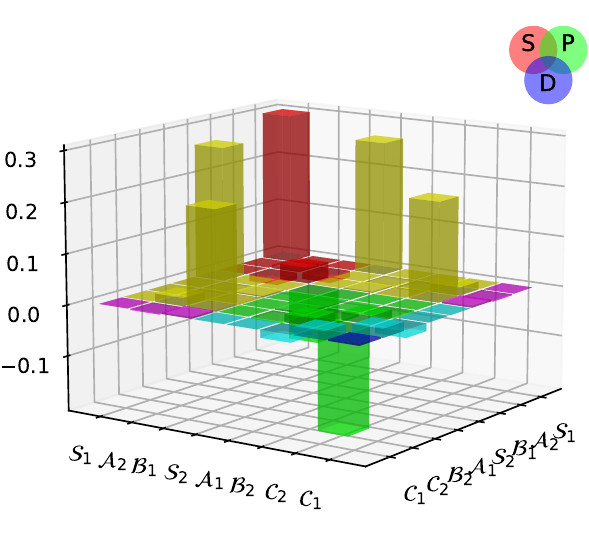}}
\caption{\label{fignormalisationAMD} Contributions of the various quark+diquark orbital angular momentum components to the canonical normalisation of the Poincar\'e-covariant proton wave function after rest-frame projection. The values drawn are listed in Table \ref{tabnormalisationAMD}.}
\end{figure}

Using the assignments in Fig.\,\ref{Legend}, the distinct contributions from each partial wave to the proton's canonical normalisation constant are depicted in Fig.\,\ref{fignormalisationAMD}. (Recall that the canonical normalisation constant is related to the $Q^2=0$ value of the charge
form factors linked to the electrically charged members of a given hadron multiplet: in this case, that is the proton Dirac form factor.) 
From Fig.\,\ref{fignormalisationAMD}, in the rest frame, one observes that the proton canonical normalisation is largely determined by $\mathsf S$-wave components, but there are significant, destructive $\mathsf P$-wave contributions and also strong, constructive $\mathsf S \otimes \mathsf P$-wave interference terms. 
There is no contribution from $\mathsf D$-wave. 
These results are consistent with those obtained in the QCD-kindred framework whose interaction is momentum-dependent \cite{Liu:2022ndb}. 

\begin{table}[!htbp]
 \caption{  Contributions of the various diquark components to the canonical normalisation .}
 \label{tabnormalisationdiquark}
 \centering
 \begin{center}
  \renewcommand\arraystretch{1.2}
  \begin{tabular}{|c|c|c|}
  \hline
  &  $SC$ &  $AV$ \\
  \hline
  $SC$ & 0.74 & 0\\
  \hline
  $AV$ & 0 & 0.26\\
  \hline
  \end{tabular}
 \end{center}
\end{table}

Working with the results in Table \ref{tabnormalisationAMD}, one arrives at SCI prediction for the contributions of the various diquark components to the canonical normalisation.  
They are listed in Table \ref{tabnormalisationdiquark}. 
While the $[u d]_{0+}$ scalar diquark (SC) is dominant, material contributions also owe to the $\{u d\}_{1+}$, $\{u u\}_{1+}$ axial-vector diquarks (AV): roughly 74\% of the proton's canonical normalisation constant is provided by the scalar diquark and the remainder owes to the axial-vector diquark. 
The QCD-kindred framework gives similar results \cite{Liu:2022ndb}. 
It is notable that, unlike the results predicted by the QCD-kindred framework, $\mathrm{SC} \otimes \mathrm{AV}$ interference contribution are absent because there are no contributions arising from the photon coupling to the exchange quark in the SCI case \cite{Wilson:2011aa}. 

\subsection{Axial charges}\label{subsecAMDaxilcharge}

\begin{figure}[!htbp]
\centerline{\includegraphics[width=3.8in]{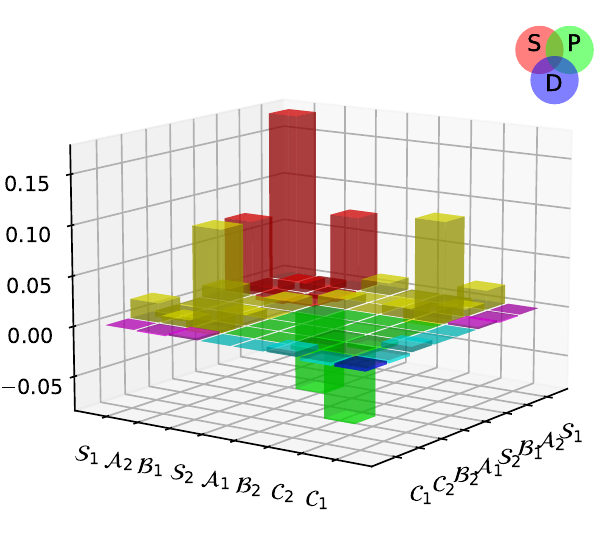}}
\caption{\label{figga0AMD} Contributions of the various quark+diquark orbital angular momentum components to $g_{A}^{(0)}$ after rest-frame projection. The values drawn are listed in Table \ref{tabaxialAMD}.}
\end{figure}

\begin{table}[t]
 \caption{Contributions of the various diquark components to the axial charge $g_{A}^{(0)} = 0.42$ -- see Table~\ref{AllCharges}.
 \label{tabaxialdiquark} }
 \centering
 \begin{center}
  \renewcommand\arraystretch{1.2}
  \begin{tabular}{|c|c|c|}
  \hline
  &  $SC$ &  $AV$ \\
  \hline
  $SC$ & 0.12 & 0.09\\
  \hline
  $AV$ & 0.09 & 0.12\\
  \hline
  \end{tabular}
 \end{center}
\end{table}

In Sec.\,\ref{SecFour}, I provided SCI predictions for the proton's axial charge along with their flavour separation at $Q^2=0$. 
Here, I will dissect the SCI predictions in order to expose the various quark+diquark orbital angular momentum contributions to these quantities, following the scheme employed for the canonical normalisation. 

The contributions of various quark+diquark orbital angular momentum components to $g_{A}^{(0)}$ are depicted in Fig.\,\ref{figga0AMD}. 
Comparing with Fig.\,\ref{fignormalisationAMD}, the axial charge $g_{A}^{(0)}$ results share a similar pattern: the contributions from $\mathsf S$-wave components dominate, but there are significant, destructive $\mathsf P$-wave contributions, as well as strong, constructive $\mathsf S \otimes \mathsf P$-wave interference terms. 
From the results listed in Table~\ref{tabaxialAMD}, one can see that, if only $\mathsf S$-wave components are considered, the value of $g_{A}^{(0)}$ is underestimated: 
$g_{A \supset S}^{(0)}/g_{A}^{(0)}=0.74$, where $g_{A \supset S}^{(0)}$ is the value of $g_{A}^{(0)}$ obtained by including only $\mathsf S$-wave components. 

Similar to the case of the canonical normalisation, drawing on the results in Table~\ref{tabaxialAMD}, I obtain predictions for the contributions of the various diquark components to the axial charge $g_{A}^{(0)}$, as shown in Table\,\ref{tabaxialdiquark}. 
It will be observed that with this formulation of the SCI the contribution of the scalar diquark ($0.118$) is approximately equal to that of the axial-vector diquark ($0.123$).  
I am currently working to understand whether this outcome is accidental or a special feature of the SCI.  As shown by a comparison between Ref.\,\cite{Chen:2022odn} (Table~1) and Table~\ref{FlavourDiagramSeparated} herein, it is not true in general.

In contrast to the results for the canonical normalisation, which measures the electric current, there is a large contribution arising from $\mathrm{SC} \otimes \mathrm{AV}$ interference when the axial current is used instead. 

\begin{figure}[!htbp]
\vspace*{2ex}

\leftline{\hspace*{4.5em}{\large{\textsf{A}}}}
\vspace*{-4ex}
\centerline{\includegraphics[width=3.8in]{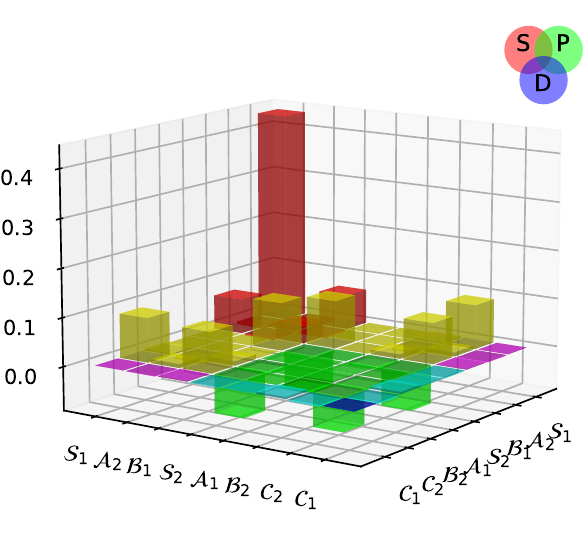}}
\vspace*{1ex}

\leftline{\hspace*{4.5em}{\large{\textsf{B}}}}
\vspace*{-4ex}
\centerline{\includegraphics[width=3.8in]{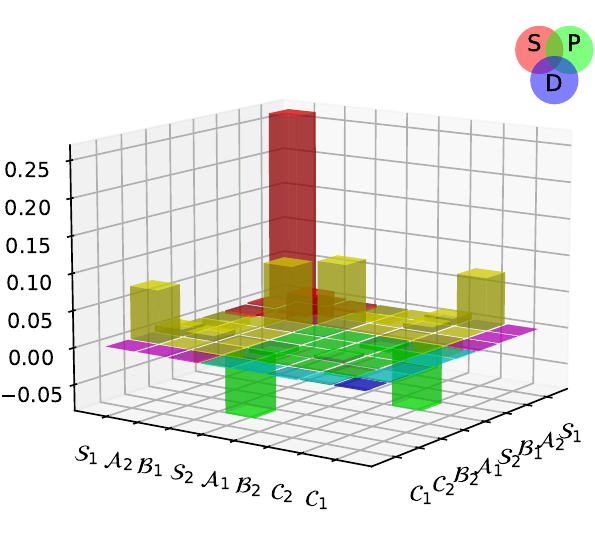}}
\caption{\label{figgaudAMD}
{\sf Panel A}.
Contributions of the various quark+diquark orbital angular momentum components to $g_{A}^{u}$ after rest-frame projection. 
{\sf Panel B}.  Contributions of the various quark+diquark orbital angular momentum components to $-g_{A}^{d}$ after rest-frame projection. The values drawn are listed in Table \ref{tabaxialAMD}.
}
\end{figure}

\begin{table}[!htbp]
 \caption{ $g_{A}^{u}$ and $g_{A}^{d}$ contributions broken into rest-frame quark+diquark orbital angular momentum components.}
 \label{tabgaud}
 \centering
 \begin{center}
  \renewcommand\arraystretch{1.2}
  \begin{tabular}{|c|c|c|c|}
  \hline
  $g_{A}^{u}$ &  $S$ &  $P$ &  $D$\\
  \hline
  $S$ & 0.60 & 0.28 & 0.00\\
  \hline
  $P$ & 0.28 & -0.33 & 0.00\\
  \hline
  $D$ & 0.00 & 0.00 & 0.00 \\
  \hline
  \hline
  $g_{A}^{d}$ &  $S$ &  $P$ &  $D$\\
  \hline
  $S$ & -0.29 & -0.15 & 0.00\\
  \hline
  $P$ & -0.15 & 0.18  & 0.00\\
  \hline
  $D$ & 0.00  & 0.00  & 0.00 \\
  \hline
  \end{tabular}
 \end{center}
\end{table}

The contributions of various quark+diquark orbital angular momentum components to the flavour-separated charges $g_{A}^{u}$ and $-g_{A}^{d}$ are depicted in Figs.\ref{figgaudAMD}. The results of $g_{A}^{u}$ and $-g_{A}^{d}$ are similar to those of $g_{A}^{(0)}$. When summing over the same set of orbital angular momentum components, one can derive the results presented in Table \ref{tabgaud}. 
It is interesting that the ratios of $g_{A}^{d}$ to $g_{A}^{u}$ for each pair of results in Table~\ref{tabgaud} are roughly the same, \textit{viz}.\ all approximately equal to the final result listed in Eq.\,\eqref{Ratioaxial}:
\begin{equation}
g_{A}^{d}/g_{A}^{u} = -0.50.
\label{Ratiodtou}
\end{equation}
At present, I am working to uncover which particular aspects of the SCI lead to this outcome. 

\section{Summary}\label{cha4review}
As a continuation and complement to the findings presented in Chapter\,\ref{chapter3}, I calculated the angular momentum decomposition of the canonical normalisation and axial charge of the proton. 

The SCI predictions for the proton's canonical normalisation are, in many ways, similar to the results obtained with the QCD-kindred framework, whose interaction is momentum-dependent: 
the proton canonical normalisation constant is dominated by $\mathsf S$-wave contributions, 
there are significant destructive $\mathsf P$-wave contributions, 
and strong $\mathsf S \otimes \mathsf P$-wave constructive interference terms. 
The SCI also predicts that roughly 74\% of the proton's canonical normalisation constant is provided by the scalar diquark and 26\% by axial-vector diquark. 
However, $\mathrm{SC} \otimes \mathrm{AV}$ interference does not contribute to the SCI canonical normalisation, an outcome which differs from the prediction of the QCD-kindred framework. 

The results for contributions from the various quark+diquark orbital angular momentum components to the proton's axial charge $g_A^{(0)}$ and its flavour separation into $g_A^{u}$ and $-g_A^{d}$ pieces are typically similar to those of canonical normalisation. 
It is interesting that the ratios of $g_A^{d}$ to $g_A^{u}$, computed from $\mathsf S$-wave, $\mathsf P$-wave and $\mathsf S \otimes \mathsf P$-wave interference, are the similar and all roughly equal to the net result, $g_{A}^{d}/g_{A\approx }^{u} = -0.50$.  
\chapter{Polarised parton distribution functions and proton spin}\label{chapter5}
\section{Introduction}\label{chap5secintro}

It is worth highlighting that the proton is the most fundamental bound-state in Nature. In isolation, it is stable; at least, the lower limit on its lifetime is many orders-of-magnitude greater than the $\sim 14$-billion-year age of the Universe. Moreover, the proton is characterised by two fundamental Poincar\'e invariant quantities: mass squared, $m_p^2$; and total angular momentum squared, $J^2=J(J+1)=3/4$. One can also incorporate parity, $P=+1$, in which case the proton is identified as a $J^P=\tfrac{1}{2}^+$ state.

As discussed above, according to contemporary theory, the proton is composed of three valence quarks: $u+u+d$, which interact based on the rules described by the Lagrangian density of quantum chromodynamics. QCD itself is a Poincar\'e-invariant quantum non-Abelian gauge field theory. It is worth emphasising that $P$ is a Poincar\'e invariant quantum number. On the other hand, every separation of $J$ into a sum of orbital angular momentum ($L$) and spin ($S$), $L+S$, is observer dependent. Therefore, there is no correlation between $P$ and $L$ in QCD and no objective (Poincar\'e-invariant) significance for $L$ and $S$ individually \cite{Brodsky:2022fqy}.

Assuming isospin symmetry, as I have done throughout, so that $u$ and $d$ quarks are mass-degenerate, then the full wave function of the $J^P=\tfrac{1}{2}^+$ proton can be 
described by a Poincar\'e-covariant four-component spinor whose complete form involves 128 distinct scalar (Poincar\'e-invariant) functions \cite{Eichmann:2009qa, Wang:2018kto}. It follows that in any observer-dependent reference frame, this wave function includes $\mathsf S$-, $\mathsf P$- and $\mathsf D$-wave orbital angular momentum components. The nature of the angular momentum represented by these components relies on the degrees-of-freedom (dof) employed to solve the proton bound-state problem. Typically, these dof vary with the resolving scale of any probe utilised to measure the property of the proton. In QCD, there is no specific scale at which the proton $J=\tfrac{1}{2}$ can be regarded simply as the sum of the spins of its valence dof \cite{EuropeanMuon:1987isl}. Similar observations can be made about $m_p^2$; namely, the distribution of proton mass among its constituents depends upon the choices of variables and frame made when solving the bound-state problem. Either or both of these may rely on the resolving scale employed to specify the problem.

As evident from the discussions in previous chapters, these remarks highlight again that there is no objective meaning to any decomposition of the proton's $J=\tfrac{1}{2}$ into subcomponents of any kind. Such a decomposition is contextual and only attains significance once choices of variables and frame are established. A convenient approach is to project Poincar\'e-covariant wave functions onto the light-front because the wave functions obtained thereby are the probability amplitudes associated with parton distribution functions (DFs) \cite{Brodsky:1989pv, Brodsky:1997de, Heinzl:2000ht}.

The selection of variables is more complex. Herein I adopt a perspective characteristic of CSMs. Namely, at the hadron scale, $\zeta_{\cal H}<m_p$, QCD bound-state problems are most efficiently solved by employing dressed-parton dof: dressed-gluons and -quarks, each of which possesses a momentum-dependent mass. As explained above, this approach is firmly founded in QCD theory and has been widely employed with phenomenological success - see, \emph{e.g}., Refs.\,\cite{Roberts:2021nhw, Binosi:2022djx, Papavassiliou:2022wrb, Ding:2022ows, Ferreira:2023fva, Roberts:2022rxm, Salme:2022eoy, Carman:2023zke} for discussions of both facets.  Notably, $\zeta_{\cal H}$ represents the scale at which all properties of a given hadron are determined by its valence quasiparticle dof \cite{Cui:2020dlm, Cui:2020tdf, Ding:2019qlr}.

\section{Results based on Faddeev equation}
\label{PFE}
As I have repeatedly highlighted, EHM is an essential characteristic of strong interactions. In the absence of Higgs boson couplings into QCD, it entails the dynamical generation of a nuclear-scale mass for the proton, $m_p \approx 1 \mathrm{GeV}$, accompanied by the formation of massless pseudoscalar Nambu-Goldstone bosons \cite{Roberts:2016vyn}. As a consequence of EHM, any quark+antiquark interaction that accurately describes meson properties also generates nonpointlike diquark correlations in multiquark systems, something which Sec.\,\ref{secBSE} discussed in detail.  This justifies the use of the quark+diquark Faddeev equation (Fig.\,\ref{figFaddeev}) in describing proton properties.  Working with this simplification offers the advantage that at most 16 (instead of 128) scalar functions are sufficient to fully express the Poincar\'e-covariant proton wave function.

\begin{table}[t]
\caption{\label{anValence}
Gegenbauer coefficients that define the hadron-scale valence quark DFs in Eq.\,\eqref{ExpandGegenbauer}.
Each entry should be divided by $10^3$.
 }
\begin{center}
\begin{tabular*}
{\hsize}
{
c@{\extracolsep{0ptplus1fil}}
c@{\extracolsep{0ptplus1fil}}
c@{\extracolsep{0ptplus1fil}}
c@{\extracolsep{0ptplus1fil}}
c@{\extracolsep{0ptplus1fil}}}\hline\hline
\multicolumn{5}{c}{$a_n^{\mathpzc{u}}$}\\\hline
1 & 2 & 3 & 4 & 5 \\
$403$ & $112$ & $7.31 $ & $-10.4$ & $-4.90\ $ \\\hline
6 & 7 & 8 & 9 & 10 \\
$-0.0474$ & $1.15$ & $0.828$ & $0.334$  & $0.0635$ \\\hline\hline
\multicolumn{5}{c}{$a_n^{\mathpzc{d}}$}\\\hline
1 & 2 & 3 & 4 & 5 \\
$482$ & $161$ & $20.8$ & $-11.5$ & $-7.11\ $ \\\hline
6 & 7 & 8 & 9 & 10 \\
$-0.430$ & $1.50$ & $1.03$ & $0.349$  & $0.0543$ \\\hline\hline
\end{tabular*}
\end{center}
\end{table}

In proceeding to a discussion of proton helicity DFs, the following predictions from the quark+diquark Faddeev equation at the hadron scale are important.
\begin{enumerate}[label=(\emph{\roman*}), ref=(\roman*)]
\item  The proton consists of both isoscalar-scalar (SC) and isovector-axial-vector (AV) diquark correlations, with the AV correlations being responsible for $\approx 35$\% of the wave function canonical normalisation \cite{Mezrag:2017znp}.  (This is the result of the QCD-kindred approach.)\label{item2i}
\item In the rest frame, the quark+diquark Faddeev wave function of the proton contains $\mathsf S$-, $\mathsf P$- and $\mathsf D$-wave orbital angular momentum components -- see Fig.\,3a in Ref.\cite{Liu:2022ndb}, which, in using the QCD-kindred approach, improves upon Fig.\,\ref{fignormalisationAMD} above.
\item Calculated unpolarised valence quark DFs in a proton with the preceding two features are reliably interpolated using a finite sum of Gegenbauer polynomials \cite{Chang:2022jri}:
\begin{equation}
\label{ExpandGegenbauer}
{\mathpzc q}(x;\zeta_{\cal H})  = n_{\mathpzc{q}} 140 x^3(1-x)^3 \left[1 + \sum_{n=1}^{10} a_n^{\mathpzc{q}} C_n^{7/2}(1-2x)\right]\,,
\end{equation}
$n_{\mathpzc u}=2n_{\mathpzc d}=2$, where the coefficients are listed in Table~\ref{anValence}.
\item Consistent with experiment \cite{ParticleDataGroup:2020ssz}, the predicted proton axial charge is $g_A=1.25(3)$ \cite{Chen:2022odn}, where the uncertainty arises from that on the masses of the SC and AV diquarks: $m_{\rm SC}\approx 0.8\,$GeV; $m_{\rm AV} \approx 0.9\,$GeV.  Furthermore, the $u$ quark fraction of the axial charge is $g_A^u/g_A=0.76(1)$; that of the $d$ quark is $g_A^d/g_A=-0.24(1)$; and $g_A^d/g_A^u=-0.32(2)$. \label{gAvalues}
\item The singlet axial charge of the proton is Ref.\cite{Ding:2022ows} (Sec.\,9):
\begin{equation}
\label{isoscalaraxialcharge}
a_0 = 0.65(2)\,.
\end{equation}
At $\zeta_{\cal H}$ the remainder of the proton spin is lodged with quark+diquark orbital angular momentum, as was illustrated in Chapter~\ref{chapter3}, using the SCI. \label{itemv}
\end{enumerate}
I stress here that because results obtained with the QCD-kindred framework are more realistic \cite{Chen:2022odn}, they are used in this analysis of proton helicity DFs instead of those obtained with the SCI in Chapter~\ref{chapter3}. 

\section{Polarised valence quark distributions at $\zeta_{\cal H}$}
In order to deliver a Faddeev equation based prediction for the polarised valence quark DFs at the hadron scale, it is necessary to extend the methods used for the unpolarised DFs in Ref.\,\cite{Chang:2022jri}, centred on the vector current, to the axial current case. Recently, a symmetry-preserving axial current suitable for use with a solution of the quark+diquark Faddeev equation has been derived \cite{Chen:2020wuq, Chen:2021guo}; but additional time will be necessary before it can be applied to the calculation of polarised valence quark DFs.

In the meantime, I employ a phenomenological approach to address the issue, similar to those used, \emph{e.g}., in Refs.\,\cite{Liu:2019vsn, Han:2021dkc}. I also utilise constraints suggested by perturbative QCD analyses \cite{Brodsky:1994kg} to formulate straightforward \emph{Ans\"atze} for the polarised DFs. It is worth enumerating the constraints imposed.
\begin{enumerate}[label=(\emph{\alph*}), ref=(\alph*)]
\item At low-$x$, there is no correlation between the helicity of the struck quark and  the helicity of the parent proton. Therefore, the ratio of polarised to unpolarised DFs must approach zero: as $x \to 0$, $\Delta {\mathpzc q}(x;\zeta_{\cal H})/{\mathpzc q}(x;\zeta_{\cal H}) \to 0$. Drawing on Regge phenomenology, I implement this by expressing $\Delta {\mathpzc q}(x;\zeta_{\cal H}) \propto x^{\delta\alpha_R}{\mathpzc q}(x;\zeta_{\cal H})$, where $\delta\alpha_R = \tfrac{1}{2}$ represents the difference between the intercepts of the vector and axial-vector meson Regge trajectories \cite{Brisudova:1999ut}.
\item At high-$x$, both the polarised and unpolarised valence quark distributions exhibit the same power-law behaviour, \emph{viz}.\ $\Delta \mathpzc{q}(x)/\mathpzc{q}(x) \to {\rm constant} \neq 0$ as $x\to 1$.
\end{enumerate}
These constraints are implemented using four distinct mappings: $\Delta \mathpzc{q}(x;\zeta_{\cal H}) \! = \! \mathpzc{s}_\mathpzc{q} r_i(x,\gamma_i^{\mathpzc{q}}) \mathpzc{q}(x;\zeta_{\cal H})$, with
$\mathpzc{s}_{\mathpzc{u}} \! = \! 1 \! = \! - \mathpzc{s}_{\mathpzc{d}}$,
\begin{subequations}
\label{ratiofunctions}
\begin{align}
r_1(x,\gamma) & = \sqrt{x}/[1+\gamma \sqrt{x}] \,, &
r_2(x,\gamma) & = \sqrt{x}/[\gamma+ \sqrt{x}] \,, \\
r_3(x,\gamma) & = \sqrt{x}/[1+ \gamma x] \,, &
r_4(x,\gamma) & = \sqrt{x}/[\gamma+ x] \,.
\end{align}
\end{subequations}
In each case, $\gamma_i^{\mathpzc{q}}$ is fixed by requiring
$\int_0^1 dx\,\Delta\mathpzc{q}(x;\zeta_{\cal H}) = g_A^{\mathpzc{q}}$.
Referring to Sec.\,\ref{PFE}-item~\ref{gAvalues}, this yields the values listed in Table~\ref{gammavalues}. Note that these profiles are in agreement with results obtained from basis light-front quantisation applied to solve a model Hamiltonian for the proton \cite{Xu:2022abw}, as well as with ongoing studies of proton DFs using a Faddeev equation with a contact interaction \cite{Cheng:2022jxe}.

\begin{table}[t]
\caption{\label{gammavalues}
Referring to Eq.\,\eqref{ratiofunctions}, central values of the mapping parameters, $\gamma^{\mathpzc{q}}$, that reproduce the given quark's contribution to the proton axial charge.
 }
\begin{center}
\begin{tabular*}
{\hsize}
{
l@{\extracolsep{0ptplus1fil}}
c@{\extracolsep{0ptplus1fil}}
c@{\extracolsep{0ptplus1fil}}
c@{\extracolsep{0ptplus1fil}}
c@{\extracolsep{0ptplus1fil}}}\hline
$i$ & 1 & 2 & 3 & 4 \\\hline
$\gamma_i^{\mathpzc{u}}$ & $0.350$ & $0.621$ & $ 0.575$ & $0.853\ $\\
$\gamma_i^{\mathpzc{d}}$ & $1.47\phantom{0} $ & $1.26\phantom{1} $ & $2.62\phantom{5} $ & $1.60\phantom{3}\ $ \\\hline
\end{tabular*}
\end{center}
\end{table}

\begin{figure}[!htbp]
\centerline{\includegraphics[width=3.4in]{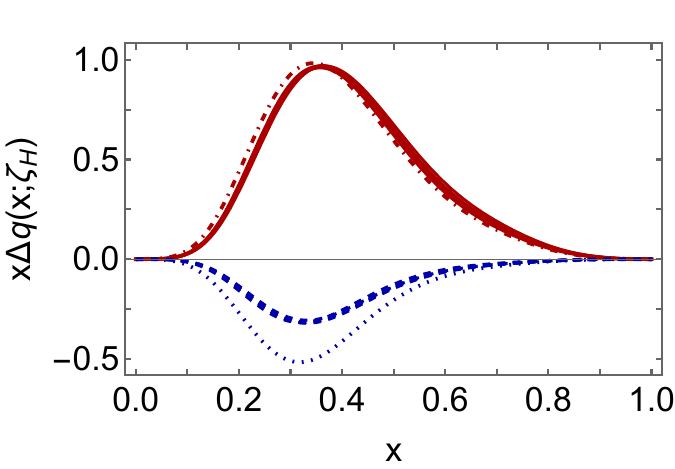}}
\caption{\label{FigPolarised}
Hadron scale polarised valence quark distributions: solid red curves -- $\mathpzc{u}$ quark; and dashed blue curves -- $\mathpzc{d}$ quark. In each case, there are five curves, \emph{viz}.\ the four produced by the mapping functions in Eq.\,\eqref{ratiofunctions}, with the coefficient values in Table~\ref{gammavalues}, and the average of these curves.
Context is provided by the unpolarised valence quark distributions: $x\mathpzc{u}(x;\zeta_{\cal H})/2$ -- dot-dashed red curve; and $[-x\mathpzc{d}(x;\zeta_{\cal H})/2]$ -- dotted blue curve.
}
\end{figure}

The resulting \emph{Ans\"atze} for the hadron-scale polarised valence quark DFs are illustrated in Fig.\,\ref{FigPolarised}, wherein they are compared with the corresponding unpolarised DFs calculated in Ref.\,\cite{Chang:2022jri} -- reproduced by Eq.\,\eqref{ExpandGegenbauer} with the coefficients listed in Table~\ref{anValence}.

At this point, considering that the $x=1$ value of any ratio of valence quark DFs is independent of the scale \cite{Holt:2010vj}, one can present an updated version of Table~1 in Ref.\,\cite{Roberts:2013mja} -- see Table~\ref{x1Update}. The general agreement between my results based on the Faddeev equation and those in the ``Faddeev'' column demonstrates that reliable estimates are provided by the simple formulae introduced in Ref.\,\cite{Roberts:2013mja} for use in analysing nucleon Faddeev wave functions to obtain $x\to 1$ values of DF ratios without the necessity of calculating the $x$-dependence of any DF.  Viewed alternately, the agreement lends support to our proposed \emph{Ans\"atze} for the polarised valence quark DF.

\begin{table}[t]
\begin{center}
\begin{tabular*}
{\hsize}
{
l@{\extracolsep{0ptplus1fil}}
c@{\extracolsep{0ptplus1fil}}
c@{\extracolsep{0ptplus1fil}}
c@{\extracolsep{0ptplus1fil}}
c@{\extracolsep{0ptplus1fil}}
c@{\extracolsep{0ptplus1fil}}}\hline
 & herein & Faddeev & SC only & SU$(4)$ & pQCD \\\hline
$\rule{0ex}{3ex}\frac{F_2^n}{F_2^p}$ & $\phantom{-}0.45(5)$ & $\phantom{-}0.49$ & $\frac{1}{4}$ & $\phantom{-}\frac{2}{3}$ &$\frac{3}{7}$   \\[1ex]
$\frac{d}{u}$ & $\phantom{-}0.23(6)$ & $\phantom{-}0.28$ & 0 & $\phantom{-}\frac{1}{2}$ & $\frac{1}{5}$ \\[1ex]
$\frac{\Delta d}{\Delta u}$ & $-0.14(3)$ & $-0.11$ & 0 & $-\frac{1}{4}$ & $\frac{1}{5}$ \\[1ex]
$\frac{\Delta u}{u}$ &  $\phantom{-}0.63(8)$ & $\phantom{-}0.65$ & 1 & $\phantom{-}\frac{2}{3}$ & 1 \\[1ex]
$\frac{\Delta d}{d}$ & $-0.38(7)$ & $-0.26$ & 0 & $-\frac{1}{3}$ & 1 \\[1ex]
$A_1^n$ &  $\phantom{-}0.15(5)$ & $\phantom{-}0.17$ & 1 & $\phantom{-}0$ &  1 \\[1ex]
$A_1^p$ & $\phantom{-}0.58(8)$ & $\phantom{-}0.59$ & 1 & $\phantom{-}\frac{5}{9}$ & 1 \\\hline
\end{tabular*}
\caption{\label{x1Update}
Predictions for $x=1$ value of the indicated quantities.
Column ``herein'' collects predictions from Ref.\,\cite{Chang:2022jri} and those obtained using the polarised DFs in Fig.\,\ref{FigPolarised}.
``Faddeev'' reproduces the DSE-1/realistic results in Ref.\,\cite{Roberts:2013mja}, obtained using simple formulae, expressed in terms of diquark appearance and mixing probabilities.
The next two columns are, respectively, results drawn from Ref.\,\protect\cite{Close:1988br} -- proton modelled as being built using an elementary scalar diquark (no AV); and Ref.\,\cite{Hughes:1999wr} -- proton described by a SU$(4)$ spin-flavour wave function.
The last column, labelled ``pQCD,'' lists predictions made in Refs.\,\protect\cite{Brodsky:1994kg,Farrar:1975yb}, which assume an SU$(4)$ spin-flavour wave function for the proton's valence-quarks and that a hard photon may interact only with a quark that possesses the same helicity as the target.  ($3/7 \approx 0.43$.)
}
\end{center}
\end{table}

Now it is pertinent to discuss the matter of helicity retention in hard scattering processes \cite{Brodsky:1994kg, Farrar:1975yb}.
If this notion holds true, then $\Delta {\mathpzc d}/{\mathpzc d} = 1=\Delta {\mathpzc u}/{\mathpzc u}$ on $x\simeq 1$ -- see Table~\ref{x1Update}-column~5.
However, these ratios remain invariant under QCD evolution (DGLAP \cite{Dokshitzer:1977sg, Gribov:1971zn, Lipatov:1974qm, Altarelli:1977zs});
such evolution cannot result in a zero in a valence quark DF;
and $\int_0^1 dx \Delta {\mathpzc d}(x;\zeta_{\cal H}) = g_A^d < 0$.
As a result, helicity retention necessitates the presence of a zero in $\Delta {\mathpzc d}(x;\zeta_{\cal H})$. Current precision data indicate that if such a zero exists, then it must occur at $x\gtrsim 0.6$ \cite{HERMES:2004zsh,COMPASS:2010hwr,CLAS:2006ozz, CLAS:2008xos, CLAS:2015otq, CLAS:2017qga,JeffersonLabHallA:2014mam,JeffersonLabHallA:2003joy, JeffersonLabHallA:2004tea}.

Since I have modelled the polarised valence quark distributions, I cannot provide a CSM argument either supporting or contradicting helicity retention. In fact, there are currently no calculations of the polarised valence quark distributions available within any nonperturbative framework that can be directly linked to QCD. Nevertheless, the mappings in Eq.\,\eqref{ratiofunctions} exclude the possibility of a zero in $\Delta {\mathpzc d}(x;\zeta_{\cal H})$. This choice is motivated by the observations that no viable direct calculation of $\Delta {\mathpzc d}(x;\zeta_{\cal H})$ yields a result with a zero on the valence quark domain -- see, e.g, Refs.\,\cite{Deur:2018roz, Xu:2022abw, Xu:2021wwj}, and phenomenological DF global fits do not provide any strong evidence for the existence of such a zero $\Delta {\mathpzc d}(x)$ \cite{Ethier:2020way}. However, a zero in $\Delta {\mathpzc d}(x;\zeta_{\cal H})$ is engineered in the model of Ref.\,\cite{Liu:2019vsn}. It can be affirmed with certainty that QCD-connected calculations of the polarised valence quark distributions are highly desirable, along with the related data on $x \gtrsim 0.6$. The latter have been obtained by the CLAS RGC Collaboration \cite{E1206109} and the E12-06-110 Collaboration \cite{Zheng:2006}, and completed analyses can reasonably be expected within a few years.

A related matter pertains to the ${\mathpzc d}/{\mathpzc u}$ (equivalently $F_2^n/F_2^p$) ratio in Table~\ref{x1Update}. Columns ``herein'', ``Faddeev'' and ``pQCD'' agree, within uncertainties. The first two are derived from calculations of the proton's Poincar\'e-covariant wave function, which incorporates scalar and axial-vector diquarks with dynamically prescribed relative strengths, as described in Sec.\,\ref{PFE}-item\,\ref{item2i}. This wave function corresponds to a structured leading-twist hadron-scale proton distribution amplitude (DA) \cite{Mezrag:2017znp, Bali:2015ykx}.
However, as the scale increases, it is anticipated that all such structure is eliminated as the DA approaches its asymptotic form and the wave function begins to exhibit SU$(4)$ spin-flavour symmetry \cite{Lepage:1980fj}.
In this case, since ${\mathpzc d}/{\mathpzc u}(x=1)$ is invariant under evolution, the ``pQCD'' prediction can be interpreted as a constraint on the relative strength of SC and AV correlations in the hadron-scale proton wave function; and that constraint is satisfied if, and only if, the Faddeev wave function possesses the properties described in Sec.\,\ref{PFE}-item\,\ref{item2i}.
Notably, this relative strength also offers an explanation \cite{Lu:2022cjx,Chang:2022jri} for modern data on $F_2^n(x)/F_2^p(x)$ \cite{JeffersonLabHallATritium:2021usd} and its extrapolation \cite{Cui:2021gzg}: on $x\simeq 1$, $F_2^n/F_2^p = 0.437(85)$.

\emph{Caveat~1}. Before proceeding further, it is important to note that all polarisation ``data'' reproduced herein were derived from analyses of experiments that employ one or another set of the available global DF fits to another body of experiments. Consequently, the ``data'' values and uncertainties are dependent on the reliability of the chosen global fit. Furthermore, there is no assurance of consistency between the newly provided experiment and the body used to produce the existing global fit. As a result, the reported ``data'' are not objective. In contrast, as I now explain, the predictions presented herein arise from an internally consistent, unified treatment of all DFs.

\section{Polarised quark distributions at $\zeta^2=3\,$GeV$^2$}
Given that $\zeta_{\cal H}$ is the scale at which the dressed valence quark dof carry all the properties of the proton, the all-orders extension of QCD evolution, as described in Refs.\,\cite{Cui:2021mom, Cui:2022bxn, Raya:2021zrz, Yin:2023dbw}, can be used to determine all proton DFs at any scale $\zeta > \zeta_{\cal H}$. This approach has been successfully used to predict proton and pion unpolarised DFs \cite{Lu:2022cjx}. More recently, it has also been successfully used to extract the pion mass distribution from available data \cite{Xu:2023bwv}. Herein, I use it to provide predictions for proton polarised DFs.

This all-orders evolution scheme is based on a single proposition \cite{Yin:2023dbw}:\\[1ex]
\hspace*{0.75em}\parbox[t]{0.95\linewidth}{{\sf P1} -- \emph{In the context of Refs.\,\cite{Grunberg:1980ja, Grunberg:1982fw}, there exists at least one effective charge, $\alpha_{1\ell}(k^2)$, which, when used to integrate the leading-order perturbative DGLAP equations, defines an evolution scheme for parton DFs that is all-orders exact}.}  
\bigskip

\noindent Such charges are not necessarily process-independent (PI); hence, not unique. Nevertheless, the possibility of a PI effective charge is not ruled out. The charge, denoted $\hat\alpha$, discussed in Refs.\,\cite{Cui:2019dwv, Cui:2020tdf, Raya:2021zrz}, has proved suitable and I adopt it herein. Using $\hat\alpha$, one can predict $\zeta_{\cal H}=0.331(2)\,$GeV. Further details on its connection with experiment and other nonperturbative extensions of QCD's running coupling can be found in Refs.\,\cite{Deur:2016tte, Deur:2022msf, Deur:2023dzc}.

\begin{figure}[htbp]
\centerline{\includegraphics[width=3.4in]{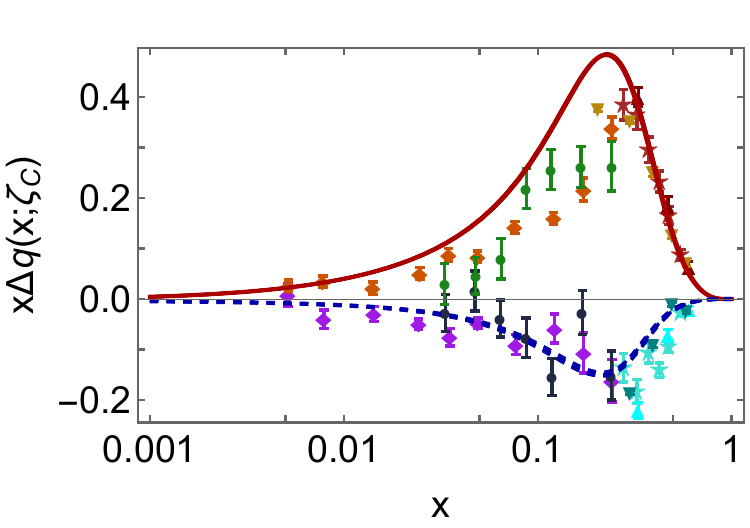}}
\caption{\label{polzetaCa}
Polarised quark DFs: $\Delta \mathpzc{u}(x;\zeta_{\rm C})$ -- solid red curves; and $\Delta \mathpzc{d}(x;\zeta_{\rm C})$ -- dashed blue curves.
Data: HERMES \cite{HERMES:2004zsh} -- circles; COMPASS \cite{COMPASS:2010hwr} -- diamonds;
filled down-triangles -- CLAS EG1 \cite{CLAS:2006ozz, CLAS:2008xos, CLAS:2015otq, CLAS:2017qga};
five-pointed stars -- E06-014 \cite{JeffersonLabHallA:2014mam};
filled up-triangles --E99-117 \cite{JeffersonLabHallA:2003joy, JeffersonLabHallA:2004tea}.
}
\end{figure}

Using the method described in Ref.\,\cite{Lu:2022cjx} and applying it to the polarised valence DFs shown in Fig.\,\ref{FigPolarised}, one obtains the $\zeta_{\rm C}=\surd 3\,$GeV polarised quark DFs drawn in Fig.\,\ref{polzetaCa}, wherein they are compared with data inferred from experiments HERMES\,\cite{HERMES:2004zsh}, COMPASS\,\cite{COMPASS:2010hwr},
CLAS EG1\,\cite{CLAS:2006ozz, CLAS:2008xos, CLAS:2015otq, CLAS:2017qga},
E06-014\,\cite{JeffersonLabHallA:2014mam} and
E99-117\,\cite{JeffersonLabHallA:2003joy, JeffersonLabHallA:2004tea}:
there is agreement on the valence quark domain for $x\gtrsim 0.2$.

\begin{figure}[t]

\leftline{\hspace*{4.5em}{{\textsf{(a)}}}}
\vspace*{-4ex}
\centerline{\includegraphics[width=3.4in]{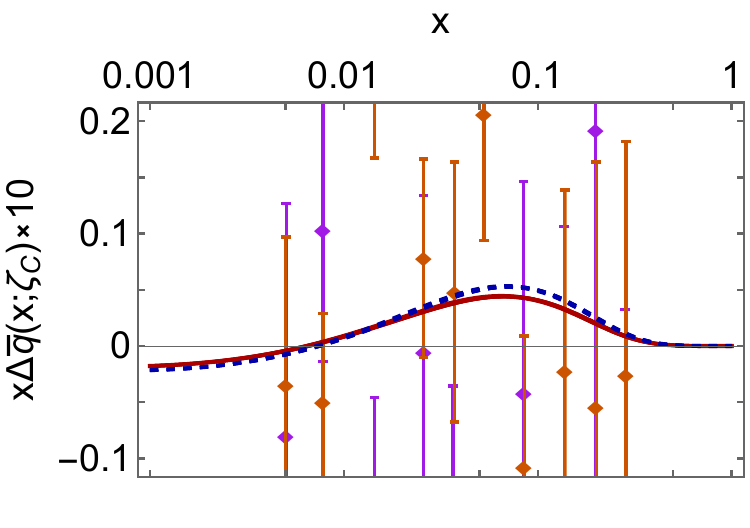}}
\vspace*{-3ex}
\leftline{\hspace*{4.5em}{{\textsf{(b)}}}}
\vspace*{-4ex}
\centerline{\includegraphics[width=3.4in]{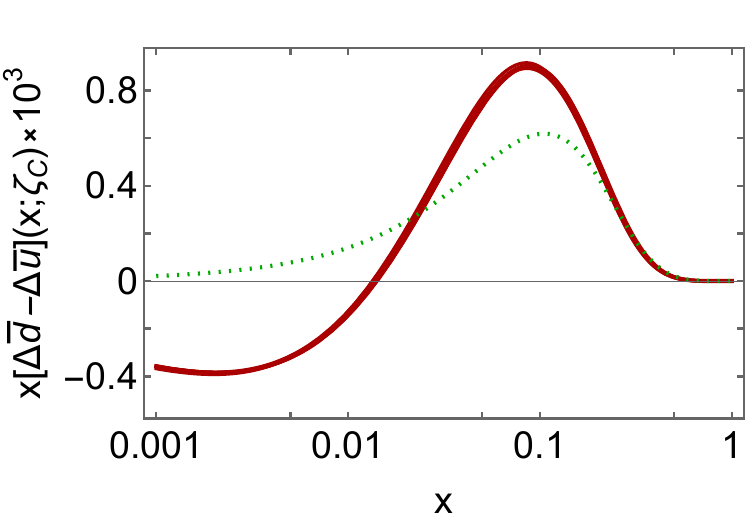}}

\caption{\label{polzetaCB}
\textsf{(a)}
Polarised antiquark DFs: $\Delta \bar{\mathpzc{u}}(x;\zeta_{\rm C})$ -- solid red curves; and $\Delta \bar{\mathpzc{d}}(x;\zeta_{\rm C})$ -- dashed blue curves.
Data from COMPASS \cite{COMPASS:2010hwr}: $\bar{\mathpzc{u}}$ -- red diamonds; and $\bar{\mathpzc{d}}$ -- purple squares.
\textsf{(b)}
$x[\Delta \bar{\mathpzc{d}}(x;\zeta_{\rm C})-\Delta \bar{\mathpzc{u}}(x;\zeta_{\rm C})]$ -- solid red curves.
For comparison: $x[\bar{\mathpzc{d}}(x;\zeta_{\rm C})- \bar{\mathpzc{u}}(x;\zeta_{\rm C})]$ -- dotted green curve.
}
\end{figure}

\begin{figure}[htbp]
\centerline{\includegraphics[width=3.4in]{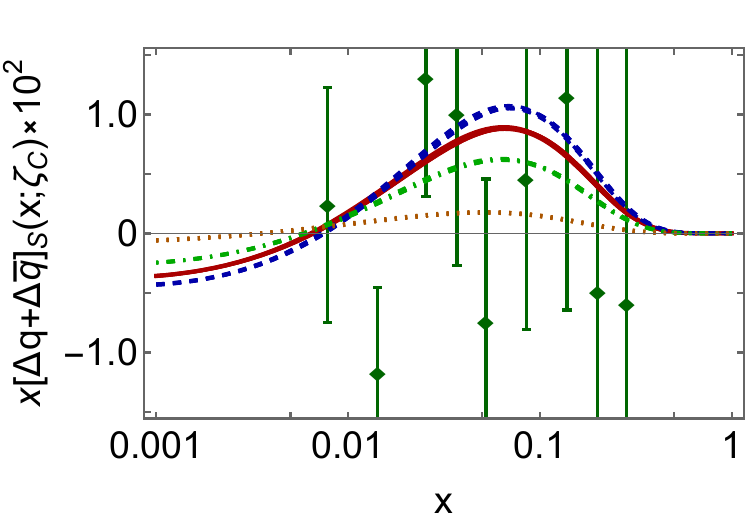}}
\caption{\label{FigPolarisedSea}
Polarised sea quark distributions at $\zeta_{\rm C}$.
Solid red curves: $x[\Delta \mathpzc{u}+\Delta\bar{\mathpzc{u}}]_S$; dashed blue curves: $x[\Delta \mathpzc{d}+\Delta \bar{\mathpzc{d}}]_S$; $x[\Delta \mathpzc{s}+\Delta\bar{\mathpzc{s}}]_S$: dot-dashed green curves; $x[\Delta \mathpzc{c}+\Delta\bar{\mathpzc{c}}]_S$: dotted orange curves.
Context is provided by values of $2x\Delta \mathpzc{s}(x;\zeta_{\rm C})$ from Ref.\,\cite{COMPASS:2010hwr}, wherein $x\Delta \mathpzc{s}(x;\zeta_{\rm C}) \approx x\Delta \bar {\mathpzc{s}}(x;\zeta_{\rm C})$.
}
\end{figure}

Referring to the COMPASS results, lying on $x\lesssim 0.2$, the collaboration's extrapolations yield $g_A^d=-0.34(5)$, $g_A^u=0.71(4)$, $g_A=g_A^u-g_A^d=1.05(6)$.
 Comparing this with Sec.\,\ref{PFE}-item\,\ref{gAvalues}, reveals agreement with the CSM value of $g_A^d$. This agreement is consistent with the match between data and my prediction for $\Delta \mathpzc{d}(x;\zeta_{\rm C})$.
On the other hand, the data tend to  fall below my prediction for $\Delta \mathpzc{u}(x;\zeta_{\rm C})$. This aligns with the observations that the COMPASS results yield a value for $g_A^u-g_A^d$ that is $0.83(5)$-times the value determined from neutron $\beta$-decay. This outcome can be attributed to the low value of $g_A^u$, which is only $0.75(3)$-times the CSM prediction.  (These statements can be made because of the negligible contribution of polarised antiquark DFs at this scale -- see Figs.\,\ref{polzetaCB} and \ref{FigPolarisedSea}.)

The polarised antiquark distributions are depicted in Fig.\,\ref{polzetaCB}a together with values from Ref.\,\cite{COMPASS:2010hwr}.
Given the scale of this image, which is set by the magnitudes of my predictions, the data exhibit large uncertainties. Consequently, they can only be used to establish reasonable boundaries for the size of these distributions.

The difference $x[\Delta \bar{\mathpzc{d}}(x;\zeta_{\rm C})-\Delta \bar{\mathpzc{u}}(x;\zeta_{\rm C})]$ is illustrated in Fig.\,\ref{polzetaCB}b and compared to the corresponding result for $x[\bar{\mathpzc{d}}(x;\zeta_{\rm C})- \bar{\mathpzc{u}}(x;\zeta_{\rm C})]$ from Refs.\,\cite{Lu:2022cjx, Chang:2022jri}, which reproduce the proton antimatter asymmetry reported in Ref.\,\cite{SeaQuest:2021zxb}. 
(This was achieved via a modest Pauli blocking factor in the gluon splitting function.) Notably, both differences exhibit the same magnitude and their trend remains similar for $x\gtrsim 0.01$: they are related by using {\sf P1}. In this case, owing to the considerable uncertainties on the available results \cite{HERMES:2004zsh,COMPASS:2010hwr},
a meaningful comparison with the data cannot be reported.

Figure~\ref{FigPolarisedSea} illustrates my predictions for all polarised sea quark DFs. Following the implementation of all-orders evolution in Ref.\,\cite{Lu:2022cjx}, one has thresholds at which the influence of heavier quarks becomes significant in evolution. This accounts for the flavour-separation amongst the polarised sea DFs. The figure includes results on $2x\Delta {\mathpzc{s}}(x;\zeta_{\rm C})$ inferred from data in Ref.\,\cite{COMPASS:2010hwr}. 
In general, the magnitude aligns with my prediction for this DF; but, again, the data uncertainties are large.  The prediction
$\int_{0.004}^{0.3} dx \,\Delta{\mathpzc{s}}_S(x;\zeta_{\rm C}) = 0.0072(1)$ is  in agreement with the inferred empirical value \cite{COMPASS:2010hwr}: $-0.01\pm 0.01\pm 0.01$.

\begin{figure}[t]
\centerline{\includegraphics[width=3.4in]{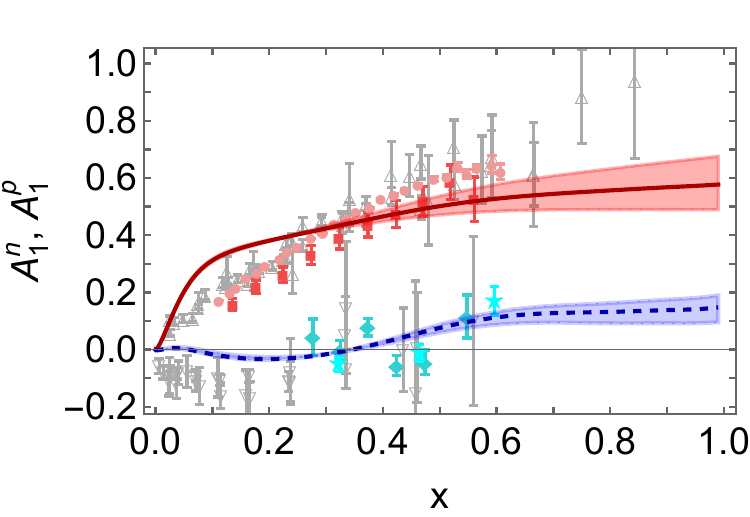}}
\caption{\label{FigA1}
Predictions for proton (solid red) and neutron (dashed blue) longitudinal spin asymmetries at $\zeta_{\rm C}$.  In each case, the associated band expresses the uncertainty deriving from Eq.\,\eqref{ratiofunctions}.
Data.
$A_1^p$:
red squares -- CLAS EG1 \cite{CLAS:2006ozz, CLAS:2008xos, CLAS:2015otq, CLAS:2017qga};
pink circles -- Ref.\,\cite{CLAS:2014qtg};
grey up-triangles -- Refs.\,\cite{E155:1999pwm, E155:2000qdr, HERMES:1998cbu, E143:1995clm, E143:1996vck, E143:1998hbs, SpinMuonSMC:1994met, SpinMuonSMC:1997mkb}.
$A_1^n$: turquoise diamonds -- E06-014 \cite{JeffersonLabHallA:2014mam};
aqua five-pointed stars -- E99-117 \cite{JeffersonLabHallA:2003joy, JeffersonLabHallA:2004tea};
grey down-triangles -- Refs.\,\cite{HERMES:1997hjr, E154:1997xfa, E154:1997ysl, E142:1996thl}.
}
\end{figure}

In Fig.\,\ref{FigA1}, I present predictions for nucleon longitudinal spin asymmetries, which are defined, for example, as in Chapter~4.7 of Ref.\,\cite{Ellis:1996mzs}.  (The contribution of $c$ quarks is practically negligible at this scale.)
For context, I also present results inferred from data collected within the past vicennium \cite{CLAS:2006ozz, CLAS:2008xos, CLAS:2015otq, CLAS:2017qga, JeffersonLabHallA:2014mam, JeffersonLabHallA:2003joy, JeffersonLabHallA:2004tea, CLAS:2014qtg}, along with selected earlier results \cite{E155:1999pwm, E155:2000qdr, HERMES:1998cbu, E143:1995clm, E143:1996vck, E143:1998hbs, SpinMuonSMC:1994met, SpinMuonSMC:1997mkb, HERMES:1997hjr, E154:1997xfa, E154:1997ysl, E142:1996thl}. The discrepancy observed between my prediction and inferences at low-$x$ may reflect known discrepancies between CSM predictions for sea quark DFs and those obtained through phenomenological fits \cite{Lu:2022cjx, Cui:2020tdf}. On the other hand, there is general agreement between my predictions and data on $x\gtrsim 0.2$. It is highly desirable to have new experiments that can provide DF information on $x\gtrsim 0.6$; particularly in relation to the previously discussed question of helicity retention. Therefore, the analyses of recently collected data obtained by the CLAS RGC Collaboration and the E12-06-110 Collaboration are much anticipated.

\section{Polarised gluon distribution at $\zeta^2=3\,$GeV$^2$}
\begin{figure}[t]

\leftline{\hspace*{4.5em}{{\textsf{(a)}}}}
\vspace*{-4ex}
\centerline{\includegraphics[width=3.4in]{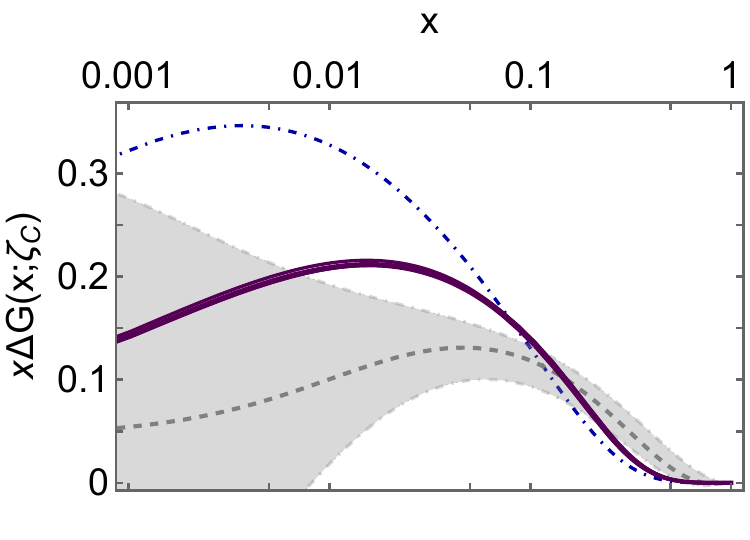}}
\vspace*{-3ex}

\leftline{\hspace*{4.5em}{{\textsf{(b)}}}}
\vspace*{-4ex}
\centerline{\includegraphics[width=3.4in]{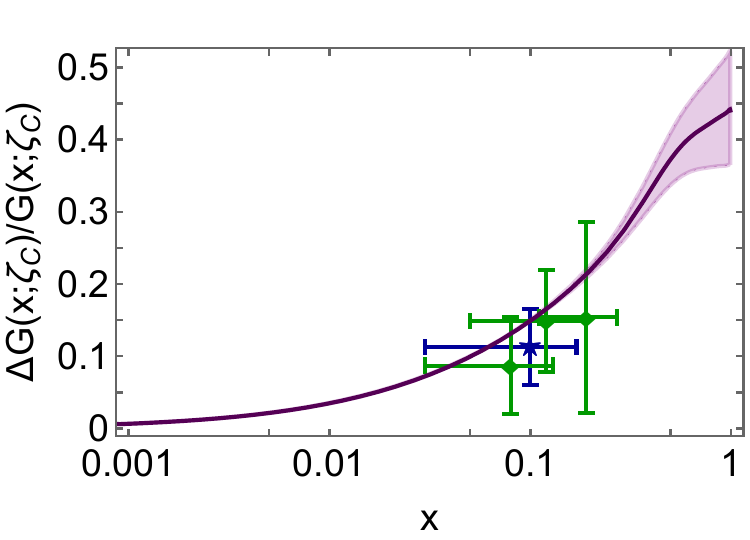}}

\caption{\label{Figgluon}
\textsf{(a)}
Polarised gluon DF: $\Delta G(x;\zeta_{\rm C})$ -- solid purple curves.
Gray dashed curve and band: typical result from phenomenological global fit \cite{deFlorian:2014yva} reported at $\zeta=10\,$GeV.
Evolved to this scale, my central prediction is the blue dot-dashed curve.
\textsf{(b)}
Polarised/unpolarised DF ratio $\Delta G(x;\zeta_{\rm C})/G(x;\zeta_{\rm C})$. The associated band expresses the uncertainty deriving from Eq.\,\eqref{ratiofunctions}.
For context, I depict values reported in Ref.\,\cite{COMPASS:2015pim}.
}
\end{figure}

Starting with the hadron-scale DFs depicted in Fig.\,\ref{FigPolarised}, the all-orders evolution scheme provides the polarised and unpolarised gluon DFs at any scale $\zeta>\zeta_{\cal H}$. My predictions for $\zeta_{\rm C}$ are shown in Fig.\,\ref{Figgluon}.

Regarding phenomenological DF fits, $\Delta G(x)$ is very poorly constrained. This is depicted by the grey band in Fig.\,\ref{Figgluon}a, taken from Ref.\,\cite{deFlorian:2014yva} at $\zeta=10\,$GeV. At this scale, my central result is represented by the dot-dashed (blue) curve. Clearly, similar to unpolarised DFs, on $x\lesssim 0.05$, my internally consistent predictions for glue (and sea) DFs have larger magnitudes compared to those derived from phenomenological fits \cite{Lu:2022cjx, Cui:2020tdf}. Notwithstanding this, on the complementary domain I obtain 
\begin{equation}
\int_{0.05}^1 dx \, \Delta G(x;\zeta=10\,{\rm GeV}) = 0.199(3)\,,
\end{equation}
in contrast to $0.19(6)$ reported in Ref.\,\cite{deFlorian:2014yva}.

My prediction for the ratio $\Delta G(x;\zeta_{\rm C})/G(x;\zeta_{\rm C})$ is shown in Fig.\,\ref{Figgluon}b. It exhibits excellent agreement with the results presented in Ref.\,\cite{COMPASS:2015pim}. 
This is highlighted, for instance, by comparing the mean value of my result on the domain covered by measurements, which is $0.167(3)$, to the value $ 0.113 \pm 0.038 \pm 0.036$ reported in Ref.\,\cite{COMPASS:2015pim}.

\section{Proton spin}
It is now pertinent to recall Eq.\,\eqref{isoscalaraxialcharge}, which records that $\approx 65$\% of the proton spin is attributed to valence quark quasiparticle degrees of freedom at the hadron scale. Importantly, this value remains scale-independent under {\sf P1} \cite{Yin:2023dbw}.

On the other hand, measurements of the proton spin are sensitive to the non-Abelian anomaly corrected combination \cite{Altarelli:1988nr}
\begin{equation}
\label{Eqa0E}
a_0^{\rm E}(\zeta) = a_0 - n_f \frac{\hat \alpha(\zeta)}{2\pi} \int_0^1 dx\,\Delta G(x;\zeta)
=: a_0 - n_f \frac{\hat \alpha(\zeta)}{2\pi} \Delta G(\zeta)\,,
\end{equation}
where $n_f$ represents the number of active quark flavours: herein, $\hat\alpha$ and {\sf P1} evolution are defined using $n_f=4$.

By utilising the result for $\Delta G(x;\zeta_{\rm C})$ from Fig.\,\ref{Figgluon} to compute the right-hand side of Eq.\,\eqref{Eqa0E}, I predict
\begin{equation}
a_0^{\rm E}(\zeta_{\rm C}) = 0.35(2)\,.
\end{equation}
This value demonstrates good agreement with that reported in Ref.\,\cite{COMPASS:2016jwv}: $0.32(7)$.

Notwithstanding the remarks in Chapter~\ref{chap5secintro}, it is common to present a light-front breakdown of the proton spin into contributions from quark and gluon spin and angular momenta:
\begin{equation}
\frac{1}{2} = \frac{1}{2} a_0 + L_q(\zeta)+\Delta G(\zeta)+L_g(\zeta)\,.
\end{equation}
This can be achieved unambiguously by exploiting the nature of the hadron scale, \emph{viz}.\
\begin{equation}
L_q(\zeta_{\cal H}) = \frac{1}{2} - \frac{1}{2} a_0 = 0.175\,,\quad
\Delta G(\zeta_{\cal H})=0=L_g(\zeta_{\cal H})\,,
\end{equation}
and subsequently applying {\sf P1} evolution, which is implemented with minor modifications of Eqs.\,(32) in Ref.\,\cite{Deur:2018roz}. In this manner, I obtain the following central values:
\begin{equation}
L_q(\zeta_{\rm C}) = -0.027\,,\;
\Delta G(\zeta_{\rm C}) = 1.31\,,\;
L_g(\zeta_{\rm C}) =-1.11\,.
\end{equation}
Clearly, while it starts with a positive value, the fraction of light-front quark angular momentum decreases steadily with increasing scale, reversing sign at a modest value of the resolving scale. Meanwhile, the increasing gluon helicity is compensated by a growth in magnitude of the light-front gluon angular momentum fraction. The asymptotic ($\zeta\to\infty$) limits are discussed elsewhere \cite{Ji:1995cu, Chen:2011gn}.

\section{Summary}
Starting with \emph{Ans\"atze} for hadron-scale proton polarised valence quark distribution functions, which were developed by using insights from perturbative QCD and constrained by solutions of a quark+diquark Faddeev equation, and supposing the existence of an effective charge which defines an evolution scheme for parton DFs (both unpolarised and polarised) that is all-orders exact, I have provided parameter-free predictions for all proton polarised DFs at the scale $\zeta_{\rm C}^2 = 3\,$GeV$^2$. In doing so, I achieved a unification of proton and pion DFs. All predictions, both pointwise behaviour and moments, show favorable agreement with results derived from data. This is exemplified by the results for polarised quark DFs [Fig.\,\ref{polzetaCa}] and nucleon longitudinal spin asymmetries [Fig.\,\ref{FigA1}].

Of particular significance is the discovery that the proton's polarised gluon DF, $\Delta G(x;\zeta_{\rm C})$, is positive and large [Fig.\,\ref{Figgluon}]. This prediction can be tested through experiments at next-generation QCD facilities \cite{Anderle:2021wcy, AbdulKhalek:2021gbh}. Moreover, the $\Delta G(x;\zeta_{\rm C})$ result allows me to predict that measurements of the proton singlet axial charge should yield a value $a_0^{\rm E}(\zeta_{\rm C}) = 0.35(2)$. This result aligns well with current data.

In the future, this analysis can be put to the test and improved by using a recently developed symmetry-preserving axial current, which is suitable for a proton described as a bound state of dressed-quark and fully interacting nonpointlike diquark degrees-of-freedom, to calculate the proton's hadron-scale polarised valence quark DFs. A first step in this direction is already in progress, using a careful treatment of a momentum-independent quark+quark interaction. The subsequent natural progression involves extending it to a study of the proton using QCD-connected Schwinger functions. Longer term objectives encompass analogous calculations that initiate with a Poincar\'e-covariant three-body treatment of the nucleon bound state problem \cite{Eichmann:2009qa, Wang:2018kto}.
\chapter{Conclusions and outlook}\label{chapter6}

\section{Key results}

The Standard Model (SM) has one known source of mass, \textit{i.e.}, the Higgs boson (HB), providing critical contributions to the evolution of the Universe. However, it is now plain that the Higgs boson is only responsible for 9 MeV out of the proton's total mass of 939 MeV, accounting for merely $1\%$ of the mass of visible material in the Universe. 
Plainly, Nature has another highly effective mechanism for generating mass, which has come to be known as EHM. 
EHM single-handedly accounts for $94\%$ of the proton's mass, with the remaining $5\%$ arising from constructive EHM + HB interference. 
The past decade of progress using CSMs has revealed the three pillars that support the EHM edifice; namely, a large dynamically generated gluon mass-scale, a process-independent effective charge, and dressed-quarks whose running masses reach a constituent-like scale at infrared momenta. 
Currently, theoretical efforts are underway to reveal the manifold and diverse manifestations of these pillars in hadron observables. Moreover, these efforts aim to emphasise the various types of measurements that can be conducted to validate the EHM paradigm. In this thesis, I exposed some of the roles played by EHM in forming the structure of the baryon octet, especially the proton, based on continuum Schwinger function methods (CSMs), especially the Dyson-Schwinger equations (DSEs). 

In this thesis, I first introduced the DSEs via a symmetry-preserving treatment of a vector × vector contact interaction (SCI), explaining: the gap equation for a dressed quark; 
the Bethe-Salpeter equation (BSE) that describes two-body scattering and bound-state problems; and the Faddeev equation appropriate for baryons treated as quark-plus-interacting-diquark bound states. 
These are the basic elements for studying baryon structure. 

Subsequently, using the SCI, predictions were delivered for octet baryon axial, induced pseudoscalar, and pseudoscalar form factors, thereby furthering progress toward a goal of unifying an array of baryon properties with analogous treatments of semileptonic decays of heavy+heavy and heavy+light pseudoscalar mesons to both pseudoscalar and vector meson final states. 
Since the approach is symmetry preserving, all mathematical and physical expressions of partial conservation of the axial current are manifest. 

The SCI results for the axial form factor, $G_A$, indicate a remarkable degree of agreement  with notions deriving from SU(3)-flavour symmetry, an outcome which can be identified as a dynamical consequence of EHM. 
Moreover, the spin-flavour structure of the Poincar\'e-covariant baryon wave functions, formulated in the presence of both flavour-antitriplet scalar diquarks and flavour-sextet axial-vector diquarks, plays a crucial role in determining all form factors. 
Taking neutral axial currents into account, SCI makes predictions for the flavour decomposition of octet baryon axial charges, which yield values for the associated SU(3) singlet, triplet, and octet axial charges. 
The findings reveal that, at the hadron scale, $\zeta_{\cal H}$, approximately 40\% of the proton's total spin is carried by valence degrees of freedom. 
Given that no other degrees of freedom exist at $\zeta_{\cal H}$, the remaining spin can be attributed to quark+diquark orbital angular momentum. 

As a continuation of the work on the proton's axial charge, I calculated the angular momentum decomposition of the proton's canonical normalisation and axial charge. 
The SCI predictions for the proton's canonical normalisation are in agreement with the results obtained using the QCD-kindred framework: the proton's canonical normalisation constant is dominated by $\mathsf S$-wave components; yet, there are also destructive $\mathsf P$-wave contributions and strong, constructive $\mathsf S \otimes \mathsf P$-wave interference terms. The results for contributions of the various quark+diquark orbital angular momentum components to the proton's singlet axial charge, $g_A^{(0)}$, and its flavour separation $g_A^{u}$ and $-g_A^{d}$ are similar to those of the canonical normalisation. It is interesting that the ratios of $g_A^{d}$ to $g_A^{u}$ for the various quark+diquark orbital angular momentum components are the roughly the same, all being $\approx -0.50$.

Continuing by formulating \emph{Ans\"atze} for proton hadron-scale polarised valence quark distribution functions (DFs), developed using insights from perturbative QCD and constrained by solutions of a quark+diquark Faddeev equation, and supposing the existence of an effective charge which defines an evolution scheme for parton DFs -– both unpolarised and polarised –- that is all-orders exact, predictions were delivered for all proton polarised DFs at the scale $\zeta_{\rm C}^2 = 3\,$GeV$^2$. The predicted DFs, both in terms of pointwise behaviour and their moments, agree favourably with results inferred from data. Of particular significance is my finding that the polarised gluon DF, $\Delta G(x;\zeta_{\rm C})$, in the proton is positive and large. This prediction can be tested through experiments at next-generation QCD facilities \cite{Anderle:2021wcy, AbdulKhalek:2021gbh}. Meanwhile, based on my result for $\Delta G(x;\zeta_{\rm C})$, I was able to predict that measurements of the proton singlet axial charge should return a value $a_0^{\rm E}(\zeta_{\rm C}) = 0.35(2)$. This result is in accord with contemporary data.

\section{Outlook}

\subsection{Nucleon Resonance Electroexcitation}
In the past decades, the CLAS Collaboration at Jefferson Laboratory in the USA have obtained a great deal of data on the electroexcitation amplitudes of nucleon resonances, including $\Delta(1232)3/2^+$, $N(1440)1/2^+$, $N(1520)3/2^-$, $N(1535)1/2^-$, \linebreak $N(1675)5/2^-$ and $\Delta(1600)3/2^+$. These data ``demand'' a comparison with sound theory predictions. The QCD-kindred framework has achieved a successful description of $\Delta(1232)3/2^+$, $N(1440)1/2^+$ and $\Delta(1600)3/2^+$ electroexcitation form factors. A unified description of the others using this framework would offer new opportunities for charting the momentum dependence of the dressed quark mass, which is one of the three pillars of EHM. So, it would play a very important role in understanding the strong interaction dynamics that govern the emergence of hadron mass.  These things are highlighted elsewhere \cite{Carman:2023zke}.

\subsection{Semileptonic decay of heavy baryons}
An extension of the analyses in Chapter.\ref{chapter3} to baryons containing one or more heavy quarks would be valuable. CP violation (CPV) has been established in the $K$, $B$ and $D$ meson systems, but not yet in analogous baryon systems. Therefore, exploring baryon CPV is one of the most important missions in both experimental and theoretical flavour physics. The semileptonic decay of heavy baryons can shed light on CPV. In addition, $\Lambda_b \rightarrow \Lambda_c e^{-} \bar{\nu}_e$ may play an important role in testing lepton flavour universality.  

\subsection{Distribution functions (DFs) of proton}
The proton's polarised DFs can help understand its spin structure, \textit{i.e.}, resolve the proton spin crisis. 
This thesis has made an important contribution, but it can be improved.  I provided an analysis of the proton's polarised DFs based on \emph{Ans\"atze}.  
Direct calculation of proton polarised DFs using wave functions calculated from the Faddeev equation is a necessary next step.

Plainly, proton DFs provide key information on proton structure, and proton one-dimensional DFs have been the focus of experiment and theory for more than fifty years.
However, little is known about proton three-dimensional DFs. 
So, drawing a three-dimensional image of the proton is a principal focus of many experimental programmes worldwide, such as the USA electron-ion collider (EIC) and the electron-ion collider in China (EicC). 
The analyses in this thesis can readily be extended to tackle the associated theory challenge.

\begin{appendices}
\chapter{Conventions and techniques}
\label{AppendixA}

\section{Euclidean metric}\label{appEuclid}

Throughout this thesis, all the work is discussed in Euclidean space. There are deep theoretical reasons for adopting this approach.  In fact, it is probable that interacting quantum field theories can only be rigorously defined in Euclidean space.  This is discussed, \textit{e.g}., in Ref.\,\cite{Ding:2022ows} (Sec.\,1).

The metric tensor in the Euclidean conventions are
\begin{equation}
a \cdot b=\delta_{\mu \nu} a_\mu b_\nu=\sum_{\mu=1}^4 a_{\mu} b_{\mu}\,,
\end{equation}
where $\delta_{\mu \nu}$ is the Kronecker delta. Hence, a space-like $Q_{\mu}$ has $Q^2>0$.

The Dirac matrices are hermitian, $\gamma_\mu^{\dagger}=\gamma_\mu$, and satisfy the algebra
\begin{equation}
\left\{\gamma_\mu, \gamma_\nu\right\}=2 \delta_{\mu \nu}\,.
\end{equation}
I define
\begin{equation}
\sigma_{\mu \nu} =\frac{i}{2}\left[\gamma_{\mu}, \gamma_{\nu}\right]\,, \quad 
\gamma_{5} =-\gamma_{1} \gamma_{2} \gamma_{3} \gamma_{4}\,,
\end{equation}
so that
\begin{equation}
\operatorname{tr}\left[\gamma_{5} \gamma_{\mu} \gamma_{\nu} \gamma_{\rho} \gamma_{\sigma}\right]=-4 \varepsilon_{\mu \nu \rho \sigma}\,, \epsilon_{1234}=1\,,
\end{equation}
where $\varepsilon_{\mu \nu \rho \sigma}$ is the completely antisymmetric Levi-Civita tensor in $\mathrm{d}=4$ dimensions. 

A Dirac-like representation of these matrices is used herein (chiral representation)
\begin{equation}
\vec{\gamma}=-i \vec{\gamma}_{M}=\left(\begin{array}{cc}
0 & -i \vec{\sigma} \\
i \vec{\sigma} & 0
\end{array}\right)\,, \quad \gamma_{4}=\gamma_{M}^{0}=\left(\begin{array}{cc}
\sigma_{0} & 0 \\
0 & -\sigma_{0}
\end{array}\right)\,,
\end{equation}
where the $2 \times 2$ Pauli matrices are
\begin{equation}
\sigma_1=\left(\begin{array}{cc}
0 & 1 \\
1 & 0
\end{array}\right)\,, \quad \sigma_2=\left(\begin{array}{cc}
0 & -i \\
i & 0
\end{array}\right)\,, \quad \sigma_3=\left(\begin{array}{cc}
1 & 0 \\
0 & -1
\end{array}\right)\,.
\end{equation}
I use
\begin{equation}
\gamma_{5}=
\left(\begin{array}{llll}
0 & 0 & 1 & 0 \\
0 & 0 & 0 & 1 \\
1 & 0 & 0 & 0 \\
0 & 1 & 0 & 0
\end{array}\right)\,.
\end{equation}

As explained in Ref.\,\cite{Roberts:1994dr} (Sec.\,2.3), one may obtain the Euclidean version of any Minkowski space expression by using the following transcription rules:
\begin{gather}
\begin{array}{ll}
\text{Configuration space} & \text{Momentum space} \\
(1) \int^{M} d^{4} x^{M} \rightarrow-i \int^{E} d^{4} x^{E} & (1) \int^{M} d^{4} k^{M} \rightarrow i \int^{E} d^{4} k^{E} \\
(2) \slashed{\partial} \rightarrow i \gamma^{E} \cdot \partial^{E} & (2) \slashed{k} \rightarrow-i \gamma^{E} \cdot k^{E} \\
(3) \slashed{A} \rightarrow-i \gamma^{E} \cdot A^{E} & (3) \slashed{A} \rightarrow-i \gamma^{E} \cdot A^{E} \\
(4) A_{\mu} B^{\mu} \rightarrow-A^{E} \cdot B^{E} & (4) k_{\mu} q^{\mu} \rightarrow-k^{E} \cdot q^{E} \\
(5) x^{\mu} \partial_{\mu} \rightarrow x^{E} \cdot \partial^{E} & (5) k_{\mu} x^{\mu} \rightarrow-k^{E} \cdot x^{E}
\end{array}
\end{gather}
where $\slashed{A}$ represents $g_{\mu \nu} \gamma_{M}^{\mu} A_{M}^{\nu}$. These transcription rules can be used as a blind implementation of an analytic continuation in the time variable, $x: x^{0} \rightarrow-i x^{4}$ with $\vec{x}^{M} \rightarrow \vec{x}^{E}$ and the same for the momentum $k$.
One also obtain $g_{\mu\nu} \rightarrow -\delta_{\mu\nu}$.

\section{Relevant expressions and relations}\label{Techn}
Here I record some of the most relevant relations used to derive SCI results and identities.

There are some useful expressions used in this thesis:
\begin{subequations}
\begin{align}   
\int \frac{d^{4} q}{(2 \pi)^{4}} \frac{1}{q^{2}+\omega}& = \frac{1}{16 \pi^{2}} \mathcal{C}^{i u}(\omega)\,, \\
\int \frac{d^{4} q}{(2 \pi)^{4}} \frac{\omega}{\left(q^{2}+\omega\right)^{2}}& =\frac{1}{16 \pi^{2}} \mathcal{C}_{1}^{i u}(\omega)\,, \\
\int \frac{d^{4} q}{(2 \pi)^{4}} \frac{1}{\left(q^{2}+\omega\right)^{2}}& =\frac{1}{16 \pi^{2}}  \overline{\mathcal{C}}_{1}^{i u}(\omega)\,.
\end{align}
\end{subequations}

Some useful integrals are listed here:
\begin{subequations}
\begin{align}
 \frac{1}{(n-1) !} \int_{0}^{\infty} d \tau \tau^{n-1} \exp (-\tau X)& = \frac{1}{X^n}\,, \\ 
 \int \frac{d^4 q}{(2 \pi)^4}(q \cdot P) \mathrm{F}\left(q^2, P^2\right)& = 0\,, \\
 \int \frac{d^4 q}{(2 \pi)^4} q_\alpha q_\beta \mathrm{F}\left(q^2\right)& = \frac{1}{4} \int \frac{d^4 q}{(2 \pi)^4} q^2 \delta_{\alpha \beta} \mathrm{~F}\left(q^2\right)\,, \\
 \int \frac{d^4 q}{(2 \pi)^4} q_\alpha q_\beta q_\mu q_\nu \mathrm{F}\left(q^2\right)&= \frac{1}{24} \int \frac{d^4 q}{(2 \pi)^4} q^4\left(\delta_{\alpha \beta} \delta_{\mu \nu}\right. \\ \nonumber
& \left.+ \delta_{\alpha \mu} \delta_{\beta \nu} +\delta_{\alpha \nu} \delta_{\beta \mu}\right) \mathrm{F}\left(q^2\right)\,.
\end{align}
\end{subequations}

Feynman parametrisations used in this thesis are
\begin{subequations}
\begin{align}
\frac{1}{D_1 D_2} & =\int_0^1 d \alpha \frac{1}{\left[(1-\alpha) D_1+\alpha D_2\right]^2}\,, \\
\frac{1}{D_1 D_2 D_3} & =\int_0^1 d \alpha \int_0^1 d \beta \frac{2 \alpha}{\left[(1-\alpha) D_1+\alpha(1-\beta) D_2+\alpha \beta D_3\right]^3}\,.
\end{align}
\end{subequations}

\section{colour and flavour coefficients}
\label{CFcoeffi}
The baryon Faddeev equations need to be augmented with appropriate colour and flavour coefficients. 
Using the colour and flavour matrices of the diquark amplitudes,
Eqs.\eqref{BSAstrucdiq}, and the quark+diquark amplitudes, Eqs. \eqref{FspaceC}, one can calculate the colour and flavour coefficients that appear in the Faddeev equation, \textit{i.e.} Eq.\eqref{FaddeevEqn}. 

The colour coefficients can be obtained by 
\begin{equation}
\label{FaddeevCol}
\frac{(\lambda_c^0)_{BA}}{\sqrt{3}}{(H_c^D)}_{AE}{(H_c^B)^T}_{EC}\frac{(\lambda_c^0)_{CD}}{\sqrt{3}}=\frac{\delta_{B A}}{\sqrt{3}} \epsilon_{A E D} \epsilon_{C E B} \frac{\delta_{C D}}{\sqrt{3}}=-2\,.
\end{equation}

For the flavour coefficients, I take the calculation of the flavour coefficient of $K_{12}$ in Eq.\eqref{KmatrixFla} as an example:
\begin{equation}
f_s^T t^{2=[us]} (t^{1=[ud]})^T f_d=\left(\begin{array}{lll}
0 & 0 & 1
\end{array}\right)\left(\begin{array}{ccc}
0 & 0 & 1\\
0 & 0 & 0\\
-1 & 0 & 0
\end{array}\right)\left(\begin{array}{ccc}
0 & 1 & 0 \\
-1 & 0 & 0 \\
0 & 0 & 0
\end{array}\right)^T\left(\begin{array}{l}
0 \\
1 \\
0
\end{array}\right)=1\,.
\end{equation}
Others can be obtained in a similar way.
\chapter{Selected octet baryons Faddeev equations}
\label{OctetFaddeev}
For notational convenience, I define
\begin{subequations}
\begin{align}
c_B^f & =\frac{g_8^2}{4 \pi^2 M_f}, \\
\sigma_B^{f,i} & =(1-\alpha) M_f^2+\alpha m_i^2-\alpha(1-\alpha) m_B^2.
\end{align}
\end{subequations}
where $g_8$ is defined in connection with Eq.\,\eqref{staticapproximation}, $f$ labels a quark flavour, $m_B$ is the baryon's mass, and $i$ is the diquark label associated with Eq.\eqref{flavourarrays}, so that $m_i$ is the mass of the associated correlation. 
Here I should note that this result is a modification of the results from Ref.\,\cite{Chen:2012qr}, which returns my result in the SU(3)-flavour symmetry limit.

\section{Proton}\label{FaddeevProton}
As previously noted, the proton's Faddeev amplitude simplifies to the form in Eqs.\,\eqref{basisoctet} by using Eq.\eqref{staticapproximation}, \textit{i.e.} 
\begin{equation}
\mathcal{S}^P(P)=s_P^1(P) \boldsymbol{I}_{\mathrm{D}}, \;
\mathcal{A}_\mu^{P i}(P)=a_{P 1}^i(P) i \gamma_5 \gamma_\mu+a_{P 2}^i(P) \gamma_5 \hat{P}_\mu, i=4,5\,,
\end{equation}
where the superscripts are diquark enumeration labels associated with Eq.\eqref{flavourarrays}. 

Similar to the analysis in Sec.\ref{ExplicitExp}, the proton's Faddeev equation is written
\begin{equation}
\label{Nucleon}
\left(\begin{array}{c}
s_1^P(P) \\
a_{P 1}^4(P) \\
a_{P 1}^5(P) \\
a_{P 2}^4(P) \\
a_{P 2}^5(P)
\end{array}\right)=\left(\begin{array}{ccccc}
\mathcal{K}_{11}^P & -\sqrt{2} \mathcal{K}_{1 4_1}^P & \mathcal{K}_{1 4_1}^P & -\sqrt{2} \mathcal{K}_{1 4_2}^P & \mathcal{K}_{1 4_{2}}^P \\
-\sqrt{2} \mathcal{K}_{4_1 1}^P & 0 & \sqrt{2} \mathcal{K}_{4_1 4_1}^P & 0 & \sqrt{2} \mathcal{K}_{4_1 4_2}^P \\
\mathcal{K}_{4_1 1}^P & \sqrt{2} \mathcal{K}_{4_1 4_1}^P & \mathcal{K}_{4_1 4_1}^P & \sqrt{2} \mathcal{K}_{4_1 4_2}^P & \mathcal{K}_{4_1 4_2}^P \\
-\sqrt{2} \mathcal{K}_{4_2 1}^P & 0 & \sqrt{2} \mathcal{K}_{4_2 4_1}^P & 0 & \sqrt{2} \mathcal{K}_{4_2 4_2}^P \\
\mathcal{K}_{4_2 1}^P & \sqrt{2} \mathcal{K}_{4_2 4_1}^P & \mathcal{K}_{4_2 4_1}^P & \sqrt{2} \mathcal{K}_{4_2 4_2}^P & \mathcal{K}_{4_2 4_2}^P
\end{array}\right)\left(\begin{array}{c}
s_1^P(P) \\
a_{P 1}^4(P) \\
a_{P 1}^5(P) \\
a_{P 2}^4(P) \\
a_{P 2}^5(P)
\end{array}\right)\,,
\end{equation}
where isospin symmetry has been used, so that the Bethe–Salpeter amplitudes for the $\{uu\}_{1^+}$ and $\{ud\}_{1^+}$ correlations are identical. 

Here it is necessary to explain the subscripts of $\mathcal{K}$: \textit{e.g.}, in $\mathcal{K}_{4_2 4_1}^P$, the label 4 is the diquark label of proton flavour structure associated with Eq.\,\eqref{flavourarrays}, and the label 1 (2) of $4_1$ ($4_2$) corresponds to the first (second) term of the axial-vector component, \textit{i.e.} $a_{P 1}^4$ ($a_{P 2}^4$). 

Using isospin symmetry, it is straightforward to show that
\begin{equation}
\label{NIspSymmetry}
a_{P j}^{4}(P)=-\sqrt{2} a_{P j}^{5}(P), j=1,2\,.
\end{equation}
Hence, Eq.\,\eqref{Nucleon} can be reduced to
\begin{equation}
\label{NucleonSimplify}
\left(\begin{array}{c}
s_1^P(P) \\
a_{P 1}^5(P) \\
a_{P 2}^5(P)
\end{array}\right)=\left(\begin{array}{ccccc}
\mathcal{K}_{11}^P & 3 \mathcal{K}_{14_1}^P & 3 \mathcal{K}_{14_2}^P \\
\mathcal{K}_{4_1 1}^P & -\mathcal{K}_{4_1 4_1}^P & -\mathcal{K}_{4_1 4_2}^P \\
\mathcal{K}_{4_2 1}^P & -\mathcal{K}_{4_2 4_1}^P & -\mathcal{K}_{4_2 4_2}^P 
\end{array}\right)\left(\begin{array}{c}
s_1^P(P) \\
a_{P 1}^5(P) \\
a_{P 2}^5(P)
\end{array}\right)\,.
\end{equation}

The entries in the proton's Faddeev kernel can be expressed as follows:
\begin{subequations}
\label{KernalProton}
\begin{align}
\mathcal{K}_{11}^P & =c_P^{u} \int_0^1 d \alpha \overline{\mathcal{C}}_1\left(\sigma_P^{u,1}\right)\left(\alpha m_P+M_u\right)(E_1-(1-\alpha)\frac{m_P}{M_u}F_1)^2\,, \\
\mathcal{K}_{1 4_1}^P & =c_P^{u} \frac{E_4}{m_4^2} \int_0^1 d \alpha \overline{\mathcal{C}}_1\left(\sigma_P^{u,4}\right)\left[\left(\left(3 M_u+\alpha m_P\right)m_4^2+2 \alpha(1-\alpha)^2 m_P^3\right) E_1\right. \\ \nonumber
& \left.-(1-\alpha)\left(\left(M_u+3 \alpha m_P\right) m_4^2+ 2(1-\alpha)^2 M_u m_P^2\right) \frac{m_P}{M_u} F_1\right]\,, \\
\mathcal{K}_{14_2}^P&=c_P^{u} \frac{E_4}{m_4^2} \int_0^1 d \alpha \overline{\mathcal{C}}_1\left(\sigma_P^{u,4}\right)\left(\alpha m_P-M_u\right)\left[(1-\alpha)^2 m_P^2-m_4^2\right]\\ \nonumber
&\times\left[E_1+(1-\alpha)\frac{m_P}{M_u}F_1\right]\,, \\
\mathcal{K}_{4_1 1}^P&=c_P^{u} \frac{E_4}{3 m_4^2} \int_0^1 d \alpha \overline{\mathcal{C}}_1\left(\sigma_P^{u,1}\right)\left(\alpha m_P+M_u\right)\left[2 m_4^2+(1-\alpha)^2 m_P^2\right]\\ \nonumber
&\times\left[E_1-(1-\alpha) \frac{m_P}{M_u} F_1\right]\,, \\
\mathcal{K}_{4_2 1}^P&=c_P^{u} \frac{E_4}{3 m_4^2} \int_0^1 d \alpha \overline{\mathcal{C}}_1\left(\sigma_P^{u,1}\right)\left(\alpha m_P+M_u\right)\left[\left(m_4^2-4(1-\alpha)^2 m_P^2\right)E_1 \right. \\ \nonumber
& \left.+(1-\alpha)\left(5 m_4^2-2(1-\alpha)^2 m_P^2\right) \frac{m_P}{M_u} F_1\right]\,, \\
\mathcal{K}_{4_1 4_1}^P&=-c_P^{u} \frac{E_4^2}{3m_4^4} \int_0^1 d \alpha \overline{\mathcal{C}}_1\left(\sigma_P^{u,4}\right) \left[\left(4 m_4^2-(1-\alpha)^2 m_P^2\right) M_u m_4^2 \right. \\ \nonumber
& \left.+\alpha (1-\alpha)^2 \left(m_4^2+2 (1-\alpha)^2 m_P^2\right) m_P^3\right]\,, \\
\mathcal{K}_{4_1 4_2}^P&=-c_P^{u} \frac{E_4^2}{3 m_4^4} \int_0^1 d \alpha \overline{\mathcal{C}}_1\left(\sigma_P^{u,4}\right)\left[(1-\alpha)^4 m_P^4+(1-\alpha)^2 m_4^2 m_P^2-2 m_4^4\right] \\ \nonumber
&\times \left(\alpha m_P-M_u\right)\,, \\
\mathcal{K}_{4_2 4_1}^P&=c_P^{u} \frac{E_4^2}{3 m_4^4} \int_0^1 d \alpha \overline{\mathcal{C}}_1\left(\sigma_P^{u,4}\right)\left[M_u\left(m_4^4-4 (1-\alpha)^2 m_4^2 m_P^2\right.\right. \\ \nonumber
& \left.\left.+6 (1-\alpha)^4 m_P^4\right) +\alpha m_P\left(-9 m_4^4+10 (1-\alpha)^2 m_4^2 m_P^2+2 (1-\alpha)^4 m_P^4\right)\right]\,, \\
\mathcal{K}_{4_2 4_2}^P&=-c_N^{u} \frac{E_4^2}{3 m_4^4} \int_0^1 d \alpha \overline{\mathcal{C}}_1\left(\sigma_P^{u,4}\right)\left[5 m_4^4-7 (1-\alpha)^2 m_4^2 m_P^2\right. \\ \nonumber
& \left.+2 (1-\alpha)^4 m_P^4\right]\left(\alpha m_P-M_u\right)\,,
\end{align}
\end{subequations}
with $E_1$, $F_1$, $E_4$ being canonically normalised Bethe-Salpeter amplitudes for diquarks corresponding to enumeration labels $i = 1, 4$ in Eq.\,\eqref{flavourarrays}. 

Introducing a parameter $\lambda$ on the right-hand side of Eq.\,\eqref{NucleonSimplify}, one obtains an eigenvalue equation of the form $K (P^2) V(P^2) =\lambda(P^2) V(P^2)$.  One repeatedly solves this equation for increasing values of $-P^2$ to obtain the function $\lambda(P^2)$.  The desired Faddeev amplitude is obtained at that lowest-magnitude value of $P^2$ for which $\lambda(P^2)=1$.  This value of $P^2$ is the baryon's mass: $\{m_B^2 = -P^2\, | \,\lambda(P^2)=1\}$.  
(The first excited state in the given channel is found at the next value of $-P^2$ for which the eigenvalue is unity \cite{Krassnigg:2003wy}, and so on.)

\section{Kernel for the $\Lambda$ baryon}\label{FaddeevLambda}
Here I explicitly specify each entry in the Faddeev equation kernel for the $\Lambda$-baryon, Eq.\eqref{eigenequLam}:
\begin{subequations}
\label{KernalLambda}
\begin{align}
\mathcal{K}_{12}^{\Lambda}&=  \frac{c_{\Lambda}^u}{2 M_u M_{u s}} \int_0^1 d \alpha \overline{\mathcal{C}}_1\left(\sigma_{\Lambda}^{u, 2}\right)\left[\alpha m_{\Lambda}+M_u\right]\left[M_u E_1 - (1-\alpha) m_{\Lambda} F_1 \right] \\ \nonumber
&\times \left[2 M_{u s} E_2 -(1-\alpha) m_{\Lambda} F_2 \right]\,, \\
\mathcal{K}_{16_1}^{\Lambda}& = \frac{c_{\Lambda}^u E_6}{M_u m_6^2} \int_0^1 d \alpha \overline{\mathcal{C}}_1\left(\sigma_{\Lambda}^{u, 6}\right)\left[\left(3 M_u m_6^2+\alpha m_{\Lambda} \left(m_6^2+2 (1-\alpha)^2 m_{\Lambda}^2\right)\right)\right. \\ \nonumber
& \left.\times M_u E_1 -(1-\alpha)\left((m_6^2+2 (1-\alpha)^2 m_{\Lambda}^2)M_u+3 \alpha m_6^2 m_{\Lambda} \right)m_{\Lambda} F_1 \right]\,, \\
\mathcal{K}_{16_2}^{\Lambda}& = \frac{c_{\Lambda}^u E_6}{M_u m_6^2} \int_0^1 d \alpha \overline{\mathcal{C}}_1\left(\sigma_{\Lambda}^{u, 6}\right)\left[\alpha m_\lambda-M_u\right]\left[M_u E_1 +(1-\alpha) m_{\Lambda}F_1\right] \\ \nonumber
&\times \left[(1-\alpha)^2m_{\Lambda}^2-m_6^2\right]\,, \\
\mathcal{K}_{21}^{\Lambda}& = \frac{c_{\Lambda}^u}{2 M_u M_{u s}} \int_0^1 d \alpha \overline{\mathcal{C}}_1\left(\sigma_{\Lambda}^{s, 1}\right)\left[\alpha m_{\Lambda}+M_s\right]\left[M_u E_1-(1-\alpha) m_{\Lambda}F_1\right] \\ \nonumber
& \times \left[2 M_{u s} E_2 -(1-\alpha)m_{\Lambda}F_2 \right]\,, \\
\mathcal{K}_{23}^{\Lambda} &= \frac{c_{\Lambda}^s}{4 M_{u s}^2} \int_0^1 d \alpha \overline{\mathcal{C}}_1\left(\sigma_{\Lambda}^{u, 2}\right)\left[2 M_{u s}E_2-(1-\alpha)m_{\Lambda}F_2\right]^2\left[\alpha m_{\Lambda}+M_u\right]\,, \\
\mathcal{K}_{2 8_1}^{\Lambda} &= \frac{c_{\Lambda}^s E_6}{2 M_{u s} m_6^2} \int_0^1 d \alpha \overline{\mathcal{C}}_1\left(\sigma_{\Lambda}^{u, 6}\right)\left[2 M_{u s} \left(\alpha m_{\Lambda}\left(m_6^2+2(1-\alpha)^2 m_{\Lambda}^2\right) \right.\right. \\ \nonumber
& \left.\left.+3 M_u m_6^2\right) E_2 -(1-\alpha)m_{\Lambda}\left(M_u\left(m_6^2+2(1-\alpha)^2 m_{\Lambda}^2\right)+3\alpha m_6^2 m_{\Lambda} \right)F_2 \right]\,, \\
\mathcal{K}_{2 8_2}^{\Lambda} &= \frac{c_{\Lambda}^s E_6}{2 M_{u s} m_6^2} \int_0^1 d \alpha \overline{\mathcal{C}}_1\left(\sigma_{\Lambda}^{u, 6}\right)\left[\alpha m_{\Lambda}-M_u\right]\left[2 M_{u s} E_2+ (1-\alpha)m_{\Lambda}F_2\right] \\ \nonumber
&\times\left[(1-\alpha)^2m_{\Lambda}^2-m_6^2\right]\,, \\
\mathcal{K}_{6_1 1}^{\Lambda} &= \frac{c_{\Lambda}^u E_6}{3 M_u m_6^2} \int_0^1 d \alpha \overline{\mathcal{C}}_1\left(\sigma_{\Lambda}^{s, 1}\right)\left[\alpha m_{\Lambda}+M_s\right]\left[M_u E_1 -(1-\alpha) m_{\Lambda}F_1 \right] \\ \nonumber
&\times\left[2 m_6^2+(1-\alpha)^2 m_{\Lambda}^2\right]\,, \\
\mathcal{K}_{6_2 1}^{\Lambda} &= \frac{c_{\Lambda}^u E_6}{3 M_u m_6^2} \int_0^1 d \alpha \overline{\mathcal{C}}_1\left(\sigma_{\Lambda}^{s, 1}\right)\left[\alpha m_{\Lambda}+M_s\right]\left[M_u\left(m_6^2-4 (1-\alpha)^2 m_{\Lambda}^2\right)E_1 \right. \\ \nonumber
& \left.+(1-\alpha)m_{\Lambda}\left(5 m_6^2-2 (1-\alpha)^2 m_{\Lambda}^2\right)F_1 \right]\,, \\
\mathcal{K}_{6_1 3}^{\Lambda} &= \frac{c_{\Lambda}^s E_6}{6 M_{u s} m_6^2} \int_0^1 d \alpha \overline{\mathcal{C}}_1\left(\sigma_{\Lambda}^{u, 2}\right)\left[\alpha m_{\Lambda}+M_u\right]\left[2 M_{u s}E_2 -(1-\alpha)m_{\Lambda}F_2 \right] \\ \nonumber
&\times \left[2 m_6^2+(1-\alpha)^2m_{\Lambda}^2\right]\,, \\
\mathcal{K}_{6_2 3}^{\Lambda} &= \frac{c_{\Lambda}^s E_6}{6 M_{u s} m_6^2} \int_0^1 d \alpha \overline{\mathcal{C}}_1\left(\sigma_{\Lambda}^{u, 2}\right)\left[2 M_{u s}\left(m_6^2-4(1-\alpha)^2 m_{\Lambda}^2\right)E_2 \right. \\ \nonumber
& \left.+(1-\alpha)m_{\Lambda}\left(5 m_6^2-2 (1-\alpha)^2 m_{\Lambda}^2\right)F_2 \right]\left[\alpha m_{\Lambda}+M_u\right]\,, \\
\mathcal{K}_{6_1 8_1}^{\Lambda} &= -\frac{c_{\Lambda}^s E_6^2}{3 m_6^4} \int_0^1 d \alpha \overline{\mathcal{C}}_1\left(\sigma_{\Lambda}^{u, 6}\right)\left[M_u m_6^2\left(4 m_6^2-(1-\alpha)^2 m_{\Lambda}^2\right)\right. \\ \nonumber
& \left.+\alpha(1-\alpha)^2 m_{\Lambda}^3\left(m_6^2+2 (1-\alpha)^2 m_{\Lambda}^2\right)\right]\,, \\
\mathcal{K}_{6_1 8_2}^{\Lambda} &= \frac{c_{\Lambda}^s E_6^2}{3 m_6^4} \int_0^1 d \alpha \overline{\mathcal{C}}_1\left(\sigma_{\Lambda}^{u, 6}\right)\left[2 m_6^4-(1-\alpha)^2m_6^2 m_{\Lambda}^2-(1-\alpha)^4m_{\Lambda}^4\right] \\ \nonumber
&\times\left[\alpha m_{\Lambda}-M_u\right]\,, \\
\mathcal{K}_{6_2 8_1}^{\Lambda} &= \frac{c_{\Lambda}^s E_6^2}{3 m_6^4} \int_0^1 d \alpha \overline{\mathcal{C}}_1\left(\sigma_{\Lambda}^{u, 6}\right)\left[M_u\left(m_6^4-4(1-\alpha)^2 m_6^2 m_{\Lambda}^2\right.\right. \\ \nonumber
& \left.\left.+6 (1-\alpha)^4 m_{\Lambda}^4\right)+ \alpha m_{\Lambda}\left(-9 m_6^4+10(1-\alpha)^2 m_6^2 m_{\Lambda}^2+2(1-\alpha)^4 m_{\Lambda}^4\right)\right]\,, \\
\mathcal{K}_{6_2 8_2}^{\Lambda} &= -\frac{c_{\Lambda}^s E_6^2}{3 m_6^4} \int_0^1 d \alpha \overline{\mathcal{C}}_1\left(\sigma_{\Lambda}^{u, 6}\right)\left[5 m_6^4-7(1-\alpha)^2 m_6^2 m_{\Lambda}^2+2 (1-\alpha)^4 m_{\Lambda}^4\right]\\ \nonumber
&\times\left[\alpha m_{\Lambda}-M_u\right]\,,
\end{align}
\end{subequations}
with $M_{u s}=(M_u M_s)/(M_u+M_s)$ and $E_{1(2)}$, $F_{1(2)}$, $E_6$ being canonically normalised Bethe-Salpeter amplitudes for diquarks corresponding to enumeration labels 
$i = 1,2,6$ in Eq.\eqref{flavourarrays}. 

The subscripts on $\mathcal{K}$ have similar meanings to those in Appendix~\ref{FaddeevProton}, \textit{e.g.}, in $\mathcal{K}_{6_1 8_2}^{\Lambda}$, the labels 6 and 8 are the diquark labels of the $\Lambda$ flavour structure associated with Eq.\,\eqref{flavourarrays}, and the label 1 (2) of $6_1$ ($8_2$) corresponds to the first (second) term of the axial-vector component, $a_{\Lambda 1}^6$ ($a_{\Lambda 2}^8$).

\section{$\Sigma^{+}$}\label{FaddeevProton}
According to the flavour structure of the $\Sigma^{+}$ in Eq.\,\eqref{SpinFlavourSigma}, one can derive its Faddeev equation directly from that of the proton by simply making the replacement $d \to s$. However, since I assumed isospin symmetry in writing the proton's Faddeev equation, this replacement is not straightforward. Hence, I provide the complete structure here. 

The Faddeev amplitude for the $\Sigma^{+}$ baryon is expressed in terms of 
\begin{equation}
\mathcal{S}^{\Sigma}=s_{\Sigma}^2(P) \mathbf{I}_{\mathrm{D}},
\quad 
\mathcal{A}_\mu^{\Sigma i}=a_{\Sigma 1}^i(P) i \gamma_5 \gamma_\mu+a_{\Sigma 2}^i(P) \gamma_5 \hat{P}_\mu,\, i=4,6\,,
\end{equation}
where the superscripts are diquark enumeration labels, Eq.\,\eqref{flavourarrays}. 
The associated Faddeev equation is
\begin{equation}
\label{FaddeevSigma}
\left(\begin{array}{c}
s_{\Sigma}^2(P) \\
a_{\Sigma 1}^4(P) \\
a_{\Sigma 1}^6(P) \\
a_{\Sigma 2}^4(P) \\
a_{\Sigma 2}^6(P)
\end{array}\right)=\left(\begin{array}{ccccc}
\mathcal{K}_{22}^{\Sigma} & -\sqrt{2} \mathcal{K}_{24_1}^{\Sigma} & \mathcal{K}_{26_1}^{\Sigma} & -\sqrt{2} \mathcal{K}_{24_2}^{\Sigma} & \mathcal{K}_{26_2}^{\Sigma} \\
-\sqrt{2} \mathcal{K}_{4_1 2}^{\Sigma} & 0 & \sqrt{2} \mathcal{K}_{4_1 6_1}^{\Sigma} & 0 & \sqrt{2} \mathcal{K}_{4_1 6_2}^{\Sigma} \\
\mathcal{K}_{6_1 2}^{\Sigma} & \sqrt{2} \mathcal{K}_{6_1 4_1}^{\Sigma} & \mathcal{K}_{6_1 6_1}^{\Sigma} & \sqrt{2} \mathcal{K}_{6_1 4_2}^{\Sigma} & \mathcal{K}_{6_1 6_2}^{\Sigma} \\
-\sqrt{2} \mathcal{K}_{4_2 2}^{\Sigma} & 0 & \sqrt{2} \mathcal{K}_{4_2 6_1}^{\Sigma} & 0 & \sqrt{2} \mathcal{K}_{4_2 6_2}^{\Sigma} \\
\mathcal{K}_{6_2 2}^{\Sigma} & \sqrt{2} \mathcal{K}_{6_2 4_1}^{\Sigma} & \mathcal{K}_{6_2 6_1}^{\Sigma} & \sqrt{2} \mathcal{K}_{6_2 4_2}^{\Sigma} & \mathcal{K}_{6_2 6_2}^{\Sigma}
\end{array}\right)\left(\begin{array}{c}
s_{\Sigma}^2(P) \\
a_{\Sigma 1}^4(P) \\
a_{\Sigma 1}^6(P) \\
a_{\Sigma 2}^4(P) \\
a_{\Sigma 2}^6(P)
\end{array}\right)\,.
\end{equation}
Regarding the subscripts of $\mathcal{K}$, \textit{e.g.}, in $\mathcal{K}_{4_1 6_2}^{\Sigma}$, the labels 4 and 6 are the diquark labels of the $\Sigma$ flavour structure associated with Eq.\,\eqref{flavourarrays}, and the label 1 (2) of $4_1$ ($6_2$) corresponds to the first (second) term of the axial-vector component, \textit{i.e.}, $a_{\Sigma 1}^4$ ($a_{\Sigma 2}^6$). 

The entries in the Faddeev kernel can be expressed as follows:
\begin{subequations}
\label{KernalSigma}
\begin{align}
\mathcal{K}_{22}^{\Sigma} & =c_{\Sigma}^{s} \int_0^1 d \alpha \overline{\mathcal{C}}_1\left(\sigma_{\Sigma}^{u,2}\right)\left(\alpha m_{\Sigma}+M_u\right)\left[E_2-(1-\alpha)\frac{m_{\Sigma}}{2 M_{us}}F_2\right]^2\,, \\
\mathcal{K}_{2 4_1}^{\Sigma} & =c_{\Sigma}^{u} \frac{E_4}{m_4^2} \int_0^1 d \alpha \overline{\mathcal{C}}_1\left(\sigma_{\Sigma}^{s,4}\right)\left[\left(3 M_s m_4^2+\alpha m_{\Sigma} (m_4^2+2 (1-\alpha)^2 m_{\Sigma}^2)\right) E_2 \right. \\ \nonumber
& \left.-(1-\alpha)\frac{m_{\Sigma}}{2 M_{us}} \left(3 \alpha m_{\Sigma} m_4^2+M_s(m_4^2+2(1-\alpha)^2 m_{\Sigma}^2)\right) F_2\right]\,, \\
\mathcal{K}_{24_2}^{\Sigma}&=c_{\Sigma}^{u} \frac{E_4}{m_4^2} \int_0^1 d \alpha \overline{\mathcal{C}}_1\left(\sigma_{\Sigma}^{s,4}\right)\left(\alpha m_{\Sigma}-M_s\right)\left[(1-\alpha)^2 m_{\Sigma}^2-m_4^2\right]\\ \nonumber
&\times \left[E_2+(1-\alpha)\frac{m_{\Sigma}}{2M_{us}}F_2\right]\,, \\
\mathcal{K}_{2 6_1}^{\Sigma} & =c_{\Sigma}^{s} \frac{E_6}{m_6^2} \int_0^1 d \alpha \overline{\mathcal{C}}_1\left(\sigma_{\Sigma}^{u,6}\right)\left[\left(3 M_u m_6^2+\alpha m_{\Sigma} (m_6^2+2 (1-\alpha)^2 m_{\Sigma}^2)\right) E_2\right. \\ \nonumber
& \left.-(1-\alpha) \frac{m_{\Sigma}}{2 M_{us}} \left(3 \alpha m_{\Sigma} m_6^2+M_u(m_6^2+2(1-\alpha)^2 m_{\Sigma}^2)\right) F_2\right]\,,\\
\mathcal{K}_{2 6_2}^{\Sigma}&=c_{\Sigma}^{s} \frac{E_6}{m_6^2} \int_0^1 d \alpha \overline{\mathcal{C}}_1\left(\sigma_{\Sigma}^{u,6}\right)\left(\alpha m_{\Sigma}-M_u\right)\left[(1-\alpha)^2 m_{\Sigma}^2-m_6^2\right]\\ \nonumber
&\times\left[E_2+(1-\alpha)\frac{m_{\Sigma}}{2M_{us}}F_2\right]\,, \\
\mathcal{K}_{4_1 2}^{\Sigma}&=c_{\Sigma}^{u} \frac{E_4}{3 m_4^2} \int_0^1 d \alpha \overline{\mathcal{C}}_1\left(\sigma_{\Sigma}^{u,2}\right)\left(\alpha m_{\Sigma}+M_u\right)\left[2 m_4^2+(1-\alpha)^2 m_{\Sigma}^2\right]\\ \nonumber
&\times\left[E_2-(1-\alpha)\frac{m_{\Sigma}}{2 M_{us}}F_2\right]\,, \\
\mathcal{K}_{4_2 2}^{\Sigma}&=c_{\Sigma}^{u} \frac{E_4}{3 m_4^2} \int_0^1 d \alpha \overline{\mathcal{C}}_1\left(\sigma_{\Sigma}^{u,2}\right)\left(\alpha m_{\Sigma}+M_u\right)\left[\left(m_4^2-4(1-\alpha)^2 m_{\Sigma}^2\right)E_2 \right. \\ \nonumber
& \left.+(1-\alpha)\frac{m_{\Sigma}}{2M_{us}}\left(5 m_4^2-2(1-\alpha)^2 m_{\Sigma}^2\right)F_2\right]\,, \\
\mathcal{K}_{6_1 2}^{\Sigma}&=c_{\Sigma}^{s} \frac{E_6}{3 m_6^2} \int_0^1 d \alpha \overline{\mathcal{C}}_1\left(\sigma_{\Sigma}^{u,2}\right)\left(\alpha m_{\Sigma}+M_u\right)\left[2 m_6^2+(1-\alpha)^2 m_{\Sigma}^2\right]\\ \nonumber
&\times\left[E_2-(1-\alpha)\frac{m_{\Sigma}}{2 M_{us}}F_2\right]\,, \\
\mathcal{K}_{6_2 2}^{\Sigma}&=c_{\Sigma}^{s} \frac{E_6}{3 m_6^2} \int_0^1 d \alpha \overline{\mathcal{C}}_1\left(\sigma_{\Sigma}^{u,2}\right)\left(\alpha m_{\Sigma}+M_u\right)\left[\left(m_6^2-4(1-\alpha)^2 m_{\Sigma}^2\right)E_2 \right. \\ \nonumber
& \left.+(1-\alpha)\frac{m_{\Sigma}}{2M_{us}}\left(5 m_6^2-2(1-\alpha)^2 m_{\Sigma}^2\right)F_2\right]\,, \\
\mathcal{K}_{4_1 6_1}^{\Sigma}&=-c_{\Sigma}^{u} \frac{E_4 E_6}{3m_4^2 m_6^2} \int_0^1 d \alpha \overline{\mathcal{C}}_1\left(\sigma_{\Sigma}^{u,6}\right) \left[M_u\left(m_4^2\left(4 m_6^2-2 (1-\alpha)^2 m_{\Sigma}^2\right)\right.\right. \\ \nonumber
& \left.\left.+(1-\alpha)^2 m_6^2 m_{\Sigma}^2\right) +\alpha (1-\alpha)^2 m_{\Sigma}^3\left(2 m_4^2-m_6^2+2 (1-\alpha)^2 m_{\Sigma}^2\right)\right]\,, \\
\mathcal{K}_{4_1 6_2}^{\Sigma}&=-c_{\Sigma}^{u} \frac{E_4 E_6}{3 m_4^2 m_6^2} \int_0^1 d \alpha \overline{\mathcal{C}}_1\left(\sigma_{\Sigma}^{u,6}\right)\left(\alpha m_{\Sigma}-M_u\right)\left[(1-\alpha)^2 m_{\Sigma}^2-m_6^2\right]\\ \nonumber
&\times \left[(1-\alpha)^2 m_{\Sigma}^2+2 m_4^2\right]\,, \\
\mathcal{K}_{4_2 6_1}^{\Sigma}&=c_{\Sigma}^{u} \frac{E_4 E_6}{3 m_4^2 m_6^2}\int_0^1 d \alpha \overline{\mathcal{C}}_1\left(\sigma_{\Sigma}^{u,6}\right)\left[M_u\left(m_4^2(m_6^2-8 (1-\alpha)^2 m_{\Sigma}^2)\right.\right. \\ \nonumber
& \left.\left.+2 (1-\alpha)^2 m_{\Sigma}^2(2 m_6^2 +3 (1-\alpha)^2 m_{\Sigma}^2)\right) +\alpha m_{\Sigma}\left(m_4^2(2 (1-\alpha)^2 m_{\Sigma}^2-9 m_6^2) \right.\right. \\ \nonumber
& \left.\left.+2 (1-\alpha)^2 m_{\Sigma}^2(4 m_6^2+(1-\alpha)^2 m_{\Sigma}^2)\right)\right]\,, \\
\mathcal{K}_{4_2 6_2}^{\Sigma}&=-c_{\Sigma}^{u} \frac{E_4 E_6}{3 m_4^2 m_6^2}\int_0^1 d \alpha \overline{\mathcal{C}}_1\left(\sigma_{\Sigma}^{u,6}\right)\left[5 m_4^2-2 (1-\alpha)^2 m_{\Sigma}^2\right]\\ \nonumber
&\times \left[m_6^2-(1-\alpha)^2 m_{\Sigma}^2\right] \left(\alpha m_{\Sigma}-M_u\right)\,, \\
\mathcal{K}_{6_1 4_1}^{\Sigma}&=-c_{\Sigma}^{u} \frac{E_4 E_6}{3m_4^2 m_6^2} \int_0^1 d \alpha \overline{\mathcal{C}}_1\left(\sigma_{\Sigma}^{s,4}\right) \left[M_s\left( m_4^2\left(4 m_6^2+ (1-\alpha)^2 m_{\Sigma}^2\right)\right.\right. \\ \nonumber
& \left.\left.-2(1-\alpha)^2 m_6^2 m_{\Sigma}^2\right) +\alpha (1-\alpha)^2 m_{\Sigma}^3\left(2 m_6^2-m_4^2+2 (1-\alpha)^2 m_{\Sigma}^2\right)\right]\,, \\
\mathcal{K}_{6_1 4_2}^{\Sigma}&=-c_{\Sigma}^{u} \frac{E_4 E_6}{3 m_4^2 m_6^2} \int_0^1 d \alpha \overline{\mathcal{C}}_1\left(\sigma_{\Sigma}^{s,4}\right)\left[(1-\alpha)^2 m_{\Sigma}^2-m_4^2\right] \\ \nonumber
&\times \left[(1-\alpha)^2 m_{\Sigma}^2+2 m_6^2\right]\left(\alpha m_{\Sigma}-M_s\right)\,, \\
\mathcal{K}_{6_2 4_1}^{\Sigma}&=c_{\Sigma}^{u} \frac{E_4 E_6}{3 m_4^2 m_6^2}\int_0^1 d \alpha \overline{\mathcal{C}}_1\left(\sigma_{\Sigma}^{s,4}\right)\left[M_s\left(m_4^2(m_6^2+4 (1-\alpha)^2 m_{\Sigma}^2)\right.\right. \\ \nonumber
& \left.\left.+2 (1-\alpha)^2 m_{\Sigma}^2(3 (1-\alpha)^2 m_{\Sigma}^2-4m_6^2 )\right) +\alpha m_{\Sigma}\left(m_4^2(8 (1-\alpha)^2 m_{\Sigma}^2-9 m_6^2) \right.\right. \\ \nonumber
& \left.\left.+2 (1-\alpha)^2 m_{\Sigma}^2(m_6^2+(1-\alpha)^2 m_{\Sigma}^2)\right)\right]\,, \\
\mathcal{K}_{6_2 4_2}^{\Sigma}&=-c_{\Sigma}^{u} \frac{E_4 E_6}{3 m_4^2 m_6^2}\int_0^1 d \alpha \overline{\mathcal{C}}_1\left(\sigma_{\Sigma}^{s,4}\right)\left[5 m_6^2-2 (1-\alpha)^2 m_{\Sigma}^2\right]\\ \nonumber
&\times \left[m_4^2-(1-\alpha)^2 m_{\Sigma}^2\right] \left(\alpha m_{\Sigma}-M_u\right)\,, \\
\mathcal{K}_{6_1 6_1}^{\Sigma}&=-c_{\Sigma}^{s} \frac{E_6^2}{3m_6^4} \int_0^1 d \alpha \overline{\mathcal{C}}_1\left(\sigma_{\Sigma}^{u,6}\right) \left[M_u m_6^2\left(4 m_6^2-(1-\alpha)^2 m_{\Sigma}^2\right) \right. \\ \nonumber
& \left.+\alpha (1-\alpha)^2 m_{\Sigma}^3\left(m_6^2+2 (1-\alpha)^2 m_{\Sigma}^2\right)\right]\,, \\
\mathcal{K}_{6_1 6_2}^{\Sigma}&=-c_{\Sigma}^{s} \frac{E_6^2}{3 m_6^4} \int_0^1 d \alpha \overline{\mathcal{C}}_1\left(\sigma_{\Sigma}^{u,6}\right)\left[(1-\alpha)^4 m_{\Sigma}^4+(1-\alpha)^2 m_6^2 m_{\Sigma}^2-2 m_6^4\right]\\ \nonumber
&\times \left(\alpha m_{\Sigma}-M_u\right)\,, \\
\mathcal{K}_{6_2 6_1}^{\Sigma}&=c_{\Sigma}^{s} \frac{E_6^2}{3 m_6^4} \int_0^1 d \alpha \overline{\mathcal{C}}_1\left(\sigma_{\Sigma}^{u,6}\right)\left[\left(m_6^4-4 (1-\alpha)^2 m_6^2 m_{\Sigma}^2+6 (1-\alpha)^4 m_{\Sigma}^4\right) \right. \\ \nonumber
& \left.\times M_u+\alpha m_{\Sigma}\left(-9 m_6^4+10 (1-\alpha)^2 m_6^2 m_{\Sigma}^2+2 (1-\alpha)^4 m_{\Sigma}^4\right)\right]\,, \\
\mathcal{K}_{6_2 6_2}^{\Sigma}&=-c_{\Sigma}^{s} \frac{E_6^2}{3 m_6^4} \int_0^1 d \alpha \overline{\mathcal{C}}_1\left(\sigma_{\Sigma}^{u,6}\right)\left[5 m_6^4-7 (1-\alpha)^2 m_6^2 m_{\Sigma}^2+2 (1-\alpha)^4 m_{\Sigma}^4\right] \\ \nonumber
& \times \left(\alpha m_{\Sigma}-M_u\right)\,,
\end{align}
\end{subequations}
with $E_{2}$, $F_{2}$, $E_{4(6)}$ being canonically normalised Bethe-Salpeter amplitudes for diquarks corresponding to enumeration labels 
$i = 2,4,6$ in Eq.\,\eqref{flavourarrays}.

\chapter{Hadron currents} \label{AppendixCurrents}

\section{Baryon currents} \label{AppendixCurrentsBaryon}

Using the propagators and amplitudes described above, one can derive the explicit expression for the baryon current drawn in Fig.\,\ref{figcurrent}.  For convenience of reference, I recall the expression in Eq.\,\eqref{FspaceC}:
\begin{equation}
\underline{\Psi}_{\Xi^0} =\Psi_{\Xi^0}^{{\mathcal{S}}_{[us]}} \mathpzc{f}_s +\Psi_{\Xi^0}^{{\mathcal{A}}_{\{us\}}}\mathpzc{f}_s
+\Psi_{\Xi^0}^{{\mathcal{A}}_{\{ss\}}}\mathpzc{f}_u\,,
\end{equation}
where the colour matrix is omitted and $\mathpzc{f}_u={\rm column}[1,0,0]$, $\mathpzc{f}_d={\rm column}[0,1,0]$, $\mathpzc{f}_s={\rm column}[0,0,1]$. The column vector is determined by $B$ and the specified diquark.  I denote the corresponding row-vector by $\bar {\mathpzc{f}}_h$, $h=u,d,s$ and also define
\begin{equation}
\underline{S} = {\rm diagonal}[S_u, S_d, S_s]\,,
\end{equation}
where the quark propagators are derived from Sec.\,\ref{secgap},

\subsection{Diagram 1}
Probe couples directly to the bystander quark, Table~\ref{DiagramLegend}, including two contributions:
\begin{equation}
J_{5(\mu)}^1(K,Q) = J_{5(\mu)}^{qS}(K,Q)+J_{5(\mu)}^{qA}(K,Q)\,.
\end{equation}
Using the notation mentioned above, 
\begin{subequations}
\label{Diagram1Explicit}
\begin{align}
J_{5(\mu)}^{qS} &  =
\int_\ell \bar\Psi^{\mathcal{S}}_{B^\prime}(P^\prime)\bar{\mathpzc{f}}_f \underline S(\ell_{+}^\prime) \underline\Gamma_{5(\mu)}^{fg}(Q) \underline S(\ell_+)\Delta^{0^+}(-\ell) {\mathpzc f}_g \Psi^{\mathcal{S}}_B(P), \\
J_{5(\mu)}^{qA} &  =
\int_\ell \bar\Psi^{\mathcal{A}}_{B^\prime \alpha}(P^\prime) \bar{\mathpzc f}_f \underline S(\ell_{+}^\prime) \underline\Gamma_{5(\mu)}^{fg}(Q) \underline S(\ell_+)\Delta_{\alpha\beta}^{1^+}(-\ell) {\mathpzc f}_g\Psi^{\mathcal{A}}_{B\beta}(P),
\end{align}
\end{subequations}
where $\ell_\pm^{(\prime)}=\ell \pm P^{(\prime)}$, the diquark propagators are given in Eqs.\,\eqref{scalarqqprop} and \eqref{avqqprop},
and $\int_\ell$ denotes the regularised four-dimensional momentum-space integral with, matching the Faddeev equation procedure, $\Lambda_{\rm uv}$ selected as the ultraviolet cutoff linked to the lightest diquark in the $B\stackrel{g\to f}{\to} B^\prime$ process.

The remaining elements in Eqs.\,\eqref{Diagram1Explicit} are
$\underline\Gamma_{5}^{fg} := {\cal T}^{fg}\Gamma_{5}^{fg}$,
$\underline\Gamma_{5\mu}^{fg} := {\cal T}^{fg}\Gamma_{5\mu}^{fg}$,
\emph{viz}.\ the dressed-quark+pseudoscalar, -quark+axial-vector vertices that express the $g\to f$ quark transition.  Their calculation is explained in Section\ref{subsecIBSE} and I adapt the results to all $g\to f$ transitions considered in this thesis.
Notably, my implementation of the SCI guarantees the following (and other) Ward-Green-Takahashi identities ($k_+=k+Q$, $\underline{\mathpzc{m}}={\rm diagonal}[m_u,m_d,m_s]$):
\begin{align}
& Q_\mu \underline\Gamma_{5\mu}^{fg}(k_+,k)
+ i \underline{\mathpzc{m}}\,\underline\Gamma_{5}^{fg}(k_+,k)
+ i \underline\Gamma_{5}^{fg}(k_+,k) \underline {\mathpzc{m}} \nonumber \\
& = \underline S^{-1}(k_+)i\gamma_5 {\cal T}^{fg}
+ i\gamma_5 {\cal T}^{fg}\underline S^{-1}(k)\,. \label{WGTIfg}
\end{align}

\subsection{Diagram 2}
There is only one term in this case, \emph{i.e}., probe strikes axial-vector diquark with dressed-quark spectator:
\begin{equation}
\begin{aligned}
&J_{5(\mu)}^2(K,Q) = J_{5(\mu)}^{A^\prime A}(K,Q) \\
& = \int_\ell \bar\Psi^{{\mathcal{A}}^\prime}_{B^\prime \alpha}(P^\prime) \bar{\mathpzc{f}}_h \underline S(\ell) \Delta_{\alpha\rho}^{1^+}(-\ell_-^\prime) \Gamma^{A^\prime A}_{5(\mu),\rho\sigma}(-\ell_-^\prime,-\ell_-) \Delta_{\sigma\beta}^{1^+}(-\ell_-) {\mathpzc{f}}_h \Psi^{\mathcal{A}}_{B\beta}(P),
\end{aligned}
\end{equation}
where $\Gamma^{A^\prime A}_{5(\mu),\rho\sigma}$ is the axial-vector diquark pseudoscalar (axial-vector) vertex.  The associated form factors need to be calculated; and for that purpose, I employed the procedure detailed in Ref.\,\cite{Chen:2021guo}.  The results are collected in Appendix~\ref{AppendixCurrentsDiquark}, with those relevant here given in Eq.\,\eqref{GammaAA}.

\subsection{Diagram 3}
\label{AppendixDiagram3}
In this case, there are two terms, \emph{i.e}., in the presence of a dressed-quark spectator, the probe strikes an axial-vector (scalar) diquark, inducing a transition to a scalar (axial-vector) diquark. Writing the former explicitly:
\begin{equation}
\begin{aligned}
&J_{5(\mu)}^3(K,Q) = J_{5(\mu)}^{SA}(K,Q) \\
& = \int_\ell \bar\Psi^{\mathcal{S}}_{B^\prime}(P^\prime) \bar{\mathpzc{f}}_h\, \underline S(\ell) \Delta^{0^+}(-\ell_-^\prime) \Gamma^{SA}_{5(\mu),\sigma}(-\ell_-^\prime,-\ell_-) \Delta_{\sigma\beta}^{1^+}(-\ell_-) {\mathpzc{f}}_h \Psi^{\mathcal{A}}_{B\beta}(P),
\end{aligned}
\end{equation}
where $\Gamma^{SA}_{5(\mu),\sigma}$ is the axial-vector$\,\to\,$scalar diquark transition vertex. Again, it is necessary to calculate the associated form factors, which I accomplished by following the procedure outlined in Ref.\,\cite{Chen:2021guo}.  The results are collected in Appendix~\ref{AppendixCurrentsDiquark}, with those relevant here given in Eq.\,\eqref{GammaSA}.  Naturally, $\Gamma^{AS}_{5(\mu),\sigma}(\ell^\prime,\ell)=\Gamma^{SA}_{5(\mu),\sigma}(\ell,\ell^\prime)$.

\subsection{Diagram 4}
Here the probe strikes the dressed-quark that is exchanged as one diquark breaks up and another is formed:
\begin{equation}
\begin{aligned}
&J_{5(\mu)}^4(K,Q) = \sum_{J_1^{P_1}, J_2^{P_2}={\mathcal{S}}, {\mathcal{A}}}
\int_\ell \int_k
\bar\Psi^{J_2^{P_2}}_{B^\prime}(P^\prime) \bar{\mathpzc{f}}_{h^\prime} \Delta^{J_2^{P_2}}(k_{qq}) \underline S(k) \underline\Gamma^{J_1^{P_1}}(\ell_{qq}) \\
& \times \left[\underline S(k_{qq}-\ell) \underline\Gamma_{5(\mu)}^{fg}(Q) \underline S(\ell_{qq}-k)\right]^{\rm T} \bar{\underline\Gamma}^{J_2^{P_2}}(-k_{qq}) \underline S(\ell) \Delta^{J_1^{P_1}}(\ell_{qq})
{\mathpzc{f}}_h\Psi^{J_1^{P_1}}_{B}(P)\,, 
\end{aligned}
\label{DiagramFour}
\end{equation}
where $(\cdot)^{\rm T}$ denotes matrix transpose,
$\bar \Gamma(K) = C^\dagger \Gamma(K)^{\rm T} C$,
and $\ell_{qq} = -\ell +P$, $k_{qq} = -k +P^\prime$.
I have omitted Lorentz indices, which can easily be reinstated once the specific transition is indicated.

Eq.\,\eqref{DiagramFour} contains four terms; but as exploited in the enumeration of Table~\ref{DiagramLegend}, symmetry relates ${\mathcal{S}}{\mathcal{A}}$ to ${\mathcal{A}}{\mathcal{S}}$; namely, there are only three distinct contributions.

It is worth highlighting here that in emulation of the SCI formulation of the Faddeev equation in Section \ref{genralStruofFad}, I have employed a variant of the so-called ``static approximation''. Consistency with this simplification is achieved by writing
\begin{align}
& \underline S(k_{qq}-\ell) \underline\Gamma_{5(\mu)}^{fg}(Q) \underline S(\ell_{qq}-k) \nonumber \\
& \to \underline \Gamma_{5(\mu)}^{fg}(Q) g_B^2\left[\frac{1}{M_f}+\frac{1}{M_g}\right]
\frac{i\gamma\cdot Q + M_f + M_g}{Q^2+(M_f+M_g)^2}\,.
\end{align}

\subsection{Diagrams 5 and 6}
In a quark--plus--interacting-diquark picture of baryons, it is typically necessary to incorporate ``seagull terms'' to guarantee the fulfillment of relevant Ward-Green-Takahashi identities \cite{Oettel:1999gc}. Those relevant to the currents in Eqs.\,\eqref{jaxdq0} and \eqref{jpsdq0} are given in Ref.\,\cite{Chen:2021guo}.  Adapted to the SCI, they read
\begin{subequations}
\begin{align}
J_{5(\mu)}^5(K,Q) & = \sum_{J_1^{P_1}, J_2^{P_2}={\mathcal{S}}, {\mathcal{A}}}
\int_\ell \int_k
\bar\Psi^{J_2^{P_2}}_{B^\prime}(P^\prime) \bar{\mathpzc{f}}_{h^\prime} \Delta^{J_2^{P_2}}(k_{qq}) \underline S(k) \chi_{5(\mu)}^{J_1^{P_1}fg}(\ell_{qq})\nonumber \\
& \times \underline S(k_{qq}-\ell)^{\rm T} \bar{\underline\Gamma}^{J_2^{P_2}}(-k_{qq})  \underline S(\ell) \Delta^{J_1^{P_1}}(\ell_{qq}) {\mathpzc{f}}_h \Psi^{J_1^{P_1}}_{B}(P)\,, \label{DiagramFive}\\
J_{5(\mu)}^6(K,Q) & = \sum_{J_1^{P_1}, J_2^{P_2}={\mathcal{S}}, {\mathcal{A}}}
\int_\ell \int_k
\bar\Psi^{J_2^{P_2}}_{B^\prime}(P^\prime) \bar {\mathpzc{f}}_{h^\prime} \Delta^{J_2^{P_2}}(k_{qq}) \underline S(k) \underline\Gamma^{J_1^{P_1}}(\ell_{qq}) \nonumber \\
& \times  \underline S(\ell_{qq}-k)^{\rm T}\bar\chi^{J_2^{P_2}fg}_{5(\mu)}(-k_{qq})  \underline S(\ell) \Delta^{J_1^{P_1}}(\ell_{qq}) {\mathpzc{f}}_h\Psi^{J_1^{P_1}}_{B}(P)\,, \label{DiagramSix}
\end{align}
\end{subequations}
where, with $m_{P_{fg}}$ denoting the mass of the $f\bar g $ pseudoscalar meson,
\begin{subequations}
\label{SeaGulls}
\begin{align}
\chi_{5\mu}^{J^{P} fg}(Q) & = - \frac{iQ_\mu}{Q^2+m_{P_{fg}}^2}
\left[\gamma_5 {\cal T}^{fg} \Gamma^{J^P}(Q) + \Gamma^{J^P}(Q) \left(\gamma_5 {\cal T}^{fg} \right)^{\rm T}\right] ,\\
i \chi_{5}^{J^{P} fg}(Q) & =
- \frac{1}{2{\mathpzc{m}}_{fg}}\frac{im_{P_{fg}}^2}{Q^2+m_{P_{fg}}^2}
\left[\gamma_5 {\cal T}^{fg} \Gamma^{J^P}(Q) + \Gamma^{J^P}(Q) \left(\gamma_5 {\cal T}^{fg} \right)^{\rm T}\right] ,\\
\bar\chi_{5\mu}^{J^{P} fg}(Q) & = - \frac{iQ_\mu}{Q^2+m_{P_{fg}}^2}
\left[\bar\Gamma^{J^P}(Q) \gamma_5 {\cal T}^{fg} + \left(\gamma_5 {\cal T}^{fg} \right)^{\rm T}\bar\Gamma^{J^P}(Q) \right] ,\\
i\bar\chi_{5}^{J^{P} fg}(Q) & = - \frac{1}{2{\mathpzc{m}}_{fg}}\frac{im_{P_{fg}}^2}{Q^2+m_{P_{fg}}^2}
\left[\bar\Gamma^{J^P}(Q) \gamma_5 {\cal T}^{fg} + \left(\gamma_5 {\cal T}^{fg} \right)^{\rm T}\bar \Gamma^{J^P}(Q) \right].
\end{align}
\end{subequations}

It is worth noting the following identity:
\begin{align}
Q_\mu \chi_{5\mu}^{J^{P} fg}(Q)  + 2i{\mathpzc{m}}_{fg} \chi_{5}^{J^{P} fg}(Q) = -i \gamma_5 {\cal T}^{fg} \Gamma^{J^P}(Q) - \Gamma^{J^P}(Q) \left(i\gamma_5 {\cal T}^{fg} \right)^{\rm T};  \label{WGTIfgseagull}
\end{align}
and the kindred relation for the conjugate seagulls.

\section{Diquark currents}
\label{AppendixCurrentsDiquark}

\begin{figure}[t]
\centerline{%
\includegraphics[width=3.8in]{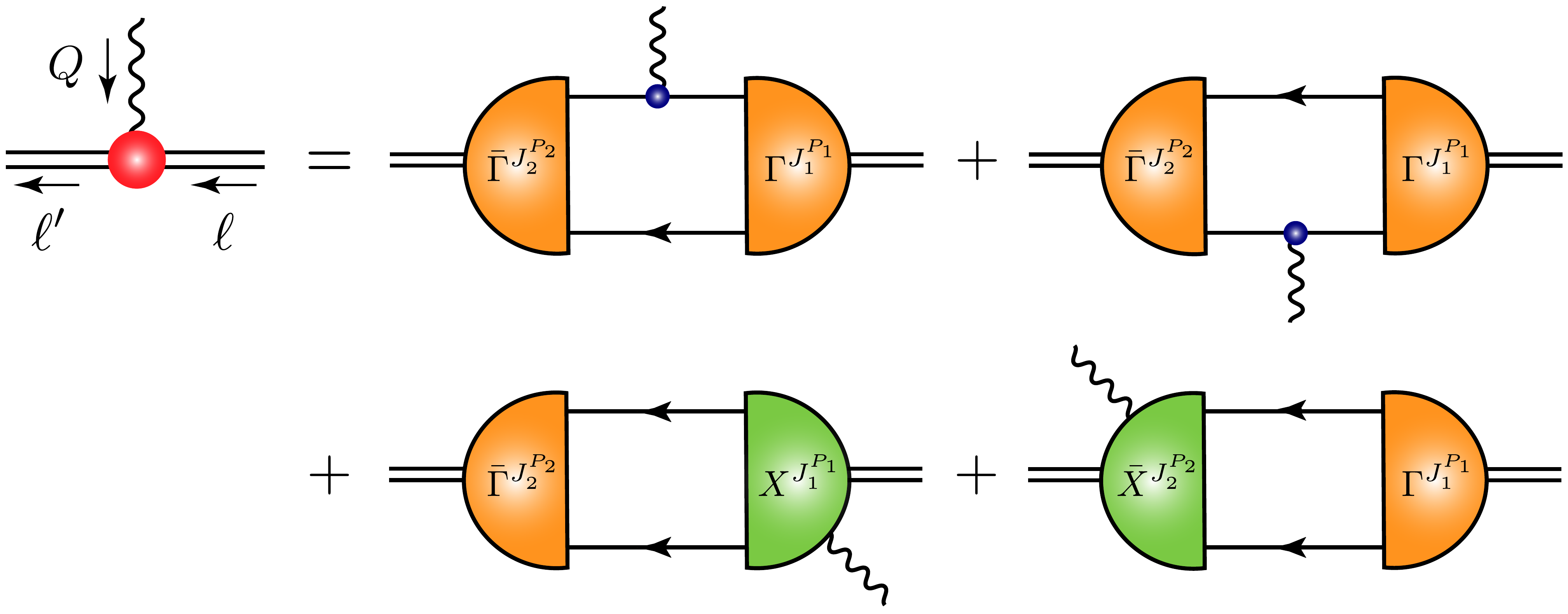}}
\caption{\label{figdqvx} Interaction vertex for the $J_1^{P_1}\to J_2^{P_2}$ diquark+probe interaction ($\ell^\prime=\ell+Q$): \emph{single line}, quark propagator; \emph{undulating  line}, pseudoscalar or axial current; $\Gamma$, diquark correlation amplitude; \emph{double line}, diquark propagator; and $\chi$, seagull interaction.}
\end{figure}

In Appendix~\ref{AppendixCurrentsBaryon}, it was demonstrated that any study of baryon axial and pseudoscalar currents utilising the quark+diquark framework of baryon structure requires knowledge of probe+diquark form factors. 
I calculated these form factors following the procedure detailed in Sec.\,III.C.4 in Ref.\,\cite{Chen:2021guo}, which employs the current illustrated in Fig.\,\ref{figdqvx}. Taking into account the systems involved, there are two form factors associated with each probe: axial-vector$\,\leftrightarrow\,$axial-vector and axial-vector$\,\leftrightarrow\,$pseudoscalar.

\subsection{Axial-vector diquark transition form factors}
Using the SCI and considering a $\{hg\}\to\{hf\}$ transition, the four diagrams depicted in Fig.\,\ref{figdqvx} can be expressed as follows:
\begin{align}
\Gamma^{AA}_{5(\mu),\rho\sigma}(\ell^\prime,\ell) & =
N_c^{\bar 3}{\rm tr}_{\rm DF}\int_t\left\{ i \bar{\underline\Gamma}_\rho^{\{h f\}}(-\ell^\prime)
\underline S(t_+^\prime)\underline\Gamma_{5(\mu)}^{fg}(Q) \underline S(t_+) i \underline\Gamma_\sigma^{\{g h\}}(\ell) \underline S(-t)^{\rm T} \right. \nonumber \\
& + i \bar{\underline\Gamma}_\rho^{\{hf\}}(-\ell^\prime) \underline S(t)  i \underline\Gamma_\sigma^{\{gh\}}(\ell) \left[ \underline S(-t_-^\prime) \underline \Gamma_{5(\mu)}^{fg}(Q) \underline S(-t_-) \right]^{\rm T} \nonumber \\
& -   \bar{\underline \Gamma}_\rho^{\{h f\}}(-\ell^\prime) \underline S(t_+^\prime) \chi_{5(\mu),\sigma}^{\{gh\}fg}(\ell) \underline S(-t)^{\rm T} \nonumber \\
& \left. -  \bar\chi_{5(\mu),\rho}^{\{hf\}fg}(-\ell^\prime) \underline S(t_+) \underline \Gamma_\sigma^{\{gh\}}(\ell) \underline S(-t)^{\rm T}\right\},
\label{GammaAA}
\end{align}
where I have made the Lorentz indices explicit, writing with reference to Eq.\,\eqref{BSAstrucdiq}, \emph{e.g}., $\underline\Gamma^{1^+}_{gh} = \underline\Gamma_\sigma^{\{g h\}}$;
$N_c^{\bar 3} = 2$ and the trace is over Dirac and flavour structure;
and $Q=\ell^\prime - \ell$, $t_\pm^{(\prime)} = t \pm \ell^{(\prime)}$.

\subsection{Axial-vector-scalar diquark transition form factors}
Similarly, for the $\{hg\} \to [hf]$ transition, the process described in Appendix~\ref{AppendixDiagram3} can be represented by the following expression:
\begin{align}
\Gamma^{SA}_{5(\mu),\sigma}(\ell^\prime,\ell) &=
N_c^{\bar 3}{\rm tr}_{\rm DF}\int_t \left\{
i \bar{\underline \Gamma}^{[hf]}(-\ell^\prime) \underline S(\ell_+^\prime) \underline\Gamma_{5(\mu)}^{fg}(Q)
\underline S(t_+) i \underline\Gamma_\sigma^{\{g h\}}(\ell) \underline S(-t)^{\rm T} \right. \nonumber \\
& + i \bar{\underline\Gamma}^{[hf]}(-\ell^\prime) \underline S(t) i \underline\Gamma_\sigma^{\{gh\}}(\ell) \left[ \underline S(-t_-^\prime) \underline \Gamma_{5(\mu)}^{fg}(Q) \underline S(-t_-) \right]^{\rm T} \nonumber \\
& -  \bar{\underline \Gamma}^{[h f]}(-\ell^\prime) \underline S(t_+^\prime) \chi_{5(\mu),\sigma}^{\{gh\}fg}(\ell) \underline S(-t)^{\rm T} \nonumber \\
& \left. -  \bar\chi_{5(\mu),\rho}^{[hf]fg}(-\ell^\prime) \underline S(t_+) \underline \Gamma_\sigma^{\{gh\}}(\ell) \underline S(-t)^{\rm T}\right\}.
\label{GammaSA}
\end{align}
As noted above, $\Gamma^{AS}_{5(\mu),\sigma}(\ell^\prime,\ell)  = \Gamma^{SA}_{5(\mu),\sigma}(\ell,\ell^\prime)$.

\subsection{Ward-Green-Takahashi identities for diquark currents}
It is worth remarking here that, by utilising Eqs.\,\eqref{WGTIfg}, \eqref{WGTIfgseagull} and kindred relations, one can readily verify the following results:
\begin{subequations}
\label{WGTIagain}
\begin{align}
0 & = Q_\mu \Gamma^{AA}_{5 \mu,\rho\sigma}(\ell^\prime,\ell) + i 2 \mathpzc{m}_{fg} \Gamma^{AA}_{5,\rho\sigma}(\ell^\prime,\ell) \,,\\
0 & = Q_\mu \Gamma^{SA}_{5 \mu,\rho}(\ell^\prime,\ell) + i 2 \mathpzc{m}_{fg}\Gamma^{SA}_{5,\rho}(\ell^\prime,\ell)\,.
\end{align}
\end{subequations}
These identities were established elsewhere \cite{Chen:2021guo}.  Being general, they can be used to constrain \emph{Ans\"atze} for the vertices involved. Nevertheless, herein, I directly calculate the SCI results.

\subsection{Probe-diquark form factors}

\begin{table}[!htbp]
\caption{\label{interpolatorcoefficientsdu}
Probe-diquark form factors for $d\to u$ transitions, which for practical purposes can be interpolated using Eq.\,\eqref{interpolator12} with the coefficients listed here.  Where written, $f=d,u$ because I assume isospin symmetry; and the absence of an entry means the coefficient is zero.
(Every $\kappa(s)$ is dimensionless; so each coefficient in Eq.\,\eqref{interpolator12} has the mass dimension necessary to cancel that of the associated $s\, ({\rm GeV}^2)$ factor.)}
\begin{center}
\begin{tabular*}
{\hsize}
{
l@{\extracolsep{0ptplus1fil}}|
c@{\extracolsep{0ptplus1fil}}
c@{\extracolsep{0ptplus1fil}}
c@{\extracolsep{0ptplus1fil}}
c@{\extracolsep{0ptplus1fil}}}\hline
$\{fd\}\to\{fu\}\ $ & $a_0\ $ & $a_1\ $ & $b_1\ $ & $b_2\ $ \\
$\kappa_{\mathpzc{p}}^{AA}$ & $0.470\ $& $\phantom{-}0.173\ $ & $0.598\ $&   \\
$\kappa_{{\mathpzc{a}}_1}^{AA}$ & $0.467\ $& $\phantom{-}0.023\ $ & $0.598\ $&   \\
$\kappa_{{\mathpzc{a}}_2}^{AA}$ & $0.470\ $& $\phantom{-}0.023\ $ & $0.598\ $&   \\
$\kappa_{{\mathpzc{a}}_3}^{AA}$ & & & & \\\hline
$\{ds\}\to\{us\}\ $ & $a_0\ $ & $a_1\ $ & $b_1\ $ & $b_2\ $ \\
$\kappa_{\mathpzc{p}}^{AA}$ & $0.492\ $& $\phantom{-}0.137\ $ & $0.567\ $&   \\
$\kappa_{{\mathpzc{a}}_1}^{AA}$ & $0.489\ $& $-0.095\ $ & $0.444\ $& $-0.129\ $  \\
$\kappa_{{\mathpzc{a}}_2}^{AA}$ & $0.492\ $& $-0.096\ $ & $0.444\ $& $-0.129\ $ \\
$\kappa_{{\mathpzc{a}}_3}^{AA}$ & & & & \\\hline
$\{ff\}\leftrightarrow [ud]\ $ & $a_0\ $ & $a_1\ $ & $b_1\ $ & $b_2\ $ \\
$\kappa_{\mathpzc{p}}^{SA}$ & $0.649\ $& $\phantom{-}0.094\ $ & $0.182\ $&   \\
$\kappa_{{\mathpzc{a}}_1}^{SA}$ & $0.587\ $& $\phantom{-}0.335\ $ & $0.781\ $& $-0.037\ $  \\
$\kappa_{{\mathpzc{a}}_2}^{SA}$ & $0.646\ $& $\phantom{-}0.327\ $ & $0.751\ $& $-0.035\ $ \\
$\kappa_{{\mathpzc{a}}_3}^{SA}$ & $0.062\ $& $\phantom{-}0.006\ $ & $0.703\ $& \\\hline
$\{(u,d) s\}\leftrightarrow [(d,u) s]\ $ & $a_0\ $ & $a_1\ $ & $b_1\ $ & $b_2\ $ \\
$\kappa_{\mathpzc{p}}^{SA}$ & $0.641\ $& $\phantom{-}0.152\ $ & $0.327\ $&   \\
$\kappa_{{\mathpzc{a}}_1}^{SA}$ & $0.571\ $& $\phantom{-}0.255\ $ & $0.686\ $& $-0.031\ $  \\
$\kappa_{{\mathpzc{a}}_2}^{SA}$ & $0.638\ $& $\phantom{-}0.254\ $ & $0.679\ $& $-0.031\ $ \\
$\kappa_{{\mathpzc{a}}_3}^{SA}$ & $0.070\ $& $\phantom{-}0.005\ $ & $0.728\ $&  \\\hline
\end{tabular*}
\end{center}
\end{table}

\begin{table}[!htbp]
\caption{\label{interpolatorcoefficientsus}
Probe-diquark form factors for $s\to u$ transitions, which can be interpolated using Eq.\,\eqref{interpolator12} with the coefficients listed here.  Where written, $f=d,u$ because I assume isospin symmetry; and the absence of an entry means the coefficient is zero.
(Every $\kappa(s)$ is dimensionless; so each coefficient in Eq.\,\eqref{interpolator12} has the mass dimension necessary to cancel that of the associated $s \, ({\rm GeV}^2)$ factor.)}
\begin{center}
\begin{tabular*}
{\hsize}
{
l@{\extracolsep{0ptplus1fil}}|
c@{\extracolsep{0ptplus1fil}}
c@{\extracolsep{0ptplus1fil}}
c@{\extracolsep{0ptplus1fil}}
c@{\extracolsep{0ptplus1fil}}}\hline
$\{fs\}\to\{fu\}\ $ & $a_0\ $ & $a_1\ $ & $b_1\ $ & $b_2\ $ \\
$\kappa_{\mathpzc{p}}^{AA}$ & $0.516\ $& $\phantom{-}0.131\ $ & $0.482\ $&   \\
$\kappa_{{\mathpzc{a}}_1}^{AA}$ & $0.480\ $& $-0.087\ $ & $0.318\ $& $-0.096\ $  \\
$\kappa_{{\mathpzc{a}}_2}^{AA}$ & $0.516\ $& $-0.093\ $ & $0.325\ $& $-0.095\ $  \\
$\kappa_{{\mathpzc{a}}_3}^{AA}$ & $0.128\ $& $-0.019\ $ & $0.416\ $& $-0.089\ $  \\\hline
$\{ss\}\to\{us\}\ $ & $a_0\ $ & $a_1\ $ & $b_1\ $ & $b_2\ $ \\
$\kappa_{\mathpzc{p}}^{AA}$ & $0.519\ $& $\phantom{-}0.113\ $ & $0.496\ $&   \\
$\kappa_{{\mathpzc{a}}_1}^{AA}$ & $0.481\ $& $\phantom{-}1.807\ $ & $4.328\ $& $\phantom{-}2.142\ $  \\
$\kappa_{{\mathpzc{a}}_2}^{AA}$ & $0.519\ $& $\phantom{-}1.877\ $ & $4.188\ $& $\phantom{-}2.083\ $  \\
$\kappa_{{\mathpzc{a}}_3}^{AA}$ & $0.076\ $& $\phantom{-}0.183\ $ & $3.090\ $& $\phantom{-}1.657\ $  \\\hline
$\{ds\}\rightarrow [ud]\ $ & $a_0\ $ & $a_1\ $ & $b_1\ $ & $b_2\ $ \\
$\kappa_{\mathpzc{p}}^{SA}$ & $0.742\ $& $\phantom{-}0.173\ $ & $0.304\ $&   \\
$\kappa_{{\mathpzc{a}}_1}^{SA}$ & $0.633\ $& $\phantom{-}0.251\ $ & $0.578\ $& $-0.024\ $  \\
$\kappa_{{\mathpzc{a}}_2}^{SA}$ & $0.712\ $& $\phantom{-}0.246\ $ & $0.552\ $& $-0.023\ $ \\
$\kappa_{{\mathpzc{a}}_3}^{SA}$ & $0.109\ $& $\phantom{-}0.006\ $ & $0.590\ $&  \\\hline
$\{ss\}\rightarrow [us]\ $ & $a_0\ $ & $a_1\ $ & $b_1\ $ & $b_2\ $ \\
$\kappa_{\mathpzc{p}}^{SA}$ & $0.691\ $& $\phantom{-}0.179\ $ & $0.376\ $&   \\
$\kappa_{{\mathpzc{a}}_1}^{SA}$ & $0.594\ $& $\phantom{-}0.199\ $ & $0.543\ $& $-0.023\ $  \\
$\kappa_{{\mathpzc{a}}_2}^{SA}$ & $0.666\ $& $\phantom{-}0.195\ $ & $0.527\ $& $-0.023\ $ \\
$\kappa_{{\mathpzc{a}}_3}^{SA}$ & $0.097\ $& $\phantom{-}0.004\ $ & $0.616\ $&  \\\hline
$\{fs\}\rightarrow [uf]\ $ & $a_0\ $ & $a_1\ $ & $b_1\ $ & $b_2\ $ \\
$\kappa_{\mathpzc{p}}^{SA}$ & $0.651\ $& $\phantom{-}0.144\ $ & $0.301\ $&   \\
$\kappa_{{\mathpzc{a}}_1}^{SA}$ & $0.626\ $& $\phantom{-}0.243\ $ & $0.580\ $& $-0.025\ $  \\
$\kappa_{{\mathpzc{a}}_2}^{SA}$ & $0.630\ $& $\phantom{-}0.238\ $ & $0.556\ $& $-0.023\ $ \\
$\kappa_{{\mathpzc{a}}_3}^{SA}$ & $0.024\ $& $\phantom{-}0.002\ $ & $0.556\ $&  \\\hline
\end{tabular*}
\end{center}
\end{table}

The expression in Eq.\,\eqref{GammaAA} yields the following explicit results:
\begin{subequations}
\begin{align}
i \Gamma^{AA}_{5,\rho\sigma}(\ell^\prime,\ell) & =
- \frac{1}{2 \mathpzc{m}_{fg}} \frac{m_{P_{fg}}^2}{Q^2+m_{P_{fg}}^2}\varepsilon_{\alpha\beta\gamma\delta}\bar\ell_\gamma Q_\delta \kappa_{{\mathpzc{p}}fg}^{AA}(Q^2)T_{\rho\alpha}^{\ell^\prime}T_{\sigma\beta}^{\ell} \\
\Gamma^{AA}_{5\mu ,\rho\sigma}(\ell^\prime,\ell) & =
-\bigg[\varepsilon_{\alpha\beta\gamma\delta}\bar\ell_\gamma Q_\delta \frac{Q_\mu}{Q^2+m_{P_{fg}}^2}\kappa_{{\mathpzc{a}}_1fg}^{AA}(Q^2)\nonumber \\
&  + \varepsilon_{\mu\alpha\beta\gamma}[\bar\ell_\gamma \kappa_{{\mathpzc{a}}_2fg}^{AA}(Q^2)
+  Q_\gamma \kappa_{{\mathpzc{a}}_3fg}^{AA}(Q^2)]\bigg]T_{\rho\alpha}^{\ell^\prime}T_{\sigma\beta}^{\ell}\,,
\end{align}
\end{subequations}
where $ \bar\ell = \ell^\prime+\ell$ and, on the domain $Q^2 \in (-m_{P_{fg}}^2, 2M_{B^\prime B}^2)$ the computed form factors $\kappa_{{\mathpzc{i}}fg}^{AA}(Q^2)$, ${\mathpzc{i}}={\mathpzc{p}}, {\mathpzc{a}}_1,{\mathpzc{a}}_2,{\mathpzc{a}}_3$, are reliably interpolated using the following function:
\begin{equation}
\label{interpolator12}
\kappa (s=Q^2) = \frac{a_0 + a_1 s}{1+ b_1 s + b_2 s^2}\,,
\end{equation}
with the coefficients listed in Tables~\ref{interpolatorcoefficientsdu} and \ref{interpolatorcoefficientsus} (charged currents) and Table~\ref{interpolatorcoefficientsneutral} (neutral currents). Note that, owing to the identities in Eqs.\,\eqref{WGTIagain}, $\kappa_{\mathpzc{p}}^{AA}(0)=\kappa_{{\mathpzc{a}}_2}^{AA}(0)$. Moreover, in the isospin symmetry limit, $m_{\{fd\}}=m_{\{fu\}}$, $f=d,u$; consequently, $\kappa_{{\mathpzc{a}}_3ud}^{AA}\equiv 0$. Furthermore, in no case considered herein does $\kappa_{{\mathpzc{a}}_3}^{AA}\neq 0$ contribute more than 1\% to any reported quantity.

\begin{table}[!htbp]
\caption{\label{interpolatorcoefficientsneutral}
Probe-diquark form factors for $g\to g$, $g=u,d,s$, neutral current transitions, which can be interpolated using Eq.\,\eqref{interpolator12} with the coefficients listed here.  Where written, $f=d,u$ because I assume isospin symmetry; and the absence of an entry means the coefficient is zero. Note that $\kappa_{{\mathpzc{a}}_3}^{AA}\equiv 0$ in this case.
(Every $\kappa(s)$ is dimensionless; so each coefficient in Eq.\,\eqref{interpolator12} has the mass dimension necessary to cancel that of the associated $s\, ({\rm GeV}^2)$ factor.)}
\begin{center}
\begin{tabular*}
{\hsize}
{
l@{\extracolsep{0ptplus1fil}}|
c@{\extracolsep{0ptplus1fil}}
c@{\extracolsep{0ptplus1fil}}
c@{\extracolsep{0ptplus1fil}}
c@{\extracolsep{0ptplus1fil}}}\hline
$\{ff\}\to\{ff\}\ $ & $a_0\ $ & $a_1\ $ & $b_1\ $ & $b_2\ $ \\
$\kappa_{\mathpzc{p}}^{AA}$ & $0.470\ $& $\phantom{-}0.173\ $ & $0.598\ $&   \\
$\kappa_{{\mathpzc{a}}_1}^{AA}$ & $0.467\ $& $\phantom{-}0.023\ $ & $0.598\ $& \\
$\kappa_{{\mathpzc{a}}_2}^{AA}$ & $0.470\ $& $\phantom{-}0.023\ $ & $0.598\ $& \\\hline
$\{ss\}\to\{ss\}\ $ & $a_0\ $ & $a_1\ $ & $b_1\ $ & $b_2\ $ \\
$\kappa_{\mathpzc{p}}^{AA}$ & $0.547\ $& $\phantom{-}0.094\ $ & $0.435\ $&   \\
$\kappa_{{\mathpzc{a}}_1}^{AA}$ & $0.475\ $& $\phantom{-}0.643\ $ & $1.878\ $& $\phantom{-}0.723\ $  \\
$\kappa_{{\mathpzc{a}}_2}^{AA}$ & $0.547\ $& $\phantom{-}0.654\ $ & $1.722\ $& $\phantom{-}0.649\ $   \\\hline
$\{fs\}\to\{fs\}\ $ & $a_0\ $ & $a_1\ $ & $b_1\ $ & $b_2\ $ \\
$\kappa_{{\mathpzc{p}}ff}^{AA}$ & $0.492\ $& $\phantom{-}0.137\ $ & $0.567\ $&   \\
$\kappa_{{\mathpzc{a}}_1ff}^{AA}$ & $0.489\ $& $-0.095\ $ & $0.444\ $& $-0.129\ $  \\
$\kappa_{{\mathpzc{a}}_2ff}^{AA}$ & $0.492\ $& $-0.096\ $ & $0.444\ $& $-0.129\ $   \\
$\kappa_{{\mathpzc{p}}ss}^{AA}$ & $0.564\ $& $\phantom{-}0.106\ $ & $0.416\ $&   \\
$\kappa_{{\mathpzc{a}}_1ss}^{AA}$ & $0.494\ $&  & $0.462\ $&  \\
$\kappa_{{\mathpzc{a}}_2ss}^{AA}$ & $0.564\ $&  & $0.469\ $&   \\\hline
$\{ud\}\leftrightarrow [ud]\ $ & $a_0\ $ & $a_1\ $ & $b_1\ $ & $b_2\ $ \\
$\kappa_{\mathpzc{p}}^{SA}$ & $0.649\ $& $\phantom{-}0.094\ $ & $0.182\ $&   \\
$\kappa_{{\mathpzc{a}}_1}^{SA}$ & $0.587\ $& $\phantom{-}0.335\ $ & $0.781\ $& $-0.037\ $  \\
$\kappa_{{\mathpzc{a}}_2}^{SA}$ & $0.646\ $& $\phantom{-}0.327\ $ & $0.751\ $& $-0.035\ $ \\
$\kappa_{{\mathpzc{a}}_3}^{SA}$ & $0.062\ $& $\phantom{-}0.006\ $ & $0.703\ $&  \\\hline
%
%
%
$\{fs\}\leftrightarrow [fs]\ $ & $a_0\ $ & $a_1\ $ & $b_1\ $ & $b_2\ $ \\
$\kappa_{{\mathpzc{p}}ff}^{SA}$ & $0.641\ $& $\phantom{-}0.152\ $ & $0.327\ $&   \\
$\kappa_{{\mathpzc{a}}_1 ff}^{SA}$ & $0.571\ $& $\phantom{-}0.255\ $ & $0.686\ $& $-0.031\ $  \\
$\kappa_{{\mathpzc{a}}_2 ff}^{SA}$ & $0.638\ $& $\phantom{-}0.254\ $ & $0.679\ $& $-0.031\ $ \\
$\kappa_{{\mathpzc{a}}_3 ff}^{SA}$ & $0.070\ $& $\phantom{-}0.005\ $ & $0.728\ $&  \\
$\kappa_{{\mathpzc{p}}ss}^{SA}$ & $0.742\ $& $\phantom{-}0.160\ $ & $0.310\ $&   \\
$\kappa_{{\mathpzc{a}}_1 ss}^{SA}$ & $0.678\ $& $\phantom{-}0.189\ $ & $0.459\ $& $-0.018\ $  \\
$\kappa_{{\mathpzc{a}}_2 ss}^{SA}$ & $0.701\ $& $\phantom{-}0.185\ $ & $0.434\ $& $-0.017\ $ \\
$\kappa_{{\mathpzc{a}}_3 ss}^{SA}$ & $0.064\ $& $\phantom{-}0.002\ $ & $0.484\ $&  \\\hline
\end{tabular*}
\end{center}
\end{table}

Turning to Eq.\,\eqref{GammaSA}, one finds:
\begin{subequations}
\begin{align}
\Gamma^{SA}_{5,\rho}(\ell^\prime,\ell)& =
T_{\rho\alpha}^\ell Q_\alpha \frac{m_{P_{fg}}^2}{Q^2+m_{P_{fg}}^2} \frac{m_{[hf]}+m_{\{gh\}}}{2{\mathpzc{m}}_{fg}}  \kappa_{{\mathpzc{p}}fg}^{SA}(Q^2) \,,\\
i \Gamma^{SA}_{5\mu ,\rho}(\ell^\prime,\ell) & =
T_{\rho\alpha}^\ell [m_{[hf]}+m_{\{gh\}}]
\bigg[ \delta_{\alpha\mu}
\kappa_{{\mathpzc{a}}_1fg}^{SA}(Q^2)- \frac{Q_\mu Q_\alpha}{Q^2+m_{P_{fg}}^2} \kappa_{{\mathpzc{a}}_2fg}^{SA}(Q^2)
  \\ \nonumber
 & + \bar\ell_\mu Q_\alpha \frac{1}{m_{\{gh\}}^2-m_{[hf]}^2} \kappa_{{\mathpzc{a}}_3fg}^{SA}(Q^2)\bigg]\,,
\end{align}
\end{subequations}
where the form factors can again be interpolated using Eq.\,\eqref{interpolator12} with the coefficients listed in Tables~\ref{interpolatorcoefficientsdu}\,--\,\ref{interpolatorcoefficientsneutral}.
\chapter{Angular momentum decomposition}
\label{AppendixD}

\begin{table}[!htbp]
\begin{center}
\caption{\label{tabnormalisationAMD}  Proton-canonical normalisation contributions broken into rest-frame quark+diquark orbital angular momentum components, defined with reference to Eqs.\,\eqref{listofOAM}.}

\renewcommand\arraystretch{1.2}
\begin{tabular}{|c|ccc|cccc|c|}
\hline
 &  $\mathcal{S}_{1}$ &  $\mathcal{A}_{2}$ &  $\mathcal{B}_{1}$ &  $\mathcal{S}_{2}$ &  $\mathcal{A}_{1}$ &  $\mathcal{B}_{2}$ &  $\mathcal{C}_{2}$ &  $\mathcal{C}_{1}$\\
\hline
$\mathcal{S}_{1}$ &  0.30 &  0.00 &  0.00 &  0.27 &  0.00 &  0.00 &  0.00 & 0.00 \\
$\mathcal{A}_{2}$ &  0.00 &  0.03 & -0.02 &  0.00 & -0.01 &  0.01 &  0.06 & 0.00 \\
$\mathcal{B}_{1}$ &  0.00 & -0.02 &  0.04 &  0.00 &  0.00 &  0.01 &  0.16 & 0.00\\
\hline
$\mathcal{S}_{2}$ &  0.27 &  0.00 &  0.00 & -0.10 &  0.00 &  0.00 &  0.00 & 0.00\\
$\mathcal{A}_{1}$ &  0.00 & -0.01 &  0.00 &  0.00 &  0.00 &  0.00 &  0.00 & 0.00 \\
$\mathcal{B}_{2}$ &  0.00 &  0.01 &  0.01 &  0.00 &  0.00 & -0.02 &  0.01 & -0.02 \\
$\mathcal{C}_{2}$ &  0.00 &  0.06 &  0.16 &  0.00 &  0.00 &  0.01 & -0.17 & -0.01 \\
\hline
$\mathcal{C}_{1}$ &  0.00 &  0.00 &  0.00 &  0.00 &  0.00 & -0.02 & -0.01 & 0.00 \\
\hline
\end{tabular}
\end{center}
\end{table}

Using the computed solutions of the Faddeev equations for the Poincar\'e-covariant baryon wave functions, evaluated in the rest frame, I computed the contributions of various quark+diquark orbital angular momentum components to the proton's canonical normalisation constant. The results are recorded in Table~\ref{tabnormalisationAMD}. It is from this table that the image in Fig.\,\ref{fignormalisationAMD} is drawn. I also calculated the kindred contributions to the proton's axial charge and its flavour separation. The results are recorded in Table~\ref{tabaxialAMD}. 

\begin{table}[!htbp]
\begin{center}
\caption{\label{tabaxialAMD} Proton axial charge contributions broken into rest-frame quark+diquark orbital angular momentum components, defined with reference to Eqs.\eqref{listofOAM}.}

\renewcommand\arraystretch{1.2}
\begin{tabular}{|c|ccc|cccc|c|}
\hline
$g_{A}^{u}$ &  $\mathcal{S}_{1}$ &  $\mathcal{A}_{2}$ &  $\mathcal{B}_{1}$ &  $\mathcal{S}_{2}$ &  $\mathcal{A}_{1}$ &  $\mathcal{B}_{2}$ &  $\mathcal{C}_{2}$ &  $\mathcal{C}_{1}$\\
\hline
$\mathcal{S}_{1}$ & 0.44  & -0.01 & 0.07  & 0.00  & 0.00  & 0.00  & 0.09  & 0.00 \\
$\mathcal{A}_{2}$ & -0.01 & 0.00  & 0.00 & 0.00  & 0.00  & -0.01 & 0.01  & 0.00 \\
$\mathcal{B}_{1}$ & 0.07  & 0.00 &  0.04 & 0.10  & 0.00  & 0.01  & 0.08 & 0.00 \\
\hline
$\mathcal{S}_{2}$ & 0.00  & 0.00 & 0.10  & -0.08 & 0.00  & -0.01 & -0.08 & 0.00 \\
$\mathcal{A}_{1}$ & 0.00  & 0.00 & 0.00  & 0.00  & 0.00  & 0.00  & 0.00  & 0.00  \\
$\mathcal{B}_{2}$ & 0.00  & -0.01 & 0.01 & -0.01 & 0.00  & 0.00  &  0.00 & 0.00\\
$\mathcal{C}_{2}$ & 0.09  & 0.01  & 0.08 & -0.08 & 0.00  &  0.00 & -0.07 & 0.00\\
\hline
$\mathcal{C}_{1}$ & 0.00  &  0.00 & 0.00 &  0.00 & 0.00  & 0.00  &  0.00 & 0.00\\
\hline
\hline
 $g_{A}^{d}$&  $\mathcal{S}_{1}$ &  $\mathcal{A}_{2}$ &  $\mathcal{B}_{1}$ &  $\mathcal{S}_{2}$ &  $\mathcal{A}_{1}$ &  $\mathcal{B}_{2}$ &  $\mathcal{C}_{2}$ &  $\mathcal{C}_{1}$\\
\hline
$\mathcal{S}_{1}$ & -0.26 & 0.00  & 0.00  & 0.01 & 0.00  & -0.01  & -0.07 & 0.00 \\
$\mathcal{A}_{2}$ & 0.00  & 0.01  & 0.00  & 0.00 & 0.00  & -0.01  & 0.00  & 0.00 \\
$\mathcal{B}_{1}$ & 0.00  & 0.00  & -0.04 & -0.09 & 0.00  & 0.00  & 0.02  & 0.00 \\
\hline
$\mathcal{S}_{2}$ & 0.01  & 0.00  & -0.09 &  0.00 & 0.00  & 0.01  & 0.08  & 0.00 \\
$\mathcal{A}_{1}$ & 0.00  & 0.00  & 0.00  & 0.00  & 0.00  & 0.00  & 0.00  & 0.00 \\
$\mathcal{B}_{2}$ & -0.01 & -0.01 & 0.00  & 0.01  & 0.00  &  0.00 & 0.00  & 0.00 \\
$\mathcal{C}_{2}$ & -0.07 & 0.00  & 0.02  & 0.08  & 0.00  & 0.00  & 0.00  &
0.00 \\
\hline
$\mathcal{C}_{1}$ & 0.00  & 0.00  & 0.00  & 0.00  & 0.00  & 0.00  & 0.00  & 0.00 \\
\hline
\hline
 $g_{A}^{(0)}$&  $\mathcal{S}_{1}$ &  $\mathcal{A}_{2}$ &  $\mathcal{B}_{1}$ &  $\mathcal{S}_{2}$ &  $\mathcal{A}_{1}$ &  $\mathcal{B}_{2}$ &  $\mathcal{C}_{2}$ &  $\mathcal{C}_{1}$\\
\hline
$\mathcal{S}_{1}$ & 0.18  & -0.01 & 0.07  & 0.01  & 0.00  & 0.01  & 0.02  & 0.00 \\
$\mathcal{A}_{2}$ & -0.01 & 0.01  & 0.00  & 0.00  & 0.00  & -0.02 & 0.01  & 0.00 \\
$\mathcal{B}_{1}$ & 0.07  & 0.00  & 0.00  & 0.01  & 0.00  & 0.01  & 0.10  & 0.00 \\
\hline
$\mathcal{S}_{2}$ & 0.01  & 0.00  & 0.01  & -0.08 & 0.00  & 0.00  & 0.00  & 0.00 \\
$\mathcal{A}_{1}$ & 0.00  & 0.00  & 0.00  & 0.00  & 0.00  & 0.00  & 0.00  & 0.00 \\
$\mathcal{B}_{2}$ & -0.01 & -0.02 & 0.01  & 0.00  & 0.00  & 0.00  & 0.00  & 0.00 \\
$\mathcal{C}_{2}$ & 0.02  & 0.01  & 0.10  & 0.00  & 0.00  & 0.00  & -0.07 & 0.00 \\
\hline
$\mathcal{C}_{1}$ & 0.00  & 0.00  & 0.00  & 0.00  & 0.00  & 0.00  & 0.00  & 0.00 \\
\hline
\end{tabular}
\end{center}
\end{table}

\end{appendices}


\backmatter
\bibliographystyle{unsrt}
\bibliography{bibfile}

\begin{thebibliography}{100}

\bibitem{Pennington:1996dy}
M.~R. Pennington.
\newblock {Calculating hadronic properties in strong QCD}.
\newblock In {\em {2nd ELFE Workshop}}, 9 1996.

\bibitem{Jaffe:2006iao}
A.~M. Jaffe.
\newblock {The Millennium Grand Challenge in Mathematics}.
\newblock {\em Not. Amer. Math. Soc.}, 53:652, 2006.

\bibitem{Capstick:2000qj}
S.~Capstick and W.~Roberts.
\newblock {Quark models of baryon masses and decays}.
\newblock {\em Prog. Part. Nucl. Phys.}, 45:S241, 2000.

\bibitem{Aznauryan:2012ba}
I.~G. Aznauryan et~al.
\newblock {Studies of Nucleon Resonance Structure in Exclusive Meson Electroproduction}.
\newblock {\em Int. J. Mod. Phys. E}, 22:1330015, 2013.

\bibitem{Crede:2013kia}
V.~Crede and W.~Roberts.
\newblock {Progress towards understanding baryon resonances}.
\newblock {\em Rept. Prog. Phys.}, 76:076301, 2013.

\bibitem{Swanson:2012zz}
E.~S. Swanson.
\newblock {Can the constituent quark model describe the pion?}
\newblock {\em Prog. Part. Nucl. Phys.}, 67:365, 2012.

\bibitem{Colangelo:2000dp}
P.~Colangelo and A.~Khodjamirian.
\newblock {QCD sum rules, a modern perspective}.
\newblock page 1495, 10 2000.

\bibitem{Brodsky:2011vgv}
S.~Brodsky, G.~de~Teramond, and M.~Karliner.
\newblock {Puzzles in Hadronic Physics and Novel Quantum Chromodynamics Phenomenology}.
\newblock {\em Ann. Rev. Nucl. Part. Sci.}, 62:2082, 2011.

\bibitem{Brodsky:2013ar}
S.~J. Brodsky, G.~F. De~T\'eramond, and H.~G. Dosch.
\newblock {Threefold Complementary Approach to Holographic QCD}.
\newblock {\em Phys. Lett. B}, 729:3, 2014.

\bibitem{Roberts:1994dr}
C.~D. Roberts and A.~G. Williams.
\newblock {Dyson-Schwinger equations and their application to hadronic physics}.
\newblock {\em Prog. Part. Nucl. Phys.}, 33:477, 1994.

\bibitem{Roberts:2000aa}
C.~D. Roberts and S.~M. Schmidt.
\newblock {Dyson-Schwinger equations: Density, temperature and continuum strong QCD}.
\newblock {\em Prog. Part. Nucl. Phys.}, 45:S1, 2000.

\bibitem{Maris:2003vk}
P.~Maris and C.~D. Roberts.
\newblock {Dyson-Schwinger equations: A Tool for hadron physics}.
\newblock {\em Int. J. Mod. Phys. E}, 12:297, 2003.

\bibitem{Bashir:2012fs}
A.~Bashir, L.~Chang, I.~C. Cloet, B.~El-Bennich, Y.-X. Liu, C.~D. Roberts, and P.~C. Tandy.
\newblock {Collective perspective on advances in Dyson-Schwinger Equation QCD}.
\newblock {\em Commun. Theor. Phys.}, 58:79, 2012.

\bibitem{Roberts:2012sv}
C.~D. Roberts.
\newblock {Strong QCD and Dyson-Schwinger Equations}.
\newblock {\em IRMA Lect. Math. Theor. Phys.}, 21:355, 2015.

\bibitem{Roberts:2015lja}
C.~D. Roberts.
\newblock {Three Lectures on Hadron Physics}.
\newblock {\em J. Phys. Conf. Ser.}, 706:022003, 2016.

\bibitem{Horn:2016rip}
T.~Horn and C.~D. Roberts.
\newblock {The pion: an enigma within the Standard Model}.
\newblock {\em J. Phys. G}, 43:073001, 2016.

\bibitem{Eichmann:2016yit}
G.~Eichmann, H.~Sanchis-Alepuz, R.~Williams, R.~Alkofer, and C.~S. Fischer.
\newblock {Baryons as relativistic three-quark bound states}.
\newblock {\em Prog. Part. Nucl. Phys.}, 91:1, 2016.

\bibitem{Burkert:2017djo}
V.~D. Burkert and C.~D. Roberts.
\newblock {Colloquium : Roper resonance: Toward a solution to the fifty year puzzle}.
\newblock {\em Rev. Mod. Phys.}, 91:011003, 2019.

\bibitem{Fischer:2018sdj}
C.~S. Fischer.
\newblock {QCD at finite temperature and chemical potential from Dyson\textendash{}Schwinger equations}.
\newblock {\em Prog. Part. Nucl. Phys.}, 105:1, 2019.

\bibitem{Qin:2020rad}
S.-x. Qin and C.~D. Roberts.
\newblock {Impressions of the Continuum Bound State Problem in QCD}.
\newblock {\em Chin. Phys. Lett.}, 37:121201, 2020.

\bibitem{Roberts:2020udq}
C.~D. Roberts and S.~M. Schmidt.
\newblock {Reflections upon the emergence of hadronic mass}.
\newblock {\em Eur. Phys. J. ST}, 229:3319, 2020.

\bibitem{Roberts:2020hiw}
C.~D. Roberts.
\newblock {Empirical Consequences of Emergent Mass}.
\newblock {\em Symmetry}, 12:1468, 2020.

\bibitem{Roberts:2021xnz}
C.~D. Roberts.
\newblock {On Mass and Matter}.
\newblock {\em AAPPS Bull.}, 31:6, 2021.

\bibitem{Roberts:2021nhw}
C.~D. Roberts, D.~G. Richards, T.~Horn, and L.~Chang.
\newblock {Insights into the emergence of mass from studies of pion and kaon structure}.
\newblock {\em Prog. Part. Nucl. Phys.}, 120:103883, 2021.

\bibitem{Binosi:2022djx}
D.~Binosi.
\newblock {Emergent Hadron Mass in Strong Dynamics}.
\newblock {\em Few Body Syst.}, 63:42, 2022.

\bibitem{Papavassiliou:2022wrb}
J.~Papavassiliou.
\newblock {Emergence of mass in the gauge sector of QCD*}.
\newblock {\em Chin. Phys. C}, 46:112001, 2022.

\bibitem{Ding:2022ows}
M.-H. Ding, C.~D. Roberts, and S.~M. Schmidt.
\newblock {Emergence of Hadron Mass and Structure}.
\newblock {\em Particles}, 6:57, 2023.

\bibitem{Ferreira:2023fva}
M.~N. Ferreira and J.~Papavassiliou.
\newblock {Gauge Sector Dynamics in QCD}.
\newblock {\em Particles}, 6:312, 2023.

\bibitem{Hofstadter:1955ae}
R.~Hofstadter and R.~W. McAllister.
\newblock {Electron Scattering From the Proton}.
\newblock {\em Phys. Rev.}, 98:217, 1955.

\bibitem{Ahrens:1988rr}
L.~A. Ahrens et~al.
\newblock {A Study of the Axial Vector Form-factor and Second Class Currents in Anti-neutrino Quasielastic Scattering}.
\newblock {\em Phys. Lett. B}, 202:284, 1988.

\bibitem{Kitagaki:1990vs}
T.~Kitagaki et~al.
\newblock {Study of neutrino d ---\ensuremath{>} mu- p p(s) and neutrino d ---\ensuremath{>} mu- Delta++ (1232) n(s) using the BNL 7-foot deuterium filled bubble chamber}.
\newblock {\em Phys. Rev. D}, 42:1331, 1990.

\bibitem{Bodek:2007vi}
A.~Bodek, S.~Avvakumov, R.~Bradford, and H.~S. Budd.
\newblock {Extraction of the axial nucleon form-factor from neutrino experiments on deuterium}.
\newblock {\em J. Phys. Conf. Ser.}, 110:082004, 2008.

\bibitem{Meyer:2016oeg}
Aaron~S. Meyer, Minerba Betancourt, Richard Gran, and Richard~J. Hill.
\newblock {Deuterium target data for precision neutrino-nucleus cross sections}.
\newblock {\em Phys. Rev. D}, 93:113015, 2016.

\bibitem{Choi:1993vt}
S.~Choi et~al.
\newblock {Axial and pseudoscalar nucleon form-factors from low-energy pion electroproduction}.
\newblock {\em Phys. Rev. Lett.}, 71:3927, 1993.

\bibitem{Bernard:1994pk}
V.~Bernard, U.~G. Meissner, and N.~Kaiser.
\newblock {Comment on `Axial and pseudoscalar nucleon form-factors from low-energy pion electroproduction.'}.
\newblock {\em Phys. Rev. Lett.}, 72:2810, 1994.

\bibitem{A1:1999kwj}
A.~Liesenfeld et~al.
\newblock {A Measurement of the axial form-factor of the nucleon by the p(e, e-prime pi+)n reaction at W = 1125-MeV}.
\newblock {\em Phys. Lett. B}, 468:20, 1999.

\bibitem{Fuchs:2003vw}
T.~Fuchs and S.~Scherer.
\newblock {Pion electroproduction, PCAC, chiral ward identities, and the axial form-factor revisited}.
\newblock {\em Phys. Rev. C}, 68:055501, 2003.

\bibitem{Hart:1977zz}
R.~D. Hart, C.~R. Cox, G.~W. Dodson, M.~Eckhause, J.~R. Kane, M.~S. Pandey, A.~M. Rushton, R.~T. Siegel, and R.~E. Welsh.
\newblock {Radiative Muon Capture in Calcium}.
\newblock {\em Phys. Rev. Lett.}, 39:399, 1977.

\bibitem{Jonkmans:1996my}
G.~Jonkmans et~al.
\newblock {Radiative muon capture on hydrogen and the induced pseudoscalar coupling}.
\newblock {\em Phys. Rev. Lett.}, 77:4512, 1996.

\bibitem{Wright:1998gi}
D.~H. Wright et~al.
\newblock {Measurement of the induced pseudoscalar coupling using radiative muon capture on hydrogen}.
\newblock {\em Phys. Rev. C}, 57:373, 1998.

\bibitem{Gorringe:2002xx}
T.~Gorringe and H.~W. Fearing.
\newblock {Induced Pseudoscalar Coupling of the Proton Weak Interaction}.
\newblock {\em Rev. Mod. Phys.}, 76:31, 2004.

\bibitem{Bardin:1980mi}
G.~Bardin, J.~Duclos, A.~Magnon, J.~Martino, A.~Richter, E.~Zavattini, A.~Bertin, M.~Piccinini, A.~Vitale, and D.~F. Measday.
\newblock {A Novel Measurement of the Muon Capture Rate in Liquid Hydrogen by the Lifetime Technique}.
\newblock {\em Nucl. Phys. A}, 352:365, 1981.

\bibitem{MuCap:2012lei}
V.~A. Andreev et~al.
\newblock {Measurement of Muon Capture on the Proton to 1\% Precision and Determination of the Pseudoscalar Coupling $g_P$}.
\newblock {\em Phys. Rev. Lett.}, 110:012504, 2013.

\bibitem{MuCap:2015boo}
V.~A. Andreev et~al.
\newblock {Measurement of the Formation Rate of Muonic Hydrogen Molecules}.
\newblock {\em Phys. Rev. C}, 91:055502, 2015.

\bibitem{UCNA:2012fhw}
M.~P. Mendenhall et~al.
\newblock {Precision measurement of the neutron $\beta$-decay asymmetry}.
\newblock {\em Phys. Rev. C}, 87:032501, 2013.

\bibitem{Mund:2012fq}
D.~Mund, B.~Maerkisch, M.~Deissenroth, J.~Krempel, M.~Schumann, H.~Abele, A.~Petoukhov, and T.~Soldner.
\newblock {Determination of the Weak Axial Vector Coupling from a Measurement of the Beta-Asymmetry Parameter A in Neutron Beta Decay}.
\newblock {\em Phys. Rev. Lett.}, 110:172502, 2013.

\bibitem{UCNA:2017obv}
M.~A.~P. Brown et~al.
\newblock {New result for the neutron $\beta$-asymmetry parameter $A_0$ from UCNA}.
\newblock {\em Phys. Rev. C}, 97:035505, 2018.

\bibitem{Darius:2017arh}
G.~Darius et~al.
\newblock {Measurement of the Electron-Antineutrino Angular Correlation in Neutron $\beta$ Decay}.
\newblock {\em Phys. Rev. Lett.}, 119:042502, 2017.

\bibitem{Castro:1977ep}
J.~J. Castro and C.~A. Dominguez.
\newblock {Upper Bound for the Induced Pseudoscalar Form-Factor in Muon Capture}.
\newblock {\em Phys. Rev. Lett.}, 39:440, 1977.

\bibitem{Bernard:2000et}
V.~Bernard, T.~R. Hemmert, and U.-G. Meissner.
\newblock {Ordinary and radiative muon capture on the proton and the pseudoscalar form-factor of the nucleon}.
\newblock {\em Nucl. Phys. A}, 686:290, 2001.

\bibitem{MuCap:2007tkq}
V.~A. Andreev et~al.
\newblock {Measurement of the rate of muon capture in hydrogen gas and determination of the proton's pseudoscalar coupling g(P)}.
\newblock {\em Phys. Rev. Lett.}, 99:032002, 2007.

\bibitem{Gaillard:1984ny}
J.~M. Gaillard and G.~Sauvage.
\newblock {HYPERON BETA DECAYS}.
\newblock {\em Ann. Rev. Nucl. Part. Sci.}, 34:351, 1984.

\bibitem{Cabibbo:2003cu}
N.~Cabibbo, E.~C. Swallow, and R.~Winston.
\newblock {Semileptonic hyperon decays}.
\newblock {\em Ann. Rev. Nucl. Part. Sci.}, 53:39, 2003.

\bibitem{Tiburzi:2008bk}
B.~C. Tiburzi and A.~Walker-Loud.
\newblock {Hyperons in Two Flavor Chiral Perturbation Theory}.
\newblock {\em Phys. Lett. B}, 669:246, 2008.

\bibitem{Alexandrou:2009qu}
C.~Alexandrou, R.~Baron, J.~Carbonell, V.~Drach, P.~Guichon, K.~Jansen, T.~Korzec, and O.~Pene.
\newblock {Low-lying baryon spectrum with two dynamical twisted mass fermions}.
\newblock {\em Phys. Rev. D}, 80:114503, 2009.

\bibitem{Munczek:1994zz}
H.~J. Munczek.
\newblock {Dynamical chiral symmetry breaking, Goldstone's theorem and the consistency of the Schwinger-Dyson and Bethe-Salpeter Equations}.
\newblock {\em Phys. Rev. D}, 52:4736, 1995.

\bibitem{Bender:1996bb}
A.~Bender, C.~D. Roberts, and L.~Von~Smekal.
\newblock {Goldstone theorem and diquark confinement beyond rainbow ladder approximation}.
\newblock {\em Phys. Lett. B}, 380:7, 1996.

\bibitem{Ward:1950xp}
J.~C. Ward.
\newblock {An Identity in Quantum Electrodynamics}.
\newblock {\em Phys. Rev.}, 78:182, 1950.

\bibitem{Green:1953te}
H.~S. Green.
\newblock {A Pre-renormalized quantum electrodynamics}.
\newblock {\em Proc. Phys. Soc. A}, 66:873, 1953.

\bibitem{Takahashi:1957xn}
Y.~Takahashi.
\newblock {On the generalized Ward identity}.
\newblock {\em Nuovo Cim.}, 6:371, 1957.

\bibitem{Takahashi:1985yz}
Y.~Takahashi.
\newblock {CANONICAL QUANTIZATION AND GENERALIZED WARD RELATIONS: FOUNDATION OF NONPERTURBATIVE APPROACH}.
\newblock In {\em {International Symposium on Quantum Field Theory}}, 6 1985.

\bibitem{Binosi:2014aea}
D.~Binosi, L.~Chang, J.~Papavassiliou, and C.~D. Roberts.
\newblock {Bridging a gap between continuum-QCD and ab initio predictions of hadron observables}.
\newblock {\em Phys. Lett. B}, 742:183, 2015.

\bibitem{Gao:2017uox}
F.~Gao, S.-X. Qin, C.~D. Roberts, and J.~Rodriguez-Quintero.
\newblock {Locating the Gribov horizon}.
\newblock {\em Phys. Rev. D}, 97:034010, 2018.

\bibitem{Cui:2019dwv}
Z.-F. Cui, J.-L. Zhang, D.~Binosi, F.~de~Soto, C.~Mezrag, J.~Papavassiliou, C.~D. Roberts, J.~Rodr\'\i{}guez-Quintero, J.~Segovia, and S.~Zafeiropoulos.
\newblock {Effective charge from lattice QCD}.
\newblock {\em Chin. Phys. C}, 44:083102, 2020.

\bibitem{Xu:2021iwv}
Z.-N. Xu, Z.-F. Cui, C.~D. Roberts, and C.~Xu.
\newblock {Heavy + light pseudoscalar meson semileptonic transitions}.
\newblock {\em Eur. Phys. J. C}, 81:1105, 2021.

\bibitem{Ebert:1996vx}
D.~Ebert, T.~Feldmann, and H.~Reinhardt.
\newblock {Extended NJL model for light and heavy mesons without q - anti-q thresholds}.
\newblock {\em Phys. Lett. B}, 388:154, 1996.

\bibitem{Gutierrez-Guerrero:2010waf}
L.~X. Gutierrez-Guerrero, A.~Bashir, I.~C. Cloet, and C.~D. Roberts.
\newblock {Pion form factor from a contact interaction}.
\newblock {\em Phys. Rev. C}, 81:065202, 2010.

\bibitem{Cui:2022fyr}
Z.-F. Cui, D.~Binosi, C.~D. Roberts, and S.~M. Schmidt.
\newblock {Hadron and light nucleus radii from electron scattering*}.
\newblock {\em Chin. Phys. C}, 46:122001, 2022.

\bibitem{Roberts:2011wy}
H.~L.~L. Roberts, A.~Bashir, L.~X. Gutierrez-Guerrero, C.~D. Roberts, and D.~J. Wilson.
\newblock {pi- and rho-mesons, and their diquark partners, from a contact interaction}.
\newblock {\em Phys. Rev. C}, 83:065206, 2011.

\bibitem{ParticleDataGroup:2020ssz}
P.~A. Zyla et~al.
\newblock {Review of Particle Physics}.
\newblock {\em PTEP}, 2020:083C01, 2020.

\bibitem{Cahill:1987qr}
R.~T. Cahill, C.~D. Roberts, and J.~Praschifka.
\newblock {Calculation of Diquark Masses in {QCD}}.
\newblock {\em Phys. Rev. D}, 36:2804, 1987.

\bibitem{Yin:2019bxe}
P.-L. Yin, C.~Chen, G.~Krein, C.~D. Roberts, J.~Segovia, and S.-S. Xu.
\newblock {Masses of ground-state mesons and baryons, including those with heavy quarks}.
\newblock {\em Phys. Rev. D}, 100:034008, 2019.

\bibitem{Yin:2021uom}
P.-L. Yin, Z.-F. Cui, C.~D. Roberts, and J.~Segovia.
\newblock {Masses of positive- and negative-parity hadron ground-states, including those with heavy quarks}.
\newblock {\em Eur. Phys. J. C}, 81:327, 2021.

\bibitem{Chen:2012txa}
C.~Chen, L.~Chang, C.~D. Roberts, S.~M. Schmidt, S.~Wan, and D.~J. Wilson.
\newblock {Features and flaws of a contact interaction treatment of the kaon}.
\newblock {\em Phys. Rev. C}, 87:045207, 2013.

\bibitem{Roberts:1994hh}
C.~D. Roberts.
\newblock {Electromagnetic pion form-factor and neutral pion decay width}.
\newblock {\em Nucl. Phys. A}, 605:475, 1996.

\bibitem{Xing:2022sor}
H.-Y. Xing, Z.-N. Xu, Z.-F. Cui, C.~D. Roberts, and C.~Xu.
\newblock {Heavy + heavy and heavy + light pseudoscalar to vector semileptonic transitions}.
\newblock {\em Eur. Phys. J. C}, 82:889, 2022.

\bibitem{Cahill:1988dx}
R.~T. Cahill, C.~D. Roberts, and J.~Praschifka.
\newblock {Baryon Structure and {QCD}}.
\newblock {\em Austral. J. Phys.}, 42:129, 1989.

\bibitem{Reinhardt:1989rw}
H.~Reinhardt.
\newblock {Hadronization of Quark Flavor Dynamics}.
\newblock {\em Phys. Lett. B}, 244:316, 1990.

\bibitem{Efimov:1990uz}
G.~V. Efimov, M.~A. Ivanov, and V.~E. Lyubovitskij.
\newblock {Quark - diquark approximation of the three quark structure of baryons in the quark confinement model}.
\newblock {\em Z. Phys. C}, 47:583, 1990.

\bibitem{Barabanov:2020jvn}
M.~Yu. Barabanov et~al.
\newblock {Diquark correlations in hadron physics: Origin, impact and evidence}.
\newblock {\em Prog. Part. Nucl. Phys.}, 116:103835, 2021.

\bibitem{Lu:2022cjx}
Y.~Lu, L.~Chang, K.~Raya, C.~D. Roberts, and J.~Rodr\'\i{}guez-Quintero.
\newblock {Proton and pion distribution functions in counterpoint}.
\newblock {\em Phys. Lett. B}, 830:137130, 2022.

\bibitem{Liu:2022ndb}
L.~Liu, C.~Chen, Y.~Lu, C.~D. Roberts, and J.~Segovia.
\newblock {Composition of low-lying J=32\ensuremath{\pm} \ensuremath{\Delta}-baryons}.
\newblock {\em Phys. Rev. D}, 105:114047, 2022.

\bibitem{Eichmann:2022zxn}
G.~Eichmann.
\newblock {Theory Introduction to Baryon Spectroscopy}.
\newblock {\em Few Body Syst.}, 63:57, 2022.

\bibitem{Eichmann:2011vu}
G.~Eichmann.
\newblock {Nucleon electromagnetic form factors from the covariant Faddeev equation}.
\newblock {\em Phys. Rev. D}, 84:014014, 2011.

\bibitem{Segovia:2015ufa}
J.~Segovia, C.~D. Roberts, and S.~M. Schmidt.
\newblock {Understanding the nucleon as a Borromean bound-state}.
\newblock {\em Phys. Lett. B}, 750:100, 2015.

\bibitem{Eichmann:2016hgl}
G.~Eichmann, C.~S. Fischer, and H.~Sanchis-Alepuz.
\newblock {Light baryons and their excitations}.
\newblock {\em Phys. Rev. D}, 94:094033, 2016.

\bibitem{Lu:2017cln}
Y.~Lu, C.~Chen, C.~D. Roberts, J.~Segovia, S.-S. Xu, and H.-S. Zong.
\newblock {Parity partners in the baryon resonance spectrum}.
\newblock {\em Phys. Rev. C}, 96:015208, 2017.

\bibitem{Chen:2017pse}
C.~Chen, B.~El-Bennich, C.~D. Roberts, S.~M. Schmidt, J.~Segovia, and S.~Wan.
\newblock {Structure of the nucleon\textquoteright{}s low-lying excitations}.
\newblock {\em Phys. Rev. D}, 97:034016, 2018.

\bibitem{Chen:2012qr}
C.~Chen, L.~Chang, C.~D. Roberts, S.~Wan, and D.~J. Wilson.
\newblock {Spectrum of hadrons with strangeness}.
\newblock {\em Few Body Syst.}, 53:293, 2012.

\bibitem{Oettel:1998bk}
M.~Oettel, G.~Hellstern, R.~Alkofer, and H.~Reinhardt.
\newblock {Octet and decuplet baryons in a covariant and confining diquark - quark model}.
\newblock {\em Phys. Rev. C}, 58:2459, 1998.

\bibitem{Cloet:2007pi}
I.~C. Cloet, A.~Krassnigg, and C.~D. Roberts.
\newblock {Dynamics, symmetries and hadron properties}.
\newblock {\em eConf}, C070910:125, 2007.

\bibitem{Segovia:2014aza}
J.~Segovia, I.~C. Cloet, C.~D. Roberts, and S.~M. Schmidt.
\newblock {Nucleon and $\Delta$ elastic and transition form factors}.
\newblock {\em Few Body Syst.}, 55:1185, 2014.

\bibitem{Roberts:2011cf}
H.~L.~L. Roberts, L.~Chang, I.~C. Cloet, and C.~D. Roberts.
\newblock {Masses of ground and excited-state hadrons}.
\newblock {\em Few Body Syst.}, 51:1, 2011.

\bibitem{Buck:1992wz}
A.~Buck, R.~Alkofer, and H.~Reinhardt.
\newblock {Baryons as bound states of diquarks and quarks in the Nambu-Jona-Lasinio model}.
\newblock {\em Phys. Lett. B}, 286:29, 1992.

\bibitem{Wilson:2011aa}
D.~J. Wilson, I.~C. Cloet, L.~Chang, and C.~D. Roberts.
\newblock {Nucleon and Roper electromagnetic elastic and transition form factors}.
\newblock {\em Phys. Rev. C}, 85:025205, 2012.

\bibitem{Chang:2012cc}
L.~Chang, C.~D. Roberts, and S.~M. Schmidt.
\newblock {Dressed-quarks and the nucleon's axial charge}.
\newblock {\em Phys. Rev. C}, 87:015203, 2013.

\bibitem{Segovia:2013rca}
J.~Segovia, C.~Chen, C.~D. Roberts, and S.~Wan.
\newblock {Insights into the gamma* N -\ensuremath{>} Delta transition}.
\newblock {\em Phys. Rev. C}, 88:032201, 2013.

\bibitem{Raya:2021pyr}
K.~Raya, L.~X. Guti\'errez-Guerrero, A.~Bashir, L.~Chang, Z.-F. Cui, Y.~Lu, C.~D. Roberts, and J.~Segovia.
\newblock {Dynamical diquarks in the ${\boldsymbol{\gamma^{(\ast)} p\to N(1535)\tfrac{1}{2}^-}}$ transition}.
\newblock {\em Eur. Phys. J. A}, 57:266, 2021.

\bibitem{Gutierrez-Guerrero:2021rsx}
L.~X. Guti\'errez-Guerrero, G.~Paredes-Torres, and A.~Bashir.
\newblock {Mesons and baryons: Parity partners}.
\newblock {\em Phys. Rev. D}, 104:094013, 2021.

\bibitem{Cheng:2022jxe}
P.~Cheng, F.~E. Serna, Z.-Q. Yao, C.~Chen, Z.-F. Cui, and C.~D. Roberts.
\newblock {Contact interaction analysis of octet baryon axial-vector and pseudoscalar form factors}.
\newblock {\em Phys. Rev. D}, 106:054031, 2022.

\bibitem{Xu:2015kta}
S.-S. Xu, C.~Chen, I.~C. Cloet, C.~D. Roberts, J.~Segovia, and H.-S. Zong.
\newblock {Contact-interaction Faddeev equation and, inter alia , proton tensor charges}.
\newblock {\em Phys. Rev. D}, 92:114034, 2015.

\bibitem{Liu:2022nku}
L.~T. Liu, C.~Chen, and C.~D. Roberts.
\newblock {Wave functions of (I,JP)=(12,32\ensuremath{\mp}) baryons}.
\newblock {\em Phys. Rev. D}, 107:014002, 2023.

\bibitem{Hecht:2002ej}
M.~B. Hecht, M.~Oettel, C.~D. Roberts, S.~M. Schmidt, P.~C. Tandy, and A.~W. Thomas.
\newblock {Nucleon mass and pion loops}.
\newblock {\em Phys. Rev. C}, 65:055204, 2002.

\bibitem{Sanchis-Alepuz:2014wea}
H.~Sanchis-Alepuz, C.~S. Fischer, and S.~Kubrak.
\newblock {Pion cloud effects on baryon masses}.
\newblock {\em Phys. Lett. B}, 733:151, 2014.

\bibitem{Garcia-Tecocoatzi:2016rcj}
H.~Garc\'\i{}a-Tecocoatzi, R.~Bijker, J.~Ferretti, and E.~Santopinto.
\newblock {Self-energies of octet and decuplet baryons due to the coupling to the baryon-meson continuum}.
\newblock {\em Eur. Phys. J. A}, 53:115, 2017.

\bibitem{Chen:2017mug}
X.~Chen, J.~Ping, C.~D. Roberts, and J.~Segovia.
\newblock {Light-meson masses in an unquenched quark model}.
\newblock {\em Phys. Rev. D}, 97:094016, 2018.

\bibitem{Julia-Diaz:2007qtz}
B.~Julia-Diaz, T.~S.~H. Lee, A.~Matsuyama, and T.~Sato.
\newblock {Dynamical coupled-channel model of $\pi N$ scattering in the $W \leq 2\,$GeV nucleon resonance region}.
\newblock {\em Phys. Rev. C}, 76:065201, 2007.

\bibitem{Suzuki:2009nj}
N.~Suzuki, B.~Julia-Diaz, H.~Kamano, T.~S.~H. Lee, A.~Matsuyama, and T.~Sato.
\newblock {Disentangling the Dynamical Origin of P-11 Nucleon Resonances}.
\newblock {\em Phys. Rev. Lett.}, 104:042302, 2010.

\bibitem{Ronchen:2012eg}
D.~Ronchen, M.~Doring, F.~Huang, H.~Haberzettl, J.~Haidenbauer, C.~Hanhart, S.~Krewald, U.~G. Meissner, and K.~Nakayama.
\newblock {Coupled-channel dynamics in the reactions $\pi N \to \pi N$, $\eta N$, $K\Lambda$, $K\Sigma$}.
\newblock {\em Eur. Phys. J. A}, 49:44, 2013.

\bibitem{Kamano:2013iva}
H.~Kamano, S.~X. Nakamura, T.~S.~H. Lee, and T.~Sato.
\newblock {Nucleon resonances within a dynamical coupled-channels model of $\pi N$ and $\gamma N$ reactions}.
\newblock {\em Phys. Rev. C}, 88:035209, 2013.

\bibitem{Pauli:1930pc}
W.~Pauli.
\newblock {Dear radioactive ladies and gentlemen}.
\newblock {\em Phys. Today}, 31N9:27, 1978.

\bibitem{Fermi:1934hr}
E.~Fermi.
\newblock {An attempt of a theory of beta radiation. 1.}
\newblock {\em Z. Phys.}, 88:161, 1934.

\bibitem{Yao:2021pyf}
Z.-Q. Yao, D.~Binosi, Z.-F. Cui, and C.~D. Roberts.
\newblock {Semileptonic $B_c \to \eta_c, J/\psi$ transitions}.
\newblock {\em Phys. Lett. B}, 818:136344, 2021.

\bibitem{Yao:2021pdy}
Z.-Q. Yao, D.~Binosi, Z.-F. Cui, and C.~D. Roberts.
\newblock {Semileptonic transitions: B(s)\textrightarrow{}\ensuremath{\pi}(K); Ds\textrightarrow{}K; D\textrightarrow{}\ensuremath{\pi},K; and K\textrightarrow{}\ensuremath{\pi}}.
\newblock {\em Phys. Lett. B}, 824:136793, 2022.

\bibitem{Mosel:2016cwa}
U.~Mosel.
\newblock {Neutrino Interactions with Nucleons and Nuclei: Importance for Long-Baseline Experiments}.
\newblock {\em Ann. Rev. Nucl. Part. Sci.}, 66:171, 2016.

\bibitem{NuSTEC:2017hzk}
L.~Alvarez-Ruso et~al.
\newblock {NuSTEC White Paper: Status and challenges of neutrino\textendash{}nucleus scattering}.
\newblock {\em Prog. Part. Nucl. Phys.}, 100:1, 2018.

\bibitem{Hill:2017wgb}
R.~J. Hill, P.~Kammel, W.~J. Marciano, and A.~Sirlin.
\newblock {Nucleon Axial Radius and Muonic Hydrogen \textemdash{} A New Analysis and Review}.
\newblock {\em Rept. Prog. Phys.}, 81:096301, 2018.

\bibitem{Novario:2020dmr}
S.~Novario, P.~Gysbers, J.~Engel, G.~Hagen, G.~R. Jansen, T.~D. Morris, P.~Navr\'atil, T.~Papenbrock, and S.~Quaglioni.
\newblock {Coupled-Cluster Calculations of Neutrinoless Double-$\beta$ Decay in $^{48}$Ca}.
\newblock {\em Phys. Rev. Lett.}, 126:182502, 2021.

\bibitem{Lovato:2020kba}
A.~Lovato, J.~Carlson, S.~Gandolfi, N.~Rocco, and R.~Schiavilla.
\newblock {Ab initio study of $\boldsymbol{(\nu_\ell,\ell^-)}$ and $\boldsymbol{(\overline{\nu}_\ell,\ell^+)}$ inclusive scattering in $^{12}$C: confronting the MiniBooNE and T2K CCQE data}.
\newblock {\em Phys. Rev. X}, 10:031068, 2020.

\bibitem{Anikin:2016teg}
I.~V. Anikin, V.~M. Braun, and N.~Offen.
\newblock {Axial form factor of the nucleon at large momentum transfers}.
\newblock {\em Phys. Rev. D}, 94:034011, 2016.

\bibitem{Chen:2020wuq}
C.~Chen, C.~S. Fischer, C.~D. Roberts, and J.~Segovia.
\newblock {Form Factors of the Nucleon Axial Current}.
\newblock {\em Phys. Lett. B}, 815:136150, 2021.

\bibitem{Chen:2021guo}
C.~Chen, C.~S. Fischer, C.~D. Roberts, and J.~Segovia.
\newblock {Nucleon axial-vector and pseudoscalar form factors and PCAC relations}.
\newblock {\em Phys. Rev. D}, 105:094022, 2022.

\bibitem{Chen:2022odn}
C.~Chen and C.~D. Roberts.
\newblock {Nucleon axial form factor at large momentum transfers}.
\newblock {\em Eur. Phys. J. A}, 58:206, 2022.

\bibitem{Alexandrou:2017hac}
C.~Alexandrou, M.~Constantinou, K.~Hadjiyiannakou, K.~Jansen, C.~Kallidonis, G.~Koutsou, and A.~Vaquero Aviles-Casco.
\newblock {Nucleon axial form factors using $N_f$ = 2 twisted mass fermions with a physical value of the pion mass}.
\newblock {\em Phys. Rev. D}, 96:054507, 2017.

\bibitem{Jang:2019vkm}
Y.-C. Jang, R.~Gupta, B.~Yoon, and T.~Bhattacharya.
\newblock {Axial Vector Form Factors from Lattice QCD that Satisfy the PCAC Relation}.
\newblock {\em Phys. Rev. Lett.}, 124:072002, 2020.

\bibitem{Faessler:2008ix}
A.~Faessler, T.~Gutsche, B.~R. Holstein, M.~A. Ivanov, J.~G. Korner, and V.~E. Lyubovitskij.
\newblock {Semileptonic decays of the light J**P = 1/2+ ground state baryon octet}.
\newblock {\em Phys. Rev. D}, 78:094005, 2008.

\bibitem{Ledwig:2014rfa}
T.~Ledwig, J.~Martin~Camalich, L.~S. Geng, and M.~J. Vicente~Vacas.
\newblock {Octet-baryon axial-vector charges and SU(3)-breaking effects in the semileptonic hyperon decays}.
\newblock {\em Phys. Rev. D}, 90:054502, 2014.

\bibitem{Yang:2015era}
G.-S. Yang and H.-C. Kim.
\newblock {Hyperon Semileptonic decay constants with flavor SU(3) symmetry breaking}.
\newblock {\em Phys. Rev. C}, 92:035206, 2015.

\bibitem{Ramalho:2015jem}
G.~Ramalho and K.~Tsushima.
\newblock {Axial form factors of the octet baryons in a covariant quark model}.
\newblock {\em Phys. Rev. D}, 94:014001, 2016.

\bibitem{Yang:2018idi}
G.-S. Yang and H.-C. Kim.
\newblock {Meson\textendash{}baryon coupling constants of the SU(3) baryons with flavor SU(3) symmetry breaking}.
\newblock {\em Phys. Lett. B}, 785:434, 2018.

\bibitem{Qi:2022sus}
R.~Qi, J.-B. Wang, G.~Li, C.-S. An, C.-R. Deng, and J.-J. Xie.
\newblock {Investigations on the flavor-dependent axial charges of the octet baryons}.
\newblock {\em Phys. Rev. C}, 105:065204, 2022.

\bibitem{Erkol:2009ev}
G.~Erkol, M.~Oka, and T.~T. Takahashi.
\newblock {Axial Charges of Octet Baryons in Two-flavor Lattice QCD}.
\newblock {\em Phys. Lett. B}, 686:36, 2010.

\bibitem{Green:2017keo}
J.~Green, N.~Hasan, S.~Meinel, M.~Engelhardt, S.~Krieg, J.~Laeuchli, J.~Negele, K.~Orginos, A.~Pochinsky, and S.~Syritsyn.
\newblock {Up, down, and strange nucleon axial form factors from lattice QCD}.
\newblock {\em Phys. Rev. D}, 95:114502, 2017.

\bibitem{Bali:2022qja}
G.~S. Bali, S.~Collins, W.~S\"ldner, and S.~Weish\"aupl.
\newblock {Leading order mesonic and baryonic SU(3) low energy constants from Nf=3 lattice QCD}.
\newblock {\em Phys. Rev. D}, 105:054516, 2022.

\bibitem{Wang:2013wk}
K.-L. Wang, Y.-X. Liu, L.~Chang, C.~D. Roberts, and S.~M. Schmidt.
\newblock {Baryon and meson screening masses}.
\newblock {\em Phys. Rev. D}, 87:074038, 2013.

\bibitem{Bedolla:2015mpa}
M.~A. Bedolla, J.~J. Cobos-Martínez, and A.~Bashir.
\newblock {Charmonia in a contact interaction}.
\newblock {\em Phys. Rev. D}, 92:054031, 2015.

\bibitem{Bedolla:2016yxq}
M.~A. Bedolla, K.~Raya, J.~J. Cobos-Mart\'\i{}nez, and A.~Bashir.
\newblock {$\eta_c$ elastic and transition form factors: Contact interaction and algebraic model}.
\newblock {\em Phys. Rev. D}, 93:094025, 2016.

\bibitem{Serna:2017nlr}
F.~E. Serna, B.~El-Bennich, and G.~Krein.
\newblock {Charmed mesons with a symmetry-preserving contact interaction}.
\newblock {\em Phys. Rev. D}, 96:014013, 2017.

\bibitem{Raya:2017ggu}
K.~Raya, M.~A. Bedolla, J.~J. Cobos-Mart\'\i{}nez, and A.~Bashir.
\newblock {Heavy quarkonia in a contact interaction and an algebraic model: mass spectrum, decay constants, charge radii and elastic and transition form factors}.
\newblock {\em Few Body Syst.}, 59:133, 2018.

\bibitem{Zhang:2020ecj}
J.-L. Zhang, Z.-F. Cui, J.~Ping, and C.~D. Roberts.
\newblock {Contact interaction analysis of pion GTMDs}.
\newblock {\em Eur. Phys. J. C}, 81:6, 2021.

\bibitem{Lu:2021sgg}
Y.~Lu, D.~Binosi, M.~Ding, C.~D. Roberts, H.-Y. Xing, and C.~Xu.
\newblock {Distribution amplitudes of light diquarks}.
\newblock {\em Eur. Phys. J. A}, 57:115, 2021.

\bibitem{Cui:2020dlm}
Z.-F. Cui, M.~Ding, F.~Gao, K.~Raya, D.~Binosi, L.~Chang, C.~D. Roberts, J.~Rodriguez-Quintero, and S.~M. Schmidt.
\newblock {Higgs modulation of emergent mass as revealed in kaon and pion parton distributions}.
\newblock {\em Eur. Phys. J. A}, 57:5, 2021.

\bibitem{Cui:2020tdf}
Z.-F. Cui, M.~Ding, F.~Gao, K.~Raya, D.~Binosi, L.~Chang, C.~D. Roberts, J.~Rodr\'\i{}guez-Quintero, and S.~M. Schmidt.
\newblock {Kaon and pion parton distributions}.
\newblock {\em Eur. Phys. J. C}, 80:1064, 2020.

\bibitem{Bernard:2001rs}
V.~Bernard, L.~Elouadrhiri, and U.-G. Meissner.
\newblock {Axial structure of the nucleon: Topical Review}.
\newblock {\em J. Phys. G}, 28:R1, 2002.

\bibitem{Baru:2011bw}
V.~Baru, C.~Hanhart, M.~Hoferichter, B.~Kubis, A.~Nogga, and D.~R. Phillips.
\newblock {Precision calculation of threshold $\pi^-d$ scattering, $\pi$N scattering lengths, and the GMO sum rule}.
\newblock {\em Nucl. Phys. A}, 872:69, 2011.

\bibitem{NavarroPerez:2016eli}
R.~Navarro~P\'erez, J.~E. Amaro, and E.~Ruiz~Arriola.
\newblock {Precise Determination of Charge Dependent Pion-Nucleon-Nucleon Coupling Constants}.
\newblock {\em Phys. Rev. C}, 95:064001, 2017.

\bibitem{Reinert:2020mcu}
P.~Reinert, H.~Krebs, and E.~Epelbaum.
\newblock {Precision determination of pion-nucleon coupling constants using effective field theory}.
\newblock {\em Phys. Rev. Lett.}, 126:092501, 2021.

\bibitem{Haidenbauer:2005zh}
J.~Haidenbauer and Ulf-G. Meissner.
\newblock {The Julich hyperon-nucleon model revisited}.
\newblock {\em Phys. Rev. C}, 72:044005, 2005.

\bibitem{Rijken:2010zzb}
T.~A. Rijken, M.~M. Nagels, and Y.~Yamamoto.
\newblock {Baryon-baryon interactions: Nijmegen extended-soft-core models}.
\newblock {\em Prog. Theor. Phys. Suppl.}, 185:14, 2010.

\bibitem{Deur:2018roz}
A.~Deur, S.~J. Brodsky, and G.~F. De~T\'eramond.
\newblock {The Spin Structure of the Nucleon}.
\newblock {\em Rept. Prog. Phys}, 82:076201, 2018.

\bibitem{Cui:2021mom}
Z.~F. Cui, Minghui Ding, J.~M. Morgado, K.~Raya, D.~Binosi, L.~Chang, J.~Papavassiliou, C.~D. Roberts, J.~Rodr\'\i{}guez-Quintero, and S.~M. Schmidt.
\newblock {Concerning pion parton distributions}.
\newblock {\em Eur. Phys. J. A}, 58:10, 2022.

\bibitem{Cui:2022bxn}
Z.~F. Cui, Minghui Ding, J.~M. Morgado, K.~Raya, D.~Binosi, L.~Chang, F.~De~Soto, C.~D. Roberts, J.~Rodr\'\i{}guez-Quintero, and S.~M. Schmidt.
\newblock {Emergence of pion parton distributions}.
\newblock {\em Phys. Rev. D}, 105:L091502, 2022.

\bibitem{Cui:2021gzg}
Z.-F. Cui, F.~Gao, D.~Binosi, L.~Chang, C.~D. Roberts, and S.~M. Schmidt.
\newblock {Valence Quark Ratio in the Proton}.
\newblock {\em Chin. Phys. Lett.}, 39:041401, 2022.

\bibitem{Chang:2022jri}
L.~Chang, F.~Gao, and C.~D. Roberts.
\newblock {Parton distributions of light quarks and antiquarks in the proton}.
\newblock {\em Phys. Lett. B}, 829:137078, 2022.

\bibitem{Chen:2018nsg}
C.~Chen, Y.~Lu, D.~Binosi, C.~D. Roberts, J.~Rodr\'\i{}guez-Quintero, and J.~Segovia.
\newblock {Nucleon-to-Roper electromagnetic transition form factors at large $Q^2$}.
\newblock {\em Phys. Rev. D}, 99:034013, 2019.

\bibitem{Lu:2019bjs}
Y.~Lu, C.~Chen, Z.-F. Cui, C.~D. Roberts, S.~M. Schmidt, J.~Segovia, and H.~S. Zong.
\newblock {Transition form factors: $\gamma^\ast + p \to \Delta(1232)$, $\Delta(1600)$}.
\newblock {\em Phys. Rev. D}, 100:034001, 2019.

\bibitem{Cui:2020rmu}
Z.-F. Cui, C.~Chen, D.~Binosi, F.~de~Soto, C.~D. Roberts, J.~Rodr\'\i{}guez-Quintero, S.~M. Schmidt, and J.~Segovia.
\newblock {Nucleon elastic form factors at accessible large spacelike momenta}.
\newblock {\em Phys. Rev. D}, 102:014043, 2020.

\bibitem{Eichmann:2009qa}
G.~Eichmann, R.~Alkofer, A.~Krassnigg, and D.~Nicmorus.
\newblock {Nucleon mass from a covariant three-quark Faddeev equation}.
\newblock {\em Phys. Rev. Lett.}, 104:201601, 2010.

\bibitem{Eichmann:2011pv}
G.~Eichmann and C.~S. Fischer.
\newblock {Nucleon axial and pseudoscalar form factors from the covariant Faddeev equation}.
\newblock {\em Eur. Phys. J. A}, 48:9, 2012.

\bibitem{Wang:2018kto}
Q.-W. Wang, S.-X. Qin, C.~D. Roberts, and S.~M. Schmidt.
\newblock {Proton tensor charges from a Poincar\'e-covariant Faddeev equation}.
\newblock {\em Phys. Rev. D}, 98:054019, 2018.

\bibitem{Qin:2019hgk}
S.-X. Qin, C.~D. Roberts, and S.~M. Schmidt.
\newblock {Spectrum of light- and heavy-baryons}.
\newblock {\em Few Body Syst.}, 60:26, 2019.

\bibitem{Li:2021qod}
Y.-S. Li, X.~Liu, and F.-S. Yu.
\newblock {Revisiting semileptonic decays of $\Lambda_{b(c)}$ supported by baryon spectroscopy}.
\newblock {\em Phys. Rev. D}, 104:013005, 2021.

\bibitem{Gell-Mann:1972iag}
M.~Gell-Mann.
\newblock {Quarks}.
\newblock {\em Acta Phys. Austriaca Suppl.}, 9:733, 1972.

\bibitem{Giannini:2015zia}
M.~M. Giannini and E.~Santopinto.
\newblock {The hypercentral Constituent Quark Model and its application to baryon properties}.
\newblock {\em Chin. J. Phys.}, 53:020301, 2015.

\bibitem{Plessas:2015mpa}
W.~Plessas.
\newblock {The constituent-quark model \textemdash{} Nowadays}.
\newblock {\em Int. J. Mod. Phys. A}, 30:1530013, 2015.

\bibitem{Brodsky:2022fqy}
S.~J. Brodsky, A.~Deur, and C.~D. Roberts.
\newblock {Artificial dynamical effects in quantum field theory}.
\newblock {\em Nat. Rev. Phys.}, 4:489, 2022.

\bibitem{Coester:1992cg}
F.~Coester.
\newblock {Null plane dynamics of particles and fields}.
\newblock {\em Prog. Part. Nucl. Phys.}, 29:1, 1992.

\bibitem{Qin:2013mta}
S.~X. Qin, L.~Chang, Y.-X. Liu, C.~D. Roberts, and S.~M. Schmidt.
\newblock {Practical corollaries of transverse Ward-Green-Takahashi identities}.
\newblock {\em Phys. Lett. B}, 722:384, 2013.

\bibitem{Qin:2014vya}
S.~X. Qin, C.~D. Roberts, and S.~M. Schmidt.
\newblock {Ward\textendash{}Green\textendash{}Takahashi identities and the axial-vector vertex}.
\newblock {\em Phys. Lett. B}, 733:202, 2014.

\bibitem{Binosi:2016rxz}
D.~Binosi, L.~Chang, J.~Papavassiliou, S.~X. Qin, and C.~D. Roberts.
\newblock {Symmetry preserving truncations of the gap and Bethe-Salpeter equations}.
\newblock {\em Phys. Rev. D}, 93:096010, 2016.

\bibitem{Qin:2020jig}
S.~X. Qin and C.~D. Roberts.
\newblock {Resolving the Bethe\textendash{}Salpeter Kernel}.
\newblock {\em Chin. Phys. Lett.}, 38:071201, 2021.

\bibitem{Anselmino:1992vg}
M.~Anselmino, E.~Predazzi, S.~Ekelin, S.~Fredriksson, and D.~B. Lichtenberg.
\newblock {Diquarks}.
\newblock {\em Rev. Mod. Phys.}, 65:1199, 1993.

\bibitem{Aznauryan:2011ub}
I.~Aznauryan, V.~D. Burkert, T.~S.~H. Lee, and V.~I. Mokeev.
\newblock {Results from the N* program at JLab}.
\newblock {\em J. Phys. Conf. Ser.}, 299:012008, 2011.

\bibitem{Edwards:2011jj}
R.~G. Edwards, J.~J. Dudek, D.~G. Richards, and S.~J. Wallace.
\newblock {Excited state baryon spectroscopy from lattice QCD}.
\newblock {\em Phys. Rev. D}, 84:074508, 2011.

\bibitem{EuropeanMuon:1987isl}
J.~Ashman et~al.
\newblock {A Measurement of the Spin Asymmetry and Determination of the Structure Function g(1) in Deep Inelastic Muon-Proton Scattering}.
\newblock {\em Phys. Lett. B}, 206:364, 1988.

\bibitem{Brodsky:1989pv}
S.~J. Brodsky and G.~P. Lepage.
\newblock {Exclusive Processes in Quantum Chromodynamics}.
\newblock {\em Adv. Ser. Direct. High Energy Phys.}, 5:93, 1989.

\bibitem{Brodsky:1997de}
S.~J. Brodsky, H.-C. Pauli, and S.~S. Pinsky.
\newblock {Quantum chromodynamics and other field theories on the light cone}.
\newblock {\em Phys. Rept.}, 301:299, 1998.

\bibitem{Heinzl:2000ht}
T.~Heinzl.
\newblock {Light cone quantization: Foundations and applications}.
\newblock {\em Lect. Notes Phys.}, 572:55, 2001.

\bibitem{Roberts:2022rxm}
C.~D. Roberts.
\newblock {Origin of the Proton Mass}.
\newblock {\em EPJ Web Conf.}, 282:01006, 2023.

\bibitem{Salme:2022eoy}
G.~Salm\`e.
\newblock {Explaining mass and spin in the visible matter: the next challenge}.
\newblock {\em J. Phys. Conf. Ser.}, 2340:012011, 2022.

\bibitem{Carman:2023zke}
D.~S. Carman, R.~W. Gothe, V.~I. Mokeev, and C.~D. Roberts.
\newblock {Nucleon Resonance Electroexcitation Amplitudes and Emergent Hadron Mass}.
\newblock {\em Particles}, 6:416, 2023.

\bibitem{Ding:2019qlr}
M.~H. Ding, K.~Raya, D.~Binosi, L.~Chang, C.~D. Roberts, and S.~M. Schmidt.
\newblock {Drawing insights from pion parton distributions}.
\newblock {\em Chin. Phys. C}, 44:031002, 2020.

\bibitem{Roberts:2016vyn}
C.~D. Roberts.
\newblock {Perspective on the origin of hadron masses}.
\newblock {\em Few Body Syst.}, 58:5, 2017.

\bibitem{Mezrag:2017znp}
C.~Mezrag, J.~Segovia, L.~Chang, and C.~D. Roberts.
\newblock {Parton distribution amplitudes: Revealing correlations within the proton and Roper}.
\newblock {\em Phys. Lett. B}, 783:263, 2018.

\bibitem{Liu:2019vsn}
T.~B. Liu, R.~S. Sufian, G.~F. de~T\'eramond, H.~G. Dosch, S.~J. Brodsky, and A.~Deur.
\newblock {Unified Description of Polarized and Unpolarized Quark Distributions in the Proton}.
\newblock {\em Phys. Rev. Lett.}, 124:082003, 2020.

\bibitem{Han:2021dkc}
C.-D. Han, G.~Xie, R.~Wang, and X.-R. Chen.
\newblock {An analysis of polarized parton distribution functions with nonlinear QCD evolution equations}.
\newblock {\em Nucl. Phys. B}, 985:116012, 2022.

\bibitem{Brodsky:1994kg}
S.~J. Brodsky, M.~Burkardt, and I.~Schmidt.
\newblock {Perturbative QCD constraints on the shape of polarized quark and gluon distributions}.
\newblock {\em Nucl. Phys. B}, 441:197, 1995.

\bibitem{Brisudova:1999ut}
M.~M. Brisudova, L.~Burakovsky, and J.~T. Goldman.
\newblock {Effective functional form of Regge trajectories}.
\newblock {\em Phys. Rev. D}, 61:054013, 2000.

\bibitem{Xu:2022abw}
S.-Q. Xu, C.~Mondal, X.-B. Zhao, Y.~Li, and J.~P. Vary.
\newblock {Nucleon spin decomposition with one dynamical gluon}.
\newblock {\em arXiv: 2209.08584}, 2022.

\bibitem{Holt:2010vj}
R.~J. Holt and C.~D. Roberts.
\newblock {Distribution Functions of the Nucleon and Pion in the Valence Region}.
\newblock {\em Rev. Mod. Phys.}, 82:2991, 2010.

\bibitem{Roberts:2013mja}
C.~D. Roberts, R.~J. Holt, and S.~M. Schmidt.
\newblock {Nucleon spin structure at very high-x}.
\newblock {\em Phys. Lett. B}, 727:249, 2013.

\bibitem{Close:1988br}
F.~E. Close and A.~W. Thomas.
\newblock {The Spin and Flavor Dependence of Parton Distribution Functions}.
\newblock {\em Phys. Lett. B}, 212:227, 1988.

\bibitem{Hughes:1999wr}
E.~W. Hughes and R.~Voss.
\newblock {Spin structure functions}.
\newblock {\em Ann. Rev. Nucl. Part. Sci.}, 49:303, 1999.

\bibitem{Farrar:1975yb}
G.~R. Farrar and D.~R. Jackson.
\newblock {Pion and Nucleon Structure Functions Near x=1}.
\newblock {\em Phys. Rev. Lett.}, 35:1416, 1975.

\bibitem{Dokshitzer:1977sg}
Y.~L. Dokshitzer.
\newblock {Calculation of the Structure Functions for Deep Inelastic Scattering and e+ e- Annihilation by Perturbation Theory in Quantum Chromodynamics.}
\newblock {\em Sov. Phys. JETP}, 46:641, 1977.

\bibitem{Gribov:1971zn}
V.~N. Gribov and L.~N. Lipatov.
\newblock {Deep inelastic electron scattering in perturbation theory}.
\newblock {\em Phys. Lett. B}, 37:78, 1971.

\bibitem{Lipatov:1974qm}
L.~N. Lipatov.
\newblock {The parton model and perturbation theory}.
\newblock {\em Yad. Fiz.}, 20:181, 1974.

\bibitem{Altarelli:1977zs}
G.~Altarelli and G.~Parisi.
\newblock {Asymptotic Freedom in Parton Language}.
\newblock {\em Nucl. Phys. B}, 126:298, 1977.

\bibitem{HERMES:2004zsh}
A.~Airapetian et~al.
\newblock {Quark helicity distributions in the nucleon for up, down, and strange quarks from semi-inclusive deep-inelastic scattering}.
\newblock {\em Phys. Rev. D}, 71:012003, 2005.

\bibitem{COMPASS:2010hwr}
M.~G. Alekseev et~al.
\newblock {Quark helicity distributions from longitudinal spin asymmetries in muon-proton and muon-deuteron scattering}.
\newblock {\em Phys. Lett. B}, 693:227, 2010.

\bibitem{CLAS:2006ozz}
K.~V. Dharmawardane et~al.
\newblock {Measurement of the x- and Q**2-dependence of the asymmetry A(1) on the nucleon}.
\newblock {\em Phys. Lett. B}, 641:11, 2006.

\bibitem{CLAS:2008xos}
Y.~Prok et~al.
\newblock {Moments of the Spin Structure Functions g**p(1) and g**d(1) for 0.05 \ensuremath{<} Q**2 \ensuremath{<} 3.0-GeV**2}.
\newblock {\em Phys. Lett. B}, 672:12, 2009.

\bibitem{CLAS:2015otq}
N.~Guler et~al.
\newblock {Precise determination of the deuteron spin structure at low to moderate $Q^2$ with CLAS and extraction of the neutron contribution}.
\newblock {\em Phys. Rev. C}, 92:055201, 2015.

\bibitem{CLAS:2017qga}
R.~Fersch et~al.
\newblock {Determination of the Proton Spin Structure Functions for $0.05 < Q^{2} < 5 GeV^{2}$ using CLAS}.
\newblock {\em Phys. Rev. C}, 96:065208, 2017.

\bibitem{JeffersonLabHallA:2014mam}
D.~S. Parno et~al.
\newblock {Precision Measurements of $A_1^n$ in the Deep Inelastic Regime}.
\newblock {\em Phys. Lett. B}, 744:309, 2015.

\bibitem{JeffersonLabHallA:2003joy}
X.~Zheng et~al.
\newblock {Precision measurement of the neutron spin asymmetry A1**N and spin flavor decomposition in the valence quark region}.
\newblock {\em Phys. Rev. Lett.}, 92:012004, 2004.

\bibitem{JeffersonLabHallA:2004tea}
X.~Zheng et~al.
\newblock {Precision measurement of the neutron spin asymmetries and spin-dependent structure functions in the valence quark region}.
\newblock {\em Phys. Rev. C}, 70:065207, 2004.

\bibitem{Xu:2021wwj}
S.-Q. Xu, C.~Mondal, J.-S. Lan, Y.~Zhao, X. B.and~Li, and J.~P. Vary.
\newblock {Nucleon structure from basis light-front quantization}.
\newblock {\em Phys. Rev. D}, 104:094036, 2021.

\bibitem{Ethier:2020way}
J.~J. Ethier and E.~R. Nocera.
\newblock {Parton Distributions in Nucleons and Nuclei}.
\newblock {\em Ann. Rev. Nucl. Part. Sci.}, 70:43, 2020.

\bibitem{E1206109}
S.~Kuhn et~al., CLAS Collaboration (E12-06-109).
\newblock {\emph{The Longitudinal Spin Structure of the Nucleon}}.

\bibitem{Zheng:2006}
X.~Zheng et~al., E12-06-110 Collaboration.
\newblock {\emph{Measurement of Neutron Spin Asymmetry A$_1^n$ in the Valence Quark Region using an 11 GeV Beam and a Polarized $^3$He Target in Hall C}}.

\bibitem{Bali:2015ykx}
G.~S. Bali et~al.
\newblock {Light-cone distribution amplitudes of the baryon octet}.
\newblock {\em JHEP}, 02:070, 2016.

\bibitem{Lepage:1980fj}
G.~P. Lepage and S.~J. Brodsky.
\newblock {Exclusive Processes in Perturbative Quantum Chromodynamics}.
\newblock {\em Phys. Rev. D}, 22:2157, 1980.

\bibitem{JeffersonLabHallATritium:2021usd}
D.~Abrams et~al.
\newblock {Measurement of the Nucleon $F^n_2/F^p_2$ Structure Function Ratio by the Jefferson Lab MARATHON Tritium/Helium-3 Deep Inelastic Scattering Experiment}.
\newblock {\em Phys. Rev. Lett.}, 128:132003, 2022.

\bibitem{Raya:2021zrz}
K.~Raya, Z.~Cui, L.~Chang, J.~M. Morgado, C.~D. Roberts, and J.~Rodriguez-Quintero.
\newblock {Revealing pion and kaon structure via generalised parton distributions *}.
\newblock {\em Chin. Phys. C}, 46:013105, 2022.

\bibitem{Yin:2023dbw}
P.-L. Yin, Y.-Z. Xu, Z.-F. Cui, C.~D. Roberts, and J.~Rodr\'\i{}guez-Quintero.
\newblock {All-Orders Evolution of Parton Distributions: Principle, Practice, and Predictions}.
\newblock {\em Chin. Phys. Lett. \emph{Express}}, 40:091201, 2023.

\bibitem{Xu:2023bwv}
Y.~Z. Xu, K.~Raya, Z.~F. Cui, C.~D. Roberts, and J.~Rodr\'\i{}guez-Quintero.
\newblock {Empirical Determination of the Pion Mass Distribution}.
\newblock {\em Chin. Phys. Lett. \emph{Express}}, 40:041201, 2023.

\bibitem{Grunberg:1980ja}
D.~Aston et~al.
\newblock {Observation of an $\omega \pi^0$ State of Mass 1.25-{GeV} Produced by Photons of Energy 20-{GeV} - 70-{GeV}}.
\newblock {\em Phys. Lett. B}, 92:211, 1980.
\newblock [Erratum: Phys.Lett.B 95, 461 (1980)].

\bibitem{Grunberg:1982fw}
G.~Grunberg.
\newblock {Renormalization Scheme Independent QCD and QED: The Method of Effective Charges}.
\newblock {\em Phys. Rev. D}, 29:2315, 1984.

\bibitem{Deur:2016tte}
A.~Deur, S.~J. Brodsky, and G.~F. de~Teramond.
\newblock {The QCD Running Coupling}.
\newblock {\em Nucl. Phys.}, 90:1, 2016.

\bibitem{Deur:2022msf}
A.~Deur, V.~Burkert, J.~P. Chen, and W.~Korsch.
\newblock {Experimental determination of the QCD effective charge $\alpha_{g_1}(Q)$}.
\newblock {\em Particles}, 5:171, 2022.

\bibitem{Deur:2023dzc}
A.~Deur, S.~J. Brodsky, and C.~D. Roberts.
\newblock {QCD Running Couplings and Effective Charges}.
\newblock {\em arXiv: 2303.00723}, 2023.

\bibitem{SeaQuest:2021zxb}
J.~Dove et~al.
\newblock {The asymmetry of antimatter in the proton}.
\newblock {\em Nature}, 590:561, 2021.
\newblock [Erratum: Nature 604, E26 (2022)].

\bibitem{CLAS:2014qtg}
Y.~Prok et~al.
\newblock {Precision measurements of $g_1$ of the proton and the deuteron with 6 GeV electrons}.
\newblock {\em Phys. Rev. C}, 90:025212, 2014.

\bibitem{E155:1999pwm}
P.~L. Anthony et~al.
\newblock {Measurement of the deuteron spin structure function g1(d)(x) for 1-(GeV/c)**2 \ensuremath{<} Q**2 \ensuremath{<} 40-(GeV/c)**2}.
\newblock {\em Phys. Lett. B}, 463:339, 1999.

\bibitem{E155:2000qdr}
P.~L Anthony et~al.
\newblock {Measurements of the Q**2 dependence of the proton and neutron spin structure functions g(1)**p and g(1)**n}.
\newblock {\em Phys. Lett. B}, 493:19, 2000.

\bibitem{HERMES:1998cbu}
A.~Airapetian et~al.
\newblock {Measurement of the proton spin structure function g1(p) with a pure hydrogen target}.
\newblock {\em Phys. Lett. B}, 442:484, 1998.

\bibitem{E143:1995clm}
K.~Abe et~al.
\newblock {Measurements of the Q**2 dependence of the proton and deuteron spin structure functions g1(p) and g1(d)}.
\newblock {\em Phys. Lett. B}, 364:61, 1995.

\bibitem{E143:1996vck}
K.~Abe et~al.
\newblock {Measurements of the proton and deuteron spin structure function g1 in the resonance region}.
\newblock {\em Phys. Rev. Lett.}, 78:815, 1997.

\bibitem{E143:1998hbs}
K.~Abe et~al.
\newblock {Measurements of the proton and deuteron spin structure functions g(1) and g(2)}.
\newblock {\em Phys. Rev. D}, 58:112003, 1998.

\bibitem{SpinMuonSMC:1994met}
D.~Adams et~al.
\newblock {Measurement of the spin dependent structure function $g_1(x)$ of the proton}.
\newblock {\em Phys. Lett. B}, 329:399, 1994.
\newblock [Erratum: Phys.Lett.B 339, 332--333 (1994)].

\bibitem{SpinMuonSMC:1997mkb}
D.~Adams et~al.
\newblock {Spin structure of the proton from polarized inclusive deep inelastic muon - proton scattering}.
\newblock {\em Phys. Rev. D}, 56:5330, 1997.

\bibitem{HERMES:1997hjr}
K.~Ackerstaff et~al.
\newblock {Measurement of the neutron spin structure function g1(n) with a polarized He-3 internal target}.
\newblock {\em Phys. Lett. B}, 404:383, 1997.

\bibitem{E154:1997xfa}
K.~Abe et~al.
\newblock {Precision determination of the neutron spin structure function g1(n)}.
\newblock {\em Phys. Rev. Lett.}, 79:26, 1997.

\bibitem{E154:1997ysl}
K.~Abe et~al.
\newblock {Next-to-leading order QCD analysis of polarized deep inelastic scattering data}.
\newblock {\em Phys. Lett. B}, 405:180, 1997.

\bibitem{E142:1996thl}
P.~L. Anthony et~al.
\newblock {Deep inelastic scattering of polarized electrons by polarized He-3 and the study of the neutron spin structure}.
\newblock {\em Phys. Rev. D}, 54:6620, 1996.

\bibitem{Ellis:1996mzs}
R.~K. Ellis, W.~J. Stirling, and B.~R. Webber.
\newblock {\em {QCD and collider physics}}, volume~8.
\newblock Cambridge University Press, 2011.

\bibitem{deFlorian:2014yva}
D.~de~Florian, R.~Sassot, M.~Stratmann, and W.~Vogelsang.
\newblock {Evidence for polarization of gluons in the proton}.
\newblock {\em Phys. Rev. Lett.}, 113:012001, 2014.

\bibitem{COMPASS:2015pim}
C.~Adolph et~al.
\newblock {Leading-order determination of the gluon polarisation from semi-inclusive deep inelastic scattering data}.
\newblock {\em Eur. Phys. J. C}, 77:209, 2017.

\bibitem{Altarelli:1988nr}
G.~Altarelli and G.~G. Ross.
\newblock {The Anomalous Gluon Contribution to Polarized Leptoproduction}.
\newblock {\em Phys. Lett. B}, 212:391, 1988.

\bibitem{COMPASS:2016jwv}
C.~Adolph et~al.
\newblock {Final COMPASS results on the deuteron spin-dependent structure function $g_1^{\rm d}$ and the Bjorken sum rule}.
\newblock {\em Phys. Lett. B}, 769:34, 2017.

\bibitem{Ji:1995cu}
X.~D. Ji, J.~Tang, and P.~Hoodbhoy.
\newblock {The spin structure of the nucleon in the asymptotic limit}.
\newblock {\em Phys. Rev. Lett.}, 76:740, 1996.

\bibitem{Chen:2011gn}
X.~S. Chen, W.~M. Sun, F.~Wang, and T.~Goldman.
\newblock {Proper identification of the gluon spin}.
\newblock {\em Phys. Lett. B}, 700:21, 2011.

\bibitem{Anderle:2021wcy}
D.~P. Anderle et~al.
\newblock {Electron-ion collider in China}.
\newblock {\em Front. Phys. (Beijing)}, 16:64701, 2021.

\bibitem{AbdulKhalek:2021gbh}
R.~Abdul~Khalek et~al.
\newblock {Science Requirements and Detector Concepts for the Electron-Ion Collider}: {EIC Yellow Report}.
\newblock {\em Nucl. Phys. A}, 1026:122447, 2022.

\bibitem{Krassnigg:2003wy}
A.~Krassnigg and C.~D. Roberts.
\newblock {DSEs, the pion, and related matters}.
\newblock {\em Fizika B}, 13:143, 2004.

\bibitem{Oettel:1999gc}
M.~Oettel, M.~Pichowsky, and L.~von Smekal.
\newblock {Current conservation in the covariant quark diquark model of the nucleon}.
\newblock {\em Eur. Phys. J. A}, 8:251, 2000.

\end{thebibliography}

\end{document}